\documentclass[11pt]{article}

\usepackage[utf8]{inputenc}
\usepackage{jheppub} 
\usepackage[usenames,dvipsnames,table]{xcolor}
\usepackage{mathtools}
\usepackage{amsmath,amssymb}
\usepackage{tikz}
\usepackage{graphicx}
\usepackage{xspace}
\usetikzlibrary{positioning}
\usetikzlibrary{arrows}
\usetikzlibrary{decorations.pathreplacing}
\usetikzlibrary{shapes.geometric}
\usetikzlibrary{calc}
\usetikzlibrary{decorations.pathmorphing}
\usetikzlibrary{shapes.misc} 
\usepackage{tikz-cd}
\usetikzlibrary{snakes}
\usepackage{pdflscape}
\usepackage{sectsty} \allsectionsfont{\boldmath} 
\usepackage[detect-all]{siunitx}
\usepackage{cancel}
\usepackage{afterpage}
\usepackage[export]{adjustbox}
\usepackage{subcaption}

\tikzstyle{ligne}=[draw, thick] 
\tikzset{bd/.style={circle, draw=black, inner sep=0pt, fill=black, minimum size=1.2mm}}

\hypersetup{hidelinks}
\hypersetup{linktoc=all}
\graphicspath{ {Figures/} } 

\DeclarePairedDelimiter\floor{\lfloor}{\rfloor}
\newcommand{\be}{\begin{eqnarray}}
\newcommand{\ee}{\end{eqnarray}}
\newcommand{\bea}{\begin{eqnarray}}
\newcommand{\eea}{\end{eqnarray}}
\newcommand{\nn}{\nonumber}
\newcommand{\bn}{\begin{enumerate}}
\newcommand{\en}{\end{enumerate}}

\def\CI{{\cal I}}

\def\r{\rho}

\def\s{\sigma}

\newcommand{\udl}[1]{\mathrm{d} #1 \,}

\newcommand{\sbfunc}[1]{s_b\left( #1\right)}

\newcommand{\Gpq}[1]{\Gamma_e\left( #1\right)}

\def\ga{\alpha}
\def\gb{\beta}

\def\Gd{\Delta}
\def\gd{\delta}

\def\gs{\sigma}

\newcommand{\bigphantomspace}{\phantom{\frac{\frac{\frac{1}{1}}{1}}{\frac{1}{1}}}}
\newcommand{\verybigphantomspace}{\phantom{\frac{\frac{\frac{\frac{1}{1}}{1}}{1}}{\frac{1}{1}}}}
\usepackage{array}

\makeatletter
\newcommand{\subalign}[1]{%
  \vcenter{%
    \Let@ \restore@math@cr \default@tag
    \baselineskip\fontdimen10 \scriptfont\tw@
    \advance\baselineskip\fontdimen12 \scriptfont\tw@
    \lineskip\thr@@\fontdimen8 \scriptfont\thr@@
    \lineskiplimit\lineskip
    \ialign{\hfil$\m@th\scriptstyle##$&$\m@th\scriptstyle{}##$\hfil\crcr
      #1\crcr
    }%
  }%
}
\makeatother
\makeatletter
\DeclareRobustCommand{\cev}[1]{%
  {\mathpalette\do@cev{#1}}%
}
\newcommand{\do@cev}[2]{%
  \vbox{\offinterlineskip
    \sbox\z@{$\m@th#1 x$}%
    \ialign{##\cr
      \hidewidth\reflectbox{$\m@th#1\vec{}\mkern4mu$}\hidewidth\cr
      \noalign{\kern-\ht\z@}
      $\m@th#1#2$\cr
    }%
  }%
}
\makeatother

\preprint{\begin{flushright} USTC-ICTS/PCFT-24-01 \\ DMUS-MP-23/15 \end{flushright}}

\title{Probing bad theories with the dualization algorithm {\fontfamily{ppl}\selectfont II} }

\author[c,d]{Simone Giacomelli,}
\author[a,b]{Chiung Hwang,}
\author[c,d,e]{Fabio Marino,}
\author[c,d]{Sara Pasquetti,}
\author[f]{Matteo Sacchi}

\affiliation[a]{Interdisciplinary Center for Theoretical Study, University of Science and Technology of China, Hefei, Anhui 230026, China}
\affiliation[b]{Peng Huanwu Center for Fundamental Theory, Hefei, Anhui 230026, China}
\affiliation[c]{Dipartimento di Fisica, Università di Milano-Bicocca,
Piazza della Scienza 3, I-20126 Milano, Italy}
\affiliation[d]{INFN, sezione di Milano-Bicocca, Piazza della Scienza 3, I-20126 Milano, Italy}
\affiliation[e]{
Department of Mathematics, University of Surrey, Guildford, GU2 7XH, UK}
\affiliation[f]{Mathematical Institute, University of Oxford, Woodstock Road, Oxford, OX2 6GG, United Kingdom}

\emailAdd{simone.giacomelli@unimib.it}
\emailAdd{chiung@ustc.edu.cn}
\emailAdd{f.marino25@campus.unimib.it}
\emailAdd{sara.pasquetti@gmail.com} 
\emailAdd{matteo.sacchi@maths.ox.ac.uk}

\abstract{
We continue our analysis of bad theories initiated in \cite{Giacomelli:2023zkk}, focusing on quiver theories with bad unitary and special unitary gauge groups in three dimensions. By extending the dualization algorithm we prove that the partition function of bad linear quivers can be written as a distribution, given by a sum of terms involving a product of delta functions times the partition function of a good quiver theory. We describe in detail the good quiver theories appearing in the partition function of the bad theory and discuss the brane interpretation of our result. We also discuss in detail the lift of these theories to 4d quivers with symplectic gauge groups, in which our results can be recovered by studying the Higgsing triggered by the expectation value for certain chiral operators. The paper is accompanied by a Mathematica file which implements the algorithm for an arbitrary unitary bad linear quiver.
}

\begin{document} 

\maketitle
\flushbottom

\section{Introduction}
\label{sec:Intro}

One of the most remarkable results in the study of quantum field theory is the discovery of dualities, which establish the equivalence of different-looking theories, at least in some range of the parameters. Often the weak coupling limit of one theory corresponds to a strong coupling limit of the dual and therefore by exploiting their equivalence we can learn about the dynamics of both theories in the strong coupling regime, where standard perturbative methods are not applicable. In this sense dualities represent a key tool to improve our understanding of nonperturbative quantum field theory. This is exemplified by Seiberg duality \cite{Seiberg:1994pq} which provides us with detailed information about the low-energy dynamics of supersymmetric quantum chromodynamics (SQCD) in four dimensions. 

By now many other examples of dualities have been found, especially in the context of supersymmetric field theories for which dualities seem to be ubiquitous. At present the challenge is to find an algorithm which allows us to establish systematically the existence of dual descriptions of a given theory. It is also desirable to find a systematic way to connect different dualities together, deriving all of them from a restricted set of fundamental dualities. In this respect an important result is \cite{Aharony:2013dha} (and its many follow up works) which shows how dualities in different dimensions can be related via compactification. In particular, this implies that dualities in lower dimension can be derived from those in higher dimension. \\

Mirror symmetry \cite{Intriligator:1996ex,Hanany:1996ie},  relating pairs of 3d $\mathcal{N}=4$ theories
with the  Higgs branch of one theory  mapped to the Coulomb branch of the other theory and vice-versa, is one of the most 
interesting and studied examples of dualities (see e.g. \cite{deBoer:1996mp,Aharony:1997bx,Kapustin:1999ha,Borokhov:2002cg}).
For $\mathcal{N}=4$ theories admitting a description in terms of a Hanany--Witten brane system \cite{Hanany:1996ie},
mirror symmetry can be regarded as a consequence  of Type IIB S-duality. Furthermore, mirror symmetry has played a crucial role in the development of nonperturbative techniques which allow us to study the quantum moduli space of theories with eight supercharges in three dimensions or higher, even in the absence of a Lagrangian description of the theory (see e.g.~\cite{Bullimore:2015lsa, Cremonesi:2015lsa, Collinucci:2016hpz, Collinucci:2017bwv, Ferlito:2017xdq, Hanany:2018uhm, Cabrera:2018jxt, Cabrera:2019izd, Bourget:2019aer, Bourget:2019rtl, Cabrera:2019dob, Bourget:2020asf, Bourget:2020mez, Beratto:2020wmn, vanBeest:2021xyt, Kang:2022zsl, Bourget:2023cgs, Bourget:2021jwo, Bourget:2022tmw, Bourget:2021csg, Bourget:2023uhe, Bourget:2024mgn, vanBeest:2020kou, VanBeest:2020kxw, Closset:2020scj, Closset:2020afy, Giacomelli:2020ryy, Closset:2021lwy, Beem:2023ofp}).

Recently, a new approach to study 3d mirror dualities, the \emph{mirror dualization algorithm}, has been developed \cite{Bottini:2021vms,Hwang:2021ulb,Comi:2022aqo}, which generalizes the construction \cite{Kapustin:1999ha} of abelian mirror pairs by local dualization to the non-abelian case. The algorithm was originally formulated for linear unitary quivers,
namely the $T_\rho^\sigma[SU(N)]$ theories  \cite{Gaiotto:2008ak} (see also \cite{Cremonesi:2014uva}) realized on linear Hanany--Witten brane setups with D3-branes suspended between NS5 and D5-branes, and implements field theoretically the local action of S-duality, which acts on each 5-brane by creating a S-duality domain wall to its right and to its left
\be
\text{NS5} \rightarrow S(\text{D5})S^{-1}\,,\qquad \text{D5} \rightarrow S(\overline{\text{NS5}})S^{-1}\,.
\ee

At the field theory level we associate to each 5-brane a \emph{QFT block}: NS5-branes are associated to a bifundamental matter block $(1,0)$, while D5-branes are associated to a fundamental matter block $(0,1)$.
S-duality acts locally on these blocks by swapping them, creating on their sides the \emph{S-wall theory}, which is realized by a variant of the $T[SU(N)]$ theory as proposed by \cite{Gaiotto:2008ak}.
The local dualizations of the QFT blocks correspond to genuine IR dualities, the \emph{basic duality moves}, which can in turn  be derived from the Seiberg-like Aharony duality \cite{Aharony:1997gp}.

The  algorithm consists of various steps: we first chop the initial quiver into QFT blocks, we then dualize them by means of the basic duality moves and finally glue them back together. Such a procedure can produce in the end a theory where some VEVs are turned on, triggering a non-trivial RG flow.
In \cite{Comi:2022aqo} it was shown that these VEVs can be extinguished using another duality move that swaps $(1,0)$ and $(0,1)$-blocks, which is the field theory counterpart of the Hanany--Witten move \cite{Hanany:1996ie}.
For good linear quivers a finite sequence of these moves is enough to reach the end of the RG flow and obtain the final mirror dual. 
In particular, starting from the $T_\rho^\sigma[SU(N)]$ theory, the algorithm  outputs the mirror dual theory with swapped partitions
\be
T_\rho^\sigma[SU(N)]\overset{\small \text{mirror}}{\longleftrightarrow} T_\sigma^\rho[SU(N)]\,.
\ee 
The algorithm then provides a purely field theoretic derivation of mirror symmetry by assuming only more fundamental Seiberg-like dualities.

The mirror algorithm can also be extended to the case of dualities involving $(p,q)$ 5-branes, which will have a new associated $(1,1)$ QFT block, and  to the full $SL(2,\mathbb{Z})$ S-duality group \cite{Comi:2022aqo}. It can also be applied to prove mirror dualities for theories with four superchagres \cite{Benvenuti:2023qtv}.

A 4d version of the algorithm has also been developed and   can be used to prove 4d mirror-like dualities.
Indeed, as shown in  \cite{Hwang:2020wpd},  the 3d $T_\rho^\sigma[SU(N)]$ family has an {\it uplift} to the family of 4d $\mathcal{N}=1$ symplectic quivers named  $E_\rho^\sigma[USp(2N)]$ (defined in Appendix \ref{app:Erhosigma_Trhosigma}). This class of 4d  theories (labelled by partitions $\rho$ and $\sigma$ of an integer $N$)  enjoy the mirror-like duality 
\be
E_\rho^\sigma[USp(2N)]\overset{\small \text{mirror}}{\longleftrightarrow} E_\sigma^\rho[USp(2N)]\,.
\ee 
In 3d, after a circle reduction and real mass deformations, they reduce to the 3d $\mathcal{N}=4$ linear quivers of the $T_\rho^\sigma[SU(N)]$ family.
The 4d version of the algorithm can then derive these mirror-like dualities among members of the $E_\rho^\sigma[USp(2N)]$ family by means of the 4d counterpart of the 3d basic duality moves which locally implement the action of S-duality. These $4d$ basic duality moves can all be derived from a Seiberg-like duality for simplectic groups, namely the Intriligator--Pouliot (IP) duality \cite{Intriligator:1995ne}.
In this case, the mirror algorithm can be extended to the full $PSL(2,\mathbb{Z})$ group \cite{Comi:2022aqo}.\\

In \cite{Giacomelli:2023zkk} we initiated the study of 3d bad theories focusing on the case of SQCD, namely a $U(N_c)$ gauge theory with $N_f$ flavors. 
This theory for $N_f < 2 N_c-1$ is known to be {\it bad} ({\it ugly} for $N_f = 2 N_c-1$) \cite{Gaiotto:2008ak},
meaning that  some monopoles operators fall below the unitarity bound and decouple in the IR.
Bad theories have a rather intricate IR dynamics and have been studied in various works \cite{Nanopoulos:2010bv,Kim:2012uz,Yaakov:2013fza,Bashkirov:2013dda,Hwang:2015wna,Hwang:2017kmk,Assel:2017jgo,Dey:2017fqs,Assel:2018exy}.
In particular  \cite{Assel:2017jgo} pointed out that 
the most singular locus in the Coulomb branch of $U(N_c)$ SQCD with $N_f$ flavors is a submanifold of codimension $\floor{N_f/2}$ rather than an isolated point and is such that the UV gauge symmetry is broken to a subgroup of lower rank: around it the Coulomb branch looks like that of SQCD with gauge group $U(\floor{N_f/2})$ and $N_f$ flavors, times some flat directions (which for 3d $\mathcal{N}=4$ theories are parametrized by free twisted hypermultiplets).

In \cite{Giacomelli:2023zkk} we applied the mirror dualization algorithm to the 4d/3d SQCD case.
Interestingly, we observed that also the 4d SQCD uplift for $N_f< 2N_c$ has a moduli space with the same feature as its 3d counterpart we just described. In this sense we will say that a 4d theory is bad.

The main result is that the 4d index/3d sphere partition function is a distribution rather than a regular function.\footnote{The 
distributional nature of indices/partition functions was first observed in \cite{Spiridonov:2014cxa}, in the case of theories with quantum deformed moduli spaces and patterns of  chiral symmetry breaking.}
Indeed, the schematic structure of the index/partition function of the 4d/3d bad SQCD is given by a sum of terms each involving a Dirac delta distribution, enforcing a particular constraint on the fugacities/FI parameters, which multiplies the 4d index/3d partition function of an interacting SCFT, the mirror of
a good SQCD of lower rank,  plus various gauge singlets which are just free fields.
In addition, there is an extra frame with no delta distribution whose interacting part is the mirror of the good SQCD $U(N_f-N_c)$ with $N_f$ hypers in 3d and $USp(2N_f-2N_c)$ with one antisymmetric and $2N_f+4$ fundamental chirals in 4d.\\

In this paper we focus on 3d $\mathcal{N}=4$ linear quivers with bad gauge nodes and on their 4d $\mathcal{N}=1$ uplifts. In order to dualize these theories we extend the mirror dualization algorithm. Indeed we will see that the $(1,0)$-$(0,1)$ blocks swap is not enough to take care of all the VEVs, but we need a new non-trivial duality move, the $(0,1)$-$(0,1)$ blocks swap, which in 3d we interpret as a non trivial property of the swap of D5-branes.
Such a duality is just a rewriting of the aforementioned result for SQCD, so when we apply it we get a sum of frames weighted (all but one) by a delta function.

The mirror dual of a bad linear quiver is then a sum of multiple frames, each of them having an interacting good linear quiver part ($E_\rho^\sigma[USp(2N)]$ in 4d and $T_\rho^\sigma[SU(N)]$ in 3d) and a free sector. Some of these frames will be also multiplied by one or more Dirac deltas, enforcing various conditions on the electric FI parameters. Moreover, the result can be presented by replacing in each frame the good interacting quiver theory with its mirror dual.
In this way we obtain the sum of electric duals of a bad quiver.

We also propose an \emph{electric dualization algorithm} which directly outputs the electric duals of a bad quiver.
The idea is to use the result we found in \cite{Giacomelli:2023zkk} for the bad SQCD and to apply it locally to each bad node.
This procedure is defined both in 4d and in 3d, and it can be described as follows. Given a bad theory, choose any ugly/bad node and carve it out from the rest of the quiver, together with the attached matter and singlets, to isolate a bad SQCD.
Then replace the carved out SQCD theory with the appropriate sum of good theories plus free sectors, as dictated by the aforementioned result for the SQCD we found in \cite{Giacomelli:2023zkk}. Finally, glue back the dualized parts to the rest of the quiver. At this point we have a collection of quivers. If some of them still contain ugly/bad nodes, iterate the procedure until all the generated frames contain only good gauge nodes.
In the end we will get a sum of good frames, multiplied by a free sector and possibly by one or more Dirac deltas: they will constitute the correct electric dual of the starting bad theory.

\begin{figure}[!ht]
	\centering
	\includegraphics[width=.75\textwidth]{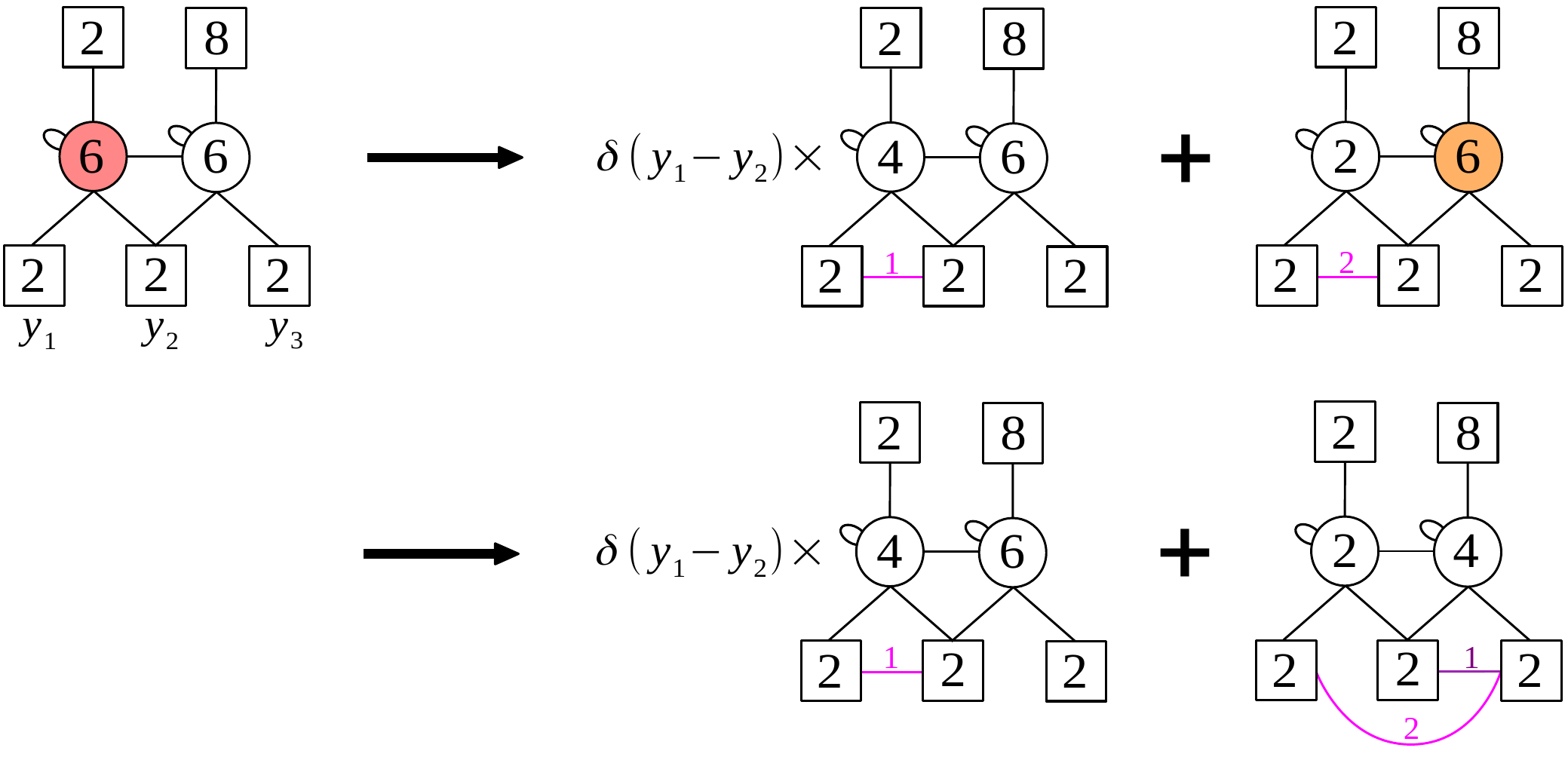}
	\caption{A schematic representation of the electric dualization of the 4d theory with gauge ranks $N_c = (3,3)$ and flavor ranks $N_f = (1,4)$. Its first node is bad and is highlighted in red. After one iteration of the electric algorithm we get two frames: the first one is good, while the second one has an ugly node, highlighted in orange, and thus needs one more iteration of the algorithm, resulting in the two quivers in the second line as the final dual theories. The numbers above the colored bifundamentals indicate their multiplicity. More details such as charges and omitted singlets can be found in Section \ref{sec:3}.}
	\label{fig:cartoon_example1_4d}
\end{figure}

This electric algorithm is illustrated in Figure \ref{fig:cartoon_example1_4d}, where the dualization of the 4d theory with gauge ranks $N_c = (3,3)$ and flavor ranks $N_f = (1,4)$ is schematically depicted. 
 This example will be discussed in detail in the main body of the paper.
 
The SQCD carving out operation, which is needed to perform the electric algorithm, is more rigorously described by a $(1,0)$-$(1,0)$ blocks swap, an IR duality coming from the aforementioned result for SQCD; in 3d it can be interpreted as a non trivial property of the swap of NS5-branes.

The mirror algorithm and the electric algorithm are equivalent.
However, the electric algorithm is perhaps faster and can be coded. Indeed this paper comes with a Mathematica notebook which, given any 3d $\mathcal{N}=4$ linear quiver with any number of bad/ugly nodes, outputs all its dual frames, specifying their FI parameters, their free sectors and the Dirac deltas they are multiplied by.\\

We are also interested in the 3-sphere partition function of bad linear quivers containing both unitary and special unitary gauge nodes.
An $SU(N)$ gauge node can be obtained from a $U(N)$ gauge node by gauging its $U(1)$ topological symmetry, which at the level of the partition function is implemented by integrating over the associated FI parameter. Under mirror symmetry the topological symmetry is mapped to a flavor symmetry, so turning a node from unitary to special unitary corresponds in the mirror dual theory to gauging some $U(1)$ flavor symmetry. One of the advantages of the dualization algorithm is that it provides the precise mapping between topological and flavor symmetries across the mirror duality so that we know exactly which $U(1)$ flavor symmetry should be gauged. This approach therefore provides an alternative to the brane locking method developed in \cite{Bourget:2021jwo}.

We first review how to do this for a good linear quiver theory with mixed unitary and special unitary gauge nodes to get its mirror dual by starting from that of the theory with unitary gauge nodes only (see also \cite{Dey:2020hfe,Bourget:2021jwo,Dey:2022eko,Dey:2023vrt} for related discussions).
We then apply the same logic to our result for the bad $U(N_c)$ SQCD to get an expression for the partition function of the bad $SU(N_c)$ SQCD. Since we are integrating over all possible values of the FI parameter, we find that this receives contribution from each of the frames that we found in the unitary case, including those that correspond to very specific values of the FI. More precisely, we find that the partition function of the bad $SU(N_c)$ SQCD is the sum of the partition functions of multiple good theories, not necessarily SQCD, none of which is multiplied by a delta now since this can be traded for the integral over the FI that we introduced. In some cases, as a result of the gauging, the free sector of the unitary theory ends up being coupled to the interacting part.

The structure of the full moduli space of the $SU(2)$ SQCD with two flavors was worked out in \cite{Assel:2018exy}, where it was shown that there are two distinct singular points of maximal codimension, i.e.~with no twisted hyper sector. The low energy effective description around each of these singular points is the same, and is given by the good SQED with two flavors. Our result for this case is compatible with \cite{Assel:2018exy} since we find that the partition function of $SU(2)$ with two flavors is the sum of two copies of the partition function of $U(1)$ with two flavors. A proposal for the full moduli space of the bad $SU(N_c)$ SQCD with $N_f$ flavors has been recently made in \cite{Bourget:2023cgs}, based on the conjecture that the quantum Coulomb branch can be obtained by a procedure called \emph{inversion} of the classical Higgs branch \cite{Grimminger:2020dmg}. It would be interesting to compare our results for the partition function of the bad $SU(N_c)$ SQCD with $N_f$ flavors with the findings of \cite{Bourget:2023cgs} for the full moduli space.\\

The rest of the paper is organized as follows.
In Section \ref{sec:QFT_Ingredients} we introduce all the preliminary 4d/3d QFT ingredients needed in the following sections. We review the basic duality moves realizing S-duality on the QFT blocks and we present the three blocks swap moves required to implement the two algorithms. 
In Section \ref{sec:3} we define the 4d mirror and electric algorithms we need in the rest of the paper. We illustrate them using the theory with gauge ranks $N_c = (3,3)$ and flavor ranks $N_f = (1,4)$ as an example.
In Section \ref{sec:example} we analyze in detail the 4d theory with $N_c = (5,3)$ and $N_f = (3,2)$ from various perspectives: we dualize the theory using both the mirror and the electric algorithms and we also study the theory adopting the Higgsing perspective, which reproduces the same answer obtained via the algorithms. 
In Section \ref{braneint} we give a Type IIB brane interpretation of the 3d $(1,0)$-$(1,0)$ and $(0,1)$-$(0,1)$ blocks swap moves. We also discuss the application of the 3d algorithms to study the fusion of two S-walls into an Identity-wall. We find that, quite remarkably, the algorithm reproduces the findings of \cite{Bottini:2021vms}. This provides a strong consistency check of our construction.
In Section \ref{badpartitions} we characterize the possible dual frames of a bad linear quiver theory, in particular we provide the rules determining the maximal and minimal dual frames, namely those having gauge groups with the largest and smallest possible rank respectively. The analysis is based entirely on the combinatorics of partitions of an integer $N$ which is specified by the linear quiver data.
Finally, in Section \ref{sec:SU} we focus on 3d $\mathcal{N}=4$ good/bad quiver theories including $SU(N)$ gauge nodes, which we obtain from theories with $U(N)$ gauge node only by gauging a subset of the $U(1)$ topological symmetries of the theory. The analysis is carried out at the level of the 3d partition function. In particular, we propose a precise method to identify unambiguously the $U(1)$ symmetries to be gauged. The main text is supplemented by various appendices reviewing conventions and previous results, and detailing some of the more technical derivations of results used in the rest of the paper.

\section{S-walls, QFT blocks, basic moves and swaps}
\label{sec:QFT_Ingredients}

In this section we introduce the class of theories we are interested in, we review the known ingredients of the dualization algorithms and introduce the new ones.

We begin by introducing the class of 3d and 4d linear quivers which we will be focusing on.
Following the notation of  \cite{Comi:2022aqo}, we will then introduce  the S-duality wall and the Identity-wall, the $(1,0)$ and $(0,1)$ QFT blocks corresponding to NS5 and D5-branes respectively, and the basic moves corresponding to their S-dualization.

We will next present the block swaps: the $(1,0)$-$(0,1)$ block swap, previously discussed in \cite{Comi:2022aqo} and which in the 3d case is the field the field theory counterpart of the  Hanany--Witten move, and the new 
 $(1,0)$-$(1,0)$ and $(0,1)$-$(0,1)$ block swaps whose  brane interpretation will be discussed in Section \ref{braneint}.
 

\subsection{Linear quivers in 3d and 4d}\label{linearquivers}

In 3d we will consider $\mathcal{N}=4$ unitary linear quivers. In this paper we will always represent them in an $\mathcal{N}=2$ notation: an $\mathcal{N}=4$ hypermultiplet contains two $\mathcal{N}=2$ chiral multiplets in conjugate representations which we represent by double lines with opposite orientations, while an $\mathcal{N}=4$ vector multiplet contains an $\mathcal{N}=2$ vector multiplet and an $\mathcal{N}=2$ adjoint chiral multiplet and we represent the latter with an arc. As usual, round nodes denote gauge symmetries while square nodes denote flavor symmetries, which in 3d will all be of unitary type.

These quivers admit a brane description in Type IIB string theory \cite{Hanany:1996ie}. If the linear quiver contains $k$ gauge groups of rank $N_i$ and the number of flavors at the $i$-th node is $F_i$ we engineer the theory by including $k+1$ NS5-branes. Between the $i$-th and $(i+1)$-th NS5’s we suspend $N_i$ D3-branes and we also include a set $F_i$ D5-branes. 

\begin{figure}[!ht]
	\centering
	\includegraphics[width=\textwidth]{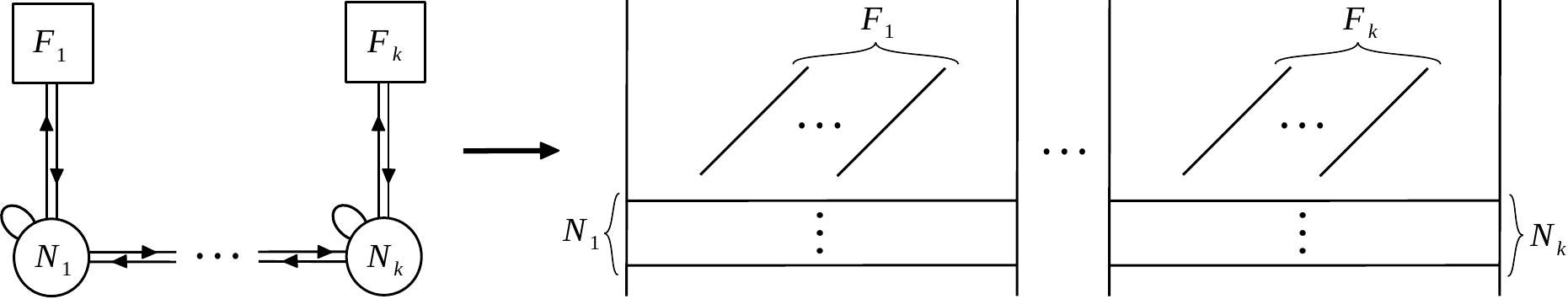}
	\caption{The 3d $\mathcal{N}=4$ $T_{\rho}^{\sigma}[SU(N)]$ theory drawn in $\mathcal{N}=2$ language and the associated Hanany--Witten brane setup. All nodes represent unitary symmetries, with round nodes being gauge symmetries and square node flavor symmetries. Each line denotes a chiral field transforming under the symmetries of the nodes that it connects, where a line between two nodes is a field in the bifundamental or antibifundamental representation depending on the orientation while an arc is a field in the adjoint representation.}
	\label{fig:HW_setup}
\end{figure}

This class of theories is usually dubbed $T_{\rho}^{\sigma}[SU(N)]$ theories \cite{Gaiotto:2008ak}, since the quiver data $\{ N_i, F_i\}$ can be encoded in the sequence of integers $\rho$ and $\sigma$ as follows:
\begin{itemize}
\item The number of flavors $F_i$ is equal to $\hat\sigma_i-\hat\sigma_{i+1}$, where we denote by $\hat\sigma_i$ the $i$-th element of $\sigma^T$, the  transpose to $\sigma$; 
\item The rank of the gauge groups $N_i$ is given by the formula 
\be\label{rank} N_i=\sum_{j>i}\rho_j-\sum_{j>i}\hat\sigma_j\,.\ee
\end{itemize}
Notice that positivity of the rank of the various gauge groups requires the constraint 
\be\label{rankcond}\rho\leq\sigma^T\quad\text{i.e.}\quad \sum_{j\leq i}\rho_j\leq\sum_{j\leq i}\hat\sigma_j\;\;\forall i\ee 
and therefore we should always restrict to $\rho$ and $\sigma$ satisfying this constraint for the quiver to make sense.

We can  bring all the D5-branes to the end of the quiver (say for definiteness to the left, even though also the other choice is perfectly fine), keeping into account the Hanany--Witten effect as the D5’s go across an NS5-brane. We end up with a set-up having $\sum_iF_i$ D5-branes, followed on their right by the $k+1$ NS5’s. 
In particular there will be a set of  $F_1$ D5-branes with  a single D3-brane ending on each of them, then there will be a set of $F_2$ D5s with   two D3-branes ending on each of them, and so on. The total number of D3-branes ending on D5’s is $N$ defined by the relation \be\label{totalN} N=\sum_i iF_i\;.\ee 
We then see that the sequence of integers $\sigma$ is actually a partition of $N$
and can be written as \be \sigma=[k^{F_k},(k-1)^{F_{k-1}},\dots,2^{F_2},1^{F_1}]\,.\ee 
The sequence of integer  $\rho$ can then be determined via \eqref{rank}.  

Before proceeding, a comment about the notation is in order. Usually the formalism described above is applied to describe good theories (that is each node is good $F_i+N_{i-1}+N_{i+1}\geq 2N_i$).
For good theories both $\rho$ and $\sigma$ are ordered partitions of $N$, with all the elements of the partition being positive integers.
For a bad theory however,  the sequence $\rho$ we get by applying \eqref{rank} is in general merely a sequence of integers whose sum is $N$. They are not ordered and are not all positive. 

In Appendix \ref{trsdef} we provide a more detailed definition of the $T_{\rho}^{\sigma}[SU(N)]$ theories by giving an expression for their $S^3_b$ partition function, which includes a set of background CS couplings for the global symmetries. These are important for various reasons, such as when discussing the variants of the same quivers but with some special unitary gauge nodes or to respect the global symmetry enhancements in the presence of a non-trivial background monopole for the global symmetries. \\

A special case of this class of theories corresponds to having the trivial partitions $\sigma=\rho=[1^N]$, in which it is customary to omit both of them and only name the theory $T[SU(N)]$. For a fixed $N$, this is the theory with the maximal global symmetry
\be 
SU(N)_X\times SU(N)_Y\times U(1)_{m_A}\,,
\ee
where $SU(N)_X$ is the manifest flavor symmetry of the quiver, $SU(N)_Y$ is enhanced from the $U(1)^{N-1}$ topological symmetry and $U(1)_{m_A}$ is the commutant of the $\mathcal{N}=2$ R-symmetry inside the $\mathcal{N}=4$ R-symmetry that manifests as a flavor symmetry in the $\mathcal{N}=2$ notation we are using. In this case, in the brane set-up we have $N$ D3-branes stretched between $N$ D5 and $N$ NS5-branes, where on each five-brane ends a single D3. This theory is special since we can recover from it any other generic $T_{\rho}^{\sigma}[SU(N)]$ theory with an RG flow triggered by giving nilpotent VEVs to the moment map operators for the $SU(N)_X$ and $SU(N)_Y$ symmetries which are labelled by $\sigma$ and $\rho$ respectively (see e.g.~\cite{Cremonesi:2014uva,Hwang:2020wpd}).

\begin{figure}[!ht]
\centering
    \includegraphics[width=0.8\textwidth]{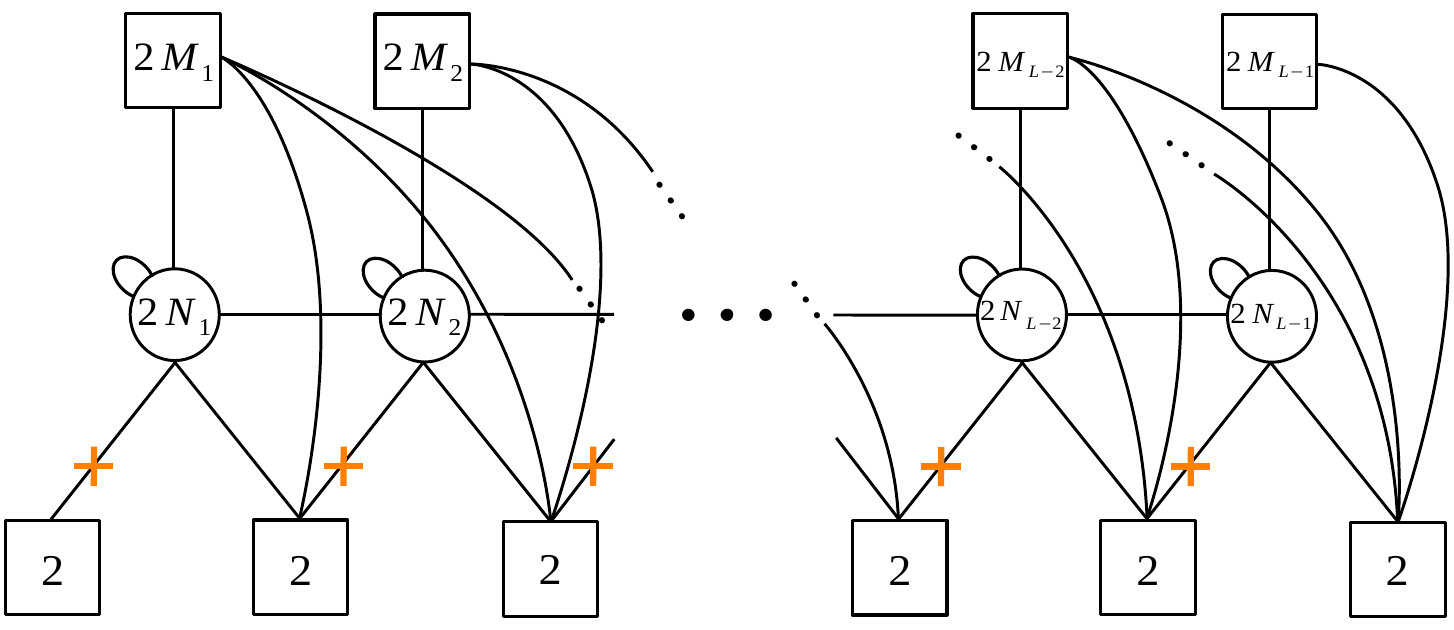}
    \caption{The 4d $\mathcal{N}=1$ $E_{\rho}^{\sigma}[USp(2N)]$ theory. All nodes represent symplectic symmetries, with round nodes being gauge symmetries and square node flavor symmetries as usual. Each line denotes a chiral field transforming under the symmetries of the nodes that it connects, where a line between two nodes is a field in the bifundamental representation while an arc is a field in the antisymmetric representation. The orange crosses represent the singlets flipping the meson built from the chiral on which they are placed (possibly dressed with some power of the antisymmetric). We use a black cross to indicate a singlet whereas   an orange cross for multiple singlets.}
    \label{fig:Generic_Erhosigma_simplified}
\end{figure}

The 3d $\mathcal{N}=4$ $T_{\rho}^{\sigma}[SU(N)]$ theories have a 4d $\mathcal{N}=1$ uplift called $E_{\rho}^{\sigma}[USp(2N)]$ \cite{Hwang:2020wpd}, where by ``uplift" we mean that the latter reduces to the former upon dimensional reduction and various real mass deformations.\footnote{Notice that this RG flow entails an enhancement of supersymmetry.} The structure of these 4d theories, summarized in Figure \ref{fig:Generic_Erhosigma_simplified}, is similar to that of their 3d counterparts and the ranks ${N_i,F_i}$ are still determined from $\sigma$ and $\rho$ with the same rule. However, there are a few differences: the groups are all symplectic rather than unitary, we have additional fundamental chiral fields forming the structure of a saw, and we have additional gauge singlet fields. In particular, there are two types of singlets:
\begin{itemize}
\item Those denoted by a cross at the $i$-th position are in number $N_{i-1}-N_i$ (where $N_0=0$) and they flip the meson constructed with the corresponding saw chirals dressed with the antisymmetric chiral up to the power $N_{i-1}-N_i-1$. 
\item Those denoted by lines connecting some of the flavor nodes. In particular, each upper flavor box is connected to all the bottom saw boxes that sit to its right. These singlets flip the minimal operator in the bifundamental representation of the two non-abelian symmetries, that is made of a sequence of chirals corresponding to the shortest path between the two boxes.
\end{itemize}
These latter singlets are in particular crucial to have an enhancement of the $SU(2)$ symmetries of the saw (see \cite{Hwang:2020wpd} for more details).
Finally, to fully specify the $E_{\rho}^{\sigma}[USp(2N)]$ theory one should also provide the superpotential. This consists of three main parts:
\begin{itemize}
\item A cubic interactions between the bifundamental and the antisymmetric chirals, which is the {\it uplift} of the superpotential we have in 3d due to $\mathcal{N}=4$  supersymmetry.
\item A cubic interaction between the chirals for each triangle of the saw.
\item The superpotential for the flipping fields discussed above.
\end{itemize}

Similarly to the 3d case, also in 4d the theory with trivial $\sigma=\rho=[1^N]$, which is denoted by $E[USp(2N)]$, is special as it has the maximal global symmetry
\be
USp(2N)_x\times USp(2N)_y\times U(1)_t\times U(1)_c
\ee
and a generic $E_{\rho}^{\sigma}[USp(2N)]$ theory can be obtained from it by giving a VEV to two operators in the antisymmetric representations of $USp(2N)_x$ and $USp(2N)_y$ whose form is fully determined in terms of $\sigma$ and $\rho$ respectively (see \cite{Hwang:2020wpd} for more details). \\

There is a notion of good/ugly/bad also for this class of 4d theories, which was introduced in \cite{Giacomelli:2023zkk} for the case of the SQCD. The condition is identical to the one we have in 3d: if all nodes are such that $F_i+N_{i-1}+N_{i+1}> 2N_i$ then the quiver is good, otherwise it is either ugly or bad depending on whether we have a strict equality of the opposite inequality for at least one node. Similarly to 3d, this implies that, while $\sigma$ is always a partition of $N$, $\rho$ is instead just a sequence of integers summing up to $N$, which for a bad theory is not necessarily ordered nor positive and hence not a partition.
As discussed in \cite{Giacomelli:2023zkk} for the SQCD case and as we will see extensively in this paper, a 4d bad theory is characterized by the fact that its index is not an ordinary function but rather a distribution expressed in terms of combinations of delta functions. Physically, this is related to the fact that the moduli space of the theory does not include the origin and hence there is no point where the full gauge symmetry is preserved. Instead, one is forced to give a VEV associated with one of the delta constraints, which Higgses the gauge group. This was discussed extensively in \cite{Giacomelli:2023zkk} for the SQCD and we will see a quiver example in Section \ref{sec:example}.

The index of the $E_{\rho}^{\sigma}[USp(2N)]$ theory in our conventions is given in Appendix \ref{ersdef}.

\subsection{S-wall and Identity-wall}\label{sec:SwallIdwall}
We now discuss the ingredient of the dualization algorithms.
First of all, we review the notions of S-wall and Identity-wall first introduced in \cite{Bottini:2021vms} (see \cite{Comi:2022aqo} for our conventions).

\begin{figure}[!ht]
	\centering
	\includegraphics[width=.9\textwidth]{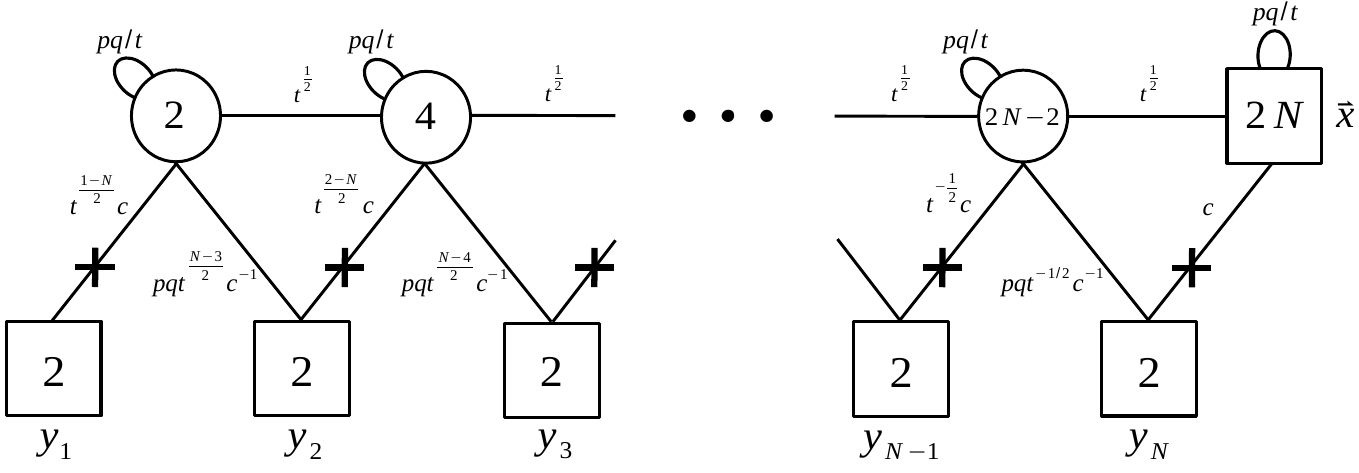}
	\caption{The 4d $FE[USp(2N)]$ theory. The combinations of the fugacities close to each line encode the charges of the corresponding chiral fields under the abelian symmetries including the R-symmetry. Close to each node we instead specify the label for the fugacities of the corresponding symmetry. For circle nodes these are gauge fugacities over which we integrate in the index, while for square nodes they are parameters on which the index depends.}
	\label{fig:FEUSp2N}
\end{figure}

\paragraph{4d case}\mbox{}\\[5pt]
The 4d $\mathsf{S}$-wall theory is the $FE[USp(2N)]$ theory introduced in \cite{Pasquetti:2019hxf}, which is defined as the $E[USp(2N)]$ theory with an additional flipping field in the antisymmetric of $USp(2N)_x$ (hence the name). This is the quiver theory depicted in Figure \ref{fig:FEUSp2N}.

\begin{figure}[!ht]
	\centering
	\includegraphics[width=.7\textwidth]{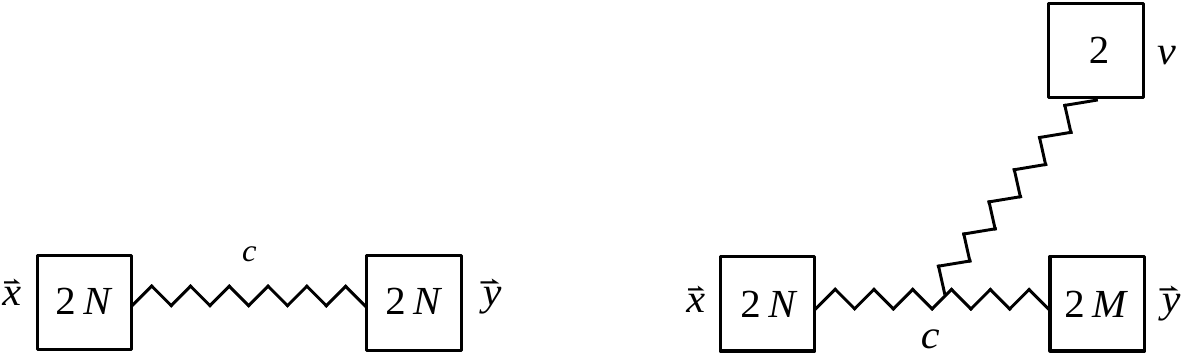}
	\caption{The 4d $\mathsf{S}$-wall theory and its asymmetric version. Close to each wiggle line we specify its $c$ fugacity, which might be identified with various combinations of the abelian symmetries fugacities when gluing the $\mathsf{S}$-wall to other blocks.}
	\label{fig:S_operator}
\end{figure}

We will represent it compactly as on the left of Figure \ref{fig:S_operator}, where we exhibit its two $USp(2N)$ symmetries. At the level of the index, the $\mathsf{S}$-wall is recursively defined as
\begingroup\allowdisplaybreaks
\begin{align}
\mathcal{I}_{\mathsf{S}}^{(N)}(\vec{x};\vec{y};t;c) &=\Gpq{pq\,c^{-2}}\Gpq{pq\,t^{-1}}^N\prod_{i<j}^N\Gpq{pq\,t^{-1}x_i^{\pm1}x_j^{\pm1}}\prod_{i=1}^N\Gpq{c\,y_N^{\pm1}x_i^{\pm1}}\nn\\
&\qquad\times\oint\udl{\vec{z}_{N-1}} \Gd_{N-1}(\vec z_{N-1}) \prod_{a=1}^{N-1}\prod_{i=1}^N\Gpq{t^{1/2}z_a^{\pm1}x_i^{\pm1}}\Gpq{pq\,t^{-1/2}c^{-1}y_N^{\pm1}z_a^{\pm1}}\nn \\
&\qquad\times\mathcal{I}_{\mathsf{S}}^{(N-1)}\left(z_1,\cdots,z_{N-1};y_1,\cdots,y_{N-1};t;t^{-1/2}c\right)\,.
\end{align}
\endgroup
It is important to remember that the S-duality group in 4d is $PSL(2,\mathbb{Z})$, and therefore the property $\mathsf{S}=\mathsf{S}^{-1}$ holds.

It is possible to turn on a deformation breaking one of the two $USp(2N)$ symmetries down to $USp(2M)\times SU(2)$ for $M<N$ (see \cite{Hwang:2020wpd,Bottini:2021vms} for more details). This gives a new theory that we call the 4d asymmetric $\mathsf{S}$-wall and that we represent compactly as on the right of Figure \ref{fig:S_operator} so to manifestly display all of its non-abelian symmetries. At the level of the index this deformation corresponds to the specialization of some of the flavor fugacities
\begin{equation}
	\mathcal{I}_{\mathsf{S}}^{(N)}(\vec{x};\vec{y},t^{\frac{N-M-1}{2}}v,\cdots,t^{-\frac{N-M-1}{2}}v;t;c)\,.
\end{equation}

\begin{figure}[!ht]
	\centering
	\includegraphics[width=.7\textwidth]{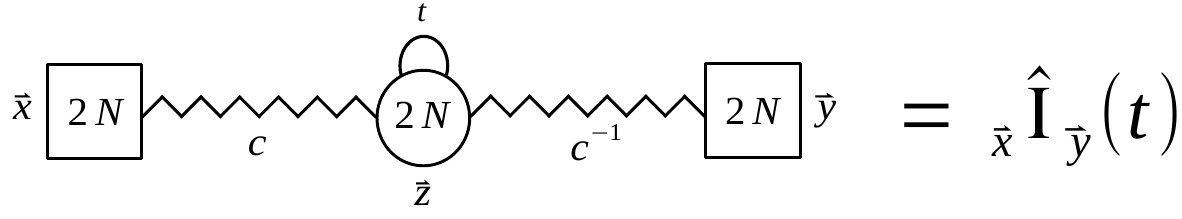}
	\caption{The 4d gluing of two $\mathsf{S}$-walls to give an Identity-wall. The gluing consists of gauging a diagonal combination of the $USp(2N)$ symmetries of the two $\mathsf{S}$-walls and adding an antisymmetric of this $USp(2N)$ symmetry that couples in the superpotential with the antisymmetric operators of the two glued blocks.}
	\label{fig:Identity_wall}
\end{figure}

If we glue two symmetric $\mathsf{S}$-walls together we get what we call a 4d Identity-wall \cite{Bottini:2021vms}, represented in Figure \ref{fig:Identity_wall}. Here and in the rest of this paper, by ``glue" we mean gauging a diagonal combination of the $USp(2N)$ symmetries of the two glued blocks. Moreover, if an antisymmetric $A$ of this $USp(2N)$ symmetry is added as in this case, this couples in the superpotential with the antisymmetric operators $\mathsf{A}_{L/R}$ of the two glued blocks (see e.g.~\cite{Comi:2022aqo})
\be\label{eq:Wgluing4d}
\mathcal{W}_{gluing}=A(\mathsf{A}_L+\mathsf{A}_R)\,.
\ee

At the level of the index we have
\begin{equation}
\oint\udl{\vec{z}_N}\Gd_N(\vec{z};t)\mathcal{I}_{\mathsf{S}}^{(N)}(\vec{x};\vec{z};t;c)\mathcal{I}_{\mathsf{S}}^{(N)}(\vec{z};\vec{y};t;c^{-1})={}_{\vec{x}}\hat{\mathbb{I}}_{\vec{y}}(t) \,,
\end{equation}
where we defined the index of the 4d Identity-wall as
\begin{equation}
{}_{\vec x}\hat{\mathbb{I}}_{\vec y}(t)=\frac{\prod_{j=1}^N 2\pi ix_j}{\Gd_N(\vec{x};t)}\sum_{\gs\in S_N}\sum_{\pm}\prod_{j=1}^N\gd\left(x_j- y_{\gs(j)}^\pm\right)\,.
\end{equation}

As explained in \cite{Bottini:2021vms}, this should be understood as a theory with a breaking of the $USp(2N)_x$ and $USp(2N)_y$ global symmetries to their diagonal combination, due to a VEV at the quantum level for an operator in their bifundamental representation. This is an example of a 4d bad quiver theory. Consider indeed gluing the two $\mathsf{S}$-walls by gauging their manifest $USp(2N)$ symmetry in Figure \ref{fig:FEUSp2N}. The antisymmetric operators $\mathsf{A}_{L/R}$ would then correspond to the antisymmetric chiral for the $USp(2N)$ symmetry of each $\mathsf{S}$-wall and the gluing superpotential \eqref{eq:Wgluing4d} would be a mass term for two out of the three antisymmetric chirals $A$, $\mathsf{A}_L$ and $\mathsf{A}_R$. After integrating the massive fields out one obtains an $E_\rho^\sigma$ theory, but the middle $USp(2N)$ gauge node would only see $2N-2<2N$ flavors and would thus be bad. The properties of this theory of its index behaving as a delta distribution and of the quantum VEV are a manifestation of its badness. In Subsection \ref{subsec:SS1} we will come back to this observation.

\begin{figure}[!ht]
	\centering
	\includegraphics[width=.7\textwidth]{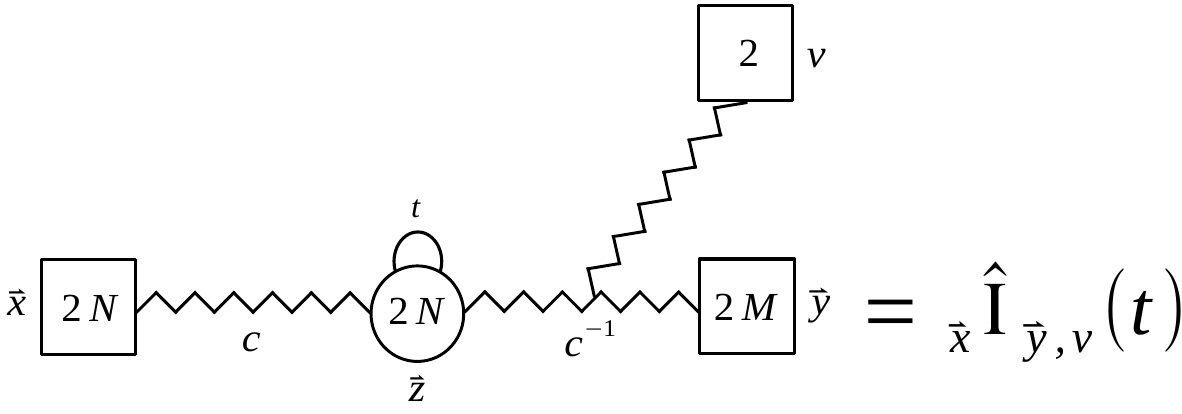}
	\caption{The 4d gluing of a symmetric and an asymmetric $\mathsf{S}$-walls to give an asymmetric Identity-wall.}
	\label{fig:Asymm_Identity_wall}
\end{figure}

If we instead glue one symmetric and one asymmetric $\mathsf{S}$-wall together we get what we call a 4d asymmetric Identity-wall, represented in Figure \ref{fig:Asymm_Identity_wall}.
\\ 
At the level of the index we have
\begin{equation}
\oint\udl{\vec{z}_N}\Gd_N(\vec{z};t)\mathcal{I}_{\mathsf{S}}^{(N)}(\vec{x};\vec{z};t;c)\mathcal{I}_{\mathsf{S}}^{(N)}(\vec{z};\vec{y},t^{\frac{N-M-1}{2}}v,\cdots,t^{-\frac{N-M-1}{2}}v;t;c^{-1})={}_{\vec{x}}\hat{\mathbb{I}}_{\vec{y},v}(t) \,,\nn\\
\end{equation}
where we defined the index of the 4d asymmetric Identity-wall as 
\begin{equation}
{}_{\vec{x}}\hat{\mathbb{I}}_{\vec{y},v}(t)=\frac{{ \prod_{j=1}^{N} 2\pi i x_j }}{\Gd_N (\vec x;t) }\left.\sum_{\sigma \in S_N,\pm}\prod_{j=1}^{N}\delta\left(x_j-y_{\sigma(j)}^{\pm1}\right)\right|_{y_{M+k}=t^{\frac{N-M+1-2k}{2}}v}\,.
\end{equation}

\begin{figure}[!ht]
	\centering
	\includegraphics[width=.8\textwidth]{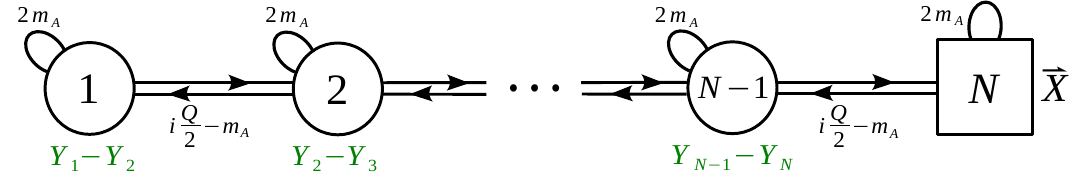}
	\caption{The $FT[U(N)]$ theory. We specify the R-charge $q_R$ and the $U(1)_{m_A}$ charge $q_A$ of each chiral field via a combination $iq_R \frac{Q}{2}+q_Am_A$ of the squashing parameter $Q=b+b^{-1}$ and of the axial mass $m_A$ on which the $S^3_b$ partition function depends. Moreover, we write in green the FI parameters on each gauge node.}
	\label{fig:FTSUN}
\end{figure}

\begin{figure}[!ht]
	\centering
	\includegraphics[width=.7\textwidth]{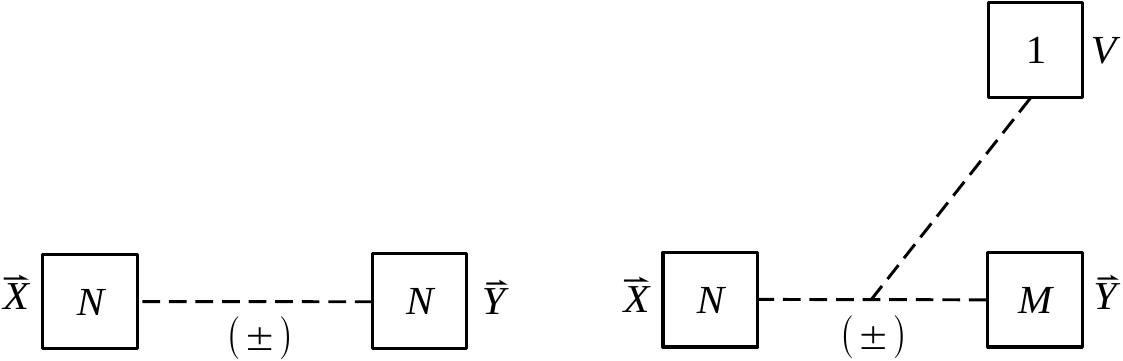}
	\caption{The 3d symmetric $\mathcal{S}^{\pm1}$-walls on the left and their asymmetric versions on the right. Close to each node we specify the label for the mass parameters of the corresponding symmetry on which the $S^3_b$ partition function depends.}
	\label{fig:S_operator_3d}
\end{figure}

\paragraph{3d case}\mbox{}\\[5pt]
In 3d we start by considering the $FT[U(N)]$ theory, which is the $T[U(N)]$ theory\footnote{The $T[U(N)]$ theory is the product of the $T[SU(N)]$ theory and the $T[U(1)]$ theory, where the latter is just a $U(1)\times U(1)$ mixed background CS theory. Its contribution to the $T[U(N)]$ partition function is given by the exponential prefactor in \eqref{eq:TUNpf}. 
The $T[U(N)]$ theory then has a $U(N)_X\times U(N)_Y\times U(1)_{m_A}$ global symmetry.} \cite{Gaiotto:2008ak} modified by adding a set of singlet chiral fields in the adjoint representation of the manifest $U(N)_X$ flavor symmetry. This can also be obtained as the dimensional reduction of the 4d $FE[USp(2N)]$ theory \cite{Pasquetti:2019hxf}. 
The $FT[U(N)]$ theory is shown in Figure \ref{fig:FTSUN}. 
\\
Its partition function is 
\begingroup\allowdisplaybreaks
\be\label{eq:TUNpf}
&&\mathcal{Z}_{FT[U(N)]}(\vec{X};\vec{Y};m_A)=\mathrm{e}^{2\pi iY_N\sum_{a=1}^NX_a}\prod_{a,b=1}^N\sbfunc{i\frac{Q}{2}-2m_A\pm(X_a-X_b)}\nn\\
&&\quad\times\int\udl{\vec{Z}^{(N-1)}_{N-1}}  
\Delta^{3d}_{N-1}\left(\vec Z^{(N-1)}\right) 
e^{-2\pi i Y_N\sum_{j=1}^{N-1}Z_j^{(N-1)}} 
\prod_{j=1}^{N-1}\prod_{a=1}^Ns_b\left(m_A\pm(Z_j^{(N-1)}-X_a)\right)\nn\\
&&\quad\times\mathcal{Z}_{FT[U(N-1)]}\left(Z_1^{(N-1)},\cdots,Z_{N-1}^{(N-1)};Y_1,\cdots,Y_{N-1};m_A\right)\,.
\ee
\endgroup
The 3d $\mathcal{S}$-wall is then defined as
\begin{equation}
\mathcal{Z}_{\mathcal{\mathcal{S}}}^{(N)}(\vec{X};\vec{Y};m_A)\equiv\mathcal{Z}_{FT[U(N)]}(\vec{X};-\vec{Y};m_A)=\mathcal{Z}_{FT[U(N)]}(-\vec{X};\vec{Y};m_A) \,.
\end{equation}
Since in 3d the S-duality group is $SL(2,\mathbb{Z})$, the generators $\mathcal{S}$ and $\mathcal{S}^{-1}$ are distinct and related as follows:
\begin{equation}
\mathcal{Z}_{\mathcal{\mathcal{S}}^{-1}}^{(N)}(\vec{X};\vec{Y};m_A)=\mathcal{Z}_{\mathcal{\mathcal{S}}}^{(N)}(\vec{X};-\vec{Y};m_A)=\mathcal{Z}_{\mathcal{\mathcal{S}}}^{(N)}(-\vec{X};\vec{Y};m_A) \,.
\end{equation}
We will represent compactly the $\mathcal{S}$-wall displaying both of its $U(N)$ global symmetries as on the left of Figure \ref{fig:S_operator_3d}, where we distinguish between $\mathcal{S}$ and $\mathcal{S}^{-1}$ by a $(+)$ and $(-)$ label respectively.

Similarly to 4d, we can obtain an asymmetric $\mathcal{S}$-wall by turning on a deformation that breaks one of the $U(N)$ symmetries down to $U(M)\times U(1)$ with $M<N$ and we represent it compactly as on the right of Figure \ref{fig:S_operator_3d}. At the level of the $S^3_b$ partition function, this deformation forces the following specialization of the real mass parameters for the broken symmetry:
\begin{equation}
Y_{M+j}=\frac{N-M+1-2j}{2}(iQ-2m_A)+V,\qquad j=1,\cdots,N-M\,.
\end{equation}

\begin{figure}[!ht]
	\centering
	\includegraphics[width=.7\textwidth]{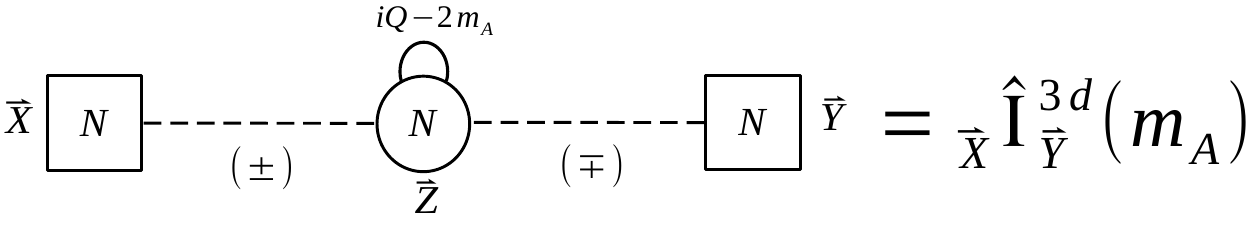}
	\caption{The 3d gluing of an $\mathcal{S}$ and an $\mathcal{S}^{-1}$-walls to give an Identity-wall. The gluing consists of gauging a diagonal combination of the $U(N)$ symmetries of the two $\mathcal{S}$-walls and adding an adjoint of this $U(N)$ symmetry that couples in the superpotential with the moment map operators of the two glued blocks.}
	\label{fig:Identity_wall_3d}
\end{figure}

Also in 3d holds the property that if we glue two symmetric $\mathcal{S}$-walls, these annihilate each other giving an Identity-wall \cite{Bottini:2021vms} (see Figure \ref{fig:Identity_wall_3d}). Again by ``glue" we mean that we gauge a diagonal combination of the $U(N)$ symmetries and we add an adjoint chiral field $\Phi$ for this new gauge node that couples in the superpotential with the moment map operators $\Phi_{L/R}$ of each glued block
\be \label{eq:Wgluing3d}
\mathcal{W}_{gluing}=\Phi(\Phi_L+\Phi_R)\,.
\ee

At the level of the partition function, we have
\begin{align}
\label{3ddeltaTSUN}
&\int\udl{\vec{Z}_N}\Delta^{3d}_N(\vec{Z};m_A)\mathcal{Z}_{\mathcal{S}}^{(N)}(\vec{Z};\vec{X};m_A)\mathcal{Z}_{\mathcal{S}}^{(N)}(\vec{Z};-\vec{Y};m_A)={}_{\vec X}\hat{\mathbb{I}}^{3d}_{\vec Y}(m_A)
\end{align}
where we defined the partition function of the 3d Identity-wall as
\begin{equation}
{}^{\phantom{}}_{\vec X}\hat{\mathbb{I}}^{3d}_{\vec Y}(m_A)=\frac{\sum_{\gs\in S_N}\prod_{j=1}^N\gd\left(X_j-Y_{\gs(j)}\right)}{\Gd^{3d}_N(\vec{X};m_A)}
\end{equation}

This is again an example of a 3d bad theory, as it can be understood if we consider gauging the manifest $U(N)$ flavor symmetry of each $\mathcal{S}$-wall, since this would give a $U(N)$ gauge node with $2N-2$ flavors. In Subsection \ref{subsec:SS1} we will recover the result \eqref{3ddeltaTSUN} from this perspective.

\begin{figure}[!ht]
	\centering
	\includegraphics[width=.7\textwidth]{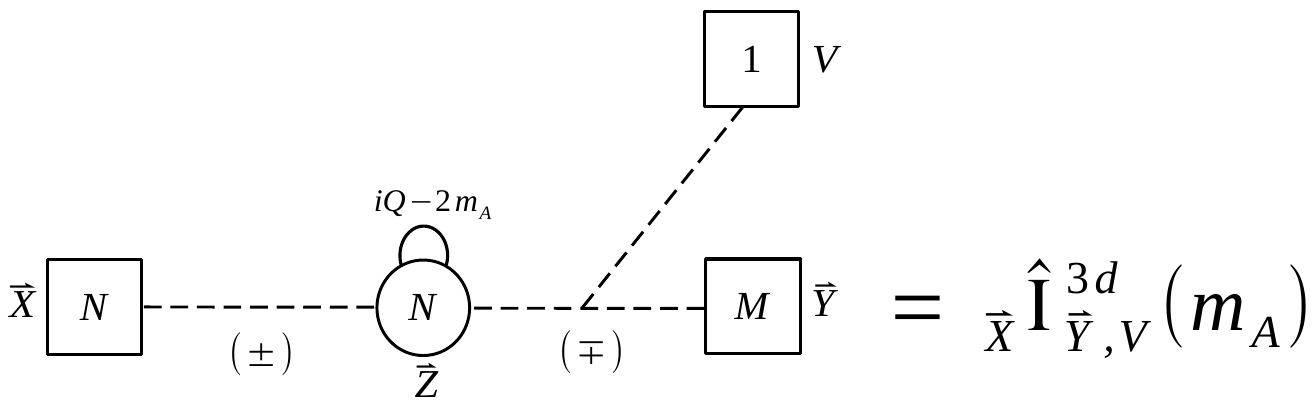}
	\caption{The 3d asymmetric Identity-wall.}
	\label{fig:Asymm_Identity_wall_3d}
\end{figure}

If we instead glue one symmetric and one asymmetric $\mathcal{S}$-wall together we get what is shown in Figure \ref{fig:Asymm_Identity_wall_3d}, where we defined the 3d asymmetric Identity-wall as
\begin{equation}
{}^{\phantom{}}_{\vec X}\hat{\mathbb{I}}^{3d}_{\vec Y,V}(m_A)=\frac{1}{\Gd^{3d}_N(\vec{X};m_A)}\left.\sum_{\sigma \in S_N}\prod_{j=1}^{N}\delta\left(X_j-Y_{\sigma(j)}\right)\right|_{Y_{M+k}=\frac{N-M+1-2k}{2}(iQ-2m_A)+V}\,.
\end{equation}

\subsection{QFT blocks}
We now recall the definitions of the QFT blocks first introduced in \cite{Hwang:2021ulb} (see \cite{Comi:2022aqo} for our conventions).

\begin{figure}[!ht]
	\centering
	\includegraphics[width=0.8\textwidth]{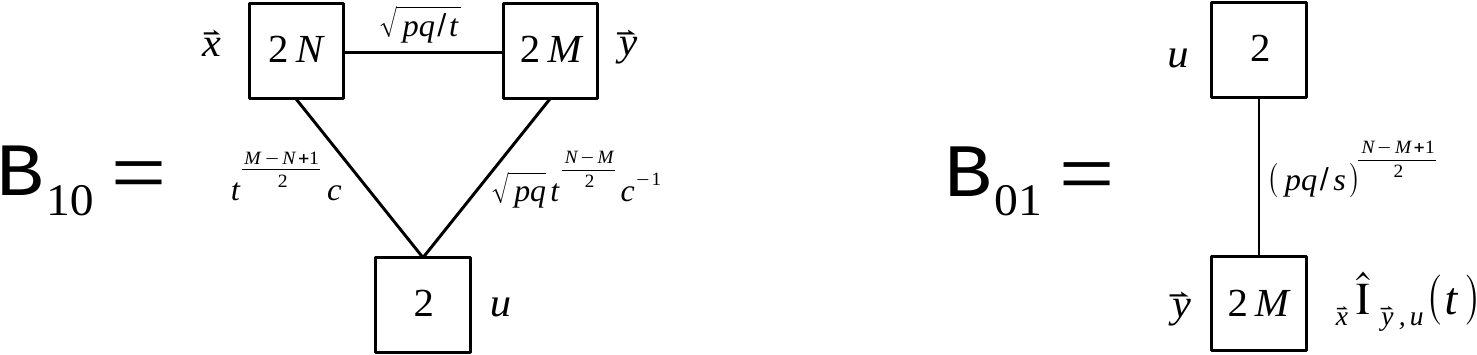}
	\caption{The 4d $\mathsf{B}_{10}$ block on the left and the $\mathsf{B}_{01}$-block on the right.}
	\label{fig:Blocks_4d}
\end{figure}

\paragraph{4d case}\mbox{}\\[5pt]
The 4d $\mathsf{B}_{10}$ and $\mathsf{B}_{01}$-blocks are defined as the theories shown in Figure \ref{fig:Blocks_4d}. The $\mathsf{B}_{10}$ block is a collection of chirals interacting with a cubic superpotential corresponding to the product of the three edges of the triangle. The $\mathsf{B}_{01}$ is also a collection of chirals, but notice that it is also multiplied by an Identity-wall.
\\
At the level of the index we have
\begin{align}
\mathcal{I}_{(1,0)}^{(N,M)}(\vec{x};\vec{y};u;t;c) & = \prod_{i=1}^N\prod_{j=1}^M \Gpq{(pq/t)^{\frac{1}{2}} x_i^\pm y_j^\pm} \prod_{i=1}^N  \Gpq{t^{\frac{M-N+1}{2}} c x_j^\pm u^\pm} \nonumber\\
& \quad\times\prod_{j=1}^M\Gpq{(pq)^{\frac{1}{2}} t^{\frac{N-M}{2}}c^{-1} y_j^\pm u^\pm} 
\end{align}
and
\be
\mathcal{I}_{(0,1)}^{(N,M)} (\vec{x};\vec{y};u;s) = 
		\prod_{j=1}^M \Gpq{(pq/s)^{\frac{N-M+1}{2}} 
		y_j^\pm u^\pm}{}_{\vec x}\hat{\mathbb{I}}_{\vec y,u}(t) \,,
\ee
where $s$ can be either $t$ or $pq/t$.
On the l.h.s.~we do not list the argument $t$ of the 4d Identity-wall, as throughout the paper  it will always be $t$.
Here we are assuming without loss of generality that $M\leq N$, since the case $M\geq N$ can be obtained just by a vertical reflection.

\begin{figure}[!ht]
	\centering
	\includegraphics[width=0.8\textwidth]{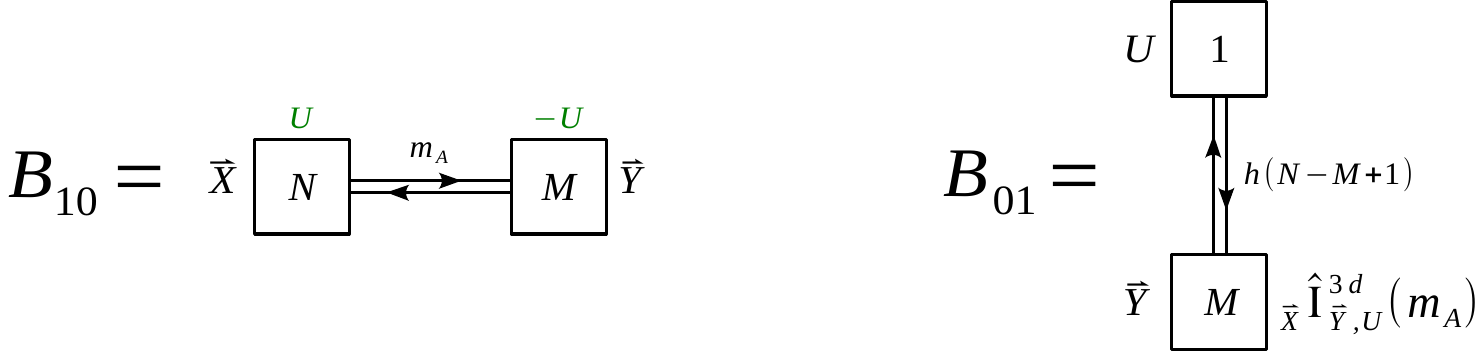}
	\caption{The 3d $\mathcal{B}_{10}$-block on the left and the $\mathcal{B}_{01}$-block on the right.}
	\label{fig:Blocks_3d}
\end{figure}

\paragraph{3d case}\mbox{}\\[5pt]
The 3d $\mathcal{B}_{10}$ and $\mathcal{B}_{01}$-blocks are similarly defined as the theories shown in Figure \ref{fig:Blocks_3d}. The $\mathcal{B}_{10}$-block is just a $U(N)\times U(M)$ bifundamental hyper with some background CS coupling between these symmetries and a $U(1)$ global symmetry that will play the role of the topological symmetry when the non-abelian symmetries are gauged in a quiver, in which case the parameter $U$ is the corresponding FI parameter. The $\mathcal{B}_{01}$-block is instead a single $U(N)$ fundamental hyper multiplied by an Identity-wall.

At the level of the partition function we have (again $M\leq N$)
\begin{equation}
\mathcal{Z}_{(1,0)}^{(N,M)}(\vec{X};\vec{Y};U;m_A) =
\mathrm{e}^{2\pi iU\left(\sum_{a=1}^NX_a-\sum_{b=1}^MY_b\right)} 
\prod_{a=1}^N\prod_{b=1}^M \sbfunc{\frac{i Q}{2}-m_A\pm(X_a-Y_b)}
\end{equation}
and 
\be
\mathcal{Z}_{(0,1)}^{(N,M)} (\vec{X};\vec{Y};U;h) = \prod_{a=1}^M \sbfunc{\frac{iQ}{2}-h(N-M+1)\pm(Y_a-U)}{}^{\phantom{}}_{\vec X}\hat{\mathbb{I}}^{3d}_{\vec Y,U}(m_A)\,,
\ee
where $h$ can be either $m_A$ or $\frac{i Q}{2}-m_A$.
On the l.h.s.~we do not list the argument $m_A$ of the 3d Identity-wall, as throughout the paper it will always be $m_A$.

\subsection{Basic S-duality moves}
Now we consider the S-dualization of the QFT blocks introduced in the previous subsection, again following the conventions of \cite{Comi:2022aqo}.
As shown in \cite{Bottini:2021vms}, the 4d and 3d basic moves are IR dualities which can be demonstrated by iterations of respectively Intriligator--Pouliot \cite{Intriligator:1995ne} and Aharony   \cite{Aharony:1997gp} dualities.

\begin{figure}[!ht]
	\includegraphics[width=\textwidth]{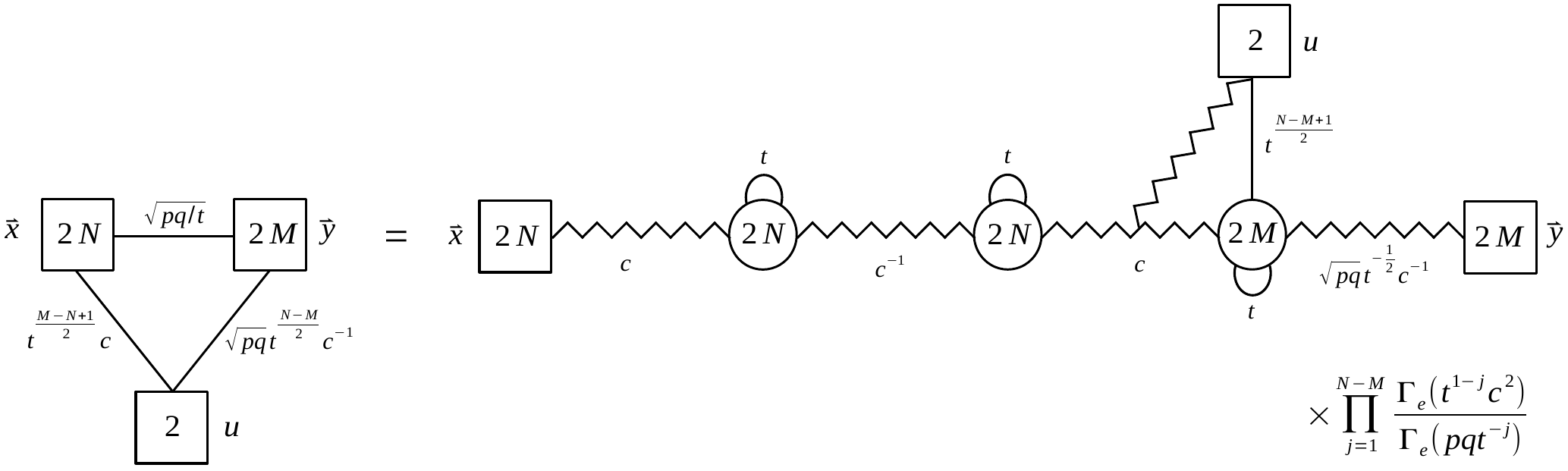}
	\caption{The $\mathsf{B}_{10}= \mathsf{S}\mathsf{B}_{01}\mathsf{S}^{-1}$ duality move. The gluing consists of gauging a diagonal combination of the $USp(2N)$ symmetries while adding an antisymmetric chiral that couples in the superpotential with the antisymmetric operators of the two glued blocks.}
	\label{fig:B10_Sdual_4d}
\end{figure}

\paragraph{4d case}\mbox{}\\[5pt]
The basic move dualizing the $\mathsf{B}_{10}$ block into the $\mathsf{B}_{01}$-block is represented in Figure \ref{fig:B10_Sdual_4d}. Again all blocks are glued by gauging a diagonal combination of their $USp(2N)$ symmetries while adding an antisymmetric chiral that couples in the superpotential with the antisymmetric operators of the two glued blocks.

Its index expression is
\begin{eqnarray}
&&\mathcal{I}_{(1,0)}^{(N,M)}\left(\vec{x};\vec{y};u;t;c\,t^{\frac{M-N}{2}}\right) =\prod_{i=1}^{N-M}\frac{\Gpq{t^{1-i}c^2}}{\Gpq{pq\,t^{-i}}}\oint \udl{\vec{z}^{\,(1)}_N}\udl{\vec{z}^{\,(2)}_M}\Gd_N\left(\vec{z}^{\,(1)};t\right)\Gd_M\left(\vec{z}^{\,(2)};t\right)\nn\\
&&\quad\qquad\times\mathcal{I}_{\mathsf{S}}^{(N)}\left(\vec{x};\vec{z}^{\,(1)};t;c\right)
{\mathcal{I}_{(0,1)}^{(N,M)}\left(\vec{z}^{\,(1)};\vec{z}^{\,(2)};u;pq/t\right)
\mathcal{I}_{\mathsf{S}}^{(M)}\left(\vec{z}^{\,(2)};\vec{y};t;(pq/t)^{\frac{1}{2}}c^{-1}\right)} \,.
\label{eq:B10_DualityMove_4d}
\end{eqnarray}

\begin{figure}[!ht]
	\includegraphics[width=\textwidth]{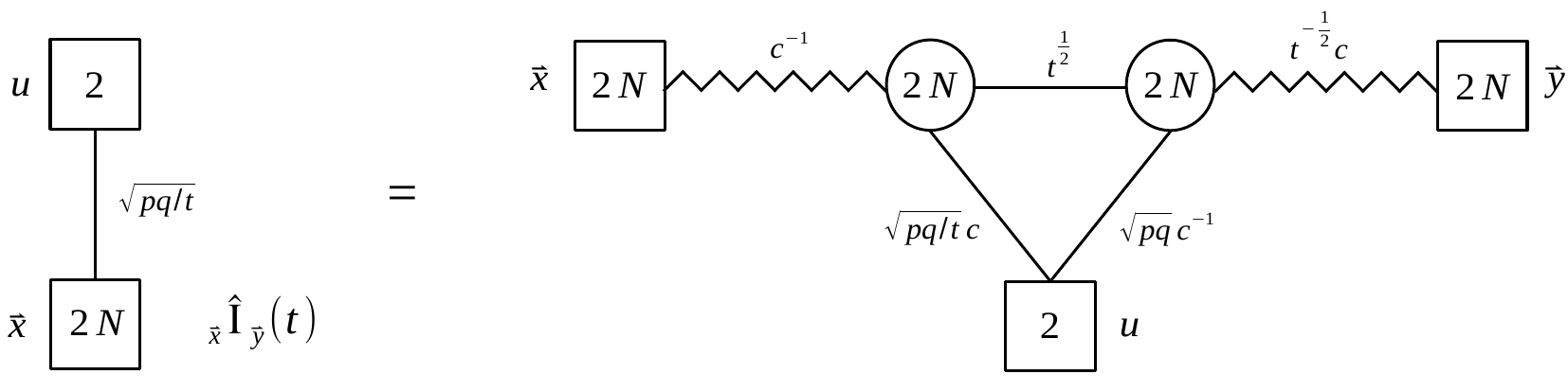}
	\caption{The $\mathsf{B}_{01}= \mathsf{S}\mathsf{B}_{10}\mathsf{S}^{-1}$ duality move. The gluing consists of gauging a diagonal combination of the $USp(2N)$ symmetries while adding a superpotential coupling the antisymmetric operators of the two $\mathsf{S}$-walls and the meson constructed from the bifundamental.}
	\label{fig:B01_Sdual_4d}
\end{figure}

On the other hand, the basic move dualizing the $\mathsf{B}_{01}$-block into the $\mathsf{B}_{10}$ block is represented in Figure \ref{fig:B01_Sdual_4d}. In this case the gluing is again done by gauging a diagonal combination of the $USp(2N)$ symmetries of the glued blocks, but since we do not add any antisymmetric chiral the superpotential consists of a coupling between the antisymmetric operators $\mathsf{A}_{L/R}$ of the two $\mathsf{S}$-walls and the meson constructed from the bifundamental $Q$
\be
\mathcal{W}_{gluing}=Q^2(\mathsf{A}_L+\mathsf{A}_R)\,.
\ee

Its index expression is
\begin{eqnarray}
&&{\mathcal{I}_{(0,1)}^{(N,N)} (\vec{x};\vec{y};u;t)=\oint \udl{\vec{w}^{\,(0)}_N}\udl{\vec{w}^{\,(1)}_N} \Gd_N(\vec{w}^{\,(0)})  \Gd_N(\vec{w}^{\,(1)})} \nn\\
&&\quad\times{\mathcal{I}_{\mathsf{S}}^{(N)}(\vec{x};\vec{w}^{\,(0)};t;c^{-1}) \mathcal{I}_{(1,0)}^{(N,N)} \left(\vec{w}^{\,(0)};\vec{w}^{\,(1)};u;pq/t;c
\right)\mathcal{I}_{\mathsf{S}}^{(N)}(\vec{w}^{\,(1)};\vec{y};t;t^{-\frac{1}{2}}c) } \,.
\label{eq:B01_DualityMove_4d}
\end{eqnarray}

\begin{figure}[!ht]
	\includegraphics[width=\textwidth]{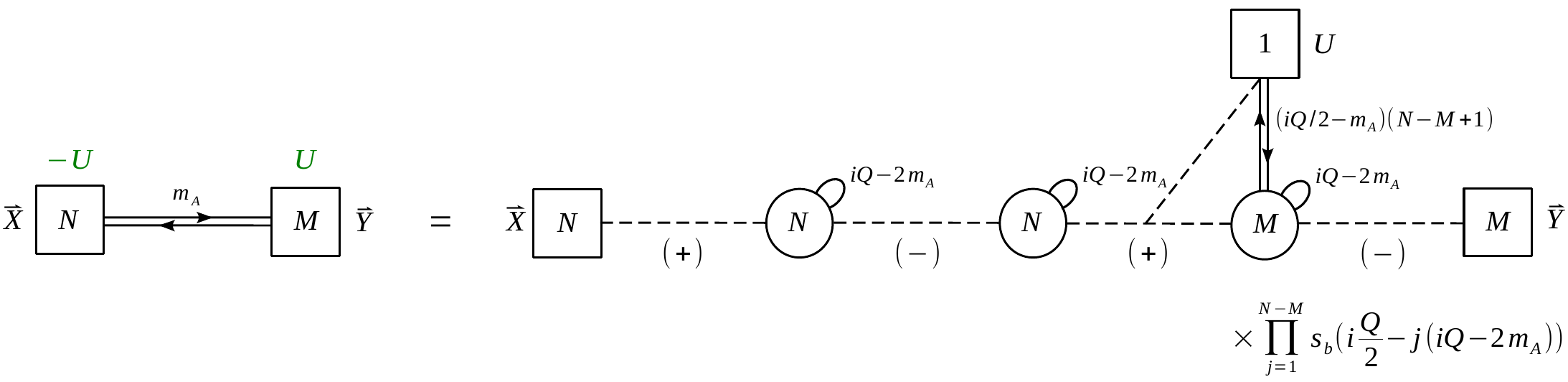}
	\caption{The $\mathcal{B}_{-10}= \mathcal{S}\mathcal{B}_{01}\mathcal{S}^{-1}$ duality move, where $\mathcal{B}_{-10}$ is the same as $\mathcal{B}_{10}$ but with the opposite sign of the FI parameters. The gluing consists of gauging a diagonal combination of the $U(N)$ symmetries and adding an adjoint of this $U(N)$ symmetry that couples in the superpotential with the moment map operators of the two glued blocks.}
	\label{fig:B10_Sdual_3d}
\end{figure}

\paragraph{3d case}\mbox{}\\[5pt]
The basic move dualizing the $\mathcal{B}_{10}$-block into the $\mathcal{B}_{01}$-block is represented in Figure \ref{fig:B10_Sdual_3d}. As before, each pair of blocks is glued by gauging a diagonal combination of their $U(N)$ symmetries  and adding an adjoint of this $U(N)$ symmetry that couples in the superpotential with the moment map operators of the two glued blocks.

Its partition function expression is
\begin{align}
&\mathcal{Z}_{(-1,0)}^{(N,M)} (\vec{X};\vec{Y};U;m_A)= \prod_{j=1}^{N-M} s_b\left(i\frac{Q}{2}-j(iQ-2m_A) \right)
\int\left(\prod_{k=1}^2\udl{\vec{Z}^{(k)}_M}\Gd_M^{3d}(\vec{Z}^{(k)};m_A)\right)\nn\\
&\qquad\times
\mathcal{Z}_{\mathcal{S}}^{(N)}\left(\vec{X};\vec{Z}^{(1)};m_A\right)
\mathcal{Z}_{(0,1)}^{(N,M)} \left(\vec{Z}^{(1)};\vec{Z}^{(2)};U;i\frac{Q}{2}-m_A\right)
\mathcal{Z}_{\mathcal{S}^{-1}}^{(M)}(\vec{Z}^{(2)};\vec{Y};m_A) \,,
\end{align}
where we used $\mathcal{Z}_{(-1,0)}^{(N,M)}(\vec{X};\vec{Y};U;m_A)=\mathcal{Z}_{(1,0)}^{(N,M)}(\vec{X};\vec{Y};-U;m_A)$.\\

\begin{figure}[!ht]
	\includegraphics[width=\textwidth]{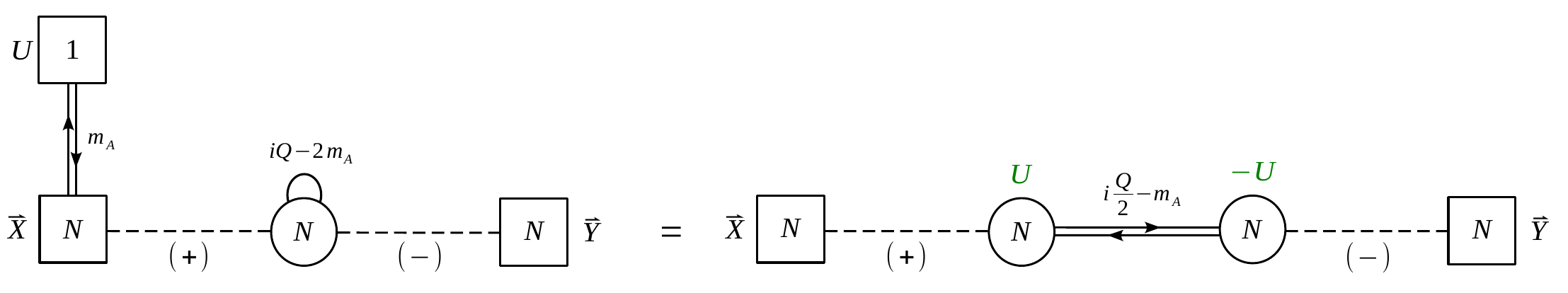}
	\caption{The $\mathcal{B}_{01}= \mathcal{S}\mathcal{B}_{10}\mathcal{S}^{-1}$ duality move. The gluing consists of gauging a diagonal combination of the $U(N)$ symmetries while adding a superpotential coupling the moment map operators of the two $\mathcal{S}$-walls and the meson constructed from the bifundamental.}
	\label{fig:B01_Sdual_3d}
\end{figure}

On the other hand, the basic move dualizing the $\mathcal{B}_{01}$-block into the $\mathcal{B}_{10}$-block is represented in Figure \ref{fig:B01_Sdual_3d}. In this case the gluing is again done by gauging a diagonal combination of the $U(N)$ symmetries of the glued blocks, but since we do not add any adjoint chiral the superpotential consists of a coupling between the moment map operators $\Phi_{L/R}$ of the two $\mathcal{S}$-walls and the meson constructed from the bifundamentals $Q$ and $\widetilde{Q}$
\be
\mathcal{W}_{gluing}=Q\widetilde{Q}(\Phi_L+\Phi_R)\,.
\ee

Its partition function expression is
\begin{align}
\mathcal{Z}_{(0,1)}^{(N)} (\vec{X};\vec{Y};U;m_A)&=\int\udl{\vec{Z}^{(1)}_N}\udl{\vec{Z}^{(2)}_N}\Gd_N^{3d}(\vec{Z}^{(1)})\Gd_N^{3d}(\vec{Z}^{(2)})\mathcal{Z}_{\mathcal{S}}^{(N)}(\vec{X};\vec{Z}^{(1)};m_A)\nonumber\\
&\quad\times\mathcal{Z}_{(1,0)}^{(N,N)}\left(\vec{Z}^{(1)};\vec{Z}^{(2)};U;i\frac{Q}{2}-m_A\right)\mathcal{Z}_{\mathcal{S}^{-1}}^{(N)}(\vec{Z}^{(2)};\vec{Y};m_A)\,.
\end{align}

\subsection{$(1,0)$-$(0,1)$ blocks swap}
\label{sec:HW-move}
In this section we present the duality move corresponding to the swap of a $(1,0)$-block and a $(0,1)$-block introduced in \cite{Comi:2022aqo}.
The 3d version of this identity, as we will see in Section \ref{braneint}, is the QFT counterpart of the Hanany--Witten (HW) move \cite{Hanany:1996ie}.

\begin{figure}[!ht]
	\includegraphics[width=\textwidth]{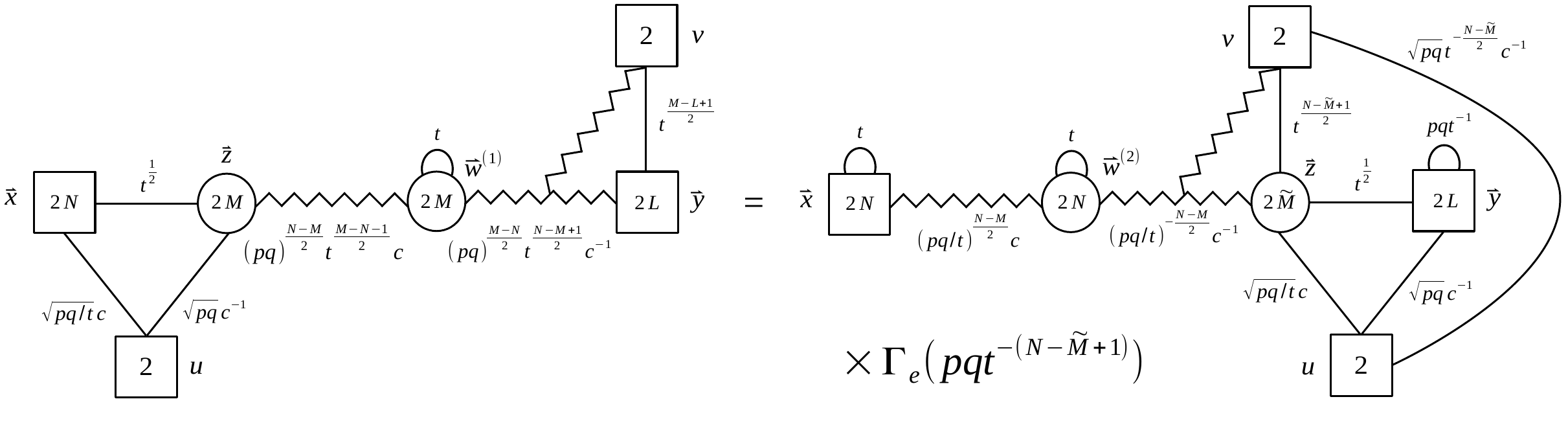}
	\caption{The swap of a $\mathsf{B}_{10}$-block and a $\mathsf{B}_{01}$-block with $N\geq\widetilde{M}\geq 0$. On the r.h.s.~the rank of the second gauge node is $\widetilde{M}=N+L-M+1$.}
\label{fig:HW_move_4d}
\end{figure}

\paragraph{4d case}\mbox{}\\[5pt]
Let us focus on the 4d theory given by a $\mathsf{B}_{10}$ $USp(2N)\times USp(2M)$ block glued to a $\mathsf{B}_{10}$ $USp(2M)\times USp(2L)$ block, with $N\geq\widetilde{M}\geq 0$ where $\widetilde{M}=N+L-M+1$.\footnote{The case $\widetilde{M}>N$, that is $L<M$, can be obtained by just using the relation in Figure \ref{fig:HW_move_4d} from right to left after moving the singlets.}
  As shown in Figure \ref{fig:HW_move_4d}, we can swap the order of the two blocks with the effect of modifying the rank of the intermediate nodes (plus some singlets).
\\

The index identity associated to this swap is
\begingroup\allowdisplaybreaks
\begin{align}
\label{eq:HW_move_4d}
    & \oint d\Vec{z}_M \,\, \Delta_M\left(\Vec{z}\right) 
    \mathcal{I}^{(N,M)}_{(1,0)}\left( \vec{x};\vec{z};u;pqt^{-1};c (pqt^{-1})^\frac{ N-M}{2}\right)
    \mathcal{I}_{(0,1)}^{(M,L)} \Big(\vec{z};\vec{y};v;pq/t\Big)
    \nonumber\\
    & \quad = 
    \Gamma_e\left( pqt^{-(N-\widetilde{M}+1)} \right) \Gamma_e\left( (pq)^{\frac{1}{2}} t^{-\frac{N-\widetilde{M}}{2}} c^{-1} u^{\pm} v^{\pm} \right)  A_N\left(\vec{x};t\right) A_L\left(\vec{y};pq/t\right) \\
    & \qquad \times 
    \oint d\Vec{z}_{\widetilde{M}}\,\, \Delta_{\widetilde{M}}\left( \Vec{z} \right) 
    \mathcal{I}_{(0,1)}^{(N,\widetilde{M})} \Big(\vec{x};\vec{z};u;pq/t\Big) \,
    \mathcal{I}^{(\widetilde{M},L)}_{(1,0)}\left( \vec{z};\vec{y};u;pqt^{-1};c (pqt^{-1})^\frac{ \widetilde{M}-L}{2}\right)\,. \nonumber
\end{align}
\endgroup

\begin{figure}[!ht]
	\includegraphics[width=\textwidth]{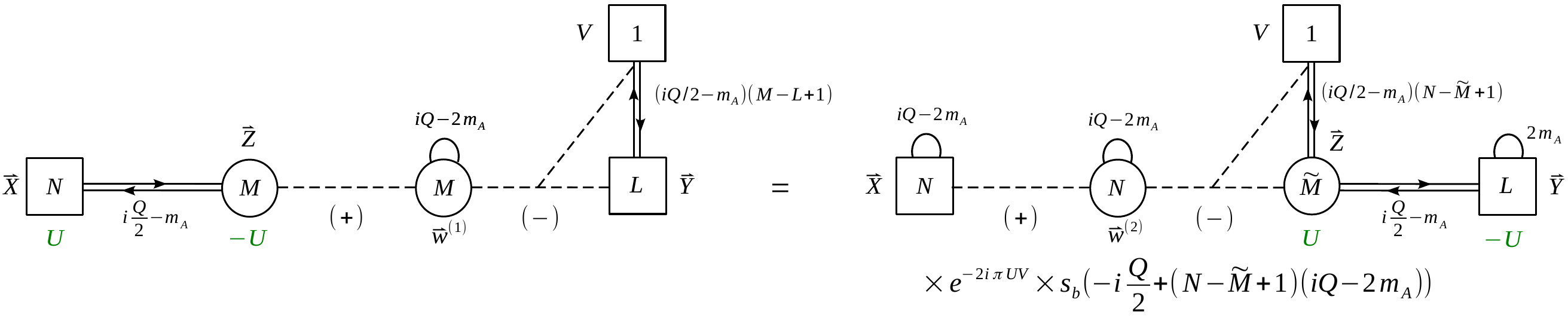}
	\caption{The swap of a $\mathcal{B}_{10}$-block and a $\mathcal{B}_{01}$-block with $N\geq\widetilde{M}\geq 0$. On the r.h.s.~the rank of the second gauge node is $\widetilde{M}=N+L-M+1$.}
\label{fig:HW_move_3d}
\end{figure}

\paragraph{3d case}\mbox{}\\[5pt]
Let us now focus on the 3d theory given by a $\mathcal{B}_{10}$ $U(N)\times U(M)$ block glued to a $\mathcal{B}_{10}$ $U(M)\times U(L)$ block, with $N\geq\widetilde{M}\geq 0$ where $\widetilde{M}=N+L-M+1$.
As shown in Figure \ref{fig:HW_move_3d}, we can swap the order of the two blocks with the effect of modifying the rank of the intermediate nodes (plus some singlets), analogously to what we did in 4d.

The partition function identity associated to this swap is
\begingroup\allowdisplaybreaks
\begin{align}
&\int\udl{\vec{Z}_M}\Gd^{3d}_M(\vec{Z})
\mathcal{Z}_{(1,0)}^{(N,M)} \left(\vec{X};\vec{Z};U;i\frac{Q}{2}-m_A \right)
\mathcal{Z}_{(0,1)}^{(M,L)} \left(\vec{Z};\vec{Y};V;\frac{iQ}{2}-m_A\right)
\nn\\
&\quad 
=
\mathrm{e}^{-2i\pi UV} s_b\left(-i\frac{Q}{2}+(N-\widetilde{M}+1)(iQ-2m_A) \right) 
A_N^{3d}\left(\vec{X};iQ-2m_A\right)
A_L^{3d}\left(\vec{Y};2m_A\right)\nn\\
&\quad\quad\times \int\udl{\vec{Z}_{\widetilde{M}}}\Gd^{3d}_{\widetilde{M}}(\vec{Z}) \,
\mathcal{Z}_{(0,1)}^{(N,\widetilde{M})} \left(\vec{X};\vec{Z};V;\frac{iQ}{2}-m_A\right) \,
\mathcal{Z}_{(1,0)}^{(\widetilde{M},L)}\left(\vec{Z};\vec{Y};U;i\frac{Q}{2}-m_A\right)\,.
\end{align}
\endgroup

\subsection{$(1,0)$-$(1,0)$ blocks swap}
\label{sec:B10-swap}
In this section we introduce a new move corresponding to the swap of two $(1,0)$-blocks.

\begin{figure}[!ht]
	\centering
	\includegraphics[width=.4\textwidth]{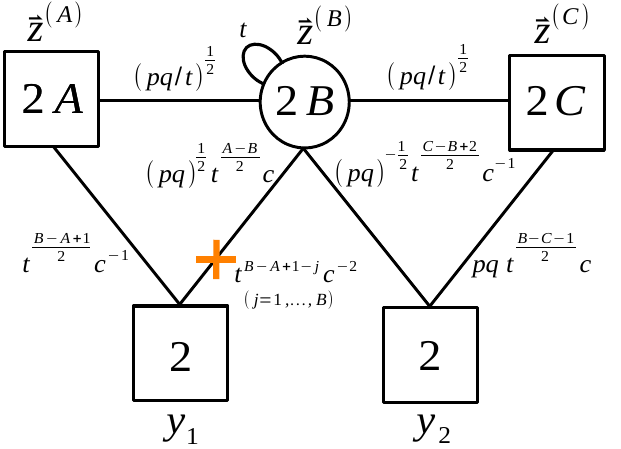}
	\caption{Two 4d $\mathsf{B}_{10}$-blocks glued together. The superpotential consists of a coupling between the $USp(2B)$ antisymmetric and the mesons constructed from the left and right bifundamentals, as well as cubic interactions for the two triangles. Notice also that we added for convenience some singlets represented with an orange cross to the left $\mathsf{B}_{10}$-block, which flip the mesons constructed from the corresponding chiral of the saw.}
		 \label{fig:B10B10_object_4d}
\end{figure}

\paragraph{4d case}\mbox{}\\[5pt]
Let us focus on the theory in  Figure \ref{fig:B10B10_object_4d}, given by two $\mathsf{B}_{10}$-blocks, a $USp(2A)\times USp(2B)$ block and an $USp(2B)\times USp(2C)$ block, glued together.

The index of this theory, which we denote by $\mathcal{I}^{(A,B,C)}_{(1,0)(1,0)}$, is \begingroup\allowdisplaybreaks
\begin{align}
	& \mathcal{I}_{(1,0)(1,0)}^{(A,B,C)}\left(\vec{z}^{\,(A)};\vec{z}^{\,(C)};y_1,y_2;t,c \right) = 
	\label{eq:B10B10_object_3d}
	\\
	& \quad = 
	\prod_{j=1}^{B}\Gamma_e\left(t^{B-A+1-j}c^{-2}\right) \nonumber\\
	& \qquad \times
	\oint \udl{\vec{z}^{\,(B)}}\Gd_N\left(\vec{z}^{\,(B)};t\right) 
	\mathcal{I}_{(1,0)}^{(A,B)}(\vec{z}^{\,(A)};\vec{z}^{\,(B)};y_1;t;c^{-1})\,
	\mathcal{I}_{(1,0)}^{(B,C)}(\vec{z}^{\,(B)};\vec{z}^{\,(C)};y_2;t;\sqrt{t/pq}\, c^{-1})
	\,, \nonumber
\end{align}
\endgroup

\begin{figure}[!ht]
	\centering
	\includegraphics[width=\textwidth]{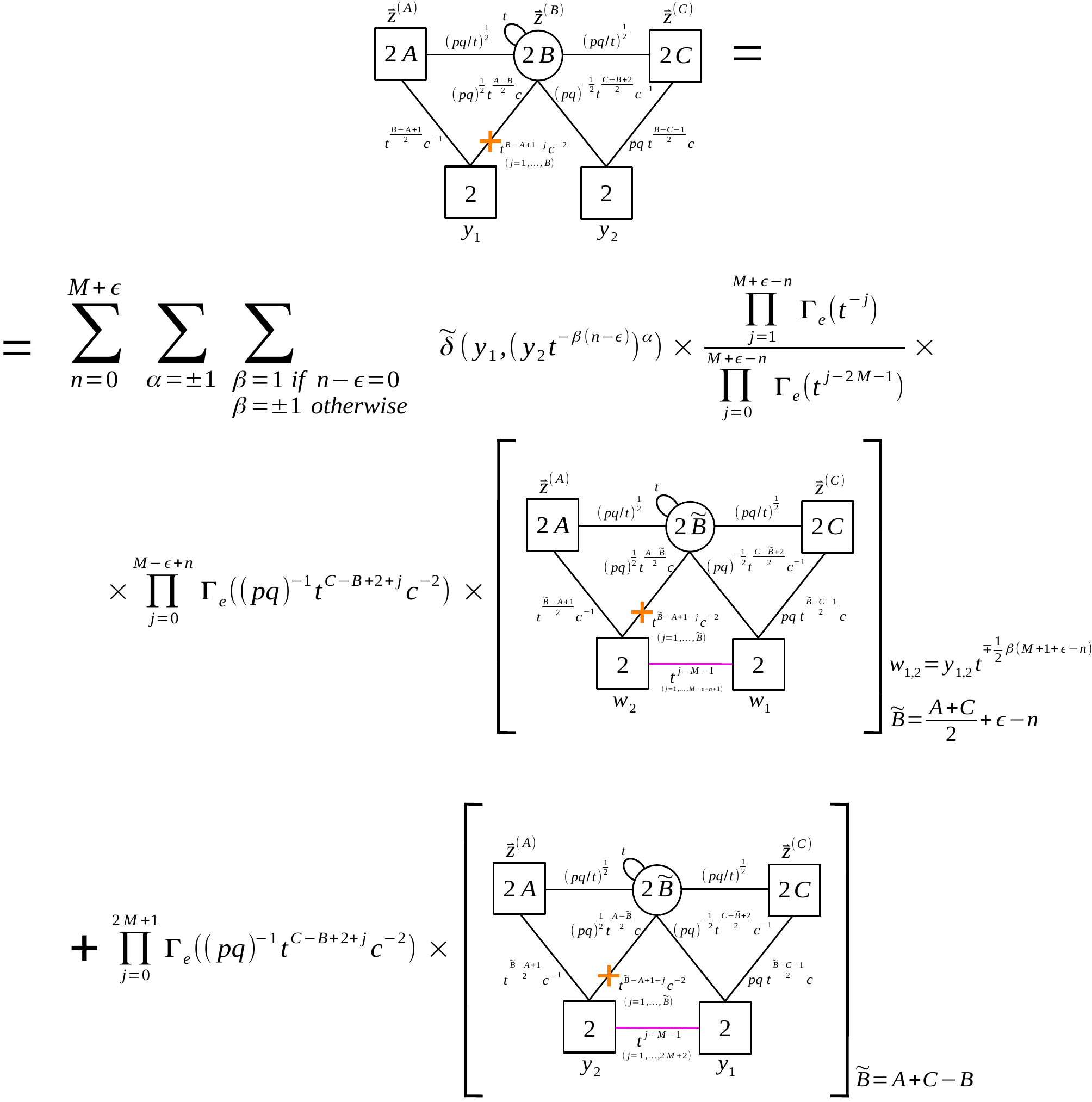}
	\caption{The 4d $\mathsf{B}_{10}$-blocks swap.}
		 \label{fig:B10_swap_4d}
\end{figure}

Using our result for the bad SQCD \eqref{eq:4d_bad_SQCD} we can prove (see Appendix \ref{app:5branes_swap_proofs}) the non-trivial relation represented in Figure \ref{fig:B10_swap_4d}, which holds for $A+C<2B$.
We interpret this relation as the fact that swapping two  $\mathsf{B}_{10}$-blocks  (notice the change of fugacities $y_1\rightarrow w_2$ and $y_2\rightarrow w_1$) is a non-trivial move.
With respect to the previous $(1,0)$-$(0,1)$ blocks swap, we not only modify the intermediate rank, but also have to consider the sum over multiple frames. 
Each frame comes with a delta function setting a constraint on the $y_1, y_2$ fugacities and there is an extra frame with no delta and generic fugacities. 
At the level of the index we have
\begingroup\allowdisplaybreaks
\begin{align}
	&\mathcal{I}_{(1,0)(1,0)}^{(A,B,C)}\left(\vec{z}^{\,(A)};\vec{z}^{\,(C)};y_1,y_2;t,c \right) = \nonumber\\
& = 
\sum_{n=0}^{M+\epsilon} 
    \sum_{\alpha=\pm 1} 
    \sum_{\substack{\beta = 1 \text{ if } n-\epsilon = 0, \\ \beta = \pm1 \text{ otherwise}}} 
    \left\{
    \tilde\delta\left(y_1,\left(y_2 t^{-(n-\epsilon)\beta}\right)^{\alpha}\right) 
    \verybigphantomspace
    \right.\nonumber\\
& \qquad \times \left.
    \frac{\prod_{j=1}^{M+\epsilon-n}\Gamma_e\left(t^{-j}\right)}{\prod_{j=0}^{M+\epsilon-n} \Gamma_e\left(t^{\, j-2M-1}\right)} 
    \prod_{j=0}^{M-\epsilon+n}\Gamma_e\left((pq)^{-1}t^{\,j-B+C+2}c^{-2}\right)
    \right. \nonumber\\
& \qquad \times \left.\left[
\mathcal{I}^{(A,\widetilde{B},C)}_{(1,0)(1,0)}\left(\vec{z}^{\,(A)};\vec{z}^{\,(C)};w_2,w_1;t;c\right)
\prod_{j=1}^{M-\epsilon+n+1}\Gamma_e\left(t^{\,j-M-1}w_1^{\pm}w_2^{\pm}\right)
\right]_{\subalign{&\widetilde{B}=\frac{A+C}{2}+\epsilon-n\\[4pt] &w_{1,2}=y_{1,2}t^{\mp\beta\left( M+1+\epsilon-n \right)}}} 
\right\}\nonumber\\
& \quad +
\prod_{j=0}^{2M+1}\Gamma_e\left((pq)^{-1}t^{\,j-B+C+2}c^{-2}\right)\nonumber\\
& \qquad\times
\left[
\mathcal{I}^{(A,\widetilde{B},C)}_{(1,0)(1,0)}\left(\vec{z}^{\,(A)};\vec{z}^{\,(C)};y_2,y_1;t;c\right)
\prod_{j=1}^{2M+2}\Gamma_e\left(t^{\,j-M-1}y_1^{\pm}y_2^{\pm}\right)
\right]_{\widetilde{B}=A+C-B}
\,, 
\label{eq:B10B10_swap_4d}
\end{align}
\endgroup
where we defined $M=B-\frac{A+C}{2}-1$ and
\begin{equation}
	\epsilon = \begin{cases}
		0 \qquad \text{if $A+C$ is even} \,,\\
		-\frac{1}{2} \quad \text{if $A+C$ is odd} \,.
	\end{cases}
\end{equation}
We also defined $\tilde{\delta}(x,y)$ as in \eqref{eq:def_delta_tilde}.
Note that the identity \eqref{eq:B10B10_swap_4d} also holds for $N_f<N_c$, although in this case the contribution without the $\delta$-function is absent.

\begin{figure}[!ht]
	\centering
	\includegraphics[width=.5\textwidth]{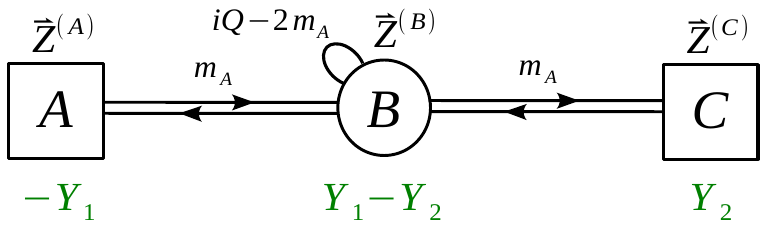}
	\caption{Two 3d $\mathcal{B}_{10}$-blocks glued together. The superpotential consists of the standard one of 3d $\mathcal{N}=4$ theories, which couples the $U(B)$ adjoint chiral to the mesons constructed from the left and right bifundamentals.}
		 \label{fig:B10B10_object_3d}
\end{figure}

\paragraph{3d case}\mbox{}\\[5pt]
Let's now consder the  3d theory in Figure \ref{fig:B10B10_object_3d}, given by two $\mathcal{B}_{10}$-blocks, a $U(A)\times U(B)$ block and an $U(2B)\times U(2C)$ block, glued together.

The partition function of this theory, which we denote by $\mathcal{Z}_{(1,0)(1,0)}^{(A,B,C)}$, is given by
\begingroup\allowdisplaybreaks
\begin{align}
	& \mathcal{Z}_{(1,0)(1,0)}^{(A,B,C)}\left( \vec{Z}^{\,(A)};\vec{Z}^{\,(C)};Y_1,Y_2;m_A \right) 
	= \nonumber\\
	& \quad = 
	\int \udl{\vec{Z}^{\,(B)}} \Gd_B^{3d}\left(\vec{Z}^{\,(B)};m_A\right) 
	\mathcal{Z}_{(1,0)}^{(A,B)}(\vec{Z}^{\,(A)};\vec{Z}^{\,(B)};-Y_1;m_A)\,
	\mathcal{Z}_{(1,0)}^{(B,C)}(\vec{Z}^{\,(B)};\vec{Z}^{\,(C)};-Y_2;m_A)
	\,.
\end{align}
\endgroup

\begin{figure}[!ht]
	\centering
	\includegraphics[width=\textwidth]{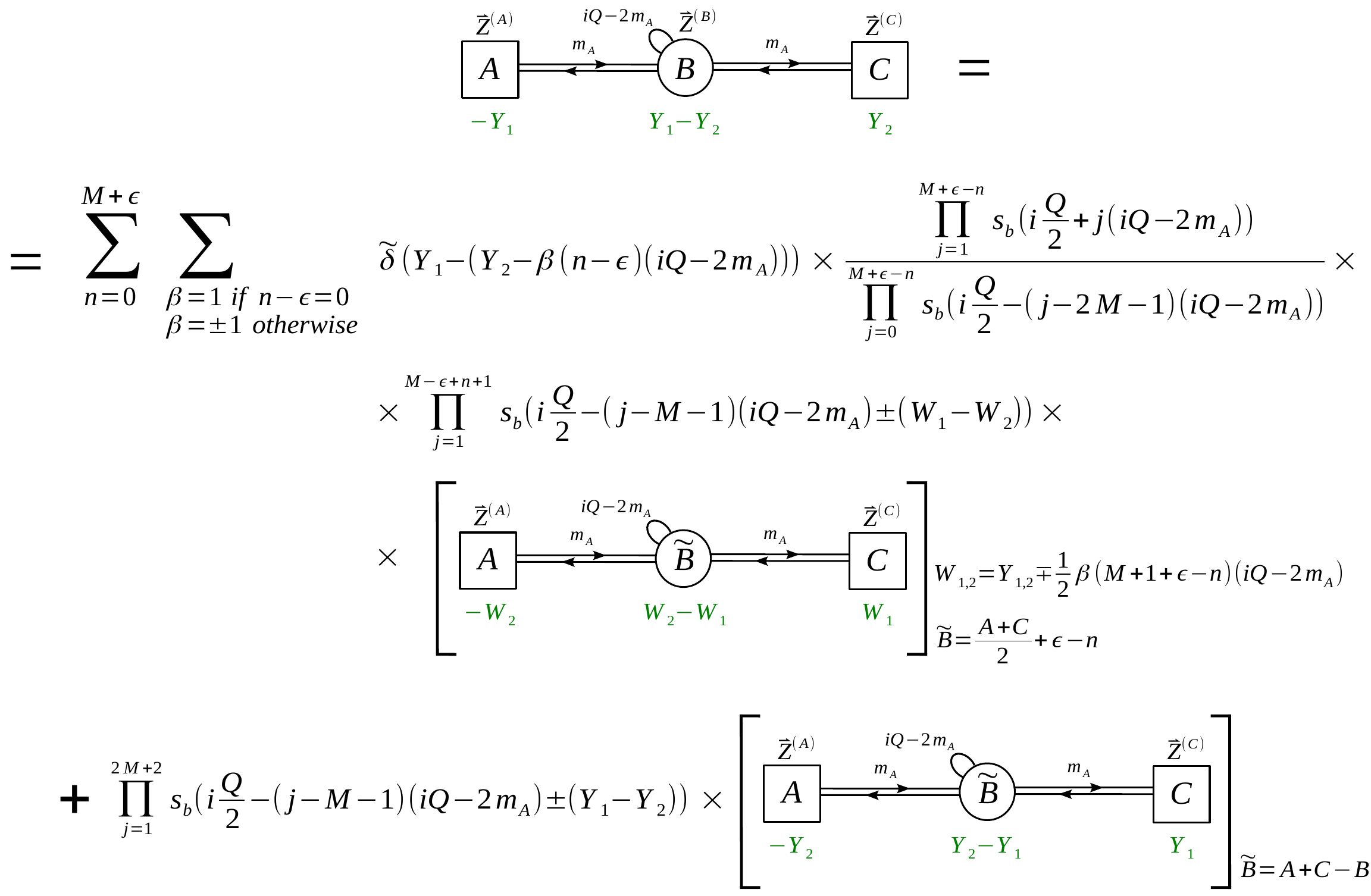}
	\caption{The 3d $\mathcal{B}_{10}$-blocks swap.}
	\label{fig:B10_swap_3d}
\end{figure}

If we take the 3d reduction of the identity \eqref{eq:B10B10_swap_4d} we get the identity, represented in Figure \ref{fig:B10_swap_3d}, which holds for $A+C<2B$.
Again, we interpret this relation as the fact that swapping two $\mathcal{B}_{10}$-blocks is a non-trivial operation. We will provide a brane interpretation of this move in Section \ref{braneint}.

At the level of the partition function we have
\begingroup\allowdisplaybreaks
\begin{align}
& \mathcal{Z}^{(A,B,C)}_{(1,0)(1,0)}\left(\vec{Z}^{\,(A)};\vec{Z}^{\,(C)};Y_1,Y_2;m_A\right)
= \nonumber\\
& = 
\sum_{n=0}^{M+\epsilon} 
    \sum_{\substack{\beta = 1 \text{ if } n-\epsilon = 0, \\ \beta = \pm1 \text{ otherwise}}} 
    \left\{
    \delta\bigg(Y_1-\left(Y_2-\beta(n-\epsilon)(iQ-2m_A)\right)\bigg) 
    \verybigphantomspace
    \right.\nonumber\\
& \qquad \times
    \frac{\prod_{j=1}^{M+\epsilon-n}s_b\left(i\frac{Q}{2}+j(iQ-2m_A)\right)}{\prod_{j=0}^{M+\epsilon-n} s_b\left(i\frac{Q}{2}-(j-2M-1)(iQ-2m_A)\right)}
    \times \nonumber\\
& \qquad \times \left[
\prod_{j=1}^{M-\epsilon+n+1}s_b\left(i\frac{Q}{2}-(j-M-1)(iQ-2m_A)\pm(W_1-W_2)\right)
\bigphantomspace\right.\nonumber\\
& \qquad\,\,\,\, \left.\left.\bigphantomspace\times
\mathcal{Z}^{(A,\widetilde{B},C)}_{(1,0)(1,0)}\left(\vec{Z}^{\,(A)};\vec{Z}^{\,(C)};W_2,W_1;m_A\right) 
\right]_{\subalign{&\widetilde{B}=\frac{A+C}{2}+\epsilon-n\\[4pt] &W_{1,2}=Y_{1,2}\mp\frac{1}{2}\beta\left( M+1+\epsilon-n \right)(iQ-2m_A)}} 
\right\}\nonumber\\[5pt]
& \quad +
\left[
\prod_{j=1}^{2M+2}s_b\left(i\frac{Q}{2}-(j-M-1)(iQ-2m_A)\pm(Y_1-Y_2)\right)
\right.\nonumber\\
& \qquad\left.\bigphantomspace\times
\mathcal{Z}^{(A,\widetilde{B},C)}_{(1,0)(1,0)}\left(\vec{Z}^{\,(A)};\vec{Z}^{\,(C)};Y_2,Y_1;m_A\right) 
\right]_{\widetilde{B}=A+C-B}
\,, 
\label{eq:B10B10_swap_3d}
\end{align}
\endgroup
where we defined $M$ and $\epsilon$ as before.
Note that the identity \eqref{eq:B10B10_swap_3d} also holds for $N_f<N_c$, although in this case the contribution without the $\delta$-function is absent.

\subsection{$(0,1)$-$(0,1)$ blocks swap}
\label{sec:B01-swap}

\begin{figure}[!ht]
	\centering
	\includegraphics[width=.6\textwidth]{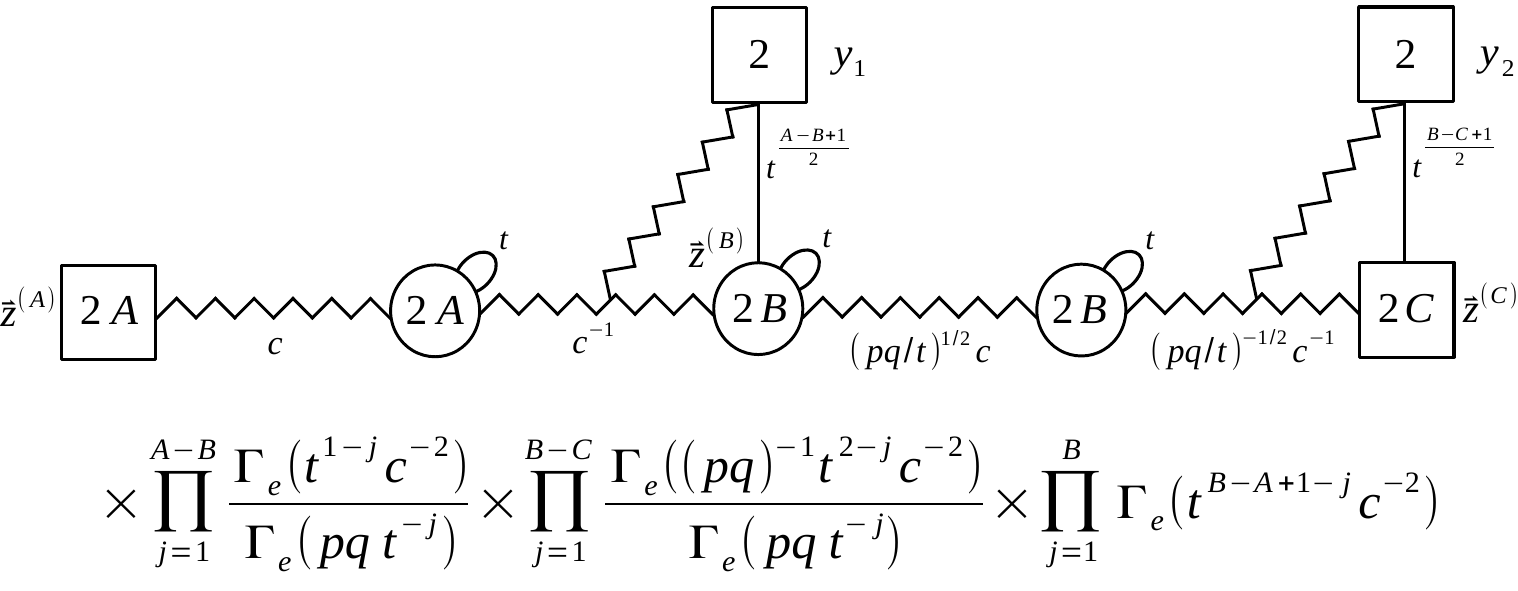}
	\caption{Two 4d $\mathsf{B}_{01}$-blocks glued together for $A \geq B \geq C$.}
		 \label{fig:B01B01_object_4d_V1}
\end{figure}

\begin{figure}[!ht]
	\centering
	\includegraphics[width=.6\textwidth]{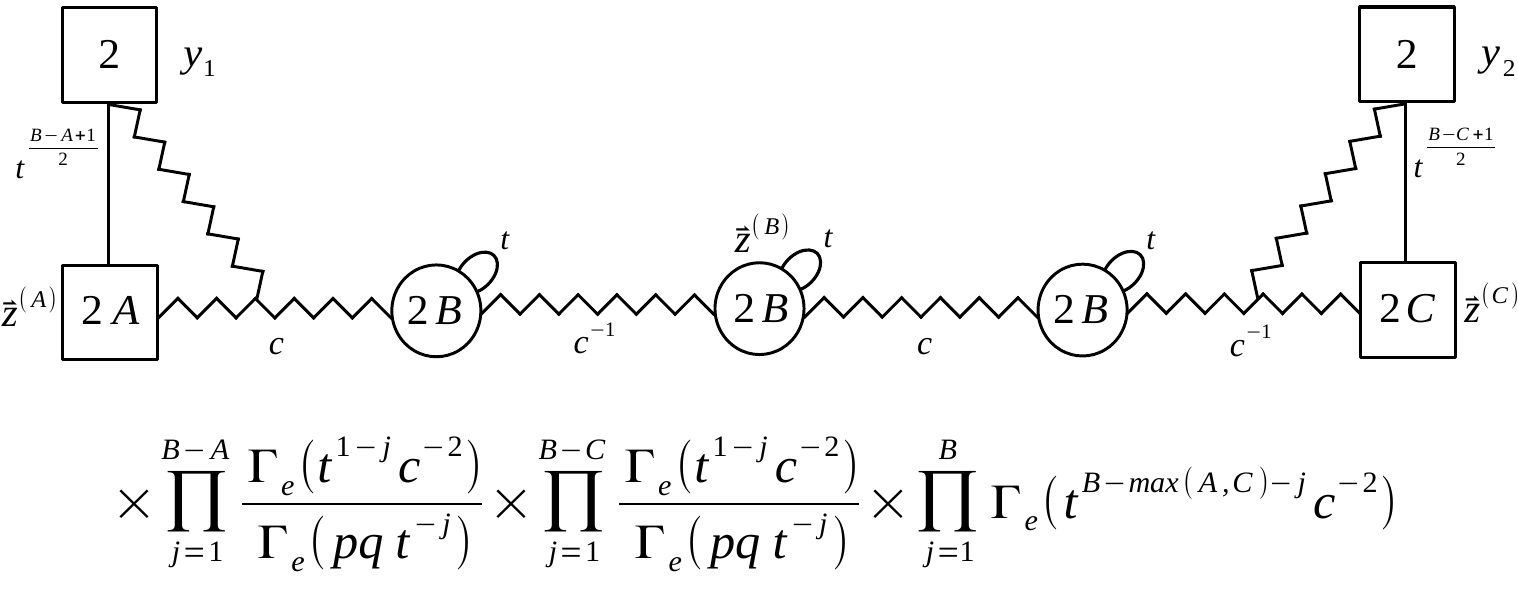}
	\caption{Two 4d $\mathsf{B}_{01}$-blocks glued together for $B \geq A\,, B \geq C$.}
		 \label{fig:B01B01_object_4d_V3}
\end{figure}

Finally, we introduce the move implementing the swap of two $(0,1)$-blocks.
\paragraph{4d case}\mbox{}\\[5pt]
We consider the theory obtained by gluing two  asymmetric $\mathsf{B}_{01}$-blocks plus some extra singlets.
We have various possibities depending on the ordering of the ranks as shown in Figures \ref{fig:B01B01_object_4d_V1} and \ref{fig:B01B01_object_4d_V3}.
The index of this theory, which we denote as $\mathcal{I}^{(A,B,C)}_{(0,1)(0,1)}$ 
for $A+C<2B$, is
\begingroup\allowdisplaybreaks
\begin{align}
& \mathcal{I}^{(A,B,C)}_{(0,1)(0,1)}\left(\vec{z}^{\,(A)};\vec{z}^{\,(C)};y_1,y_2;t;c\right) = \label{eq:B01B01_object_4d}\\
& = F(A,B,C;t;c) \oint 
\udl{\vec{z}^{\,(B)}} \Gd_B\left(\vec{z}^{\,(B)};t\right)
\mathcal{I}_{(0,1)}^{(A,B)} (\vec{z}^{\,(A)};\vec{z}^{\,(B)};y_1;pq/t)\,
\mathcal{I}_{(0,1)}^{(B,C)} (\vec{z}^{\,(B)};\vec{z}^{\,(C)};y_2;pq/t) \,,\nonumber
\end{align}
\endgroup
where we collected inside $F(A,B,C;t;c)$ some singlets depending on the ranks $A,B,C$
\begin{align}
	& F(A,B,C;t;c) = \label{Ffactor} \\ 
	& \, =
	\begin{cases}
	\prod_{j=1}^{A-B}\frac{\Gamma_e\left(t^{1-j}c^{-2}\right)}{\Gamma_e\left(pqt^{-j}\right)}
	\prod_{j=1}^{B-C}\frac{\Gamma_e\left((pq)^{-1}t^{2-j}c^{-2}\right)}{\Gamma_e\left(pqt^{-j}\right)}
	\prod_{j=1}^{B}\Gamma_e\left(t^{B-A+1-j}c^{-2}\right)
	\quad\,\,\text{if } A \geq B \geq C \,,	\\[9pt]
	\prod_{j=1}^{B-A}\frac{\Gamma_e\left(t^{1-j}c^{-2}\right)}{\Gamma_e\left(pqt^{-j}\right)}
	\prod_{j=1}^{B-C}\frac{\Gamma_e\left(t^{1-j}c^{-2}\right)}{\Gamma_e\left(pqt^{-j}\right)}
	\prod_{j=1}^{B}\Gamma_e\left(t^{B-max(A,C)-j}c^{-2}\right)
	\!\!\qquad\text{if } B \geq A\,, B \geq C \,.
	\end{cases} \nonumber
\end{align}

\begin{figure}[!p]
	\centering
	\includegraphics[width=\textwidth,center]{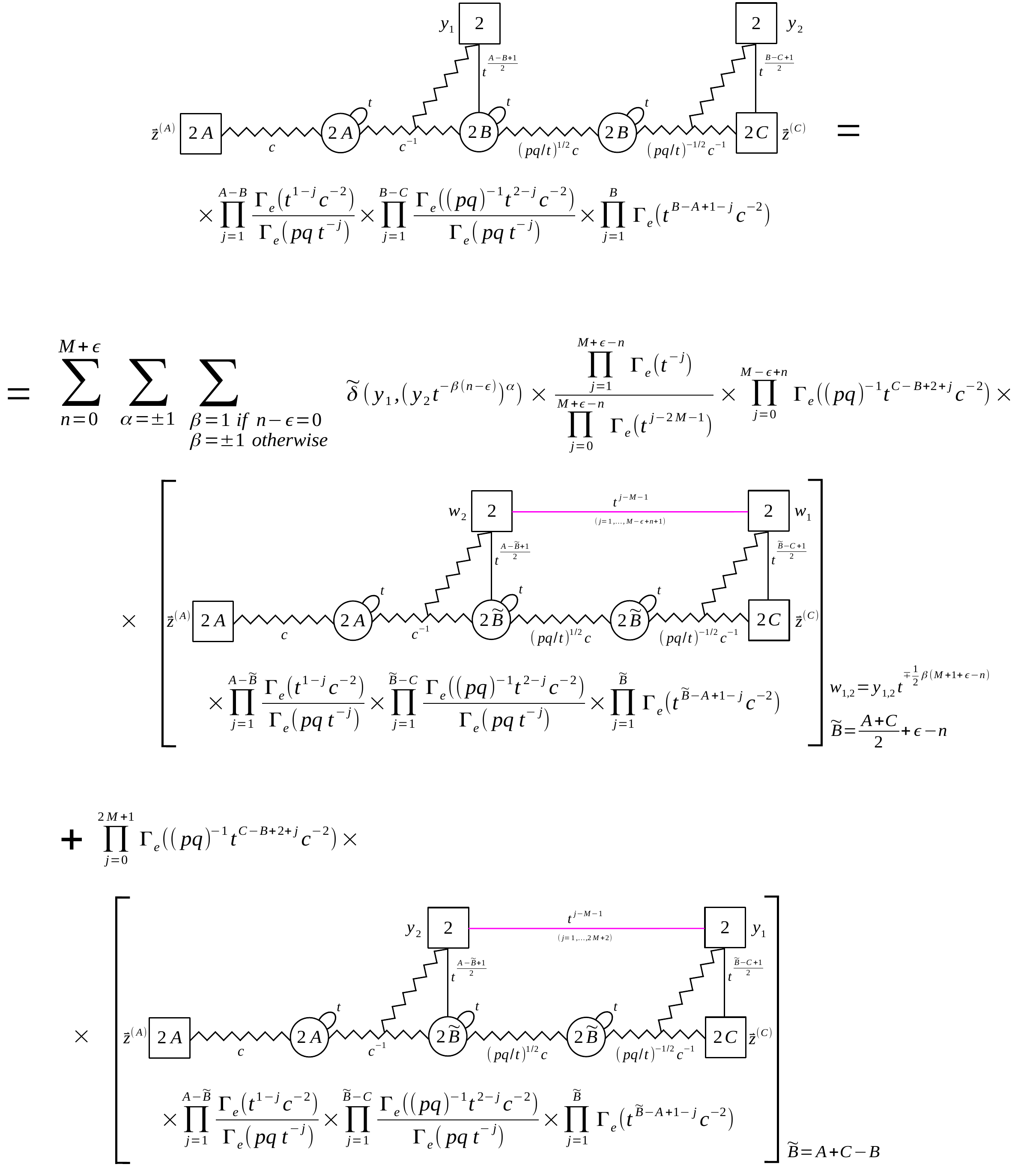}
	\caption{The 4d $\mathsf{B}_{01}$-blocks swap for $A\geq B \geq C$ and $A+C<2B$.}
		 \label{fig:B01_swap_4d_V1}
\end{figure}

By taking the $\mathsf{S}$-dual of the $(1,0)$-$(1,0)$ move of the previous section  (that is by $\mathsf{S}$-dualizing each  $(1,0)$ block frame by frame), multiplying each frame on the left and right hand side  by an $\mathsf{S}$-wall  and using that $\mathsf{S}^2=1$ we can prove  the non-trivial $(0,1)$-$(0,1)$ swap  represented in Figure \ref{fig:B01_swap_4d_V1}.\footnote{
The situation where $C \geq B \geq A$ can be obtained from a vertical reflection of the one with $A \geq B \geq C$ and thus does not correspond to an independent case. A similar relation holds for $ B \ge A$, $ B \ge C$ and $A+C<2B$. However, this case is a bit more subtle since in the intermediate frames the orientation of the asymmetric walls might change in a way that is similar to the difference between Figure \ref{fig:B01B01_object_4d_V1} and \ref{fig:B01B01_object_4d_V3}.}
We interpret this relation as the fact the swapping two  $\mathsf{B}_{01}$-blocks  (notice the change of fugacities $y_1\rightarrow w_2$ and $y_2\rightarrow w_1$) is a non-trivial move.
Also in this case we have to consider the sum over multiple frames. Moreover, we produce a set of singlets transforming under $USp(2)_{w_1}\times USp(2)_{w_2}$. 

At the level of the index we have for any $A,B,C$ such that $A+C<2B$
\begingroup\allowdisplaybreaks
\begin{align}
& \mathcal{I}^{(A,B,C)}_{(0,1)(0,1)}\left(\vec{z}^{\,(A)};\vec{z}^{\,(C)};y_1,y_2;t;c\right)
= \nonumber\\
& = 
\sum_{n=0}^{M+\epsilon} 
    \sum_{\alpha=\pm 1} 
    \sum_{\substack{\beta = 1 \text{ if } n-\epsilon = 0, \\ \beta = \pm1 \text{ otherwise}}} 
    \left\{
    \tilde\delta\left(y_1,\left(y_2 t^{-(n-\epsilon)\beta}\right)^{\alpha}\right) 
    \verybigphantomspace
    \right.\nonumber\\
& \qquad \times \left.
    \frac{\prod_{j=1}^{M+\epsilon-n}\Gamma_e\left(t^{-j}\right)}{\prod_{j=0}^{M+\epsilon-n} \Gamma_e\left(t^{\, j-2M-1}\right)} 
    \prod_{j=0}^{M-\epsilon+n}\Gamma_e\left((pq)^{-1}t^{\,j-B+C+2}c^{-2}\right)
    \right. \nonumber\\
& \qquad \times \left.\left[
\mathcal{I}^{(A,\widetilde{B},C)}_{(0,1)(0,1)}\left(\vec{z}^{\,(A)};\vec{z}^{\,(C)};w_2,w_1;t;c\right)
\prod_{j=1}^{M-\epsilon+n+1}\Gamma_e\left(t^{\,j-M-1}w_1^{\pm}w_2^{\pm}\right)
\right]_{\subalign{&\widetilde{B}=\frac{A+C}{2}+\epsilon-n\\[4pt] &w_{1,2}=y_{1,2}t^{\mp\beta\left( M+1+\epsilon-n \right)}}} 
\right\}\nonumber\\
& \quad +
\prod_{j=0}^{2M+1}\Gamma_e\left((pq)^{-1}t^{\,j-B+C+2}c^{-2}\right)\nonumber\\
& \qquad\times
\left[
\mathcal{I}^{(A,\widetilde{B},C)}_{(0,1)(0,1)}\left(\vec{z}^{\,(A)};\vec{z}^{\,(C)};y_2,y_1;t;c\right)
\prod_{j=1}^{2M+2}\Gamma_e\left(t^{\,j-M-1}y_1^{\pm}y_2^{\pm}\right)
\right]_{\widetilde{B}=A+C-B}
\,, 
\label{eq:D5_swap_4d}
\end{align}
\endgroup
where we defined $M$ and $\epsilon$ as before.
Note that the identity \eqref{eq:D5_swap_4d} also holds for $N_f<N_c$, although in this case the contribution without the $\delta$-function is absent.

\begin{figure}[!ht]
	\centering
	\includegraphics[width=.8\textwidth]{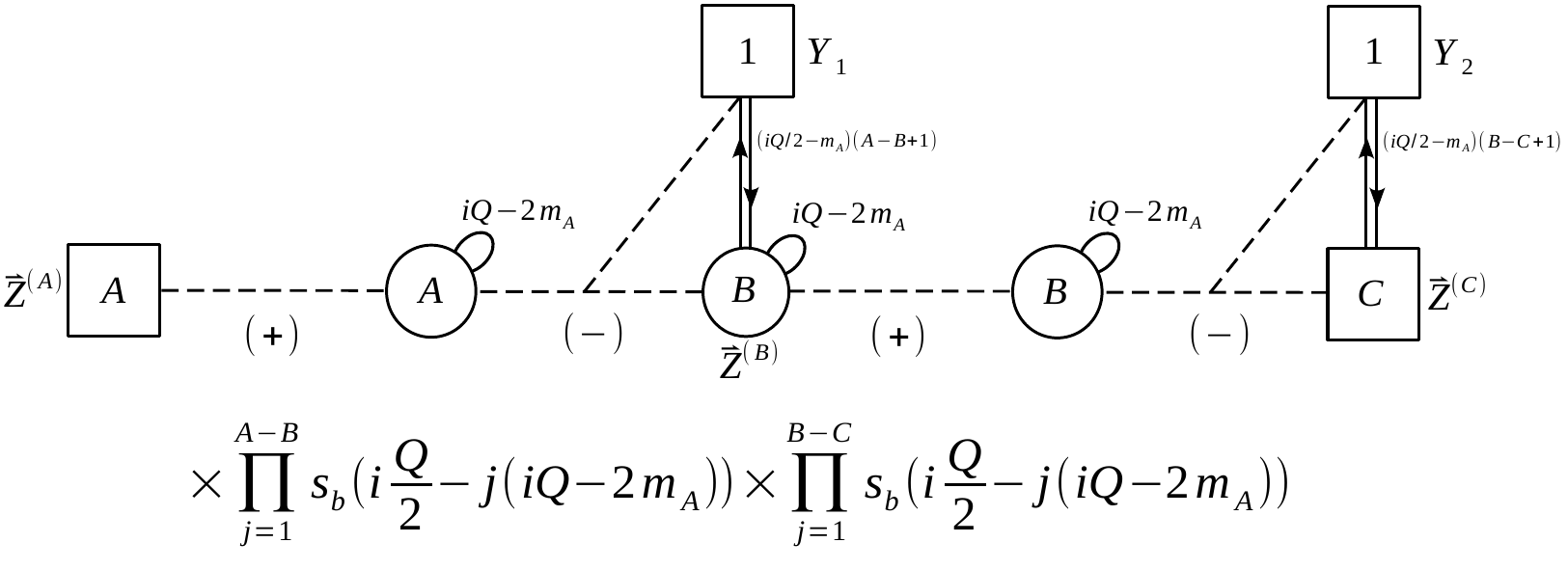}
	\caption{Two 3d $\mathcal{B}_{01}$-blocks glued together for $A \ge B \ge C$.}
		 \label{fig:B01B01_object_3d}
\end{figure}
\begin{figure}[!p]
	\centering
	\includegraphics[width=1\textwidth,center]{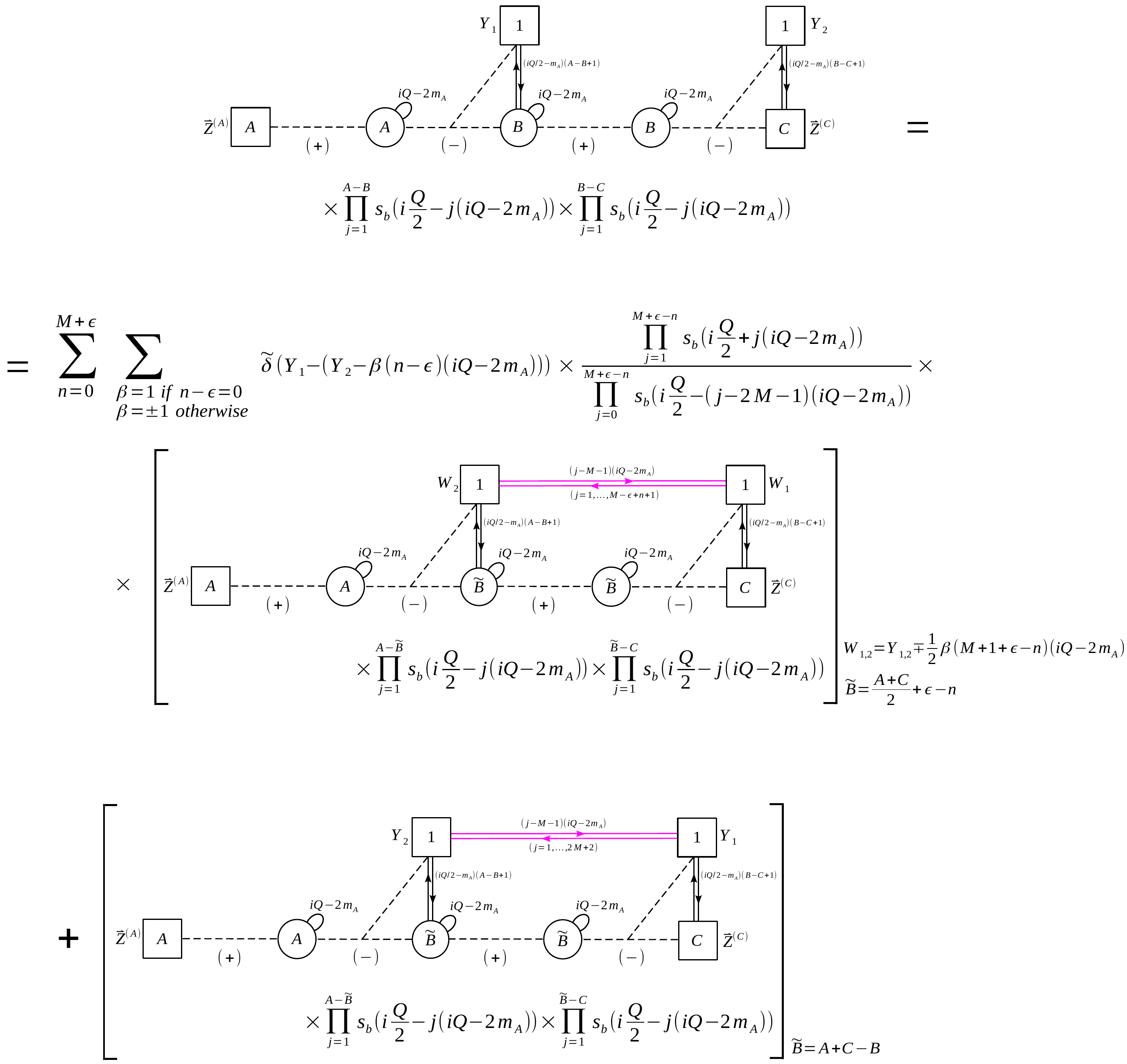}
	\caption{The 3d $\mathcal{B}_{01}$-blocks swap for $A \ge B \ge C$ and $A+C<2B$.}
		 \label{fig:B01_swap_3d_V1}
\end{figure}

\paragraph{3d case}\mbox{}\\[5pt]
Finally we consider the 3d counterpart of the above discussion. We introduce the theory  in Figure \ref{fig:B01B01_object_3d} obtained by gluing two $\mathcal{B}_{01}$-blocks plus some extra singlets (for $A \ge B \ge C$).
The partition function of this theory, which we denote by $\mathcal{Z}_{(0,1)(0,1)}^{(A,B,C)}$, is given by
\begingroup\allowdisplaybreaks
\begin{align}
& \mathcal{Z}^{(A,B,C)}_{(0,1)(0,1)}\left(\vec{Z}^{\,(A)};\vec{Z}^{\,(C)};Y_1,Y_2;m_A\right) = \nonumber\\
& \quad = 
\prod_{j=1}^{|A-B|}s_b\left(i\frac{Q}{2}-j(iQ-2m_A)\right)
\prod_{j=1}^{|B-C|}s_b\left(i\frac{Q}{2}-j(iQ-2m_A)\right) \\
& \quad \quad
\times \int \udl{\vec{Z}^{\,(B)}} \Gd_B^{3d}\left(\vec{Z}^{\,(B)};m_A\right) 
\mathcal{Z}_{(0,1)}^{(A,B)} \left(\vec{Z}^{\,(A)};\vec{Z}^{\,(B)};Y_1;\frac{iQ}{2}-m_A\right)\nonumber\\
& \quad \qquad \times
\mathcal{Z}_{(0,1)}^{(B,C)} \left(\vec{Z}^{\,(B)};\vec{Z}^{\,(C)};Y_2;\frac{iQ}{2}-m_A\right)\,.\nonumber
\end{align}
\endgroup
By dimensional reduction of the identity \eqref{eq:D5_swap_4d} we obtain the partition function identity 
valid for any $A,B,C$ such that $A+C<2B$\footnote{Notice that the singlets  in \eqref{Ffactor} do not contribute to the  3d partition function as they are charged under  $U(1)_c$ for which we activate a real mass deformation.}
\begingroup\allowdisplaybreaks
\begin{align}
& \mathcal{Z}^{(A,B,C)}_{(0,1)(0,1)}\left(\vec{Z}^{\,(A)};\vec{Z}^{\,(C)};Y_1,Y_2;m_A\right)
= \nonumber\\
& = 
\sum_{n=0}^{M+\epsilon} 
    \sum_{\substack{\beta = 1 \text{ if } n-\epsilon = 0, \\ \beta = \pm1 \text{ otherwise}}} 
    \left\{
    \delta\bigg(Y_1-\left(Y_2-\beta(n-\epsilon)(iQ-2m_A)\right)\bigg) 
    \verybigphantomspace
    \right.\nonumber\\
& \qquad \left.\times
    \frac{\prod_{j=1}^{M+\epsilon-n}s_b\left(i\frac{Q}{2}+j(iQ-2m_A)\right)}{\prod_{j=0}^{M+\epsilon-n} s_b\left(i\frac{Q}{2}-(j-2M-1)(iQ-2m_A)\right)}
    \right. \nonumber\\
& \qquad \times \left[
\prod_{j=1}^{M-\epsilon+n+1}s_b\left(i\frac{Q}{2}-(j-M-1)(iQ-2m_A)\pm(W_1-W_2)\right)
\bigphantomspace\right.\nonumber\\
& \qquad\left.\left.\bigphantomspace\times
\mathcal{Z}^{(A,\widetilde{B},C)}_{(0,1)(0,1)}\left(\vec{Z}^{\,(A)};\vec{Z}^{\,(C)};W_2,W_1;m_A\right) 
\right]_{\subalign{&\widetilde{B}=\frac{A+C}{2}+\epsilon-n\\[4pt] &W_{1,2}=Y_{1,2}\mp\frac{1}{2}\beta\left( M+1+\epsilon-n \right)(iQ-2m_A)}} 
\right\}\nonumber\\
& \quad +
\left[
\prod_{j=1}^{2M+2}s_b\left(i\frac{Q}{2}-(j-M-1)(iQ-2m_A)\pm(Y_1-Y_2)\right)
\right.\nonumber\\
& \qquad\left.\bigphantomspace\times
\mathcal{Z}^{(A,\widetilde{B},C)}_{(0,1)(0,1)}\left(\vec{Z}^{\,(A)};\vec{Z}^{\,(C)};Y_2,Y_1;m_A\right) 
\right]_{\widetilde{B}=A+C-B}
\,, 
\label{eq:D5_swap_3d}
\end{align}
\endgroup
where we defined $M$ and $\epsilon$ as before.
We   interpret this relation as the fact that swapping two $\mathcal{B}_{01}$-blocks is a non-trivial operation as depicted for  $A \ge B \ge C$ in Figure \ref{fig:B01_swap_3d_V1}. 
A similar relation holds for  $B\ge A,\,B\ge C$. 
Finally, note that the identity \eqref{eq:D5_swap_3d} also holds for $N_f<N_c$, although in this case the contribution without the $\delta$-function is absent.

\clearpage
\section{The mirror and the electric dualization algorithms }
\label{sec:3}

\subsection{The mirror  dualization algorithm}
\label{sec:magnetic dual}

In this section we review the mirror dualization algorithm introduced in \cite{Hwang:2021ulb,Comi:2022aqo} for 3d and 4d good linear quiver theories and extend it to the case of  3d and 4d bad linear quivers.

The algorithm consists of the following steps:
\begin{enumerate}
\item We chop the quiver into $(1,0)$ and $(0,1)$-blocks by ungauging the gauge nodes.  
\item We dualize each block using the basic  duality moves.
\item  We glue back the dualized blocks by restoring the original gauge symmetries.
\item   We follow the RG flow triggered by VEVs that can be generated in the previous step.
This can be done systematically by iterating the $(1,0)$-$(0,1)$ blocks swap  (and also $(0,1)$-$(0,1)$ blocks swap  for bad quivers).
\end{enumerate}
We now focus more on the fourth step involving the study of the RG flows.

For good quivers this can be implemented with various approaches as shown in \cite{Comi:2022aqo}. A very convenient approach consists in applying the $(1,0)$-$(0,1)$ swap reviewed in Section \ref{sec:QFT_Ingredients}. Indeed the presence of VEVs can be detected from  the fact that after the third step some asymmetric $(0,1)$-blocks are left.
The $(1,0)$-$(0,1)$ swap has the effect of reducing the asymmetry of the $(0,1)$-blocks and thus can be iterated to implement the propagation of the VEV. For good quiver theories it is always possible to perform enough $(1,0)$-$(0,1)$ swaps to reach configurations involving only symmetric  $(0,1)$-blocks, meaning that they contain only symmetric Identity-walls which can be trivially implemented by identifying some of the gauge nodes and disappear leading  to the final mirror dual quiver.
 
This procedure in 3d has a brane counterpart. Asymmetric  $(0,1)$-blocks correspond to D5-branes with a non-zero net number of D3-branes. To read out the mirror quiver from the brane set-up,  we need to perform a series of Hanany--Witten (HW) moves, corresponding to  $(1,0)$-$(0,1)$ swaps,  to arrive at a configuration of D5-branes with zero  net number of D3-branes which correspond to having only symmetric  $(0,1)$-blocks. The important fact is that for good theories, after the action of $\mathcal{S}$-duality, an appropriate  sequence of HW moves, is sufficient to reach the final $\mathcal{S}$-dual brane configuration from which we can read out the mirror dual quiver theory.
 
The situation for bad theories is more intricate. In particular, after the block dualization, the simple iteration of the 
$(1,0)$-$(0,1)$  moves is not enough to get rid of all the asymmetric $(0,1)$-blocks and extinguish all the VEVs. In \cite{Giacomelli:2023zkk} the effect of these VEVs in the bad SQCD case has been carefully analyzed. Schematically the structure of the 4d index/3d partition function of the bad SQCD is given by a sum of frames each involving a Dirac delta distribution, enforcing a particular constraint on the  parameters, which multiplies the index/partition function of an interacting good theory and various singlets including a set of free fields.   In addition, there is an extra frame with no delta distribution.
In the previous section we introduced the new non-trivial   $(1,0)$-$(1,0)$ and  $(0,1)$-$(0,1)$ swap moves which, as discussed in Appendix \ref{app:5branes_swap_proofs}, can be derived from the SQCD result.
These new moves basically encapsulate the subtle analysis of the collision of singularities performed in  \cite{Giacomelli:2023zkk} 
and, as we are going to see, by means of these new moves we can implement all the VEVs in bad quiver theories.
So the upgraded version of the algorithm for bad theories includes the  $(0,1)$-$(0,1)$ moves in the fourth step.
In Section \ref{braneint} we will discuss the brane counterpart of this QFT analysis.\\

\begin{figure}[!ht]
	\centering
	\includegraphics[width=.35\textwidth]{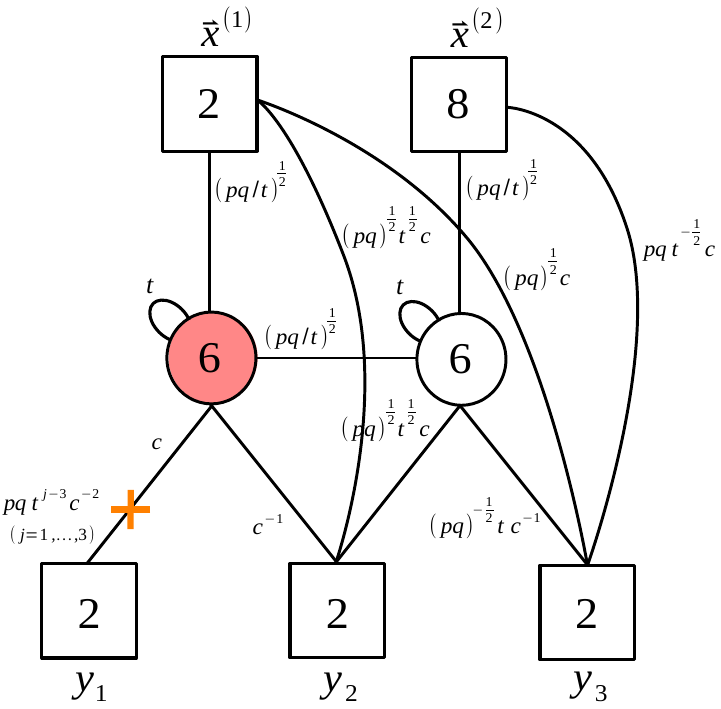}
	\caption{The 4d linear quiver corresponding to the  $E_{[2,4,3]}^{[2,2,2,2,1]}[USp(18)]$ theory.
		The red node is bad. The $SU(2)_{x_i}\times SU(2)_{y_j}$ singlets and the flipping fields (the crosses) are introduced following the  $E_{\rho}^{\sigma}[USp(2N)]$ conventions explained in Appendix \ref{app:Erhosigma_Trhosigma}.}
	\label{fig:original_theory_4d_example1}
\end{figure}

Let us give a simple example of application of the mirror dualization algorithm to a bad linear quiver. The example we consider is the case with gauge ranks $N_c = (3,3)$ and flavor ranks $N_f = (1,4)$ 
corresponding  to the partition $\sigma=[2,2,2,2,1]$ and sequence $\rho=[2,4,3]$.
We show the analysis in 4d, but the 3d steps are completely analogous. 
The 4d bad theory is shown in Figure \ref{fig:original_theory_4d_example1}.

\begin{figure}[!ht]
	\centering
	\includegraphics[width=\textwidth]{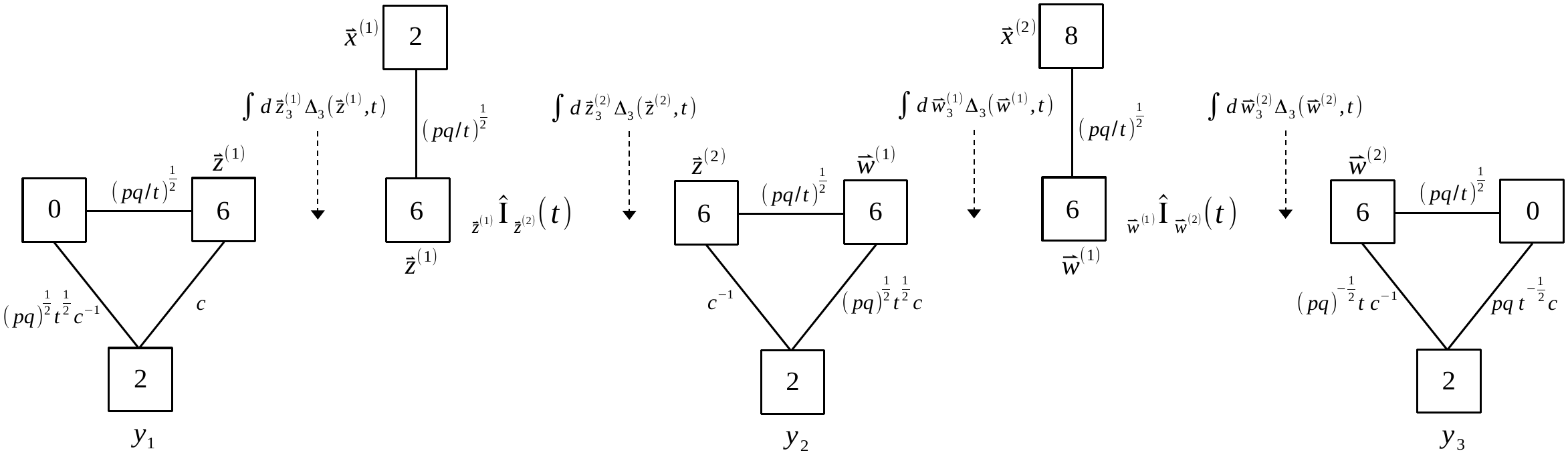}
	\caption{The decomposition into QFT blocks of the theory in Figure \ref{fig:original_theory_4d_example1}. We indicate the gaugings that should be restored to obtain the original theory by indicating their effect on the index.}
	\label{fig:chop_into_blocks_example1}
\end{figure}

Following the algorithm, we first chop the quiver into five QFT blocks by ungauging the gauge nodes as depicted in Figure \ref{fig:chop_into_blocks_example1}.
We then $\mathsf S$-dualize each block  using the basic moves and glue them back. The resulting quiver is given in the first line of Figure \ref{fig:EMdualization_step1_example1}.

At this point we notice the presence of asymmetric $\mathsf{B}_{01}$-blocks. We then perform a sequence of $(1,0)$-$(0,1)$ swaps. For example we move the leftmost $\mathsf{B}_{01}$-block to the right across three $\mathsf{B}_{10}$-blocks and move the rightmost $\mathsf{B}_{01}$-block to the left across one $\mathsf{B}_{10}$-block, as indicated by the orange arrows in the first line of Figure \ref{fig:EMdualization_step1_example1}. After these  $(1,0)$-$(0,1)$ moves, the leftmost $\mathsf{B}_{01}$-block is now symmetric, but  we still have one asymmetric $\mathsf{B}_{01}$-block left.  We stress that any other sequence of  $(1,0)$-$(0,1)$ swaps would leave some  asymmetric $\mathsf{B}_{01}$-block  behind. This is due to the fact that our quiver is bad.

\begin{figure}[!p]
	\centering
	\includegraphics[width=\textwidth,center]{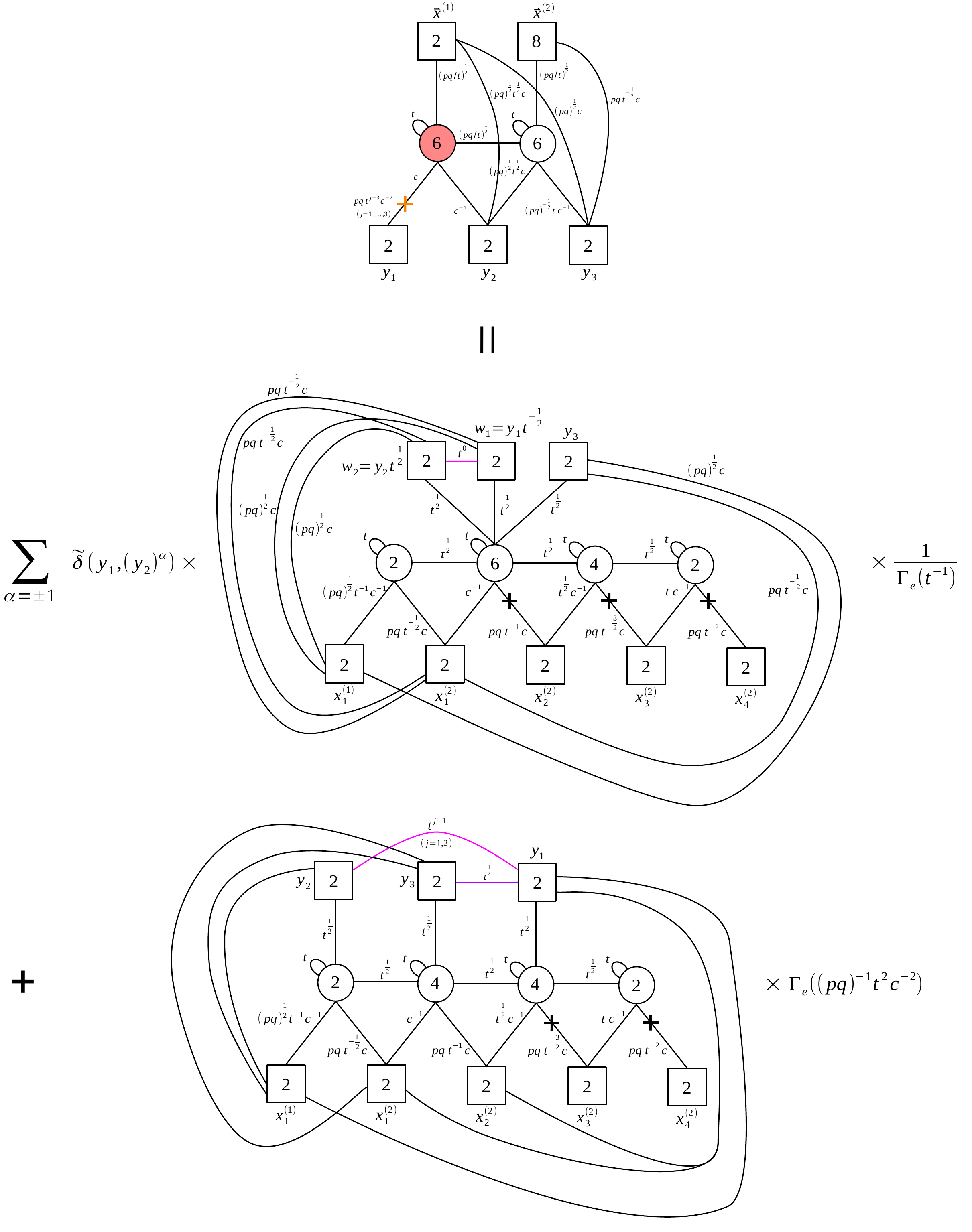}
	\caption{The comprehensive result of the mirror dual  of the 4d bad linear quiver  $E_{[2,4,3]}^{[2,2,2,2,1]}[USp(18)]$.
	The  frame (A) gives the good quiver $E_{[2,2,2,2,1]}^{[4,3,2]}[USp(18)]$ plus a free sector and a delta constraint, while the frame (B) gives the good quiver $E_{[2,2,2,2,1]}^{[3,3,3]}[USp(18)]$ plus a free sector.}
	\label{fig:allDuals_EM_4d_example1}
\end{figure}

As we have seen in the previous section, in presence of asymmetric $\mathsf{B}_{01}$-blocks the new $(0,1)$-$(0,1)$ swap in \eqref{eq:D5_swap_4d} is non-trivial and we can apply it.
In this case  we obtain two frames (A) and (B). The first frame has a delta $\tilde{\delta}(y_1,y_2)$ enforcing a condition on the $y_i$ fugacities.
Since in both (A) and (B) we still have asymmetric $\mathsf{B}_{01}$-blocks, we continue implementing $(1,0)$-$(0,1)$ and $(0,1)$-$(0,1)$  swaps
until we have only symmetric $\mathsf{B}_{01}$-blocks, as shown in Figures \ref{fig:EMdualization_step2A_example1} and \ref{fig:EMdualization_step2B_example1} for the intermediate theories (A) and (B) respectively.   In the end, as summarized in Figure \ref{fig:allDuals_EM_4d_example1}, we find that the mirror dual consists of two frames.
Each frame has an interacting part given by a good quiver theory of the form of $E_\rho^\sigma[USp(2N)]$ families defined in  Appendix \ref{app:Erhosigma_Trhosigma} and a free sector.
In particular the quivers in the  two frames appearing in Figure  \ref{fig:allDuals_EM_4d_example1} correspond to  the  $E_{[2,2,2,2,1]}^{[4,3,2]}[USp(18)]$ and $E_{[2,2,2,2,1]}^{[3,3,3]}[USp(18)]$ theories respectively.  In Section \ref{braneint} we will discuss the 3d version of this mirror dualization  together with its brane interpretation.

We then propose that for generic 4d and 3d bad quivers the fourth step of the mirror dualization algorithm can be implemented by a suitable of  combination of   $(1,0)$-$(0,1)$ and $(0,1)$-$(0,1)$ moves which  take us to a collection of frames where only symmetric $(0,1)$-blocks are present.
Each frame  has an interacting part given by a good  $E_\rho^\sigma[USp(2N)]$ or $T_\rho^\sigma[SU(N)]$ quiver theory and a free sector. Some of these frames come with a delta constraint on the parameters, so the index/partition function of bad quivers is a distribution.

\begin{figure}[!p]
	\centering
	\includegraphics[width=.8\textwidth,center]{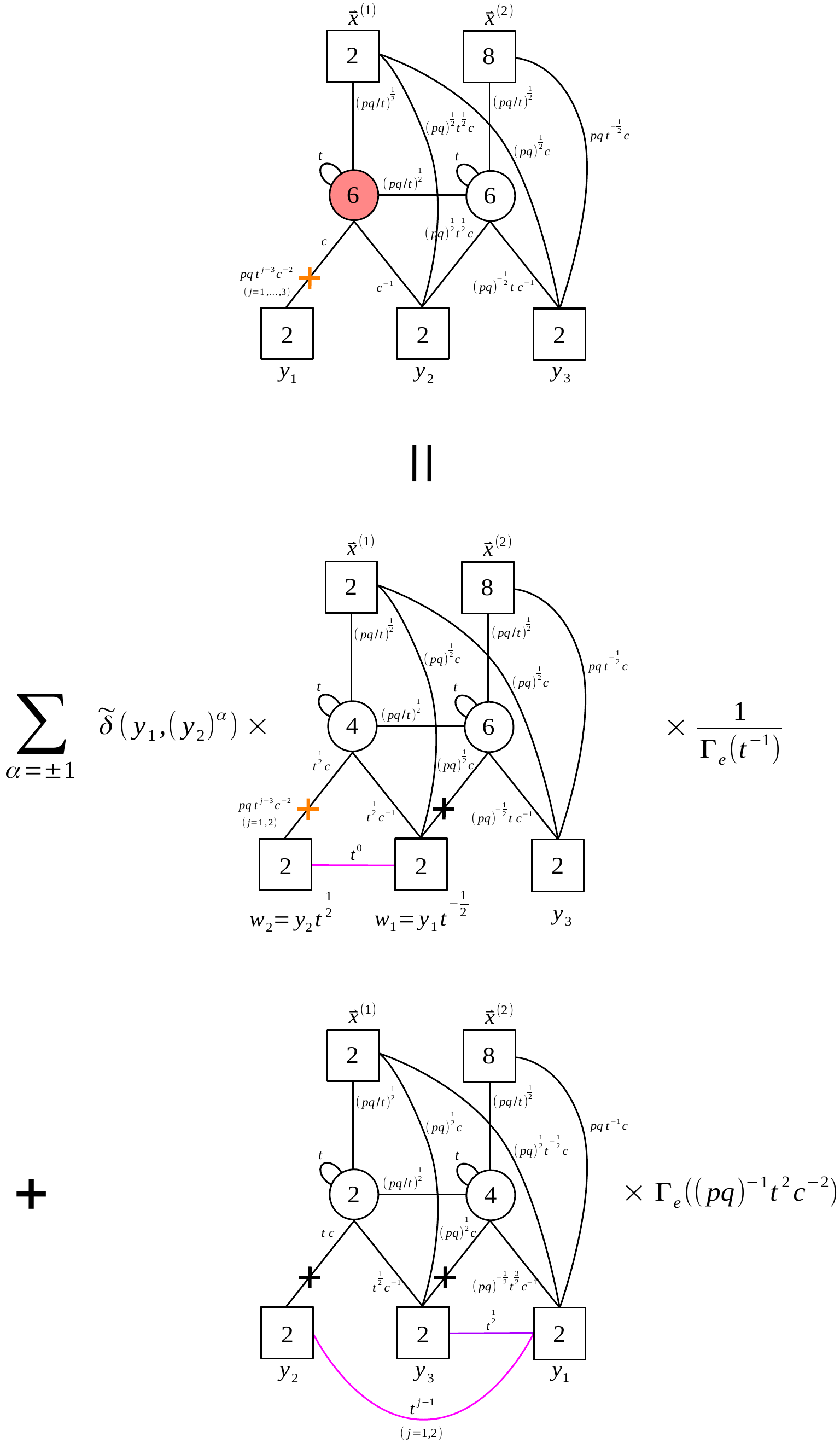}
	\caption{The comprehensive result of the electric dual of the 4d bad linear quiver  $E_{[2,4,3]}^{[2,2,2,2,1]}[USp(18)]$.}
	\label{fig:allDuals_EE_4d_example1}
\end{figure}

It is also interesting to present the result by replacing the good interacting quiver theory in each frame with its own mirror dual by employing the well-known mirror symmetry of good linear quivers.
In our example, such theories obtained by taking another step of mirror symmetry, which we call electric quivers, are given by $E_{[4,3,2]}^{[2,2,2,2,1]}[USp(18)]$ and $E_{[3,3,3]}^{[2,2,2,2,1]}[USp(18)]$, respectively, labeled by swapped partitions.
Hence, as  shown in Figure \ref{fig:allDuals_EE_4d_example1}, we can give an alternative
 {\it electric dual} description of our bad quiver as a sum of good electric quivers.\footnote{As we shall see in Section \ref{sec:SU}, in certain situations we should take the mirror dual of the free sector as well.}
The corresponding index identity is as follows:
\begingroup\allowdisplaybreaks
\begin{align}\label{eq:index_EE_bad_example_Nc33_Nf14}
	& \mathcal{I}_{E_{[2,4,3]}^{[2,2,2,2,1]}[USp(18)]}\left(\vec{x};\vec{y};t;c\right) =	\sum_{\alpha=\pm 1}
    \left[
    \tilde\delta\left(y_1,\left(y_2\right)^{\alpha}\right)
    \times
	\frac{1}{\Gamma_e\left(t^{-1}\right)}
	\times
	\Gamma_e\left(w_1^{\pm}w_2^{\pm}\right)
	\times
	\right.\nonumber\\
	& \qquad\qquad\!\left.
	\times
	\mathcal{I}_{E_{[3,3,3]}^{[2,2,2,2,1]}[USp(18)]}\left(\vec{x};\{w_2,w_1,y_3\};t;t^{\frac{1}{2}}c\right)
	\right]_{w_{1,2}=y_{1,2}t^{\mp\frac{1}{2}}}
	\\\nonumber
	& \quad + 
	\Gamma_e\left((pq)^{-1}t^{2}c^{-2}\right)
	\times
	\prod_{j=1}^{2}\Gamma_e\left(t^{j-1}y_1^{\pm}y_2^{\pm}\right)
	\times
	\Gamma_e\left(t^{\frac{1}{2}}y_1^{\pm}y_3^{\pm}\right)
	\times
	\mathcal{I}_{E_{[4,3,2]}^{[2,2,2,2,1]}[USp(18)]}\left(\vec{x};\{y_2,y_3,y_1\};t;tc\right) \,.
\end{align}
\endgroup

\begin{landscape}
	\begin{figure}
	\includegraphics[scale=.28,center]{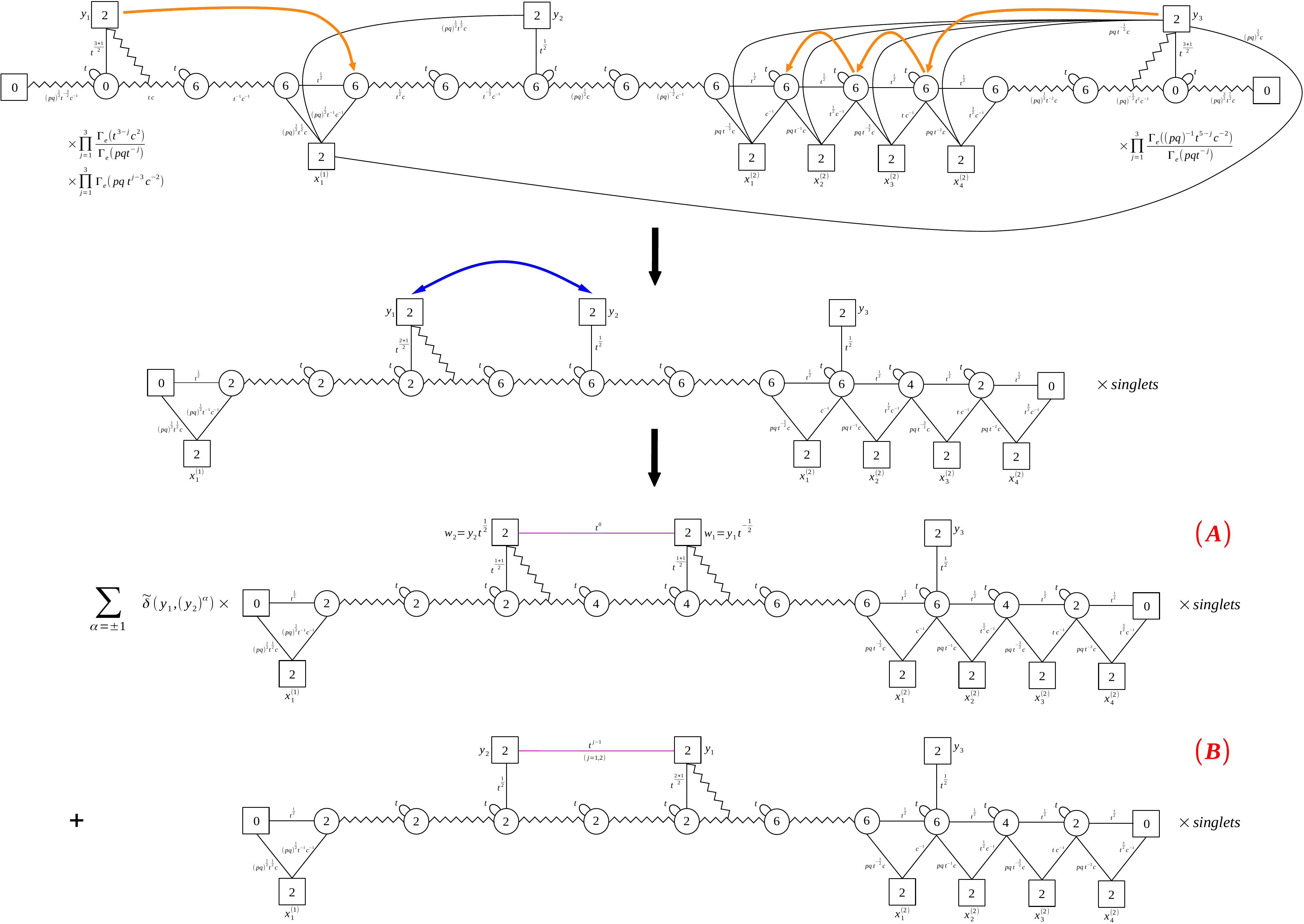}
	\caption{The initial sequence of $(1,0)$-$(0,1)$ swaps, denoted by the orange arrows, and the $(0,1)$-$(0,1)$ swap denoted by the blue arrow,
	 applied to the $\mathsf{S}$-dual of the theory in Figure \ref{fig:original_theory_4d_example1}.}
		 \label{fig:EMdualization_step1_example1}
	\end{figure}
	\begin{figure}
	\includegraphics[scale=.28,center]{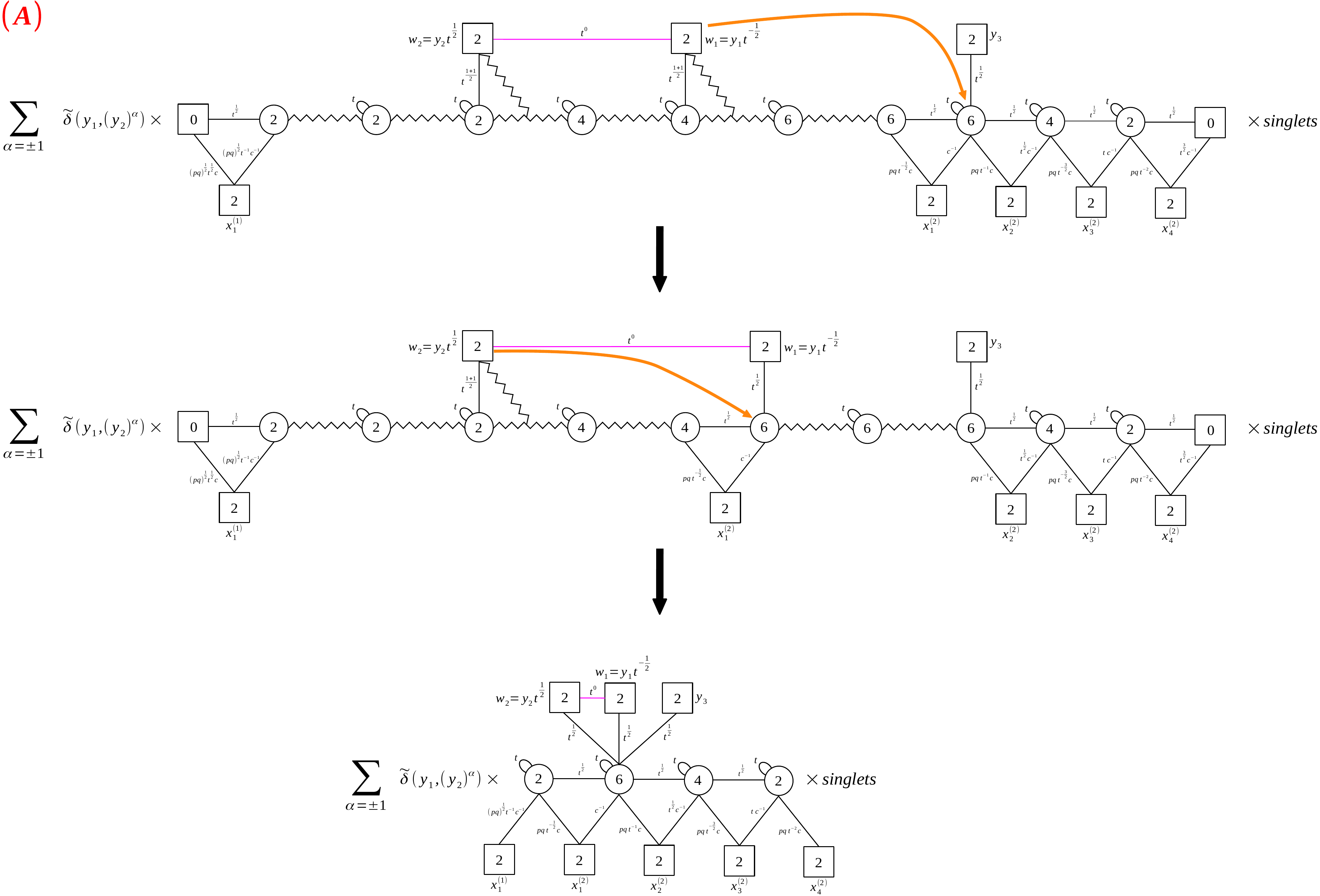}
	\caption{The following sequence of $(1,0)$-$(0,1)$ and $(0,1)$-$(0,1)$ swaps applied to the intermediate theory (A) in Figure \ref{fig:EMdualization_step1_example1}.}
		 \label{fig:EMdualization_step2A_example1}
	\end{figure}
	\begin{figure}
	\includegraphics[scale=.28,center]{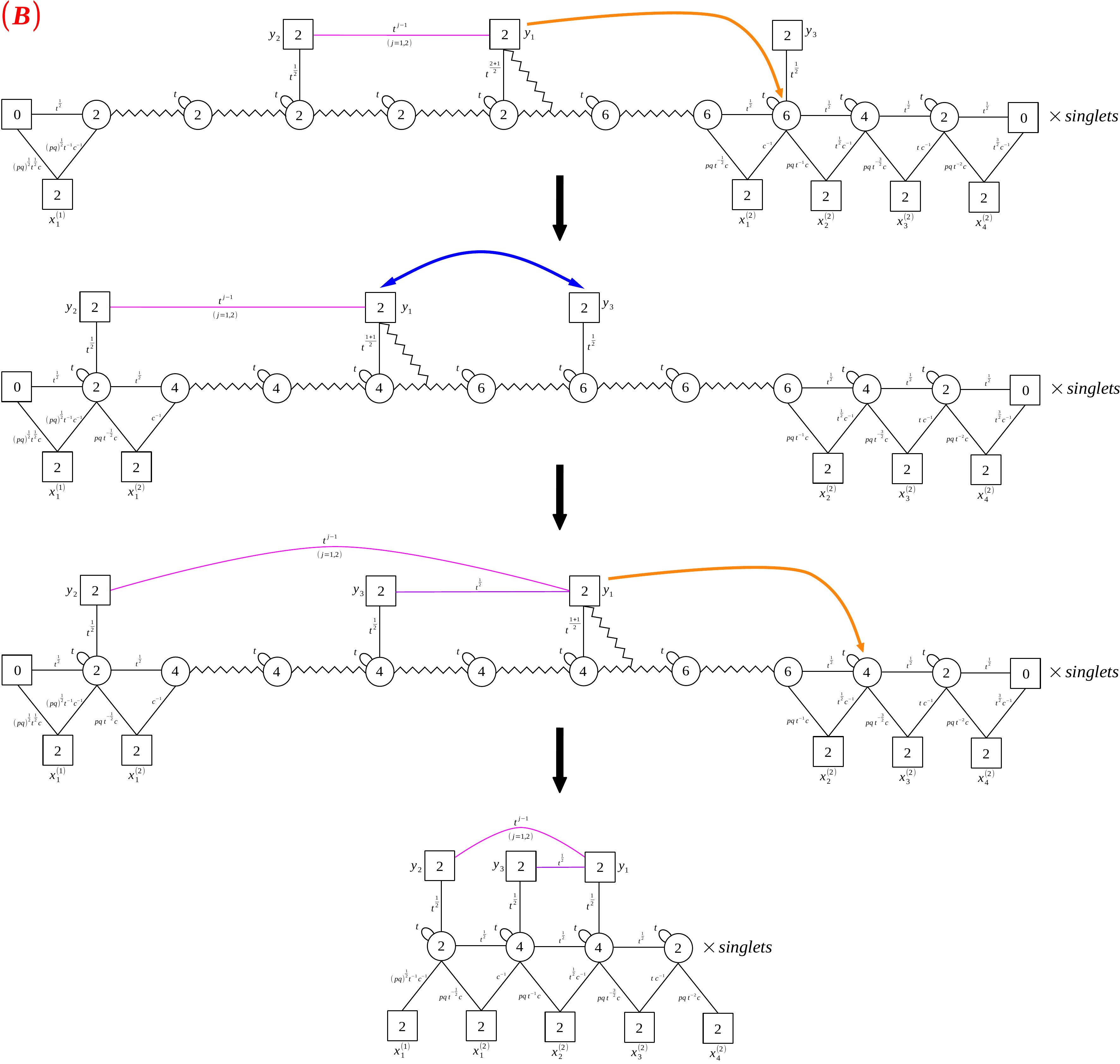}
\caption{The following sequence of $(1,0)$-$(0,1)$ and $(0,1)$-$(0,1)$ swaps applied to the intermediate theory (B) in Figure \ref{fig:EMdualization_step1_example1}.}
		 \label{fig:EMdualization_step2B_example1}
	\end{figure}
\end{landscape}

\clearpage
\subsection{The electric dualization algorithm}
\label{sec:electric dual}

In this section we  propose an {\it electric } dualization  algorithm which generates the collection of good frames in the electric dual of a bad quiver theory. The output of this algorithm is equivalent to the result we can obtain by first running the mirror dualization algorithm 
and then taking  in each frame  the mirror dual of the good quiver theory representing the corresponding interacting part. 

\begin{figure}[!ht]
	\centering
	\includegraphics[width=\textwidth]{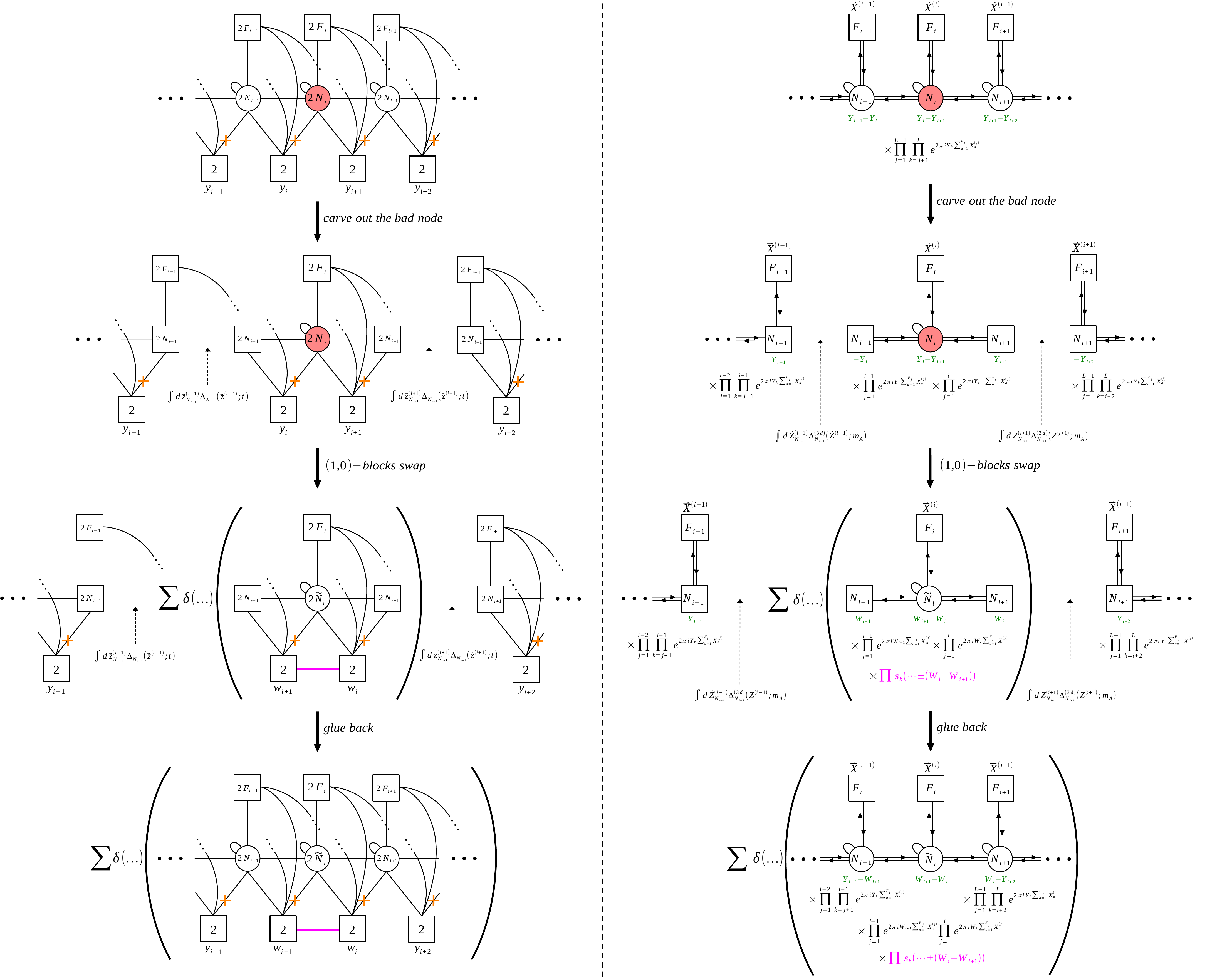}
	\caption{The three steps of the electric dualization algorithm. For simplicity we omit some of the singlets.}
	\label{fig:Generic_EE_dualization_4d}
\end{figure}

\begin{figure}[!ht]
	\centering
	\includegraphics[width=\textwidth]{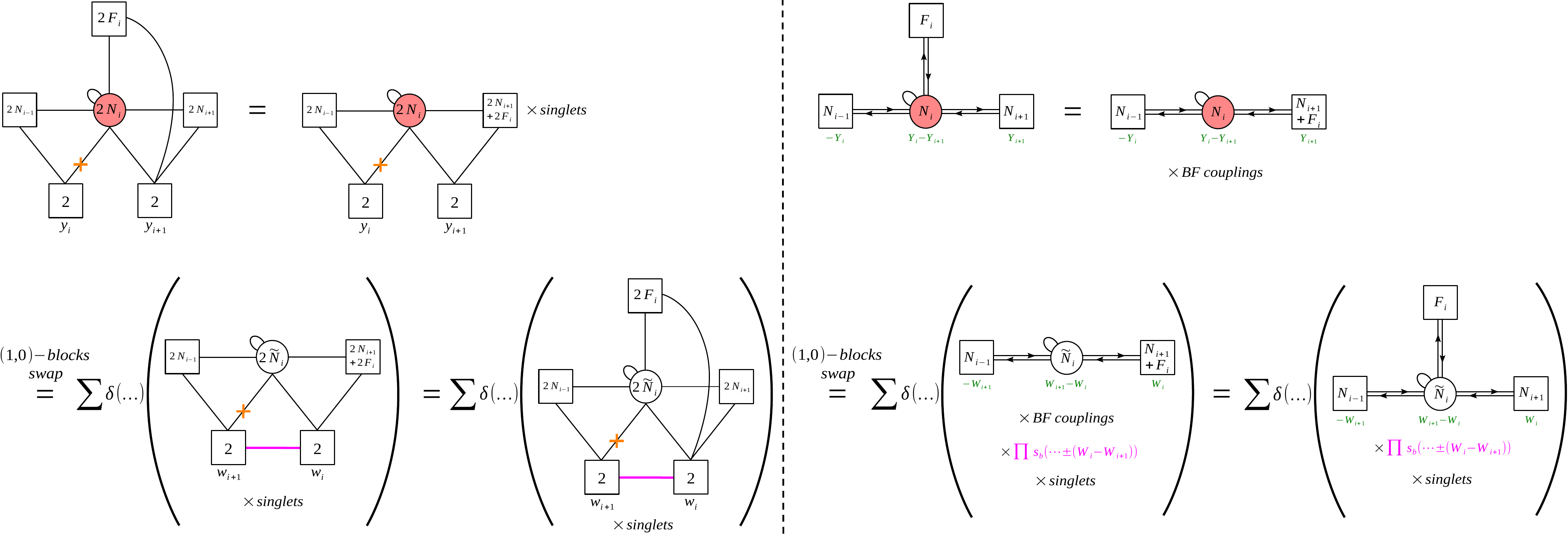}
	\caption{The SQCD electric dualization applied to a bad node of a quiver, which can be understood as a $(1,0)$-$(1,0)$ swap.}
	\label{fig:B10_swap_rewritten}
\end{figure}

The electric algorithm consists of the following steps:
\begin{enumerate}
	\item Given a bad theory, choose any ugly/bad node and carve it out from the rest of the quiver, together with the attached matter and singlets,  to isolate a bad SQCD (as shown in the first step of Figure \ref{fig:Generic_EE_dualization_4d}).
	\item Replace the carved out SQCD theory with the appropriate  sum of good theories plus free sectors
	(as shown in the second step of Figure \ref{fig:Generic_EE_dualization_4d}).
	This corresponds to using the $(1,0)$-$(1,0)$ swap as shown in Figure \ref{fig:B10_swap_rewritten}.
	\item Glue back  the dualized parts (as shown in the third step of Figure \ref{fig:Generic_EE_dualization_4d}).
	We now have a collection of quivers. 
	\item If some of the quivers generated at the previous step  still contain ugly/bad nodes, iterate the procedure
	until the all the generated frames contain only good gauge nodes.  
	\end{enumerate}

The first step in Figure \ref{fig:Generic_EE_dualization_4d} shows how to  carve out a bad SQCD from a bad quiver.
Putting unnecessary singlets aside for the moment, the carved out SQCD  essentially consists of two joint $(1,0)$-blocks
 once we combine the flavour node in the middle with the right node as shown in the first equality of Figure \ref{fig:B10_swap_rewritten}. We can then use the $(1,0)$-$(1,0)$ swap as in the second equality of Figure \ref{fig:B10_swap_rewritten}. 
 
 Once we glue black as shown in the third step of Figure \ref{fig:Generic_EE_dualization_4d} we generate a collection of frames 
where the rank of the node we focused on now varies in the range
\begin{align}
\tilde N_i = \mathrm{max}(0,\tilde F_i-N_i), \dots, \floor*{F^{\text{eff}}_i/2} \,,
\end{align}
where $F^{\text{eff}}_i$ is the effective number of flavors defined by $F^{\text{eff}}_i = F_i+N_{i-1}+N_{i+1}$.
So we have now tamed the badness of this node, but we need to check that
there no other bad nodes in the quiver, in particular notice we need to check whether  new bad nodes have appeared because of this replacement.
If the theory contains more bad nodes, the procedure needs to be repeated until we are left only with good nodes\footnote{
Similarly to the dualization of SQCD  discussed in \cite{Giacomelli:2023zkk}, the algorithm may produce additional ``unphysical'' frames that must be discarded. A physical frame is required to satisfy two consistency conditions. First, the system of delta functions associated with the frame must admit a solution for generic values of 
$m$ and $Q$, as in \cite{Giacomelli:2023zkk}. Second, the singlets introduced in the frame must remain non-singular upon imposing the delta function constraints. If a singlet becomes singular, the corresponding frame should be interpreted as a singular limit of a physical frame and thus be discarded.
}.\\

\begin{figure}[!ht]
	\centering
	\includegraphics[width=.95\textwidth]{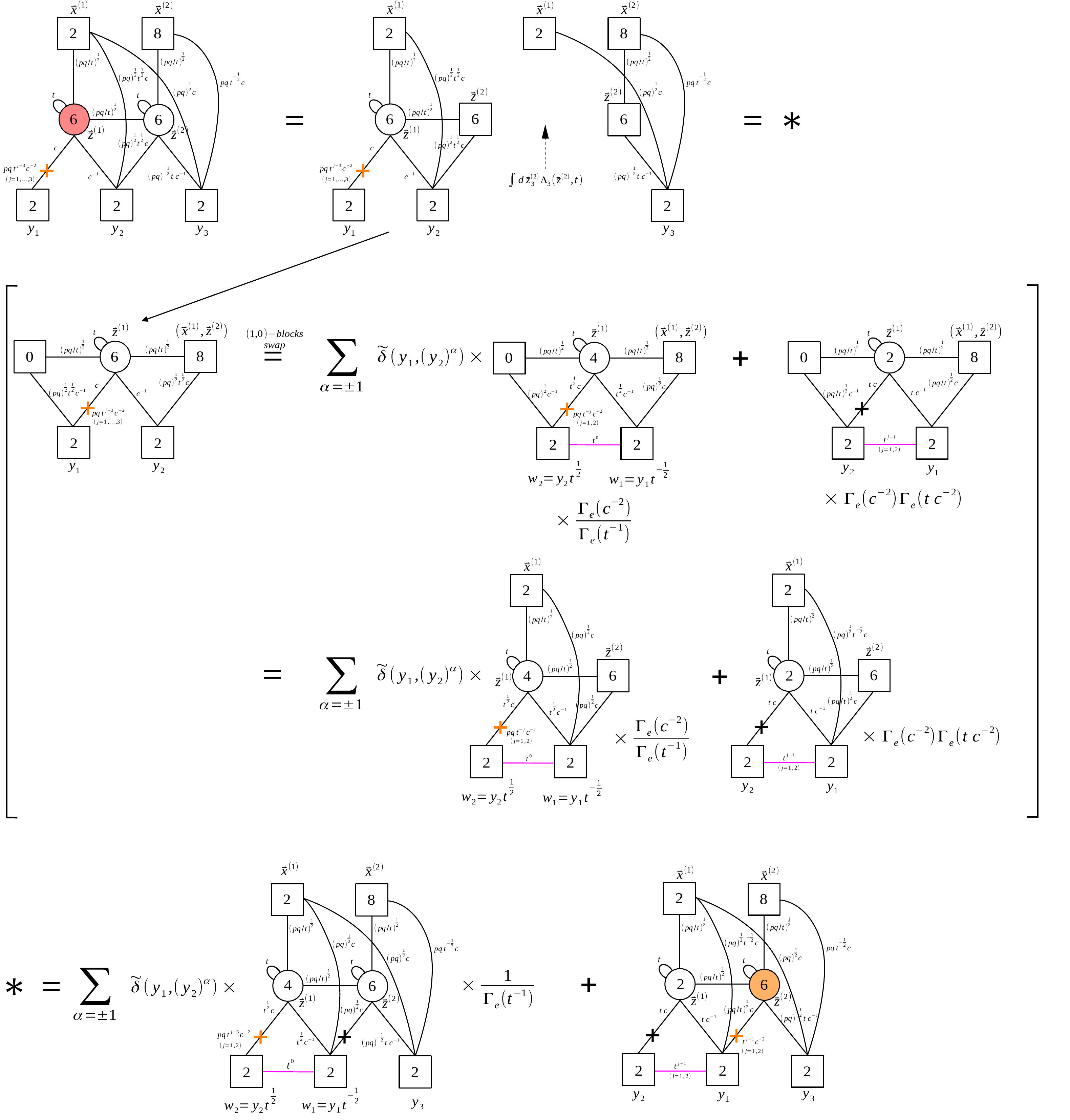}
	\caption{The first iteration of the electric dualization of the $E_{[2,4,3]}^{[2,2,2,2,1]}[USp(18)]$ theory. In the first line we carve out 
	a bad SQCD. The  flavor $USp(2)_{x^{(1)}}$ node is drew twice to indicate that the two separated parts of the quiver both have one singlet attached to it. As in the rest of the paper, bad nodes are colored in red, while ugly nodes are colored in orange.}
	\label{fig:EEdualization_explaination_1}
\end{figure}

We are now going to implement the electric algorithm on the example we studied in the previous section, the 
 $E_{[2,4,3]}^{[2,2,2,2,1]}[USp(18)]$ theory.  We  begin by carving out a bad SQCD with $N_c=3$ and $N_f=4$ as shown
  in the first step of Figure \ref{fig:EEdualization_explaination_1}.
As shown in the third line, we use the $(1,0)$-$(1,0)$  swap \eqref{eq:B10B10_swap_4d}
and  glue back to obtain the sum of the quiver theories in the last line.

\begin{figure}[!ht]
	\centering
	\includegraphics[width=0.8\textwidth]{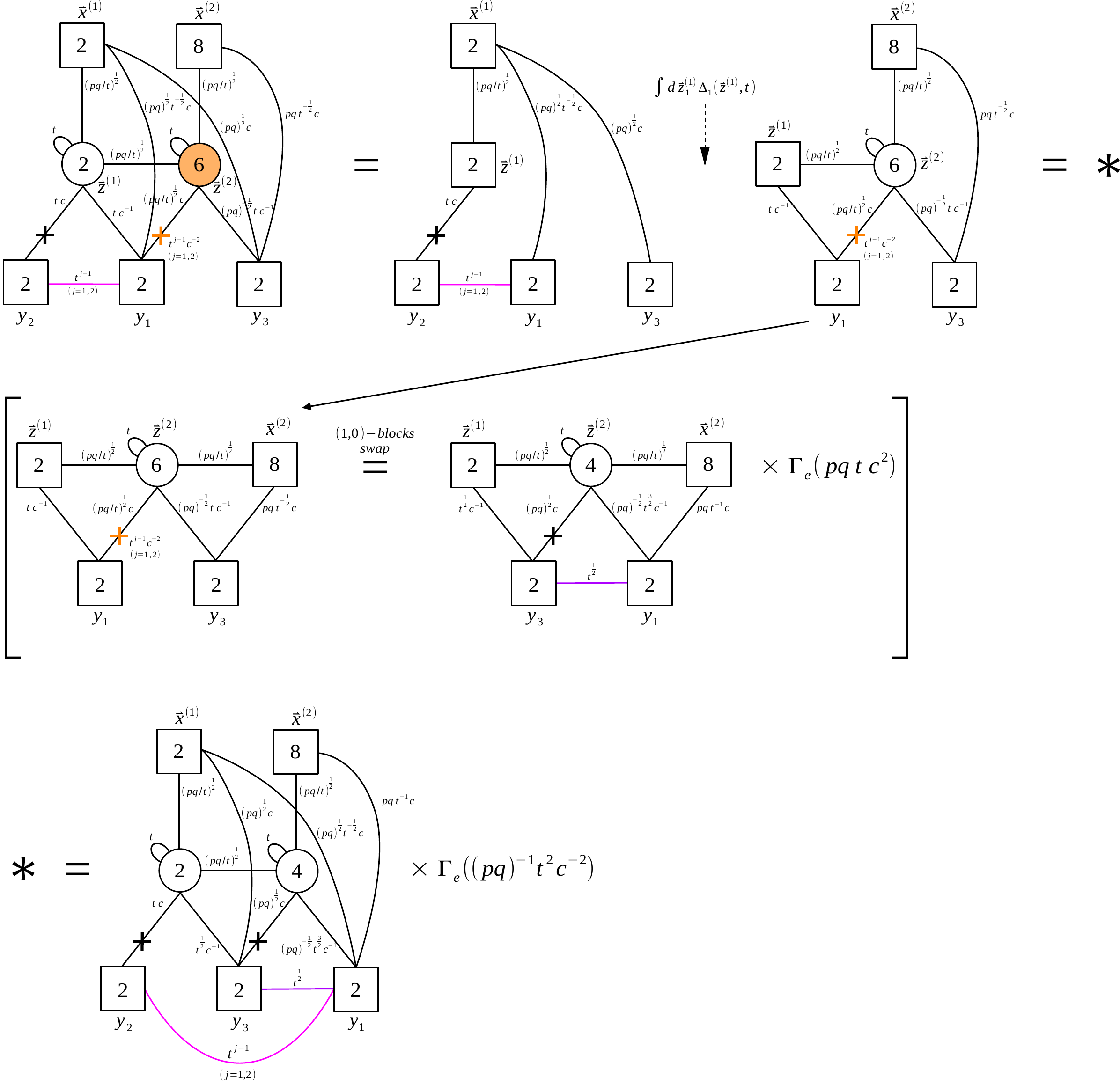}
	\caption{The second iteration of the electric dualization of the  $E_{[2,4,3]}^{[2,2,2,2,1]}[USp(18)]$ theory.}
	\label{fig:EEdualization_explaination_2}
\end{figure}

While the first  quiver is good and it is exactly the first dual frame in Figure \ref{fig:allDuals_EE_4d_example1}, the second theory has an ugly node
and we need to iterate the procedure.
As  shown in Figure \ref{fig:EEdualization_explaination_2} we thus carve out the ugly SQCD with $N_c=3$ and $N_f=5$, apply the 
$(1,0)$-$(1,0)$ swap and glue back.
The resulting quiver in the last line is exactly the second dual theory in Figure \ref{fig:allDuals_EE_4d_example1}. 


Again, one can also manipulate the 3d $\mathcal N=4$ linear quivers in the same way using the 3d version of the $(1,0)$-$(1,0)$ swap move or simply reduce the 4d result to 3d and do some deformation to obtain the 3d dual. In Section \ref{braneint} we will discuss the 3d version of this mirror dualization  together with its brane interpretation.\\

One might be concerned that the output of the electric dualization algorithm might change if we change the 
order in which  we dualize the bad nodes. It turns out the result is the same regardless  of the order. 
In particular for the frames associated with delta distributions, we will see in Section \ref{higgsing} that each frame can be obtained as a result
of a series of  Higgsings triggered by VEVs for certain dressed mesonic operators in 4d or monopoles in 3d of the original theory. We will demonstrate the Higgs mechanism at the index level as a pinching of the contour that occurs when limits on the fugacities corresponding to the delta constraints of a specific frame are taken, and we will see that such Higgsing of the index does not depend on the order of the limits; namely, it does not require a specific order of the RG flows or dualizations. 

Therefore, the electric dualization algorithm introduced in this subsection can be  implemented by just choosing a convenient bad node to start with and applying the $(1,0)$-$(1,0)$ swap. In this sense, the electric  dualization algorithm is in practice simpler than the mirror dualization algorithm of the previous section, which after the block dualization requires a  combination of  $(1,0)$-$(0,1)$ and $(0,1)$-$(0,1)$ swaps although they give exactly the same result.\\

In addition, the electric  dualization algorithm can be easily automatized into a computer code. We have attached to the paper a Mathematica code which, for any 3d linear quiver, generates the list of its electric dual frames, specifying their FI parameters, the associated delta constraints and the singlets.\\

Another important  property of the electric  dualization algorithm is that it does not need to be restricted to linear quivers and can be applied to other types of quivers such as star-shaped quivers
which we plan to investigate in the future from this perspective.

\clearpage

\section{An example in 4d with $N_c=(5,3)$ and $N_f=(3,2)$}\label{sec:example}

\begin{figure}[!ht]
	\centering
	\includegraphics[width=0.4\textwidth]{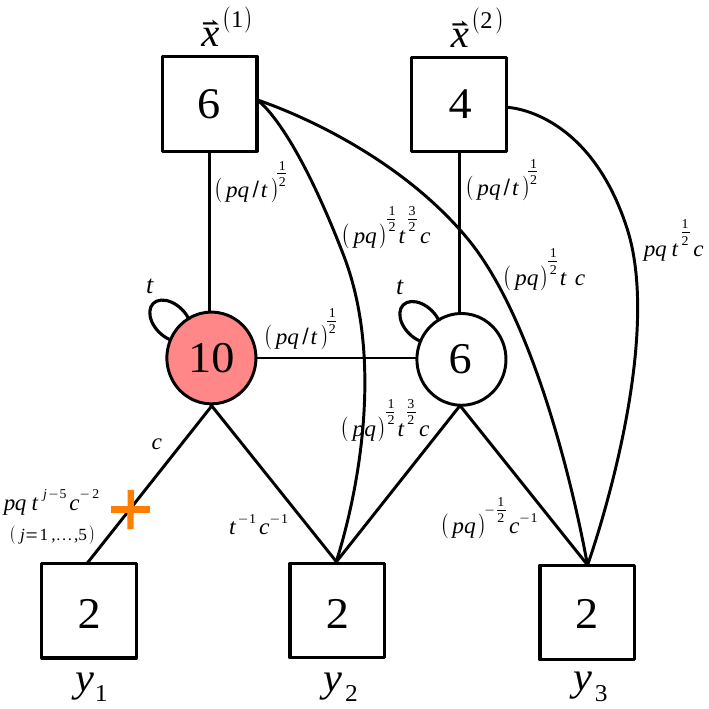}
	\caption{The 4d linear quiver with $N_c=(5,3)$ and $N_f=(3,2)$ corresponding to the ${E_{[0,4,3]}^{[2,2,1,1,1]}[USp(14)]}$ theory.	}
	\label{fig:original_theory_4d}
\end{figure}

In this section we will study the 4d theory with $N_c=(5,3)$ and $N_f=(3,2)$, with additional singlets shown  in Figure \ref{fig:original_theory_4d}, corresponding to the 
${E_{[0,4,3]}^{[2,2,1,1,1]}[USp(14)]}$ theory, whose first gauge node is bad.
In the first instance we will dualize it using the mirror dualization algorithm, as explained in Section \ref{sec:magnetic dual}. Then we will dualize it by means of the electric dualization algorithm, as explained in Section \ref{sec:electric dual}, and show how the results of the two algorithms agree. Finally, we will present a third derivation from a Higgsing perspective, along the lines of a similar analysis discussed in \cite{Giacomelli:2023zkk} for the bad SQCD. 

\subsection{The mirror dualization}
\begin{figure}
	\includegraphics[scale=.22,center]{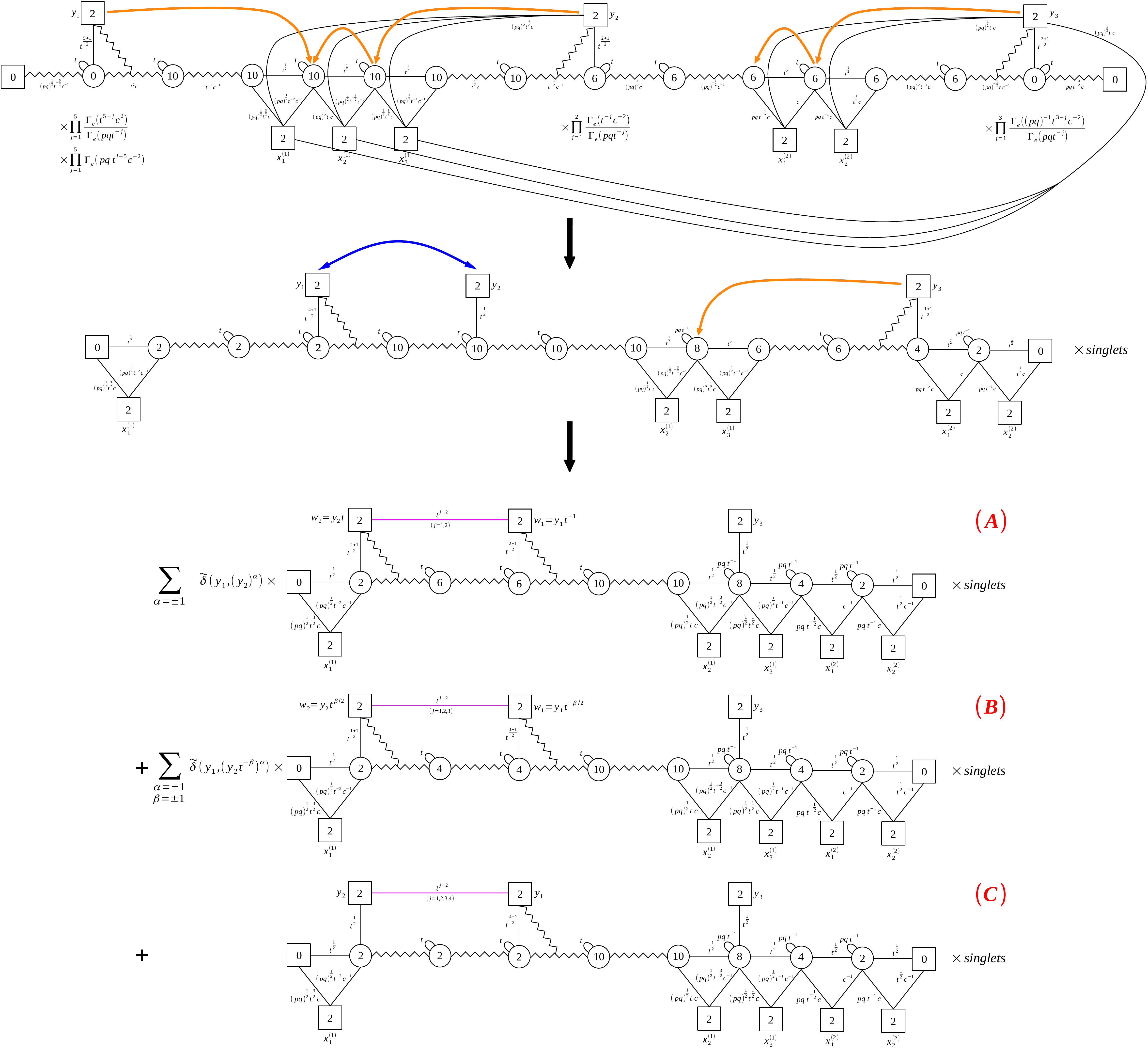}
	\caption{The first quiver is the result we get after applying the basic $\mathsf{S}$-duality moves to the original theory in Figure \ref{fig:original_theory_4d}. We apply the   $(0,1)$-$(0,1)$ swaps as indicated by  the orange arrows and we flow to the second quiver, where we did not explicitly write all the singlets to avoid clutter (and we will keep doing this until the end of the calculation). In this second quiver we also added where needed a symmetric Identity-wall in order to recognize the structure on which we apply the  $(0,1)$-$(0,1)$ swap. In the next iterations these Identity-wall insertions will be intended and not explicit. After we apply the  $(0,1)$-$(0,1)$ swap on the quiver in the second line we get the three contributions here labelled as (A), (B) and (C).}
		 \label{fig:EMdualization_step0}
\end{figure}

\begin{figure}
	\includegraphics[width=\textwidth,center]{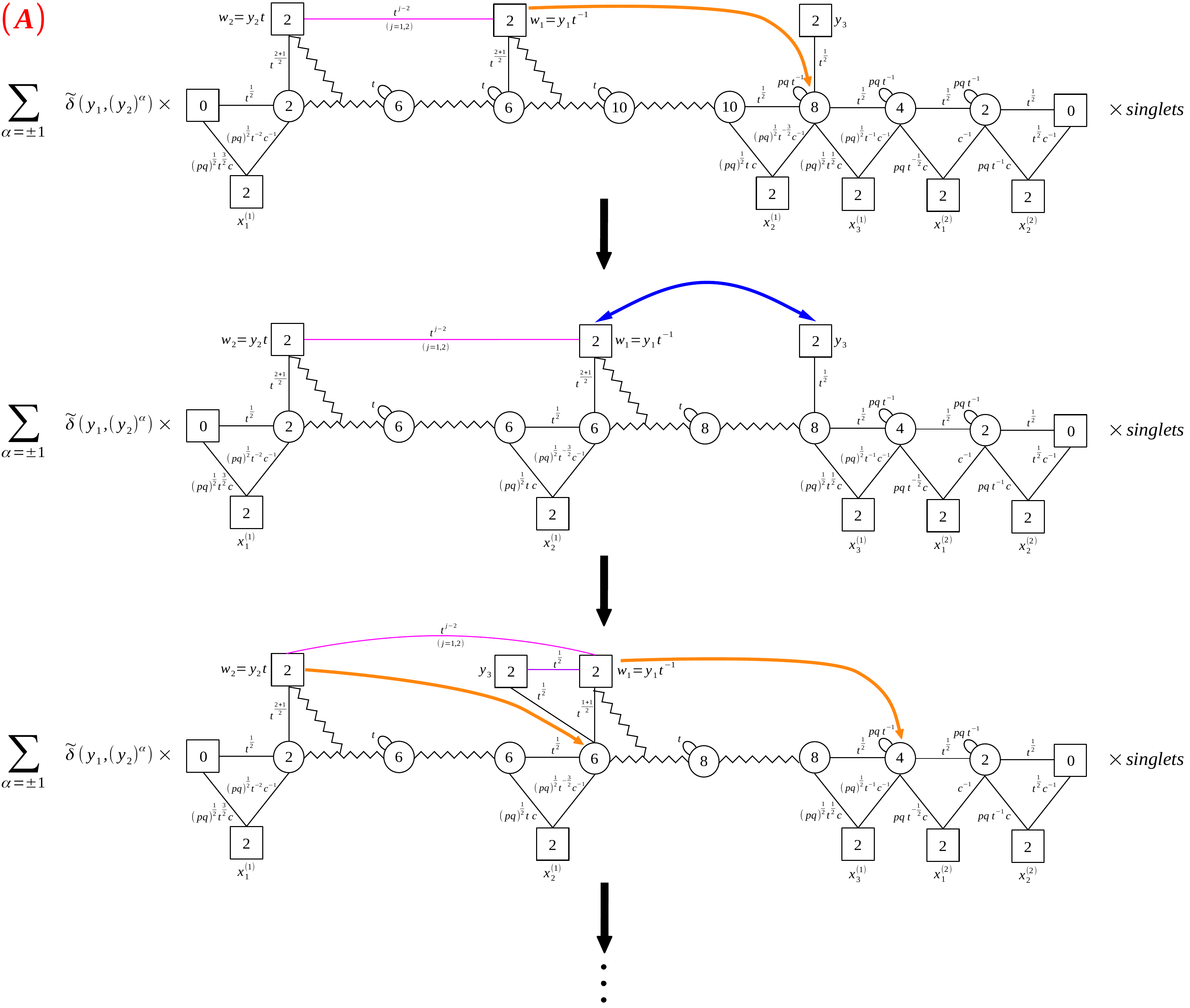}
	\caption{On theory (A) of Figure \ref{fig:EMdualization_step0} we first apply one $(1,0)$-$(0,1)$ swap, then a $(0,1)$-$(0,1)$ swap and then two $(1,0)$-$(0,1)$ swaps again.}
		 \label{fig:EMdualization_step1A}
\end{figure}

\begin{figure}[p]
	\includegraphics[width=\textwidth,center]{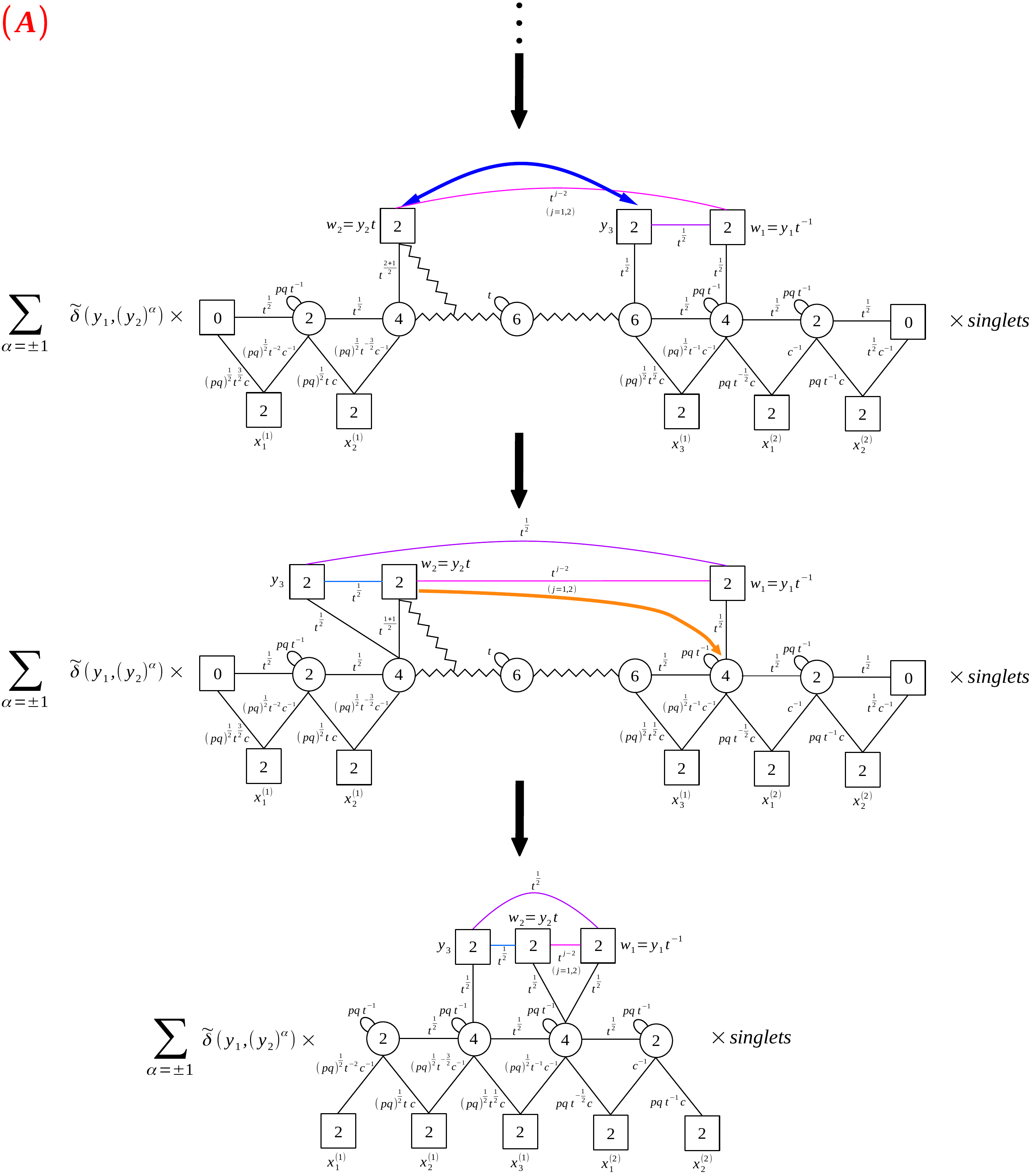}
	\caption{After the duality moves implemented on (A) in Figure  \ref{fig:EMdualization_step1A}, we perform one more $(0,1)$-$(0,1)$ swap followed by another $(1,0)$-$(0,1)$ swap and we get the frame shown at the bottom.}
	\label{fig:EMdualization_step2A}
\end{figure}

\begin{figure}
	\includegraphics[width=\textwidth,center]{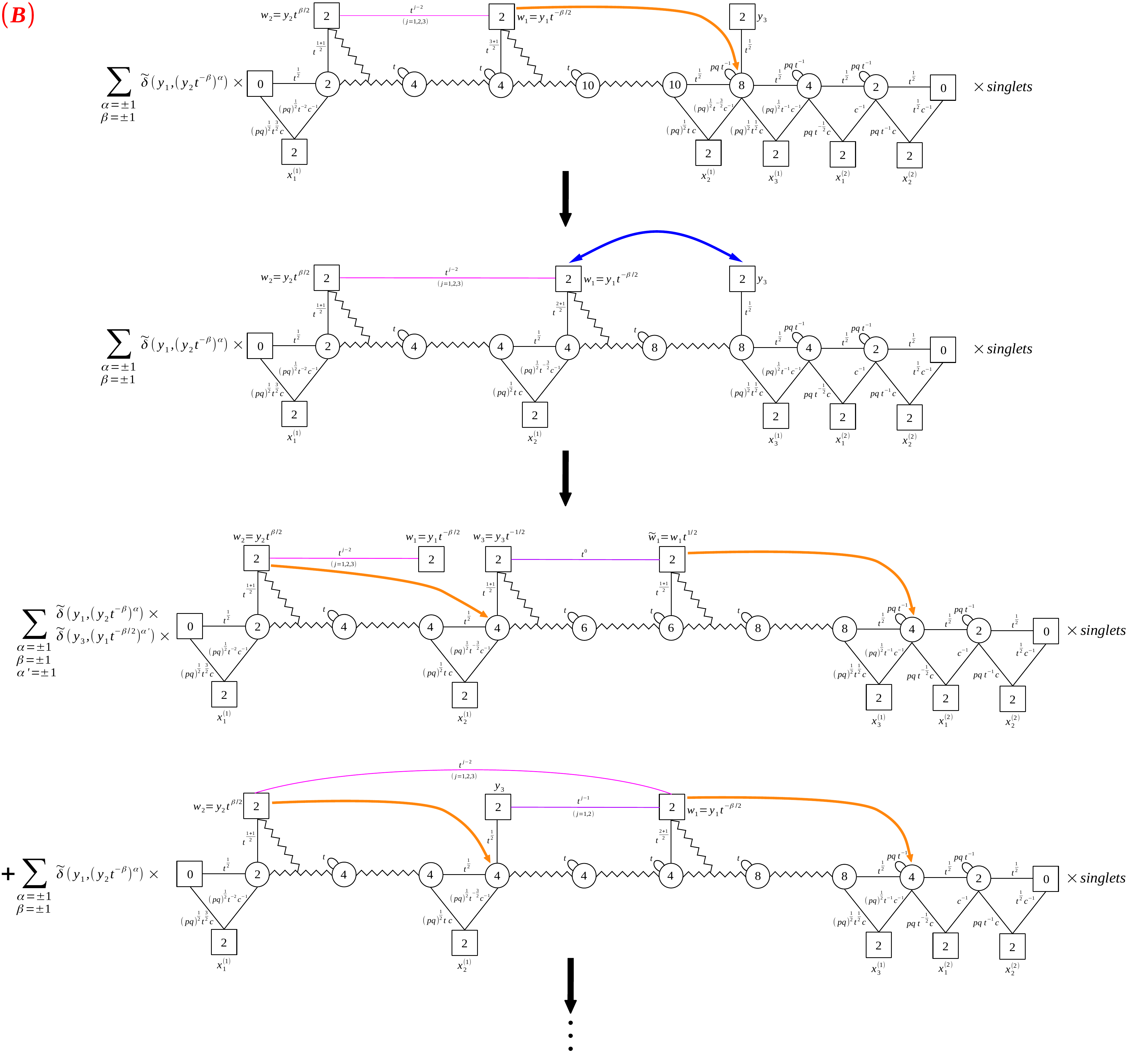}
	\caption{On theory (B) of Figure \ref{fig:EMdualization_step0} we first apply one $(1,0)$-$(0,1)$ swap and then a $(0,1)$-$(0,1)$ swap. We get the two frames shown in the bottom part of this picture. Applying on them two $(1,0)$-$(0,1)$ swaps as suggested by the orange arrows, we flow to the theories in Figure \ref{fig:EMdualization_step2B}.}
	\label{fig:EMdualization_step1B}
\end{figure}

\begin{figure}[p]
	\includegraphics[width=\textwidth,center]{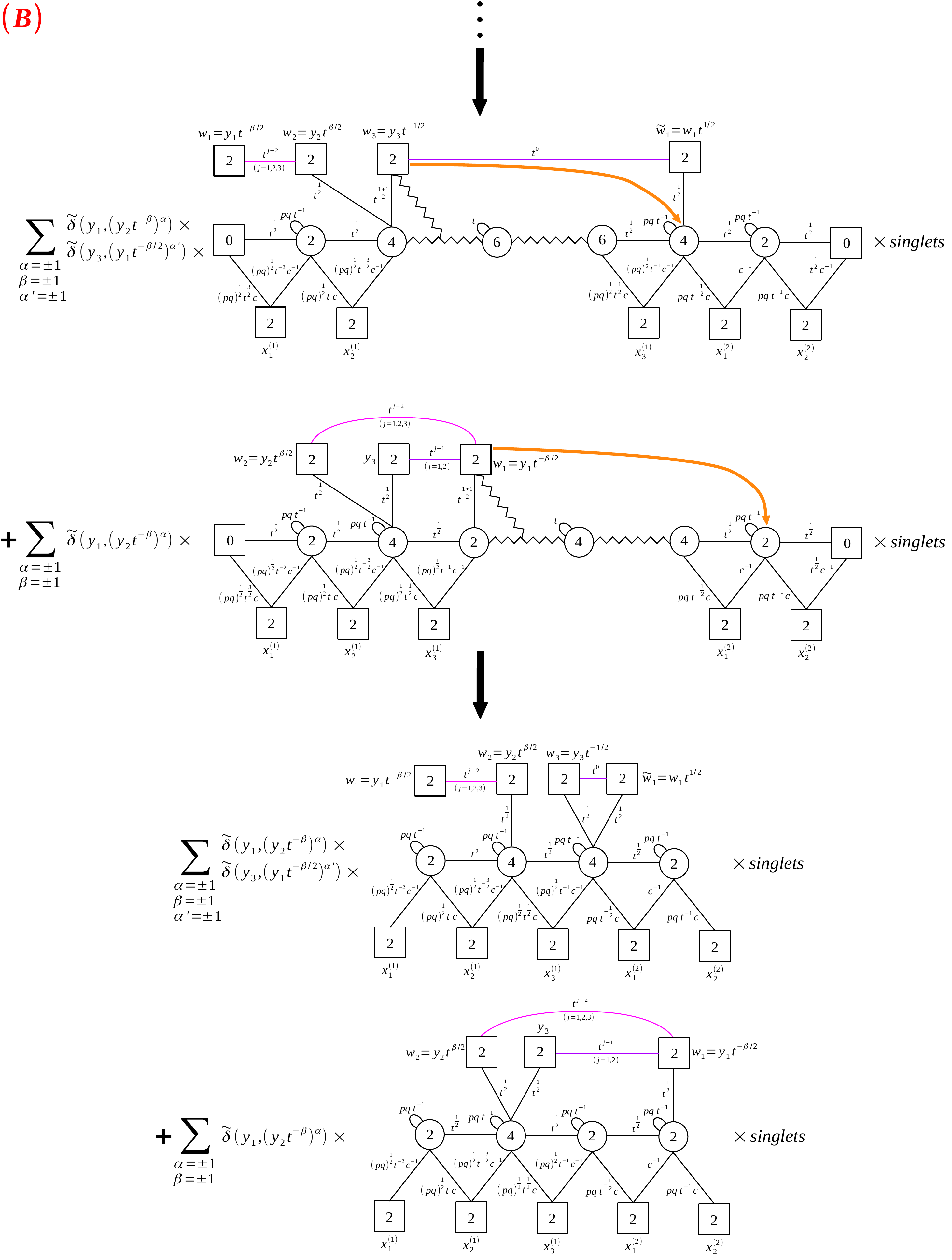}
	\caption{After performing the $(1,0)$-$(0,1)$ swaps depicted in Figure \ref{fig:EMdualization_step1B}, we need one more $(1,0)$-$(0,1)$ swap on both theories and we finally get the two frames shown at the bottom.}
	\label{fig:EMdualization_step2B}
\end{figure}

\begin{figure}
	\includegraphics[width=\textwidth,center]{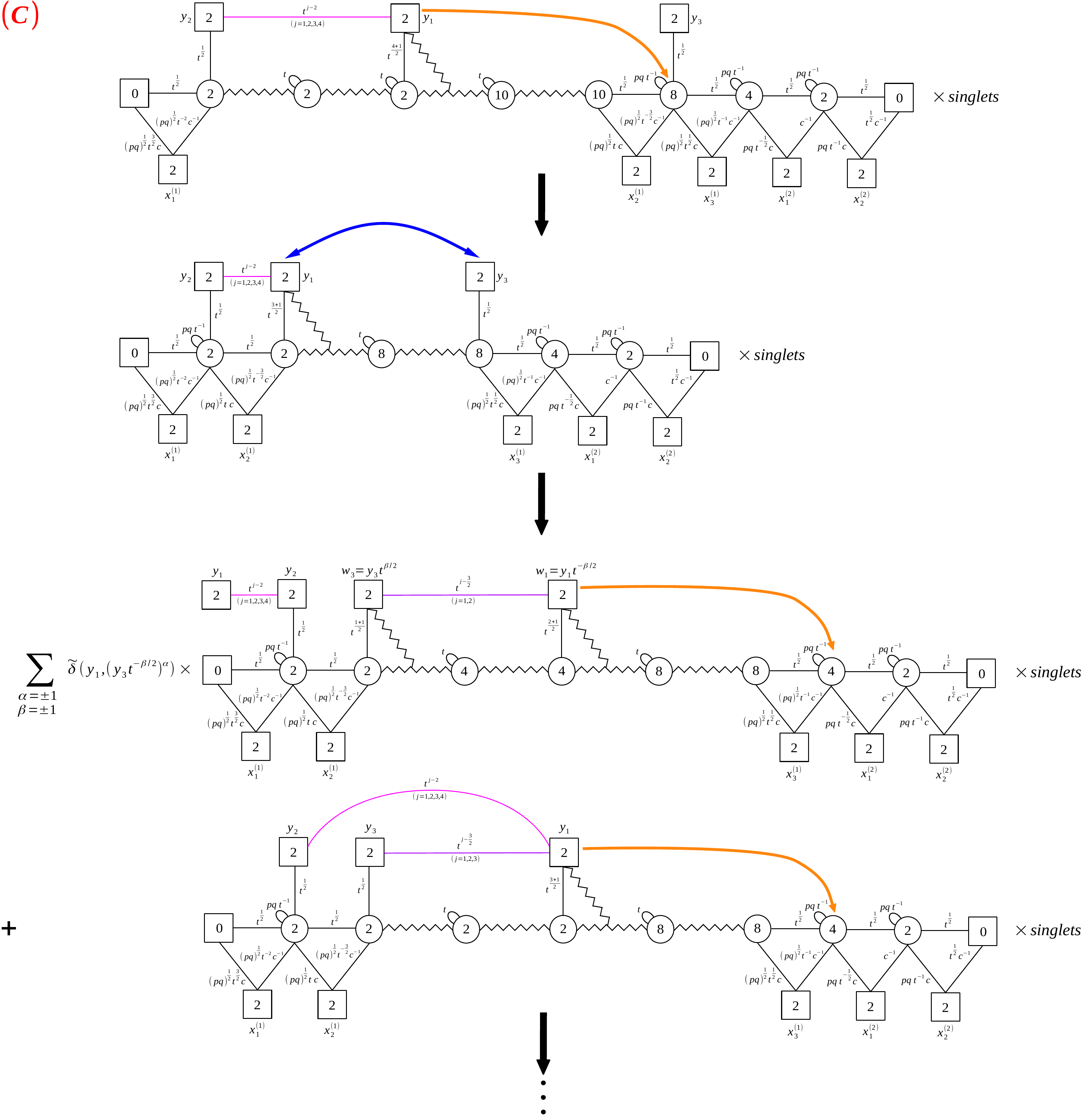}
	\caption{On theory (C) of Figure \ref{fig:EMdualization_step0} we first apply one $(1,0)$-$(0,1)$ swap and then a $(0,1)$-$(0,1)$ swap. We get the two frames shown in the bottom part of this picture. Applying on them one $(1,0)$-$(0,1)$ swap as suggested by the orange arrows, we flow to the theories in Figure \ref{fig:EMdualization_step2C}.}
	\label{fig:EMdualization_step1C}
\end{figure}

\begin{figure}[p]
	\includegraphics[width=\textwidth,center]{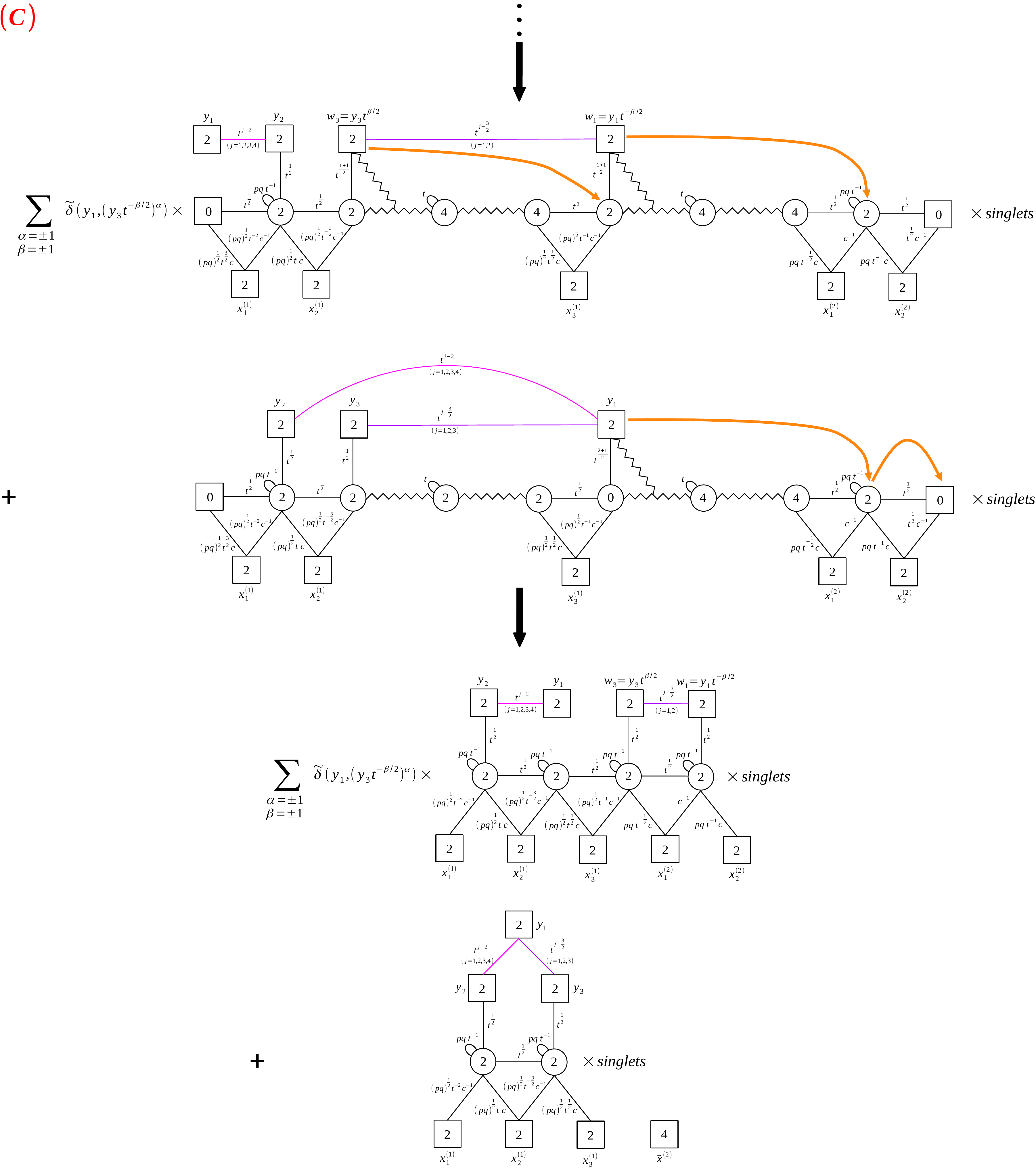}
	\caption{After performing the $(1,0)$-$(0,1)$ swaps depicted in Figure \ref{fig:EMdualization_step1C}, we need two more $(1,0)$-$(0,1)$ swaps on both theories and we finally get the two frames shown at the bottom.}
	\label{fig:EMdualization_step2C}
\end{figure}

\begin{figure}[p]
	\centering
	\includegraphics[width=1.2\textwidth,center]{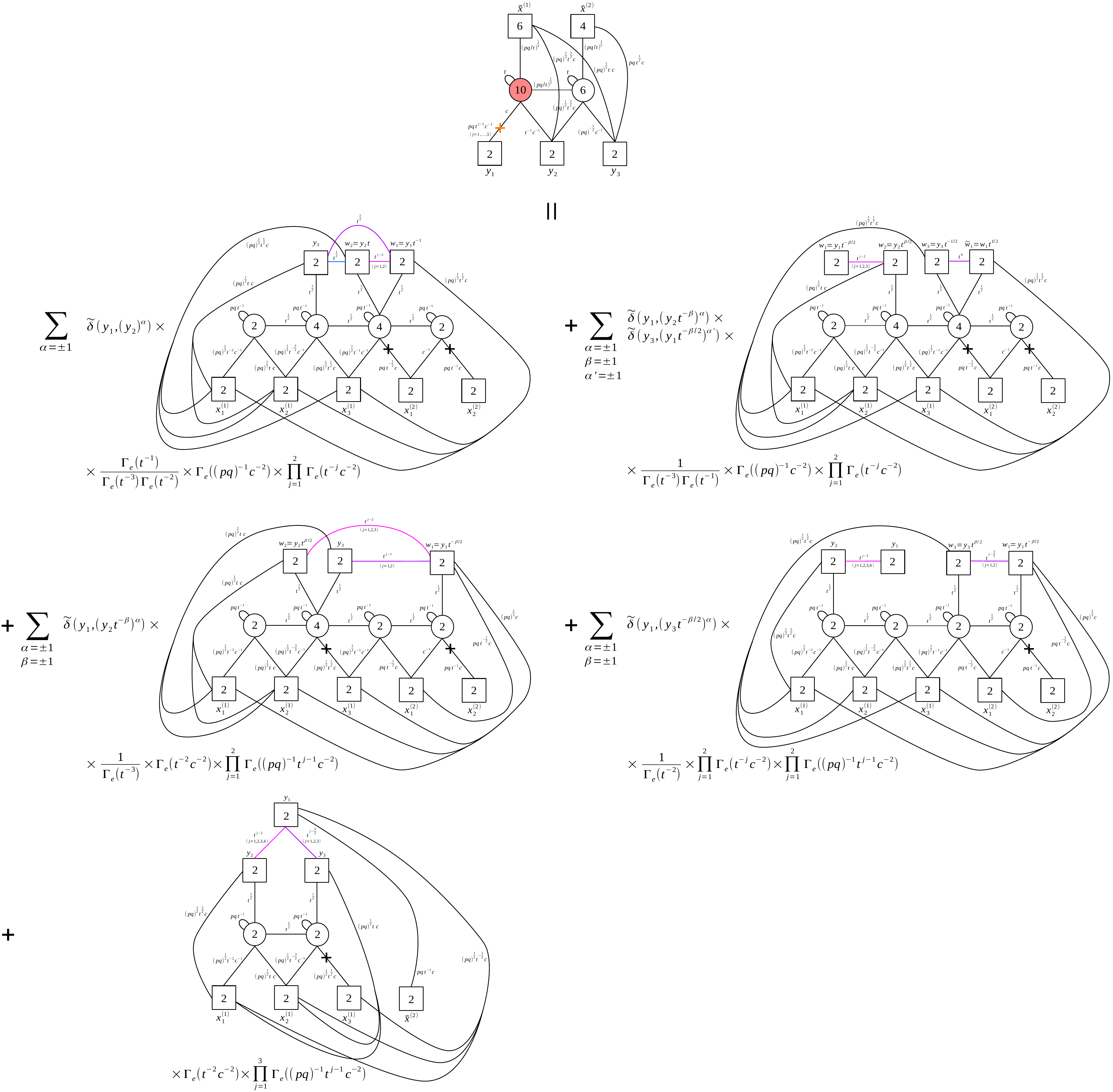}
	\caption{The comprehensive result of the mirror dualization of the 4d bad linear quiver ${E_{[0,4,3]}^{[2,2,1,1,1]}[USp(14)]}$.}
	\label{fig:allDuals_EM_4d}
\end{figure}

Given the theory in Figure \ref{fig:original_theory_4d}, we can dualize it using the mirror dualization algorithm, as explained in Section \ref{sec:magnetic dual}. 
The result of the first three steps of the mirror algorithm, decomposition into $(1,0)$ and $(0,1)$ QFT blocks, dualization of each QFT block by the basic duality moves and gluing back, 
is shown on top of Figure \ref{fig:EMdualization_step0}. We then focus on the fourth step of the algorithm, consisting in the study of the RG flow triggered by the  VEVs.
We choose to first perform some $(1,0)$-$(0,1)$ swaps (as shown by the orange arrows in the first quiver of Figure \ref{fig:EMdualization_step0}) and then we use the $(0,1)$-$(0,1)$ swap (as shown by the blue arrows in the second quiver of Figure \ref{fig:EMdualization_step0}), which produces the three frames  (A), (B) and (C) and depicted on the bottom part of Figure \ref{fig:EMdualization_step0}.\footnote{All the possible patterns of $(1,0)$-$(0,1)$ and $(0,1)$-$(0,1)$ blocks swap moves give the same result. Some of them may lead to subtle situations, where one has to refine the fugacities in order to see all the poles appearing in the index (see the discussion in \cite{Giacomelli:2023zkk}). The specific pattern chosen for this example, however, is straightforward and does not involve any further subtlety.} 
In each of these frames we still have asymmetric $\mathsf{B}_{01}$-blocks, so we need continue implementing block swaps to follow the RG flow and reach the final  dual fames.
\begin{itemize}
	\item[(A)] For the quiver (A) we apply the sequence of $(1,0)$-$(0,1)$ swaps and two $(0,1)$-$(0,1)$ swaps summarized in Figures \ref{fig:EMdualization_step1A} and \ref{fig:EMdualization_step2A}, the final result being the single frame with one delta shown on the bottom of Figure \ref{fig:EMdualization_step2A}.
	\item[(B)] For the quiver (B) we apply the sequence of $(1,0)$-$(0,1)$ swaps and one $(0,1)$-$(0,1)$ swap summarized in Figures \ref{fig:EMdualization_step1B} and \ref{fig:EMdualization_step2B}, the final result being the two frames on the bottom of Figure \ref{fig:EMdualization_step2B}: the first has two deltas, while the second has just one.
	\item[(C)] For the quiver (C) we apply the sequence of $(1,0)$-$(0,1)$ swaps and one $(0,1)$-$(0,1)$ swap summarized in Figures \ref{fig:EMdualization_step1C} and \ref{fig:EMdualization_step2C}, the final result being the two frames on the bottom of Figure \ref{fig:EMdualization_step2C}: the first has one delta, while the second has no delta and is a Wess--Zumino theory.
\end{itemize}
Performing these manipulations at the level of the index allows us to also keep track of all the singlets. The result, shown in Figure \ref{fig:allDuals_EM_4d}, has the structure of a sum of five frames,
with interacting parts given by $E^{\rho}_{\sigma}[USp(14)]$ theories with $\sigma=[2,2,1,1,1]$ and respectively  $\rho=[3,2,2]$, $\rho=[3,2,2]$, $\rho=[3,3,1]$, $\rho=[4,2,1]$ and $\rho=[4,3]$.

Notice that  the first and second frames in Figure \ref{fig:allDuals_EM_4d} have the same interacting part but  differ for the collection of deltas and  singlets. This phenomenon is not peculiar of this specific example, but happens in many other cases. The first four frames contain one or two deltas, while the last one has none and thus corresponds to generic non-trivial values of the fugacities. In all the frames having at least one delta there are some redefinitions of the flavor Cartans (the $w_i$ are the redefined $y_i$). Indeed, as shown in details in \cite{Giacomelli:2023zkk}, this redefinition is needed to properly repackage all the fields into doublets of the $SU(2)$ symmetries.

\clearpage
\subsection{The electric dualization}
We now apply the electric dualization algorithm to the bad quiver $E_{[0,4,3]}^{[2,2,1,1,1]}[USp(14)]$ in Figure \ref{fig:original_theory_4d}.

\begin{figure}[p]
	\centering
	\includegraphics[width=.8\textwidth,center]{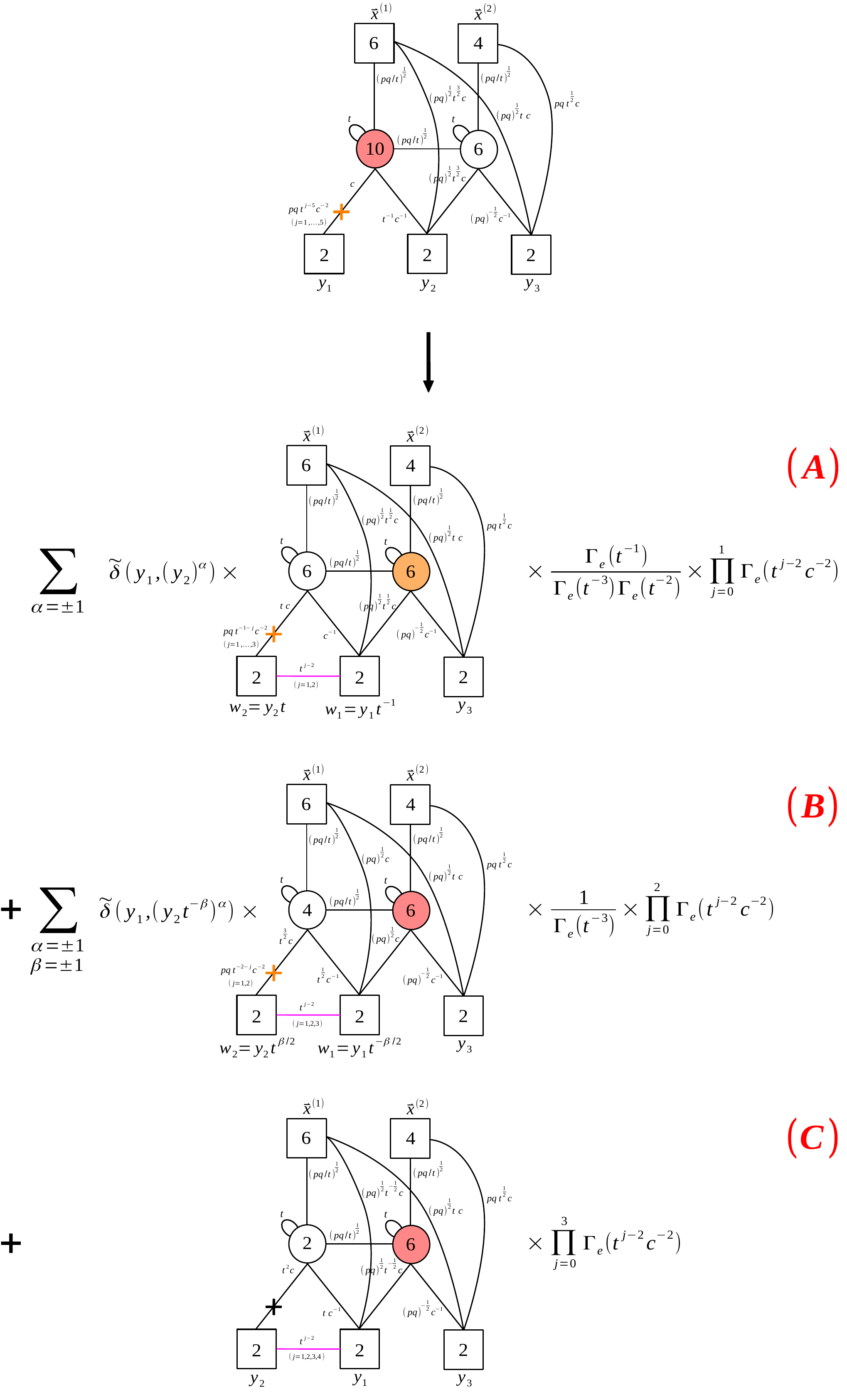}
	\caption{The local dualization of the bad node of the original theory. The three theories we obtain (labelled as (A), (B) and (C)) are not good yet, so we have to iterate this procedure. As in the rest of the paper, bad nodes are colored in red and ugly nodes are colored in orange.}
	\label{fig:EE_dualization_step0}
\end{figure}

We  first carve out the bad SQCD with $N_c=5$ and $N_f=6$, use the $(1,0)$-$(1,0)$  swap in \eqref{eq:B10B10_swap_4d}
and  glue back to obtain the sum of frames (A), (B) and (C) in Figure \ref{fig:EE_dualization_step0}.  
Notice that frame (C) has generic $y_i$ fugacities since it has no  delta constraints.

\begin{figure}[p]
	\centering
	\includegraphics[width=.95\textwidth,center]{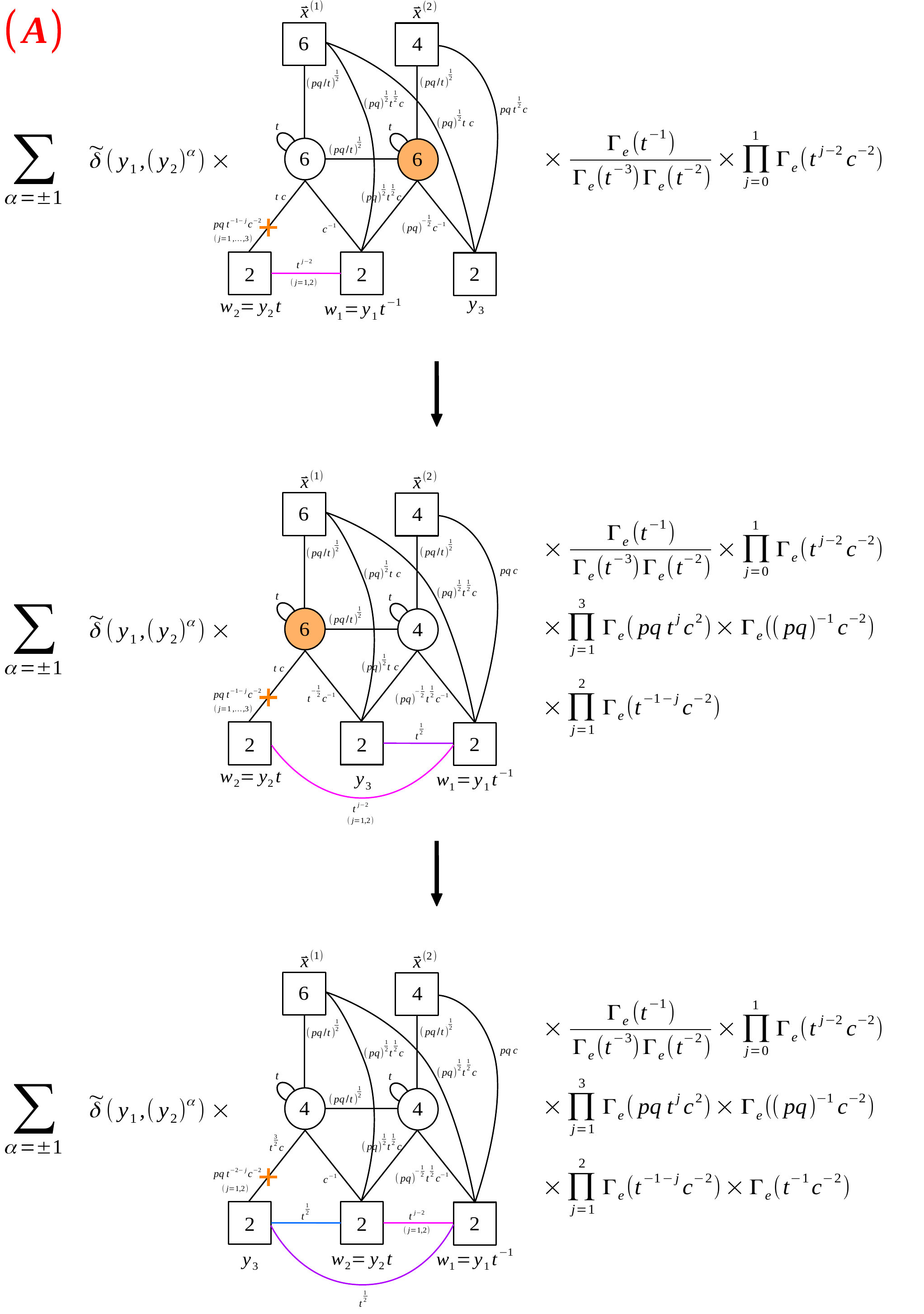}
	\caption{Theory (A) needs two more local dualization of its ugly nodes to become good.}
	\label{fig:EE_dualization_step1A}
\end{figure}

Frame (A) contains an ugly node, so we carve out the $N_c=3$ and $N_f=5$ SQCD and apply the $(1,0)$-$(1,0)$  swap in \eqref{eq:B10B10_swap_4d}.
As shown in Figure \ref{fig:EE_dualization_step1A}, the result still contains un ugly node so we need one more iteration to arrive at the final good frame. 
Notice that these two iterations, involving the dualization of an  ugly node, do not introduce further delta functions.

\begin{figure}[p]
	\centering
	\includegraphics[width=1\textwidth,center]{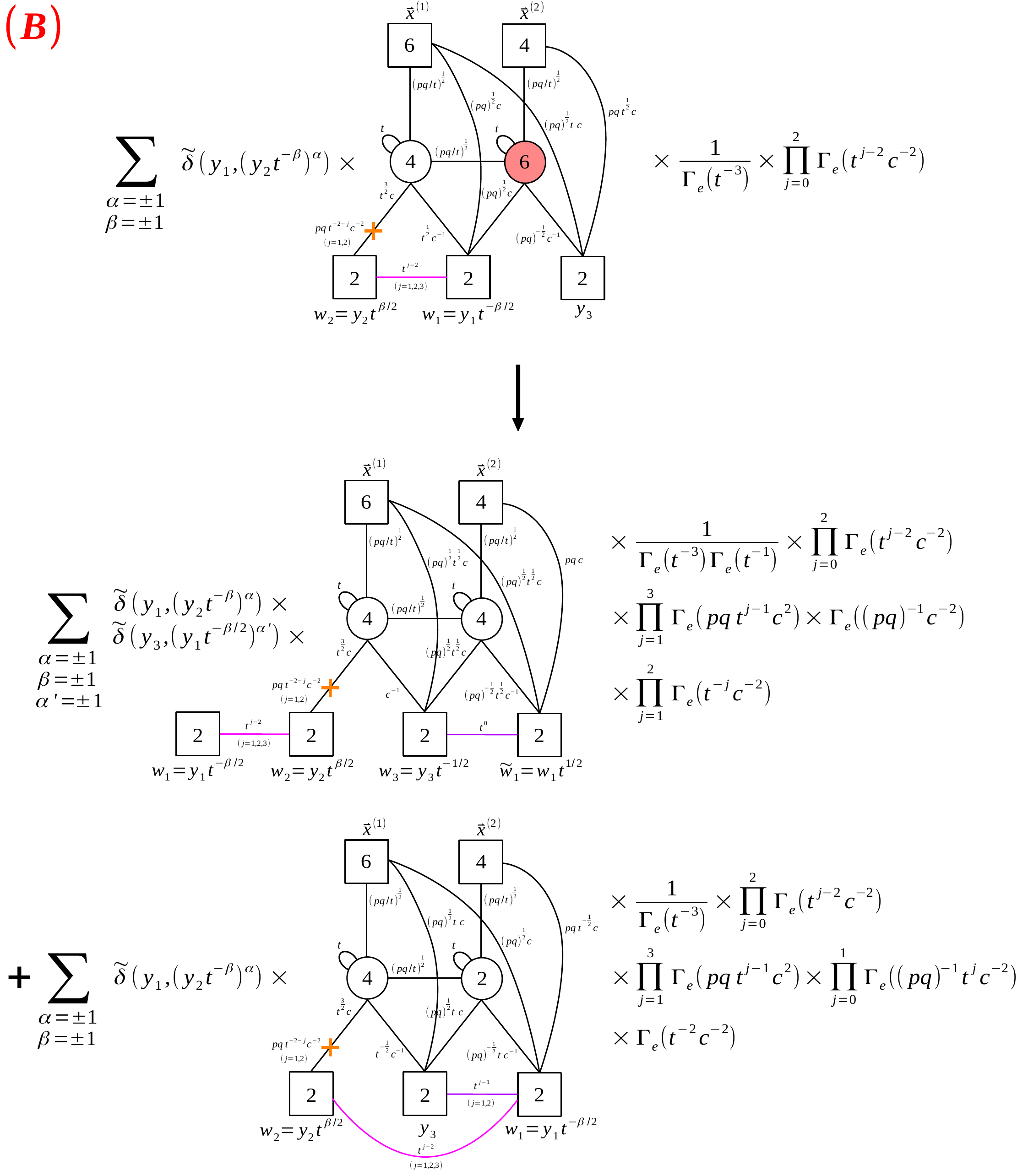}
	\caption{Theory (B) needs just one more local dualization of its bad node to produce a sum of two good frames.}
	\label{fig:EE_dualization_step1B}
\end{figure}

Frame (B) contains a bad node, so we carve out the $N_c=3$ and $N_4=5$ SQCD and apply the $(1,0)$-$(1,0)$  swap in Figure \eqref{eq:B10B10_swap_4d}
and we obtain the sum of the  two good frames shown in \ref{fig:EE_dualization_step1B}. 
Notice that in  this process we generate one extra delta function so one of the two final frames will have two delta functions.

\begin{figure}[p]
	\centering
	\includegraphics[width=1\textwidth,center]{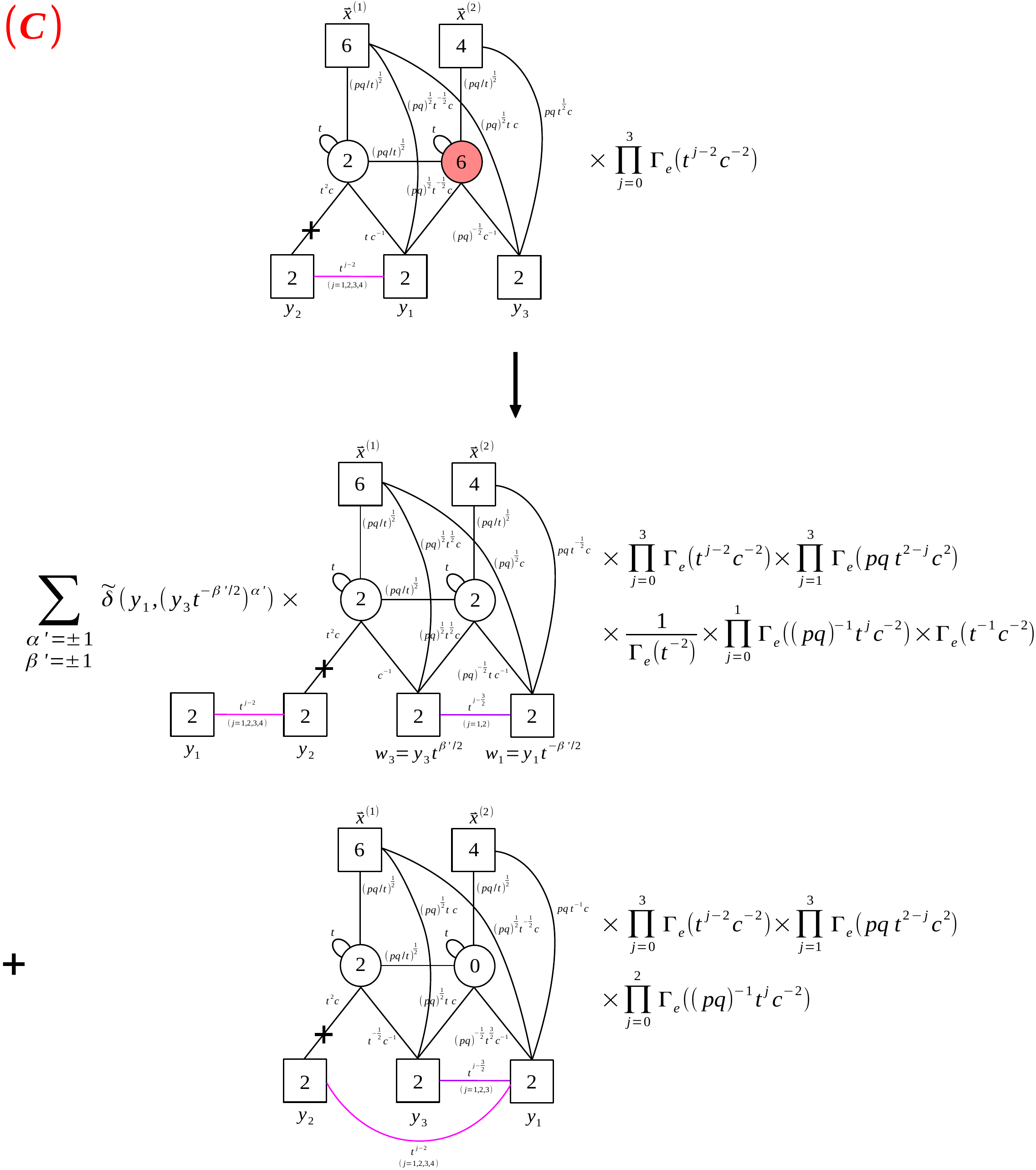}
	\caption{Theory (C) needs just one more local dualization of its bad node to produce a sum of two good frames.}
	\label{fig:EE_dualization_step1C}
\end{figure}

Also frame (C) contains a bad node, so we carve out the $N_c=3$ and $N_4=3$ SQCD and apply the $(1,0)$-$(1,0)$  swap in \eqref{eq:B10B10_swap_4d}
and we obtain the sum of the  two good frames shown in  \ref{fig:EE_dualization_step1C}.
Notice that in  this process we generate one  delta function so one of the two final frames will have one delta functions.

\begin{figure}[p]
	\centering
	\includegraphics[width=.97\textwidth,center]{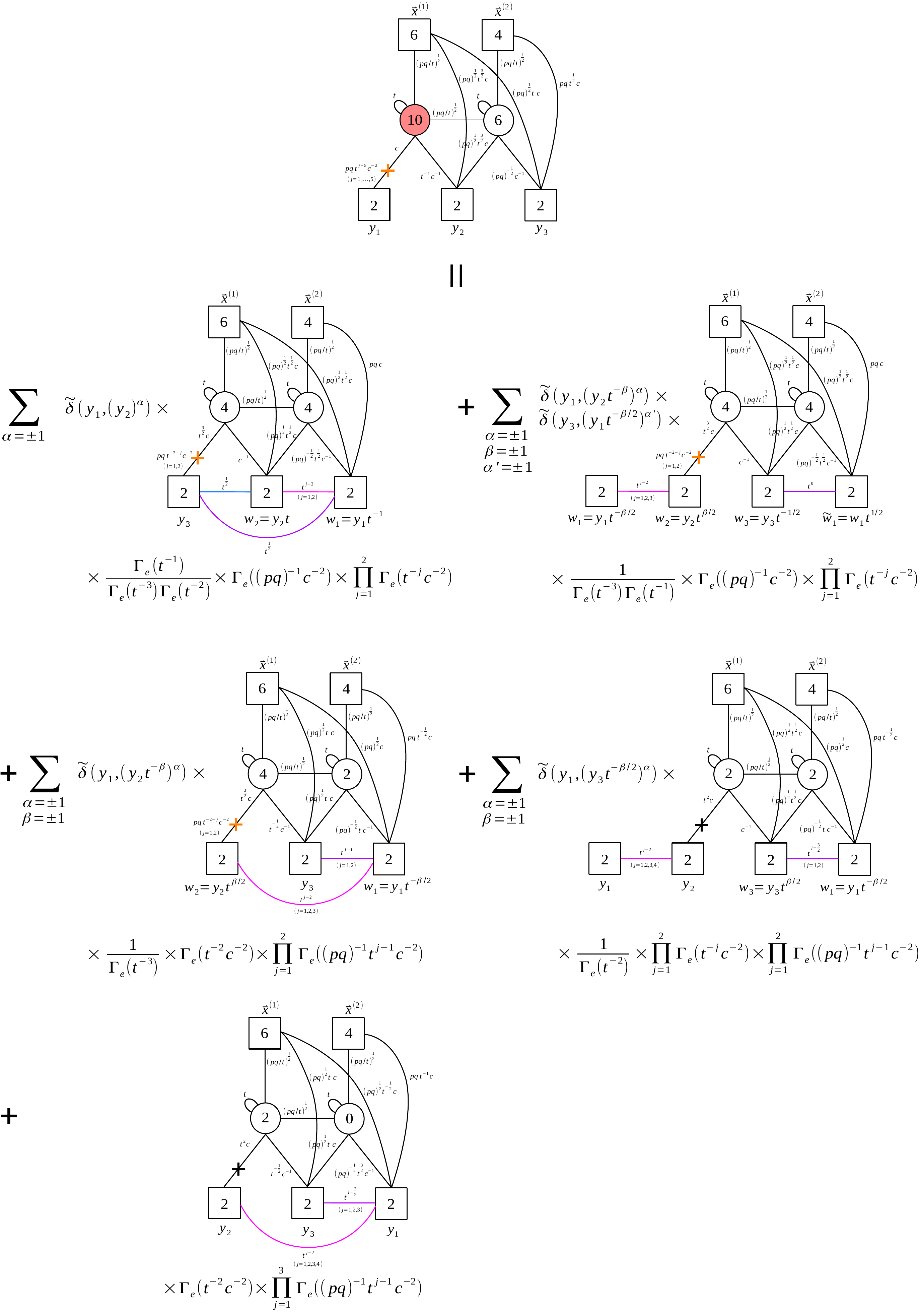}
	\caption{Summary of the result of the electric dualization of the 4d bad linear quiver ${E_{[0,4,3]}^{[2,2,1,1,1]}[USp(14)]}$.}
	\label{fig:allDuals_EE_4d}
\end{figure}

The final result is shown in Figure \ref{fig:allDuals_EE_4d} and its index is given by 
\begingroup\allowdisplaybreaks
\begin{align}\label{eq:index_EE_bad_example_Nc53_Nf32}
	& \mathcal{I}_{E_{[0,4,3]}^{[2,2,1,1,1]}[USp(14)]} = \nonumber\\[7pt]
	& \quad = 
	\sum_{\alpha=\pm 1}
    \left[
    \tilde\delta\left(y_1,\left(y_2\right)^{\alpha}\right)
    \times
	\frac{\Gamma_e\left(t^{-1}\right)}{\Gamma_e\left(t^{-3}\right)\Gamma_e\left(t^{-2}\right)}
	\Gamma_e\left((pq)^{-1}c^{-2}\right)
	\prod_{j=1}^{2}\Gamma_e\left(t^{-j}c^{-2}\right)
	\times\right.\nonumber\\
	& \qquad\qquad\quad\times
	\prod_{j=1}^{2} \Gamma_e\left(t^{j-2}w_1^{\pm}w_2^{\pm}\right)
	\times
	\Gamma_e\left(t^{\frac{1}{2}}w_1^{\pm}y_3^{\pm}\right)
	\times
	\Gamma_e\left(t^{\frac{1}{2}}w_2^{\pm}y_3^{\pm}\right)
	\times\nonumber\\
	& \qquad\qquad\!\left.\bigphantomspace
	\times
	\mathcal{I}_{E_{[3,2,2]}^{[2,2,1,1,1]}[USp(14)]}\left(\vec{x};\{y_3,w_2,w_1\};t;t^{\frac{3}{2}}c\right)
	\right]_{w_{1,2}=y_{1,2}t^{\mp1}}
	\nonumber\\
	& \qquad + 
	\sum_{\substack{\alpha=\pm 1 \\ \beta=\pm 1}}
	\sum_{\alpha'=\pm 1}
	\left[
    \tilde\delta\left(y_1,\left(y_2 t^{-\beta}\right)^{\alpha}\right)
    \times
    \tilde\delta\left(y_3,\left(y_1 t^{-\frac{1}{2}\beta}\right)^{\alpha'}\right)
    \times\bigphantomspace\right.\nonumber\\
    & \qquad\qquad\qquad\qquad\times
	\frac{1}{\Gamma_e\left(t^{-3}\right)\Gamma_e\left(t^{-1}\right)}
	\Gamma_e\left((pq)^{-1}c^{-2}\right)
	\prod_{j=1}^{2}\Gamma_e\left(t^{-j}c^{-2}\right)
	\times\nonumber\\
	& \qquad\qquad\qquad\qquad\times
	\prod_{j=1}^{3} \Gamma_e\left(t^{j-2}w_1^{\pm}w_2^{\pm}\right)
	\times
	\Gamma_e\left(\widetilde{w}_1^{\pm}w_3^{\pm}\right)
	\times\nonumber\\
	& \qquad\qquad\qquad\quad\!\left.\bigphantomspace
	\times
	\mathcal{I}_{E_{[3,2,2]}^{[2,2,1,1,1]}[USp(14)]}\left(\vec{x};\{w_2,w_3,\widetilde{w}_1\};t;t^{\frac{3}{2}}c\right)
	\right]_{\subalign{ & w_{1,2}=y_{1,2}t^{\mp\frac{1}{2}\beta} \\ & w_3=y_3t^{-\frac{1}{2}} \,,\, \widetilde{w}_1=w_1 t^{\frac{1}{2}}}}
	\nonumber\\
	& \qquad + 
	\sum_{\substack{\alpha=\pm 1 \\ \beta=\pm 1}}
	\left[
    \tilde\delta\left(y_1,\left(y_2 t^{-\beta}\right)^{\alpha}\right)
    \times
	\frac{1}{\Gamma_e\left(t^{-3}\right)}
	\Gamma_e\left(t^{-2}c^{-2}\right)
	\prod_{j=1}^{2}\Gamma_e\left((pq)^{-1}t^{j-1}c^{-2}\right)
	\times\right.\nonumber\\
	& \qquad\qquad\quad\times
	\prod_{j=1}^{3} \Gamma_e\left(t^{j-2}w_1^{\pm}w_2^{\pm}\right)
	\times
	\prod_{j=1}^{2} \Gamma_e\left(t^{j-1}w_1^{\pm}y_3^{\pm}\right)
	\times\nonumber\\
	& \qquad\qquad\!\left.\bigphantomspace
	\times
	\mathcal{I}_{E_{[3,3,1]}^{[2,2,1,1,1]}[USp(14)]}\left(\vec{x};\{w_2,y_3,w_1\};t;t^{\frac{3}{2}}c\right)
	\right]_{w_{1,2}=y_{1,2}t^{\mp\frac{1}{2}\beta}}
	\nonumber\\
	& \qquad + 
	\sum_{\substack{\alpha=\pm 1 \\ \beta=\pm 1}}
	\left[
    \tilde\delta\left(y_1,\left(y_3 t^{-\frac{1}{2}\beta}\right)^{\alpha}\right)
    \times
	\frac{1}{\Gamma_e\left(t^{-2}\right)}
	\prod_{j=1}^{2}\Gamma_e\left(t^{-j}c^{-2}\right)
	\prod_{j=1}^{2}\Gamma_e\left((pq)^{-1}t^{j-1}c^{-2}\right)
	\times\right.\nonumber\\
	& \qquad\qquad\quad\times
	\prod_{j=1}^{4} \Gamma_e\left(t^{j-2}y_1^{\pm}y_2^{\pm}\right)
	\times
	\prod_{j=1}^{2} \Gamma_e\left(t^{j-\frac{3}{2}}w_1^{\pm}w_3^{\pm}\right)
	\times\nonumber\\
	& \qquad\qquad\!\left.\bigphantomspace
	\times
	\mathcal{I}_{E_{[4,2,1]}^{[2,2,1,1,1]}[USp(14)]}\left(\vec{x};\{y_2,w_3,w_1\};t;t^{2}c\right)
	\right]_{w_{1,3}=y_{1,3}t^{\mp\frac{1}{2}\beta}}
	\nonumber\\
	& \qquad + 
	\Gamma_e\left(t^{-2}c^{-2}\right)
	\prod_{j=1}^{3}\Gamma_e\left((pq)^{-1}t^{j-1}c^{-2}\right)
	\times
	\prod_{j=1}^{4} \Gamma_e\left(t^{j-2}y_1^{\pm}y_2^{\pm}\right)
	\times
	\prod_{j=1}^{3} \Gamma_e\left(t^{j-\frac{3}{2}}y_1^{\pm}y_3^{\pm}\right)
	\times	\nonumber\\
	& \qquad\quad\times
	\mathcal{I}_{E_{[4,3,0]}^{[2,2,1,1,1]}[USp(14)]}\left(\vec{x};\{y_2,y_3,y_1\};t;t^{2}c\right)
	\,.
\end{align}
\endgroup

\clearpage
\subsection{The Higgsing perspective}\label{higgsing}
In this subsection we discuss how to recover the result for the index of a bad theory by studying Higgsings triggered by VEVs for suitable operators. We will follow the same strategy outlined in \cite{Giacomelli:2023zkk}, where it was shown how the constraint imposed by the delta functions that appear in the expression \eqref{eq:4d_bad_SQCD} for the index of the bad SQCD can be interpreted as encoding a VEV for a dressed meson which Higgses the theory down to the corresponding good SQCD. Here we will show how the same analysis can be applied to a bad quiver theory, by focusing on the example of the ${E_{[0,4,3]}^{[2,2,1,1,1]}[USp(14)]}$ theory. Although the analysis of \cite{Giacomelli:2023zkk} was carried out both at the level of the index and of the classical equations of motion, here we will only focus on the index perspective which is simpler especially for quiver theories.

\begin{figure}[h]
	\centering
	\includegraphics[width=0.27\textwidth]{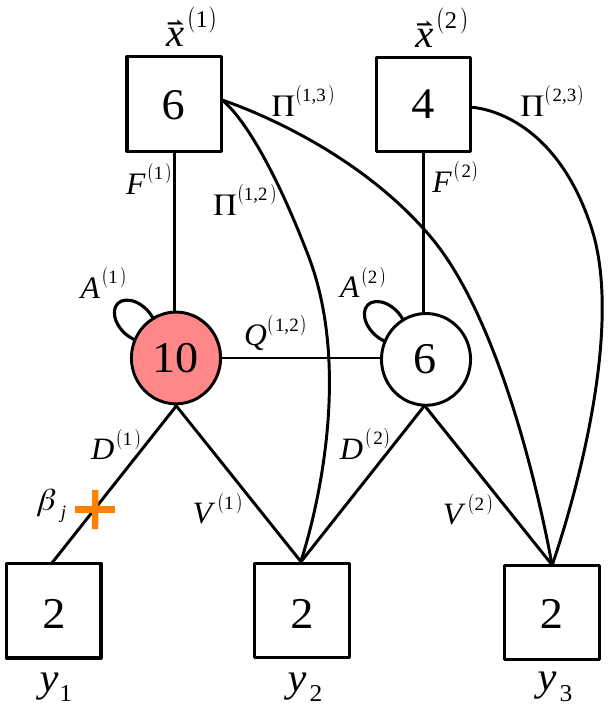}
	\caption{The naming convention of the fields in the $E_{[0,4,3]}^{[2,2,1,1,1]}$ quiver. }
	\label{fig:original_theory_4d_fields_names}
\end{figure}

We denote the various chiral fields of the theory as in Figure \ref{fig:original_theory_4d_fields_names}. Explicitly, the full index of the theory reads
\begin{align}\label{eq:indNc53Nf32}
&\mathcal{I}_{E_{[0,4,3]}^{[2,2,1,1,1]}[USp(14)]}(\vec{x}^{(1)},\vec{x}^{(2)};y_1,y_2,y_3;t;c) = \nn\\
&\times\prod_{j=1}^5\underbrace{\Gpq{pq\,t^{j-5}c^{-2}}}_{\gb_j}\prod_{\ga=1}^3\underbrace{\Gpq{(pq)^{\frac{1}{2}}t^{\frac{3}{2}}cx^{(1)\pm}_\ga y_2^{\pm}}}_{\pi^{(1,2)}}\underbrace{\Gpq{(pq)^{\frac{1}{2}}t\,cx^{(1)\pm}_\ga y_3^{\pm}}}_{\pi^{(1,3)}}\prod_{\gb=1}^2\underbrace{\Gpq{(pq)^{\frac{1}{2}}t^{\frac{1}{2}}cx^{(2)\pm}_\gb y_3^{\pm}}}_{\pi^{(2,3)}}\nn\\
&\times \oint \udl{\vec{z}^{(1)}_5}\udl{\vec{z}^{(2)}_3}\underbrace{\Gd_5(\vec{z}^{(1)};t)}_{\text{vec}^{(1)}+A^{(1)}}\underbrace{\Gd_3(\vec{z}^{(2)};t)}_{\text{vec}^{(2)}+A^{(2)}}\prod_{a=1}^5\prod_{i=1}^3\underbrace{\Gpq{(pq)^{\frac{1}{2}}t^{-\frac{1}{2}}z^{(1)\pm}_az^{(2)\pm}_i}}_{Q^{(1,2)}}\nn\\
&\times \prod_{a=1}^5\prod_{\ga=1}^3\underbrace{\Gpq{(pq)^{\frac{1}{2}}t^{-\frac{1}{2}}z^{(1)\pm}_ax^{(1)\pm}_\ga}}_{F^{(1)}}\prod_{i=1}^3\prod_{\gb=1}^2\underbrace{\Gpq{(pq)^{\frac{1}{2}}t^{-\frac{1}{2}}z^{(2)\pm}_ix^{(2)\pm}_\gb}}_{F^{(2)}}\nn\\
&\times \prod_{a=1}^5\underbrace{\Gpq{c\,z^{(1)\pm}_ay_1^\pm}}_{D^{(1)}}\underbrace{\Gpq{t^{-1}c^{-1}z^{(1)\pm}_ay_2^\pm}}_{V^{(1)}} \prod_{i=1}^3\underbrace{\Gpq{(pq)^{\frac{1}{2}}t^{\frac{3}{2}}c\,z^{(2)\pm}_iy_2^\pm}}_{D^{(2)}}\underbrace{\Gpq{(pq)^{-\frac{1}{2}}c^{-1}z^{(2)\pm}_iy_3^\pm}}_{V^{(2)}}\,,
\end{align}
where we are highlighting the contribution of each chiral field appearing in Figure \ref{fig:original_theory_4d_fields_names}.

\subsubsection*{Frame I}

We begin by showing how we can recover the theory (A) in Figure \ref{fig:EE_dualization_step0} via Higgsing.
Theory (A) comes with a delta constraint  on the index \eqref{eq:indNc53Nf32}\footnote{The case $y_1=y_2^{-1}$ works similarly.}
\begin{equation}\label{eq:constrA}
y_1=y_2\,.
\end{equation}
When this constraint is implemented we can see that the following combination of gamma functions:
\begin{equation}\label{eq:gammas}
\Gpq{c\,z^{(1)}_4y_1}\Gpq{t\,z^{(1)-1}_4z^{(1)}_5}\Gpq{t^{-1}c^{-1}z^{(1)-1}_5y_1^{-1}}
\end{equation}
provide sets of poles that pinch the integration contour of the variables $z^{(1)}_4,\,z^{(1)}_5$ at the points
\begin{equation}
z^{(1)}_4=c^{-1}y_1^{-1}\,,\qquad z^{(1)}_5=t^{-1}c^{-1}y_1^{-1}\,.
\end{equation}
Following \cite{Gaiotto:2012xa}, we then take the residue of the index at these poles.\footnote{There are also other combinations of gamma functions that lead to similar poles but for a different combination of gauge fugacities that corresponds to an action of the Weyl group of the $USp(10)$ gauge symmetry. Thus, the residues at these other poles give the same result, which compensates with the difference of the dimensions of the Weyl group that appear in the integration measure before and after the Higgsing.} This implements at the level of the index the Higgsing $USp(10)\to USp(6)$ of the left gauge node implemented by the VEV for the operator composed by the chirals whose index contribution is \eqref{eq:gammas}, which is a component of the dressed meson (see \cite{Bajeot:2023gyl,Giacomelli:2023zkk} for a more detailed analysis of similar VEVs)
\begin{equation}
\langle D^{(1)}A^{(1)}V^{(1)} \rangle \neq 0\,.
\end{equation}

After taking the residue and simplifying the contribution of massive fields (which corresponds to using the property $\Gpq{x}\Gpq{pq\,x^{-1}}=1$), we obtain exactly the index of the theory (A) in Figure \ref{fig:EE_dualization_step0} including the singlets, plus a divergent factor of $\Gpq{1}$ which has been argued in \cite{Giacomelli:2023zkk} to correspond to the delta function that we are evaluating at the origin when we impose the constraint \eqref{eq:constrA}. 

As discussed in the previous section, from this theory, one can then obtain the theory in the first frame of Figure \ref{fig:allDuals_EE_4d} by dualizing twice the ugly nodes. Since these dualizations do not introduce further delta functions, these flows are not associated to further Higgsings and we do not need to discuss them again here.  Indeed, as discussed in \cite{Giacomelli:2023zkk},  the frame with generic fugacities of the SQCD cannot be reached with a VEV since it is not associated to any delta.

\subsubsection*{Frames II and III}

Let us now consider the delta constraint
\begin{equation}
y_1=y_2 t^{-1}\,,
\end{equation}
which is associated to the theory (B) in Figure \ref{fig:EE_dualization_step0}. In this case, the corresponding VEV is
\begin{equation}\label{eq:VEV1}
\langle D^{(1)}\left(A^{(1)}\right)^2V^{(1)}\rangle\neq 0\,,
\end{equation}
which Higgses the left gauge node  of the theory in Figure \ref{fig:original_theory_4d_fields_names} as $USp(10)\to USp(4)$. Indeed, if we focus on the following combination of gamma functions:
\begin{equation}
\Gpq{c\,z^{(1)}_3y_1}\Gpq{t\,z^{(1)-1}_3z^{(1)}_4}\Gpq{t\,z^{(1)-1}_4z^{(1)}_5}\Gpq{t^{-1}c^{-1}z^{(1)-1}_5y_1^{-1}}
\end{equation}
we observe that their poles pinch the integration contour at the point
\begin{equation}
z^{(1)}_3=c^{-1}y_1^{-1}\,,\qquad z^{(1)}_4=t^{-1}c^{-1}y_1^{-1}\,,\qquad z^{(1)}_5=t^{-2}c^{-1}y_1^{-1}\,.
\end{equation}
After taking the residue, we obtain exactly the index of the theory (B) in Figure \ref{fig:EE_dualization_step0} including the singlets and a $\Gpq{1}$ corresponding to the delta function evaluated at the origin.

As shown in the previous section, at this point we need to dualize the bad node of the theory (B). This process produces the second and the third frames 
in Figure \ref{fig:allDuals_EE_4d}. The  second frame of Figure \ref{fig:allDuals_EE_4d} comes with an extra delta function
imposing the further constraint
\begin{equation}
y_3=y_1t^{-\frac{1}{2}}\,.
\end{equation}
This corresponds to the VEV 
\begin{equation}\label{eq:VEV2}
\langle D^{(2)}V^{(2)}\rangle \neq0\,,
\end{equation}
as it can be understood by looking at the gamma functions of the theory (B) in Figure \ref{fig:EE_dualization_step0}
\begin{equation}
\Gpq{(pq)^{\frac{1}{2}}t^{\frac{1}{2}}c\,z^{(2)-1}_3y_1^{-1}}\Gpq{(pq)^{-\frac{1}{2}}t^{-\frac{1}{2}}c^{-1}z^{(2)}_3y_1}\,,
\end{equation}
which provide two sets of poles that pinch the integration contour at the point
\begin{equation}
z^{(2)}_3=(pq)^{\frac{1}{2}}t^{\frac{1}{2}}c\,y_1^{-1}\,.
\end{equation}
Taking the residue at this pole implements the Higgsing $USp(6)\to USp(4)$ of the right gauge node and yields as a result the index of the theory in the second frame of Figure \ref{fig:allDuals_EE_4d}, including the singlets and a $\Gpq{1}^2$ corresponding to the two deltas evaluated at the origin.

We stress here that this second frame can be equivalently obtained by studying the two VEVs \eqref{eq:VEV1} and \eqref{eq:VEV2} in the opposite order, or in other words the RG flows triggered by them commute. This is because the Higgsing induced by the first VEV \eqref{eq:VEV1} does not affect the chiral fields involved in the second VEV \eqref{eq:VEV2}, as it can be understood from the fact that the charges of the fields $D^{(2)}$ and $V^{(2)}$ involved in the second VEV \eqref{eq:VEV2} are identical in the original theory in Figure \ref{fig:original_theory_4d_fields_names} and in the theory (B) of Figure \ref{fig:EE_dualization_step0} obtained after studying the first VEV \eqref{eq:VEV1}.

The third frame has no extra delta function and does not correspond to any further VEV so we do need to discuss it here.

\subsubsection*{Frame IV (and V)}

We conclude by discussing the Higgsing that leads to the fourth frame in Figure \ref{fig:allDuals_EE_4d}. This  case is special because it can be obtained by studying a VEV for a new type of operator that is not present in the SQCD analyzed in \cite{Giacomelli:2023zkk} and is only possible for quiver theories. 

The delta constraint leading to frame four is
\begin{equation}\label{eq:constrFIV}
y_1=y_2t^{-\frac{1}{2}}
\end{equation}
and corresponds to a VEV for one compontent of a ``long meson" constructed as a sequence of chirals that runs from one end of the quiver to the other
\begin{equation}
\langle D^{(1)}A^{(1)}Q^{(1,2)}V^{(2)}\rangle\neq 0\,.
\end{equation}
Accordingly, if we consider the combination of gamma functions
\begin{equation}
\Gpq{c\,z^{(1)}_4y_1}\Gpq{t\,z^{(1)-1}_4z^{(1)}_5}\Gpq{(pq)^{\frac{1}{2}}t^{-\frac{1}{2}}z^{(1)-1}_5z^{(2)}_3}\Gpq{(pq)^{-\frac{1}{2}}t^{-\frac{1}{2}}c^{-1}z^{(2)-1}_3y_1^{-1}}\,,
\end{equation}
these provide sets of poles that pinch the integration contour at the point
\begin{equation}
z^{(1)}_4=c^{-1}y_1^{-1}\,,\qquad z^{(1)}_5=t^{-1}c^{-1}y_1^{-1}\,,\qquad z^{(2)}_3=(pq)^{-\frac{1}{2}}t^{-\frac{1}{2}}c^{-1}y_1^{-1}\,.
\end{equation}

\begin{figure}[!ht]
	\centering
	\includegraphics[width=0.5\textwidth]{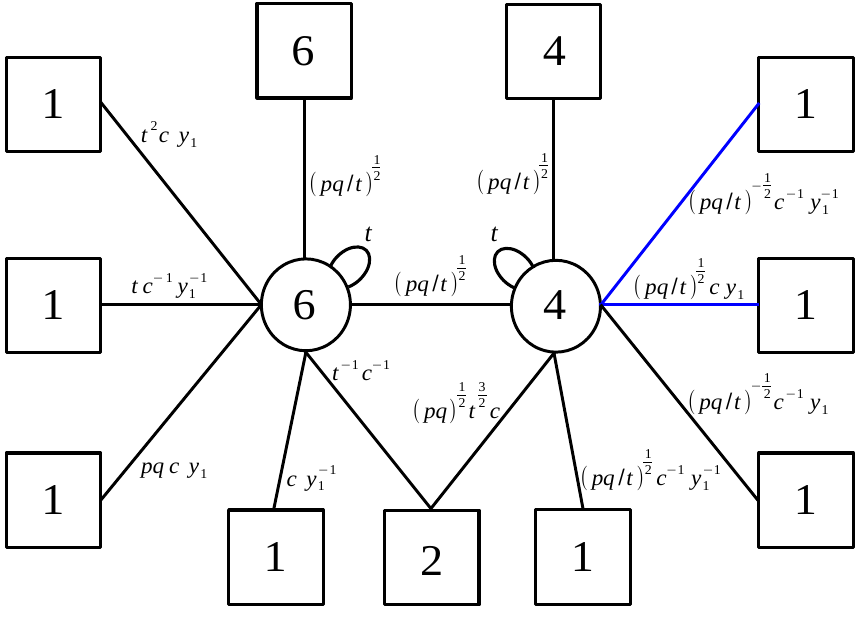}
	\caption{The intermediate theory obtained after Higgsing the linear quiver with $N_c=(5,3),\,N_f=(3,2)$ induced by a VEV for a long meson, where for simplicity we are omitting the gauge singlets. We color in blue the fields that still have a VEV. }
	\label{fig:higgsing_hybrid}
\end{figure}

Taking the residue at these poles leads to a theory that does not take the conventional form of an $E_\rho^\sigma$ theory. Such a theory is summarized in Figure \ref{fig:higgsing_hybrid}, where for simplicity we are omitting the gauge singlets. In this theory there are still some fields that are taking a VEV, which we are denoting in blue in the picture. Indeed, their contribution to the index is given by
\begin{equation}
\Gpq{(pq)^{\frac{1}{2}}t^{-\frac{1}{2}}c\,z^{(2)}_2y_1}\Gpq{(pq)^{-\frac{1}{2}}t^{\frac{1}{2}}c^{-1}z^{(2)-1}_2y_1^{-1}}
\end{equation}
and these gamma functions provide two sets of poles that pinch the integration contour at the point
\begin{equation}
z^{(2)}_2=(pq)^{-\frac{1}{2}}t^{\frac{1}{2}}c^{-1}y_1^{-1}\,.
\end{equation}
Taking the residue at this pole implements the Higgsing $USp(4)\to USp(2)$ of the right gauge node. 

\begin{figure}[!ht]
	\centering
	\includegraphics[width=0.4\textwidth]{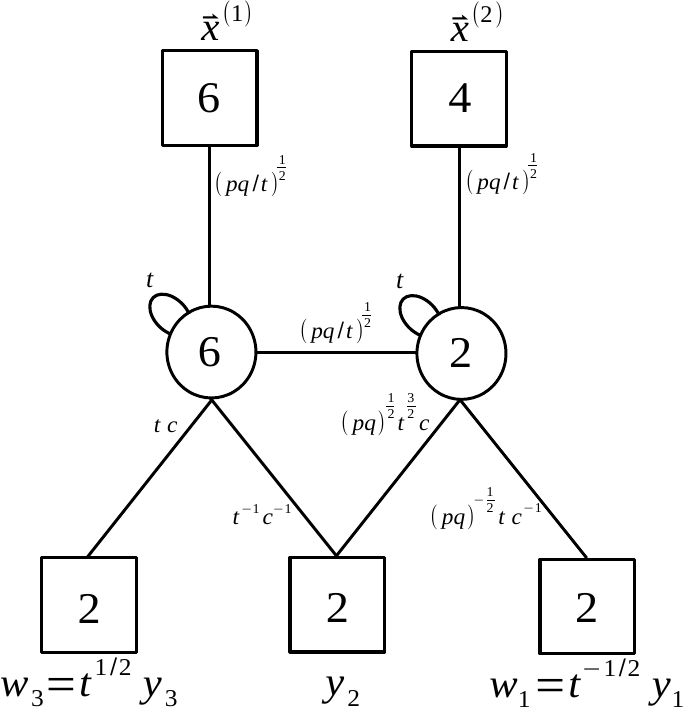}
	\caption{The theory resulting after the second Higgsing of the linear quiver with $N_c=(5,3),\,N_f=(3,2)$, where for simplicity we are omitting the gauge singlets. }
	\label{fig:higgsing_higgsing_hybrid}
\end{figure}

The result of this Higgsing is now an $E_\rho^\sigma$ theory, which we depict in Figure \ref{fig:higgsing_higgsing_hybrid} omitting again the gauge singlet fields for simplicity. However, this is still a bad theory, since the left node is effectively an SQCD with $N_c=3,\,N_f=4$.
When we dualize the bad node we produce two frames, one with an extra delta and one without the extra delta.

Let's focus on the second one which reduces the rank of this node to $N_c=1$. Keeping track of all the singlet fields, one recovers the theory in the fourth frame of Figure \ref{fig:allDuals_EE_4d} including a $\Gpq{1}$ corresponding to the delta function evaluated at the origin.

The other frame, with extra delta,  would lead to an additional frame for the bad quiver $N_c=(5,3),\,N_f=(3,2)$, which corresponds to a singularity of higher order since it would carry two delta functions. However, we can argue that such frame does not correspond to any extra singularity in the index of the bad quiver $N_c=(5,3),\,N_f=(3,2)$. One way to see this is to apply the identity of the partition functions corresponding to such possible additional frame to the theory in Figure \ref{fig:higgsing_higgsing_hybrid}. If we do so, we notice that among the singlet fields there is a contribution of the form $\Gpq{pq}$ that cancels the divergence of the extra delta function. Hence, there is no extra singularity when the constraint due to the second gamma function is imposed. Equivalently, one could study the Higgsing of the theory in Figure \ref{fig:higgsing_higgsing_hybrid} after imposing the constraint of this second gamma function and notice that in the result one only finds $\Gpq{1}$ rather than $\Gpq{1}^2$ as expected for a frame with two delta functions. We stress that for this to happen it is crucial that we introduce the redefined fugacities $w_i$ and reconstruct the free fields in the bifundamental representation of pairs of the corresponding $SU(2)_{w_i}$ symmetries, since one of these fields carries a $\Gpq{1}$. In other words, such a divergent contribution should be associated to one of the free fields and not to an extra delta singularity.

An alternative way to understand the absence of this additional extra frame is to study a different RG flow to reach the theory in the fourth frame of Figure \ref{fig:allDuals_EE_4d}, which follows more closely the electric dualization we performed in the previous subsection. Indeed, in our previous derivation we reached this frame starting from theory (C) in Figure \ref{fig:EE_dualization_step0}, which  is the generic fugacity frame with no delta which does not correspond to any VEV.
 However, as we discussed previously for the third frame, the RG flows that we are studying are expected to commute. 
 
 Hence, we can start from  theory (C) in Figure \ref{fig:EE_dualization_step0} and then  impose the constraint \eqref{eq:constrFIV}. This corresponds to turning on a VEV for a meson of the right node dressed once
\begin{equation}
\langle D^{(2)}A^{(2)}V^{(2)}\rangle\neq 0	\,.
\end{equation}
Indeed, the corresponding combination of gamma functions after imposing the constraint
\begin{equation}
\Gpq{(pq)^{\frac{1}{2}}t^{-\frac{1}{2}}c\,z^{(2)-1}_2y_1}\Gpq{t\,z^{(2)}_2z^{(2)-1}_3}\Gpq{(pq)^{-\frac{1}{2}}t^{-\frac{1}{2}}c^{-1}z^{(2)}_3y_1^{-1}}
\end{equation}
provides sets of poles that pinch the integration contour at the points
\begin{equation}
z^{(2)}_2=(pq)^{\frac{1}{2}}t^{-\frac{1}{2}}c\,y_1\,,\qquad z^{(2)}_3=(pq)^{\frac{1}{2}}t^{\frac{1}{2}}c\,y_1\,.
\end{equation}
Taking the residue at such poles implements the Higgsing $USp(6)\to USp(2)$ of the right gauge node triggered by the VEV for the dressed meson. The result is exactly the fourth frame in \ref{fig:allDuals_EE_4d}, including the singlets and a $\Gpq{1}$ corresponding to the delta function evaluated at the origin, as expected. This alternative derivation makes it clear that the previously mentioned possible extra frame is not actually present.

Finally, the fifth frame in Figure \ref{fig:allDuals_EE_4d} cannot be reached via Higgsing, since it is the generic fugacity frame
with no delta and  does not correspond to any VEV so we do not need to discuss it here.

\section{The algorithms in 3d and branes}\label{braneint}

In this section we  discuss how the 3d QFT blocks swap moves are realized in the brane picture. Using these results, one can also find the brane interpretation of each step of the 3d mirror and electric  dualization  algorithm.
In particular we will consider a specific example with $N_c = (3,3)$ and $N_f = (1,4)$ and explain the brane configurations realizing the mirror and electric dualization of this quiver theory.

\subsection{The brane interpretation of the swaps of the QFT blocks}
\label{sec:brane}
We begin this section by giving a brane interpretation of the dualities swapping pairs of  3d QFT blocks discussed in Section \ref{sec:QFT_Ingredients}.
The $(1,0)$-$(0,1)$ swap in 3d, as explained in \cite{Giacomelli:2023zkk}, is nothing but the Hanany--Witten brane transition effect \cite{Hanany:1996ie}, which generates or annihilates a D3-brane in between when NS5 and D5-branes cross each other.\\

\begin{figure}[!ht]
	\centering
	\includegraphics[width=\textwidth]{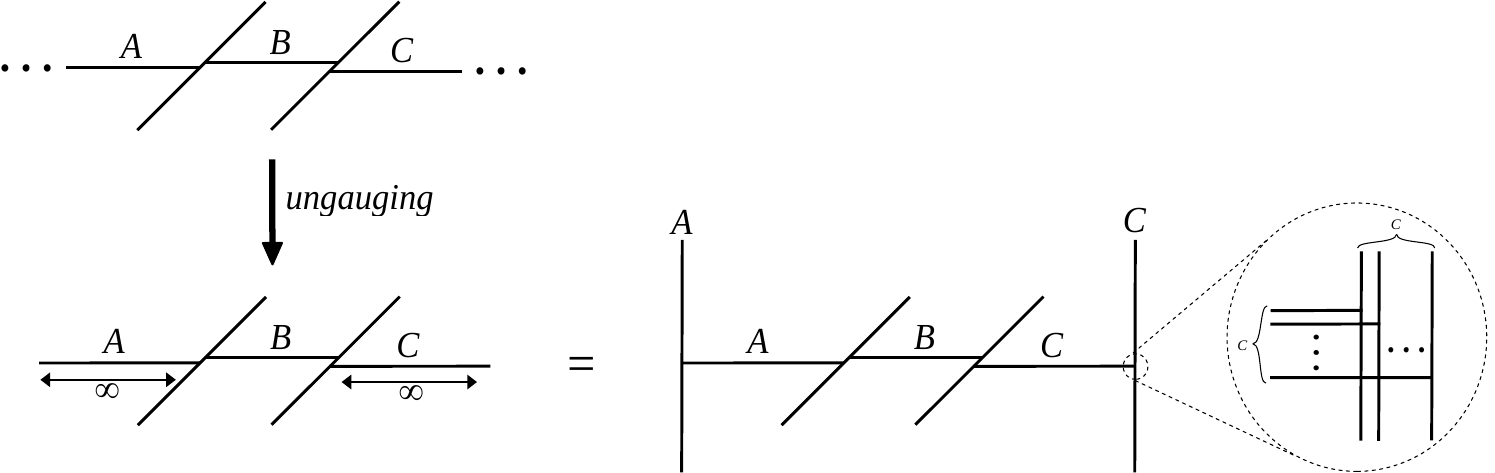}
	\caption{The brane setup corresponding to two consecutive 3d $(0,1)$-blocks and their decoupling from a quiver by ungauging.  The letters $A$ and $C$ over the vertical lines denote the number of NS5-branes while the letter placed abobe the horizontal lines denote the number of D3's.  The zoomed sector on the right displays how the D3's are attached to the stack of NS5's.}
	\label{fig:branes_two_01blocks_ungauging}
\end{figure}

Let's consider the $(0,1)$-$(0,1)$ swap in \eqref{eq:D5_swap_3d}, or equivalently in Figure \ref{fig:B01_swap_3d_V1}, which is one of the main ingredients in the mirror dualization algorithm. From the Type IIB string theory perspective, the 3d $(0,1)$-block can be realized as a D5-brane sandwiched between two stacks of D3-branes. Two consecutive $(0,1)$-blocks appearing in a generic quiver correspond to the brane configuration  shown in the first line of Figure \ref{fig:branes_two_01blocks_ungauging}.
In this set-up, we can let  the left and the right D3s stretch infinitely with the  effect of ungauging the left-most and the right-most gauge nodes of the two $(0,1)$-blocks in the field theory language. Then we can replace the set of  infinitely long D3's by a set of D3s ending on the same number of NS5-branes as shown in the second line of Figure \ref{fig:branes_two_01blocks_ungauging}, which is the brane set-up corresponding to the left hand side of the $(0,1)$-$(0,1)$ swap in Figure \ref{fig:B01_swap_3d_V1}.

As we argued from the field theory perspective in Section \ref{sec:B01-swap}, this is nothing but the brane configuration realizing the mirror of an SQCD. In particular, if $B-C > A-B$, the corresponding SQCD is a bad/ugly theory, which has multiple mirror duals depending on the FI parameter as we have studied in Part I \cite{Giacomelli:2023zkk}. In Part I, we have learned that each dual frame can be determined at the brane level by the number of decoupled D3's escaping to infinity along the parallel direction of two D5-branes. The same idea applies here, and one can find the brane picture of the 3d $(0,1)$-$(0,1)$ swap.\\

\begin{figure}[!ht]
	\centering
	\includegraphics[width=\textwidth]{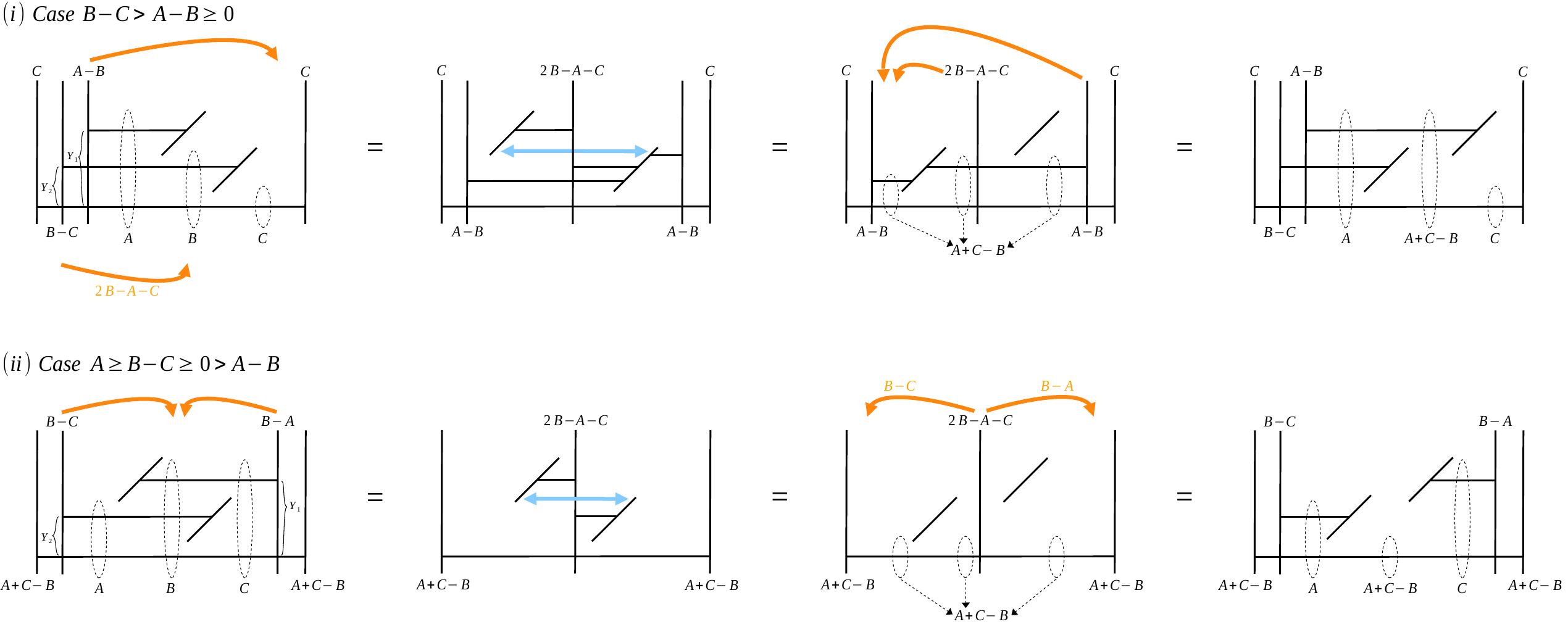}
	\caption{The brane interpretation of the swapping of two 3d $(0,1)$-blocks when $Y_1-Y_2 \neq 0$. The orange arrows indicate the rearrangement of NS5-branes, performed via $(1,0)$-$(0,1)$ swaps that correspond to HW moves. The light blue arrows indicate the swap of two D5-branes. The labels under dashed ellipses denote the number of D3-branes.}
	\label{fig:branes_01swap_nonZeroFI}
\end{figure}

First let us consider the brane configurations in Figure \ref{fig:branes_01swap_nonZeroFI}, where we allow generic vertical positions of two D5-branes, denoted by $Y_1$ and $Y_2$, which are real masses for the flavors associated with the D5's in the field theory description.
The difference between the two real masses, $Y_1-Y_2$, corresponds to the FI parameter in the original SQCD.

\begin{enumerate}
\item[(i)] Let's consider first the $B-C > A-B \geq 0$ case (the $U(B)$ node is bad/ugly since $2 B > A+C$).
To read the gauge theory describing the low energy dynamics, we should move D5's in such a way that the number of D3's ending on each side of them is the same; namely, each D5 should have vanishing net D3 number. 
For this purpose, out of the $A$ NS5-branes on the left, we first move $A-B$ of them to the right of the two D5's and we bring $2 B-A-C$ of them in between the two D5's, obtaining the second configuration in Figure \ref{fig:branes_01swap_nonZeroFI}. Now one can see that once we swap the two D5's, passing through the set of NS5's in between, the D3-branes suspended between the D5's and the NS5's in the middle are completely annihilated, and we obtain the third configuration, where every D5-branes has the same number of D3's on both sides. Indeed, such configuration is unique if $Y_1 - Y_2 \neq 0$ and leads to a quiver gauge description that is the mirror dual of the $U(A+C-B)$ SQCD with $A+C$ flavors.

Moreover, once we move the NS5-branes back to the original location, we obtain the fourth configuration, from which the last term on the right hand side of Figure \ref{fig:B01_swap_3d_V1} can be read. Note that this configuration can be directly obtained from the first configuration by swapping the two D5-branes without a collision because they are separated along the vertical direction. 

\item[(ii)] Similarly, when $A \geq B-C \geq 0 > A-B$ (where again the $U(B)$ node is bad/ugly since $2 B > A+C$) we can move $B-C$ NS5's from the left and $B-A$ from the right to the interval in between the two  D5's and swap the two D5-branes to obtain a configuration with a vanishing net D3 numbers for the two D5-branes. Once we move the NS5-branes back to the original location, we obtain the fourth configuration, which can also be obtained from the first configuration simply by swapping the two D5-branes.
\end{enumerate}
The case $0 > B-C > A-B$ is equivalent to the first case once $A$ and $C$ are exchanged.
The last configuration compatible with the bad node condition $2 B > A+C$ is $B-C > A$, which however requires $Y_1-Y_2 = 0$ that we are about to discuss.\\

If $Y_1-Y_2 = 0$, the configurations with vanishing net D3 numbers for D5-branes are not unique, which is why there are multiple dual frames when the FI parameter of the original SQCD vanishes.\footnote{In order to compare with the brane set-up, one should actually look at the deformation parameters of the vortex partition function on $\mathbb{R}^2_b\times S^1$ rather than the $S^3_b$ partition function that we used so far. The former can be obtained from factorization of the latter \cite{Pasquetti:2011fj,Beem:2012mb}, where the omega deformation parameter $b$ is related to the squashing parameter $Q=b+b^{-1}$. The delta constraint appearing in the vortex partition function when the deformation parameters are turned off $b=m_A=0$ reads $Y_1-Y_2 = 0$, which is what we consider in the brane set-up.} In this case, we can't naively swap the D5-branes due to their collision. Instead, now D3-branes can be suspended between the two D5-branes since the D5's are placed at the same vertical location. Then some of these D3-branes can move along the D5's and escape to infinity, allowing multiple possibilities for brane configurations with vanishing net D3 numbers for D5-branes. Such escaping D3's correspond to decoupled (twisted) hypers in the field theory description, and the remaining brane configuration provides the interacting fixed point in the IR.

\begin{figure}[!ht]
	\centering
	\includegraphics[width=\textwidth]{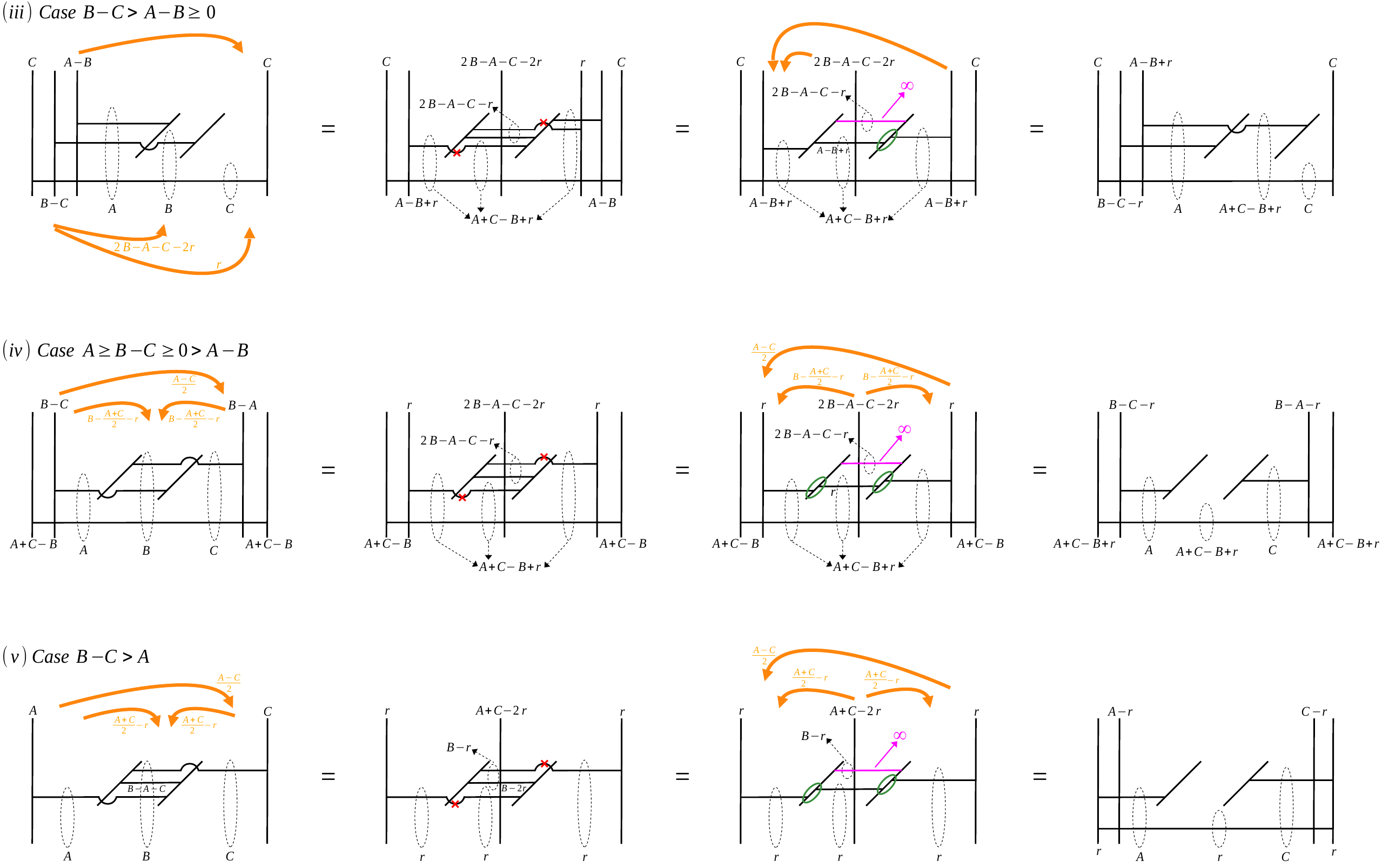}
	\caption{The brane interpretation of the swapping of two 3d $(0,1)$-blocks when $Y_1-Y_2 = 0$. The orange arrows indicate the rearrangement of NS5-branes, performed via $(1,0)$-$(0,1)$ swaps that correspond to HW moves. The red crosses represent the splitting of a D3-brane into two segments, while the green circles represent the combination of two D3 segments. The pink arrows indicate the operation of sending a stack of D3-branes to infinity. The labels under dashed ellipses indicate the number of D3-branes there contained.}
	\label{fig:branes_01swap_zeroFI}
\end{figure}

Let us look at the brane picture in Figure \ref{fig:branes_01swap_zeroFI}, where we place the two D5-branes at the same vertical location.

\begin{enumerate}
	\item[(iii)] Consider first the  $B-C > A-B \geq 0$ case. Among the $A$  NS5-branes on the left, we move $A-B+r$ of them to the right of the two D5's and $2 B-A-C-2 r$ in between the two D5's, obtaining the second configuration in Figure \ref{fig:branes_01swap_zeroFI}. Since a D3-brane crossing over a D5 can be split into two pieces terminating on each side of the D5, the second configuration is equivalent to the third configuration, where $2 B-A-C-r$ D3-branes suspended between the two D5's can be moved infinitely far away along the D5's. Eventually, every D5-brane in the remaining configuration has the same number of D3's on each side as desired and provides a quiver gauge theory describing the low energy physics, which is the mirror dual of the $U(A+C-B+r)$ SQCD with $A+C$ flavors. One can easily read off the maximum value of $r$ by requiring the number of D3-branes suspended between the two D5-branes to be non-negative; thus, $r \leq \floor*{\frac{2 B-A-C}{2}}$, i.e., $A+C-B+r \leq \floor*{\frac{A+C}{2}}$. Finally, once we move the NS5-branes back to the original location, we obtain the fourth configuration, exactly providing the terms on the right hand side of Figure \ref{fig:B01_swap_3d_V1} accompanied by the delta constraints for $r = 1, \dots, \floor*{\frac{2 B-A-C}{2}}$. In fact, one can also consider the configurations corresponding to $r = -(A+C-B), \dots, 0$ regarding negative $r$ as moving the branes in the other direction. However, these configurations do not need to be counted separately because they are obtained as limits of the configuration with $Y_1-Y_2 \neq 0$ that we have already considered.

Interestingly, before the decoupling of D3's, the left D5-brane is originally connected to NS5-branes on the right by D3-branes while the right D5-brane is connected to NS5-branes on the left (see the second configuration in Figure \ref{fig:branes_01swap_zeroFI}). On the other hand, after the decoupling of D3's, the left D5-brane is now connected to NS5-branes on the left while the right D5-brane is connected to NS5-branes on the right. In this way, although there is no real swap of the D5's, the roles of the two D5-branes are interchanged after the decoupling of D3's. This is similar to the case with $Y_1-Y_2 \neq 0$, where the two D5's are actually swapped. Indeed, if we take $r = 0$ in the fourth configuration, it looks exactly the same as the $Y_1-Y_2 \neq 0$ configuration with swapped D5's, apart from the fact that they now have the same vertical position. This is consistent with what we observe on the field theory side, where the fugacities of two $(0,1)$-blocks are exchanged after the dualization, which is why we call this duality move the $(0,1)$-$(0,1)$ swap.

\item[(iv-v)] The discussion for the other two cases,  $A \geq B-C \geq 0 > A-B$ and  $B-C>A$ (which now is allowed because the D5-branes are at the same  height) is similar.
\end{enumerate}

\begin{figure}[!ht]
	\centering
	\includegraphics[width=\textwidth]{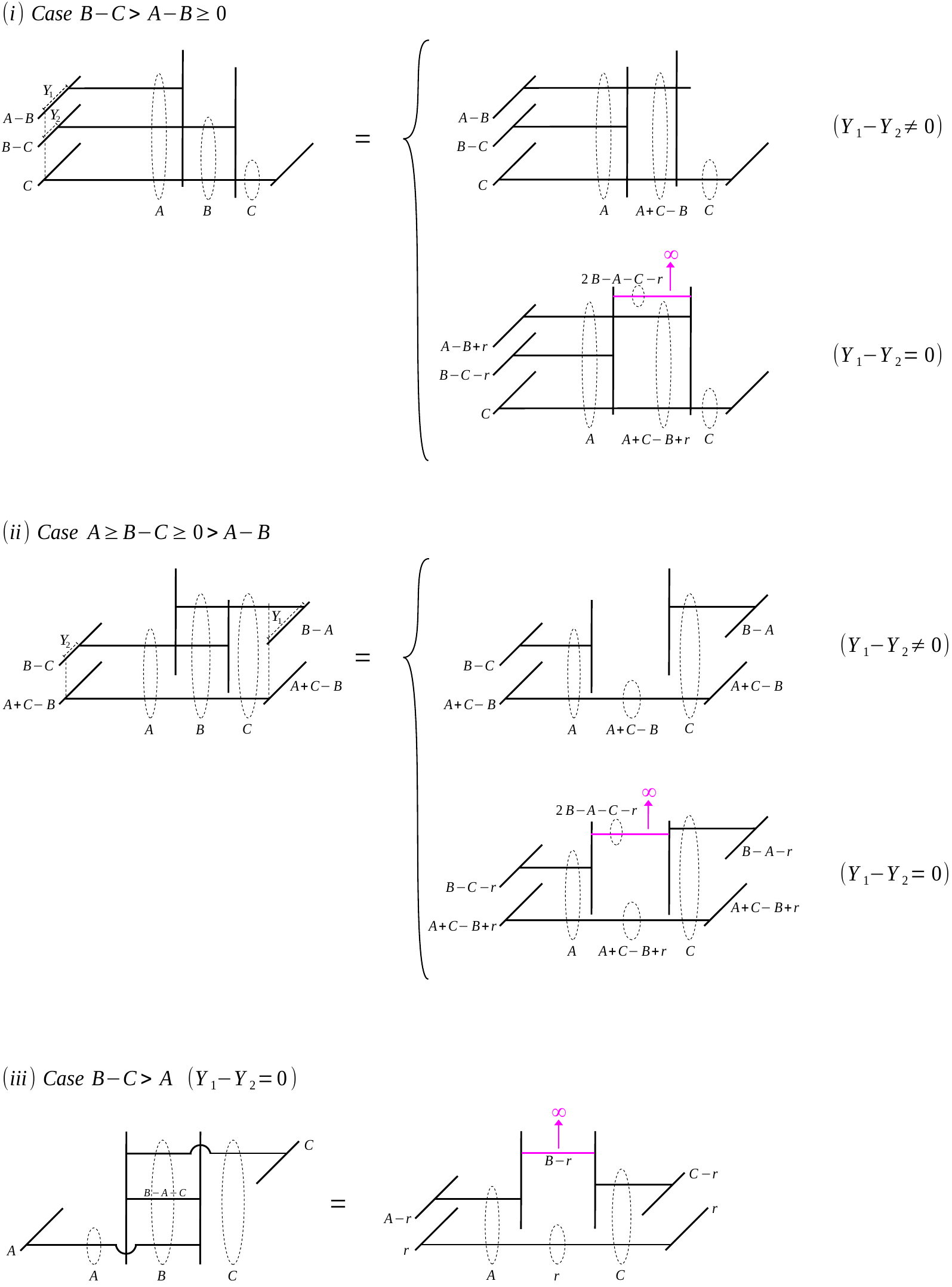}
	\caption{The brane interpretation of the swapping of two 3d $(1,0)$-blocks.}
	\label{fig:branes_10swap}
\end{figure}

The brane interpretation of the $(1,0)$-$(1,0)$ swap can be obtained as shown in Figure \ref{fig:branes_10swap}
via the  $S$-dualization of the brane configuration for the $(0,1)$-$(0,1)$ swap discussed above. 
If the FI parameter is nonzero, we can simply swap the NS5-branes, leading to the two consecutive $(1,0)$-blocks with the middle gauge node of rank $A+C-B$. If the FI parameter is zero, some of the D3's can escape to infinity, and the remaining configuration corresponds to the $U(A+C-B+r)$ SQCD with $A+C$ flavors, which is equivalent to the two consecutive $(1,0)$-blocks with the middle gauge node of rank $A+C-B+r$.

\subsection{The 3d mirror algorithm and the $(0,1)$-$(0,1)$ swap}

In this subsection, we discuss the mirror  dualization of the 3d $\mathcal N=4$ quiver example with $N_c = (3,3)$ and $N_f = (1,4)$ and the corresponding brane interpretation, especially using the result for the $(0,1)$-$(0,1)$ swap obtained in the previous subsection. 

\begin{landscape}
\begin{figure}[p]
	\centering
	\includegraphics[scale=0.28,center]{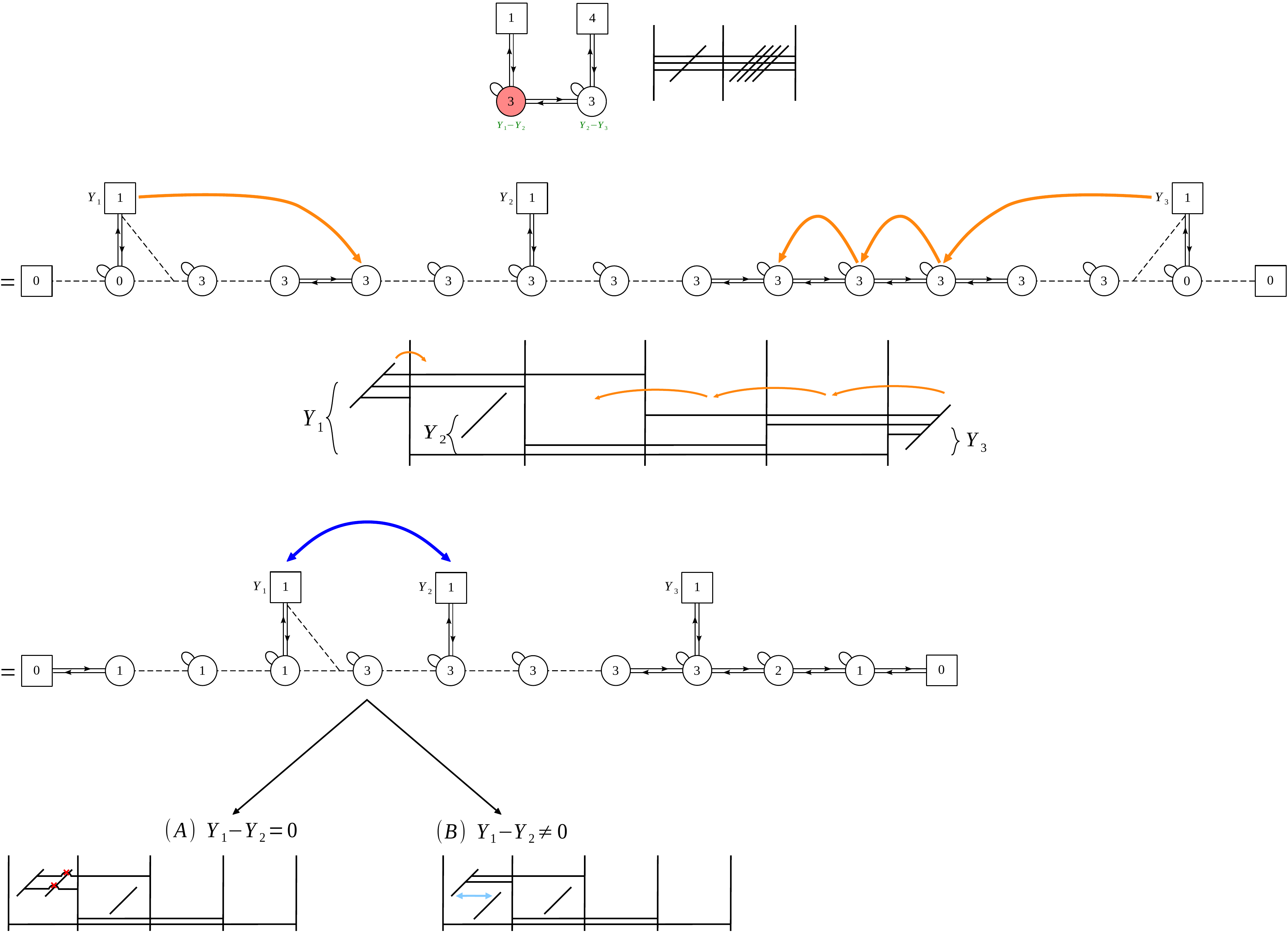}
	\caption{The brane interpretation of the mirror dualization of the quiver gauge theory with $N_c = (3,3)$ and $N_f = (1,4)$. The initial brane configuration and the subsequent steps before swapping the $(0,1)$-blocks are illustrated.}
	\label{fig:3d_EM_branes_example1}
\end{figure}
\end{landscape}

The initial brane set-up for the example is shown in the first line of Figure \ref{fig:3d_EM_branes_example1}. One can decompose the given quiver into the basic QFT blocks (see the second line), $S$-dualize each of them and recombine them (see the third line). The corresponding brane configuration is also given below the recombined quvier. As usual, one has to move D5-branes in such a way that the net number of D3's attached to each D5 vanishes. For this purpose, first we need to carry out a series of $(1,0)$-$(0,1)$ swaps as shown in the figure. In good theories, the $(1,0)$-$(0,1)$ swaps are enough to reach the frame where each D5 has zero net number of D3-branes, from which we can easily read out the corresponding quiver gauge theory. However, for bad theories, the $(1,0)$-$(0,1)$ swaps alone are not able to take us to such (good) frames regardless of which sequence of the $(1,0)$-$(0,1)$ swaps we perform. Indeed, the last quiver in Figure \ref{fig:3d_EM_branes_example1} still has an asymmetric $S$-wall associated to a D5-brane with non-zero net number of D3-branes. Hence, we need the $(0,1)$-$(0,1)$ swap as well.

As we argued in the previous subsection, the $(0,1)$-$(0,1)$ swap is realized in the brane set-up in two different ways,
(A) and (B),  depending on the relative real mass, i.e.~the vertical distance between the two D5-branes associated with the $(0,1)$-blocks (see the last brane configurations in Figure \ref{fig:3d_EM_branes_example1}). \\

\begin{figure}[p]
	\centering
	\includegraphics[width=\textwidth]{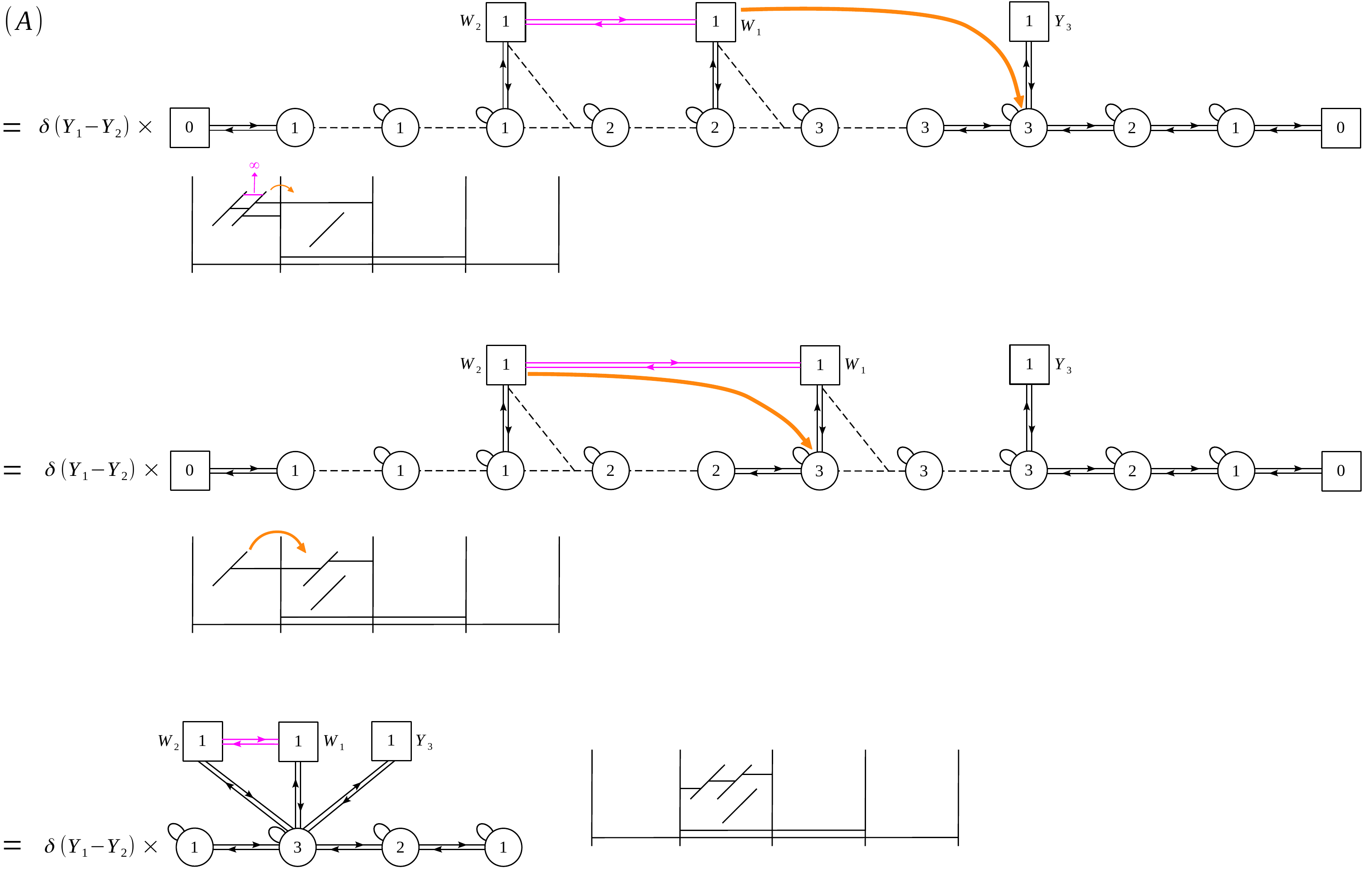}
	\caption{The brane interpretation of the mirror dualization of the quiver gauge theory with $N_c = (3,3)$ and $N_f = (1,4)$. The brane configurations after swapping the $(0,1)$-blocks with $Y_1-Y_2 = 0$ are illustrated.}
	\label{fig:3d_EM_branes_example1_A}
\end{figure}

In the case (A) with two D5's having the same vertical location, i.e.~$Y_1-Y_2 = 0$, the $(0,1)$-$(0,1)$ swap corresponds to splitting a D3-brane passing through a D5 into two pieces: one suspended between the two D5's and the other suspended between the D5 and the NS5. Subsequently, the D3 segment suspended between the two D5's escapes to infinity and contributes a decoupled singlet to the field theory. This mechanism is illustrated in Figure \ref{fig:3d_EM_branes_example1_A}, where the decoupled D3 is colored in pink. We can proceed by performing a couple more $(1,0)$-$(0,1)$ swaps as shown in the figure and obtain the final dual quiver in the last line. Remember that this brane manipulation is available because $Y_1-Y_2 = 0$, which is consistent with the field theory result that the corresponding dual frame is accompanied by the same delta constraint $Y_1-Y_2 = 0$ on the FI parameters. The manipulations that we have just described are the counterparts at the level of the branes of the field theory flow (A) in Figure \ref{fig:EMdualization_step2A_example1}.\\

\begin{figure}[p]
	\centering
	\includegraphics[width=\textwidth]{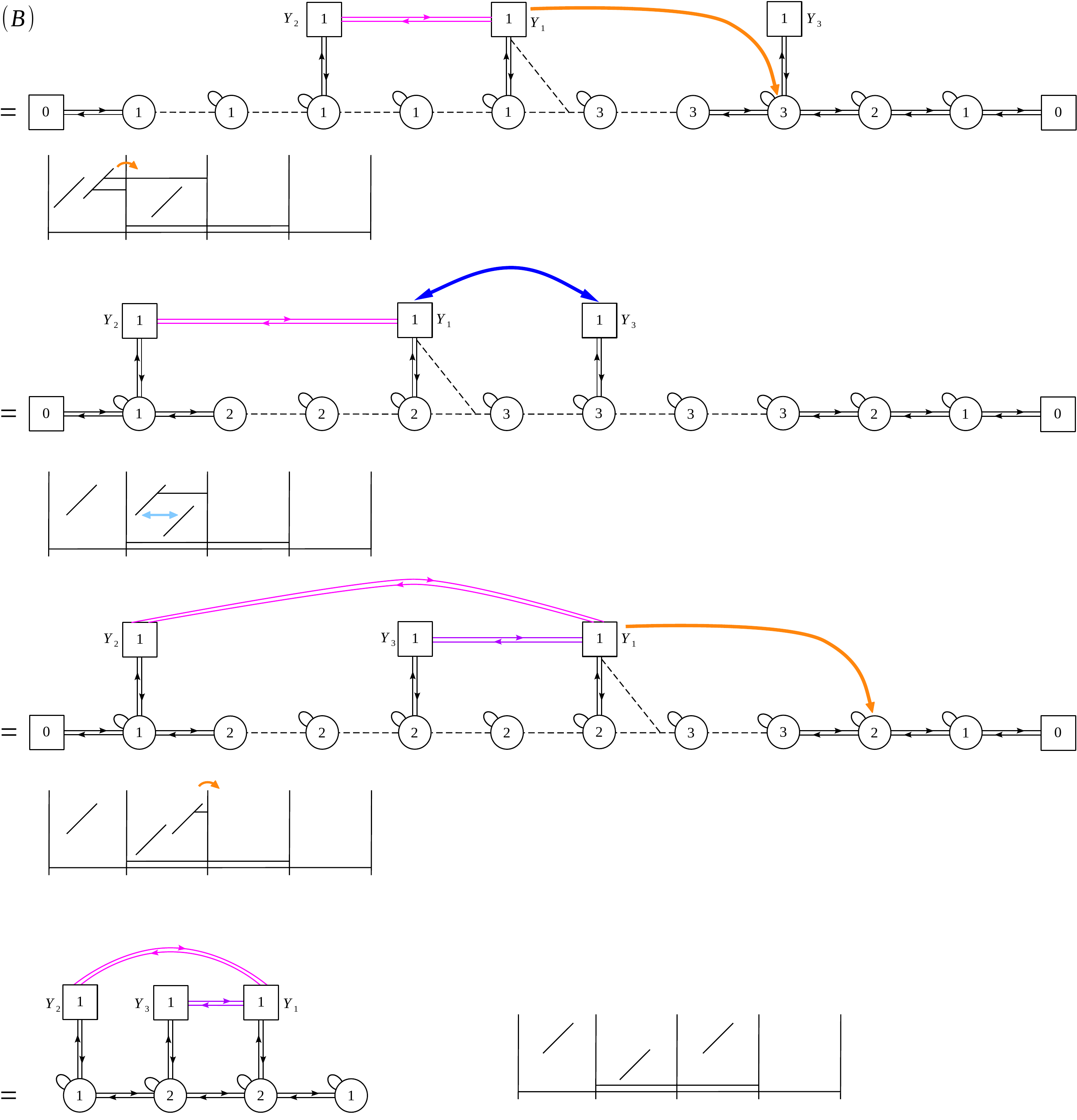}
	\caption{The brane interpretation of the mirror dualization of the quiver gauge theory with $N_c = (3,3)$ and $N_f = (1,4)$. The brane configurations after swapping the $(0,1)$-blocks with $Y_1-Y_2 \neq 0$ are illustrated.}
	\label{fig:3d_EM_branes_example1_B}
\end{figure}

In the case (B), where the two D5's are separated along the vertical direction, i.e.~$Y_1 - Y_2 \neq 0$, the corresponding $(0,1)$-$(0,1)$ swap is indeed realized by swapping the two D5's avoiding any collisions (see Figure \ref{fig:3d_EM_branes_example1_B}). Then we can further perform the extra steps of the $(1,0)$-$(0,1)$ swap and the $(0,1)$-$(0,1)$ swap displayed in the figure to obtain the final quiver in the last line. The manipulations that we have just described are the counterparts at the level of the branes of the field theory flow (B) in Figure \ref{fig:EMdualization_step2B_example1}.\\

\begin{figure}
	\centering
	\includegraphics[width=\textwidth,center]{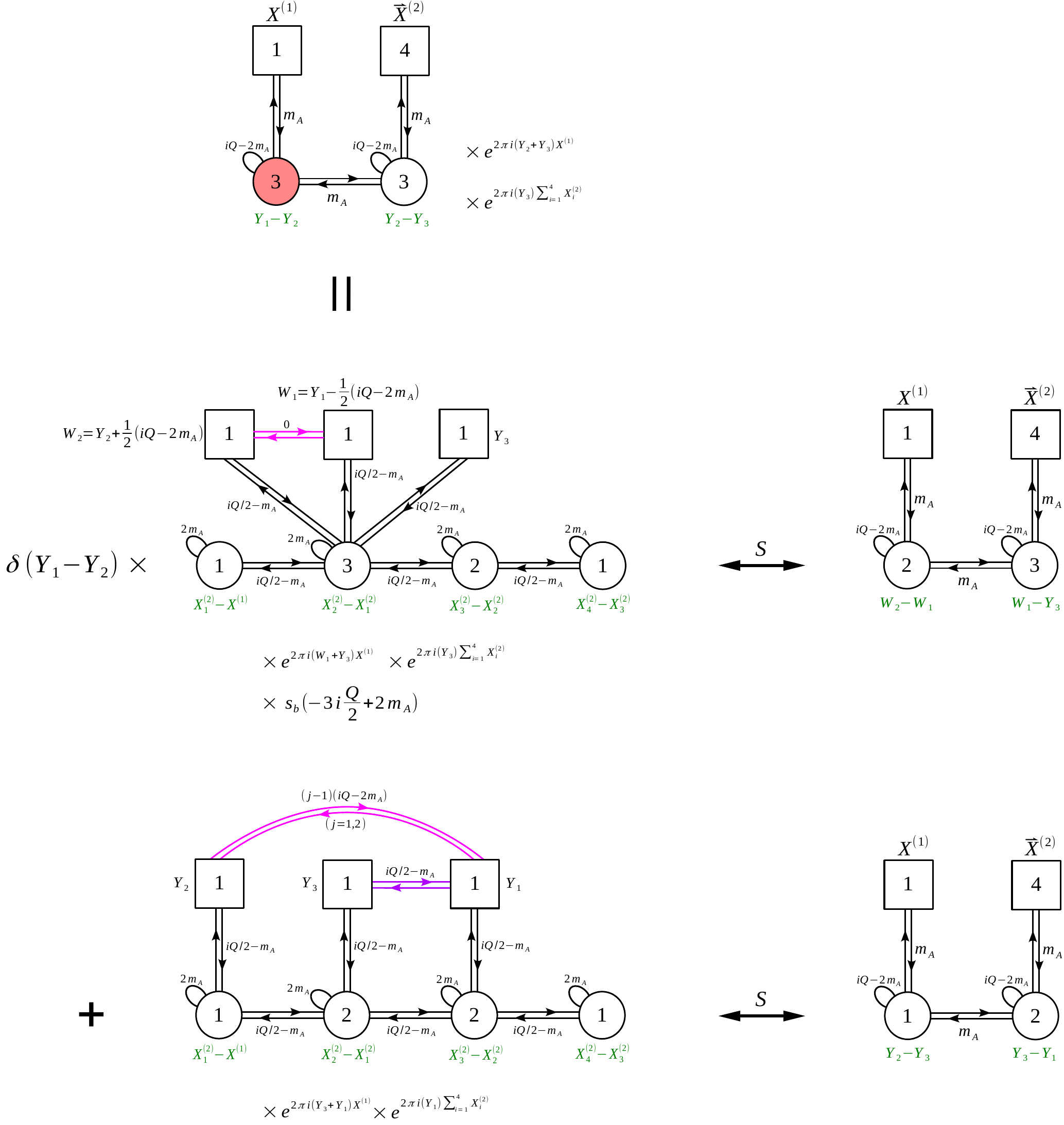}
	\caption{The comprehensive result of the mirror/electric dualization of the 3d bad linear quiver having $N_c=(3,3)$ and $N_f=(1,4)$. On the right we did not include the gauge singlets and the BF couplings. }
	\label{fig:allDuals_3d_example1}
\end{figure}

In summary, the complete result is given in Figure \ref{fig:allDuals_3d_example1}, which corresponds to the following identity between $S^3_b$ partition functions: 
\begingroup\allowdisplaybreaks
\begin{align}
	& \mathcal{Z}^{3d}_{T_{[2,4,3]}^{[2,2,2,2,1]}[SU(9)]}\left(\vec{X};\left\{Y_1,Y_2,Y_3\right\};m_A\right) = \nonumber\\[7pt]
	& \quad = 
	\left[
	\delta\left(Y_1-Y_2\right)
	\times
	s_b\left(-3i\frac{Q}{2}+2m_A\right)
	\times
	s_b\left(i\frac{Q}{2}\pm(W_1-W_2)\right)
	\times
	\right.
	\nonumber\\
	& \qquad\quad \left. \times 
	\mathcal{Z}^{3d}_{T_{[3,3,3]}^{[2,2,2,2,1]}[SU(9)]}\left(\vec{X};\left\{W_2,W_1,Y_3\right\};m_A\right)
	\right]_{W_{1,2}=Y_{1,2}\mp\frac{1}{2}(iQ-2m_A)}
	\nonumber\\[5pt]
	& \quad + 
	\prod_{j=1}^2 s_b\left(i\frac{Q}{2}-(j-1)(iQ-2m_A)\pm(Y_1-Y_3)\right)
	\times
	\nonumber\\
	& \qquad\times 
	 s_b\left(i\frac{Q}{2}-\frac{1}{2}(iQ-2m_A)\pm(Y_1-Y_2)\right)
	 \times
	\nonumber\\
	& \qquad \times
	\mathcal{Z}_{T_{[4,3,2]}^{[2,2,2,2,1]}[SU(9)]}\left(\vec{X};\{Y_2,Y_3,Y_1\};m_A\right) \,.
\end{align}
\endgroup

\clearpage
\subsection{The 3d electric  algorithm and the $(1,0)$-$(1,0)$ swap}

Now we explain the 3d electric  dualization algorithm using the $(1,0)$-$(1,0)$ swap and the corresponding brane picture. As we have seen in the 4d case, the (electric) dual frames can be easily obtained by applying the bad SQCD result in \cite{Giacomelli:2023zkk}, which is equivalent to the $(1,0)$-$(1,0)$ swap, on each bad node in a given quiver. In Section \ref{sec:brane} we have observed that the $(1,0)$-$(1,0)$ swap is realized in two different ways in the brane picture depending on the FI parameter of the bad node.

\begin{figure}[tbp]
	\centering
	\includegraphics[width=0.9\textwidth]{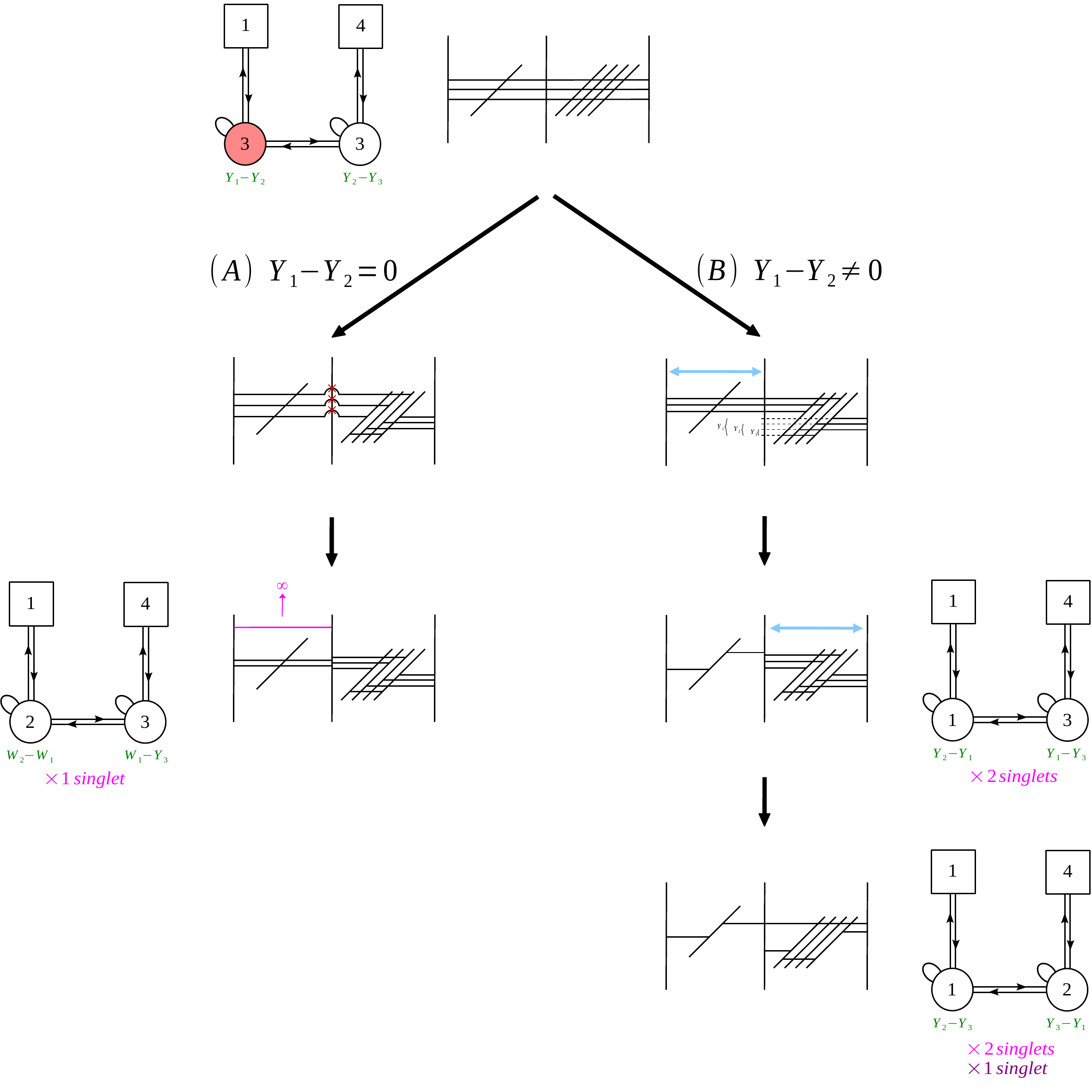}
	\caption{The brane interpretation of the electric dualization of the quiver gauge theory with $N_c = (3,3)$ and $N_f = (1,4)$.}
	\label{fig:3d_EE_branes_example1}
\end{figure}

As an example, we consider again the case with $N_c = (3,3)$ and $N_f = (1,4)$, which is illustrated in Figure \ref{fig:3d_EE_branes_example1}.
Let us first consider the case $Y_1-Y_2 = 0$, which is displayed in the left column of the figure. In this case, analogously to the $(0,1)$-$(0,1)$ swap, the $(1,0)$-$(1,0)$ swap corresponds to splitting a D3 passing through an NS5 into two pieces: one suspended between the two NS5's and the other suspended between the NS5 and the D5. Subsequently, one of the D3 segments suspended between the two NS5's escapes to infinity, resulting in the first electric dual frame associated with the delta constraint $Y_1-Y_2 = 0$.

On the other hand, if $Y_1 - Y_2 \neq 0$, the corresponding $(1,0)$-$(1,0)$ swap amounts to swapping the two NS5's avoiding any collisions, as shown in the right column of Figure \ref{fig:3d_EE_branes_example1}. At this stage, we still have a bad node and can swap the second (originally the first) NS5 and the third, resulting in the second electric dual frame with no delta constraint.

The two dual frames obtained via electric dualization using the $(1,0)$-$(1,0)$ swap exactly coincide with the frames obtained by $S$-dualizing each mirror  frame in the previous subsection. Furthermore, as already mentioned in Section \ref{sec:electric dual}, the electric  dualization does not need to be restricted to linear quivers and can be applied to other types of quivers such as a star-shaped, which we relegate to a future work.

\subsection{Application: $\mathcal{S}$-walls fusion to the Identity-wall}\label{subsec:SS1}

In this section we are going to consider the  linear quiver theory $\mathcal{T}$ without flavors.

\begin{figure}[!ht]
	\centering
	\includegraphics[width=.5\textwidth]{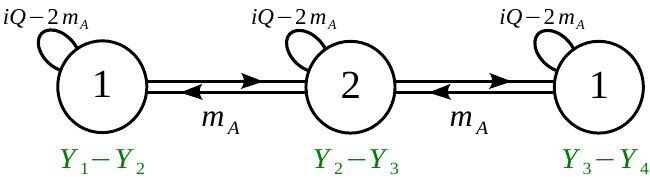}
	\caption{The theory without flavors $\mathcal{T}$. Notice this corresponds to gluing two $\mathcal{S}$-walls for $N=2$.}
	\label{fig:example_121}
\end{figure}

The central node is bad and by running the electric dualization algorithm, using the mathematica code, we obtain
two frames each with two Dirac-deltas and various chirals. By using the delta constraints and simplifying 
the massive chirals we are left with the following result for the partition function:
\begingroup\allowdisplaybreaks
\begin{align}
	& \mathcal{Z}^{3d}_\mathcal{T} = \left\{
	\delta\left(Y_1-Y_3\right) \delta\left(Y_2-Y_4\right)+	\delta\left(Y_2-Y_3\right) \delta\left(Y_1-Y_4\right)\right\}
	\nonumber\\\times
      &	s_b\left(i\frac{Q}{2}\pm (Y_1-Y_2)\right)s_b\left(-i\frac{Q}{2}+2m\right)^2 s_b\left(-i\frac{Q}{2}+2m\pm (Y_1-Y_2)\right)
	\end{align}
\endgroup

This quiver theory $\mathcal{T}$ actually coincides with the quiver obtained by gluing two $\mathcal{S}$-walls for $N=2$ as explained in Section \ref{sec:SwallIdwall}, via gauging of their manifest flavor symmetries.
Indeed by renaming the mass parameters as
\be
Y_1\to Y_1\,,\quad Y_2\to Y_2\,,\quad Y_3\to X_2\,,\quad Y_4\to X_1\,,\quad m\to i\frac{Q}{2}-m_A
\ee
we find that
\begin{align}
&\mathcal{Z}^{3d}_\mathcal{T} =\int\udl{\vec{Z}_2}\Delta^{3d}_2(\vec{Z};m_A)\mathcal{Z}_{\mathcal{S}}^{(2)}(\vec{Z};\vec{X};m_A)\mathcal{Z}_{\mathcal{S}}^{(2)}(\vec{Z};-\vec{Y};m_A)={}_{\vec X}\hat{\mathbb{I}}^{3d}_{\vec Y}(m_A)
\end{align}
with the 3d Identity-wall\footnote{In this section the focus is on showing that the partition function of the theory $\mathcal{T}$, computed using the dualization algorithm, is equivalent to the Identity-wall, in agreement with the results of \cite{Bottini:2021vms}. Our result can however also be interpreted as the statement that in each dual frame the interacting part is trivial and therefore the effective dual theory is a collection of free hypermultiplets, whose number is equal to the rank of the gauge group of the theory, namely four in the case of the theory $\mathcal{T}$. We will discuss more in detail this point in Section \ref{maxpart}.}
\begin{equation}
{}^{\phantom{}}_{\vec X}\hat{\mathbb{I}}^{3d}_{\vec Y}(m_A)=\frac{\sum_{\gs\in S_2}\prod_{j=1}^2\gd\left(X_j-Y_{\gs(j)}\right)}{\Gd^{3d}_2(\vec{X};m_A)}
\end{equation}
in accordance with the result presented in Section \ref{sec:SwallIdwall}.

Similarly, for generic $N$, we can obtain the fusion to Identity of two $\mathcal{S}$-walls using the electric dualization algorithm. 
The fusion to Identity was demonstrated in \cite{Bottini:2021vms} by applying iteratively the Intriligator--Pouliot duality \cite{Intriligator:1995ne}. So this offers another interesting consistency check of the dualization algorithm.


\section{Expected frames and partitions}\label{badpartitions}

The goal of this section is to identify the interacting sector of the various dual frames of a bad linear quiver as $T_{\rho}^{\sigma}(SU(N))$ theories and to provide a  rule to identify the maximal and minimal frames, which we will define in the following.

As we mentioned at beginning of Section \ref{linearquivers}, the notation  $T_{\rho}^{\sigma}(SU(N))$  is usually referred to good theories, for which $\rho$ and $\sigma$ are ordered partitions of $N$, with all the elements of the partition being positive integers. For a bad theory however, the $\rho$ we get by applying \eqref{rank} is in general merely a sequence of integers whose sum is $N$. They are not ordered and are not all positive. By abuse of notation, we will still refer to such $\rho$ as a partition in this section. Our goal is to extract from this sequence of integers the actual ordered partitions describing the various dual frames of the theory.

\subsection{Higgs branch flows and the $\sigma$ partition}\label{sigmapar}

\begin{figure}[!ht]
	\centering
	\includegraphics[width=.4\textwidth]{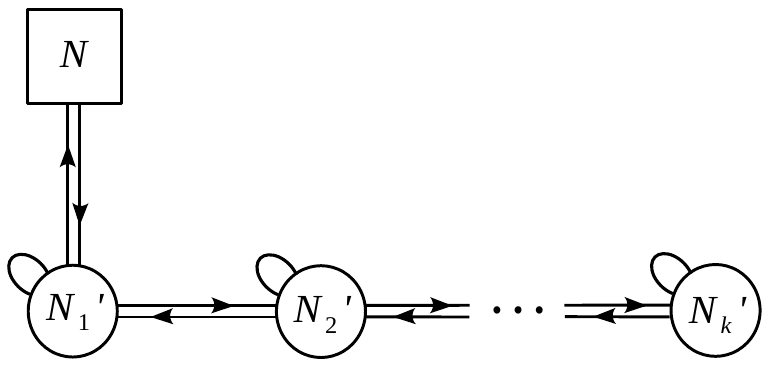}
	\caption{A new quiver obtained after moving D5-branes to the left in the brane system of Figure \ref{fig:HW_setup} with the help of Hanany--Witten moves and introducing extra D5-branes.}
	\label{newquiver}
	\end{figure}

As explained in Section \ref{linearquivers}, given a generic $T_{\rho}^{\sigma}(SU(N))$  quiver,
with associated brane set up depicted in Figure \ref{fig:HW_setup},  it can be useful  to move all the D5-branes to the left of the quiver. In the process of moving the D5-branes we change the number of D3’s in between consecutive NS5-branes, which becomes 
\be\label{rank22} N_i’=N_i+\sum_{j>i}(j-i)F_j.\ee 
As we explained this set-up contains multiple D3-branes terminating on the same D5. We can then introduce 
extra D5-branes and let each D3 end on a different D5 on the left. This removes any constraints from the s-rule and produces the auxiliary  quiver in Figure \ref{newquiver}.

This auxiliary  quiver is again a $T_{\rho}^{\sigma}[SU(N)]$ theory where $N$ and $\rho$ are as above while $\sigma$ has been replaced by the maximal partition $\sigma=[1^N]$. The quiver in Figure \ref{newquiver} can be Higgsed down
to the original quiver,  by turning on a nilpotent expectation value for the meson in the adjoint of the $SU(N)_X$ global symmetry rotating the flavors on the left.\footnote{In Section \ref{linearquivers} we explained that the $T_{\rho}^{\sigma}[SU(N)]$ theory can be obtained from the $T[SU(N)]$ theory by turning on nilpotent VEVs for the $SU(N)_X$ and $SU(N)_Y$ global symmetries labelled by $\sigma$ and $\rho$, respectively. Such an RG flow can be split into one from $T[SU(N)]$ to $T_{\rho}^{[1^N]}[SU(N)]$ triggered only by the nilpotent VEV for $SU(N)_Y$ labelled by $\rho$ and the RG flow we are considering here from $T_{\rho}^{[1^N]}[SU(N)]$ to $T_{\rho}^{\sigma}[SU(N)]$ triggered by the nilpotent VEV for $SU(N)_X$ labelled by $\sigma$.} In conclusion, the effect of the Higgsing we have just described is to change the partition $\sigma$ while leaving all other data unchanged.

So far we have parametrized our quivers with the integers $\{N_i,F_i\}$, but we can equivalently use the excess numbers $e_i$ instead of the ranks $N_i$. These are defined as $$e_i=N_{i-1}+N_{i+1}+F_i-2N_i$$ and are positive if and only if the $i$-th gauge node is good. Notice that the data $\{e_i,F_i\}$ uniquely specify the quiver since $$-C_i^jN_j=e_i-F_i,$$ where $C_i^j$ denotes the Cartan matrix of the quiver (in our case $A_{k-1}$) which is invertible. 
Crucially for our purposes, we can observe that the excess numbers as defined above are not changed by Hanany--Witten moves and consequently they are also invariant under the Higgsing we have just described. Notice that one can easily prove that the original and the auxiliary  quivers have the same excess numbers by using \eqref{rank22}. At this stage we can further observe that the number of dual frames for a bad node depends only on the value of its (negative) excess number, modulo a subtlety we will discuss momentarily. Actually we have the stronger result that, upon performing a local dualization, the change in rank of the gauge group of a bad node also depends only on its excess number.
More specifically, the local dualization replaces the negative excess number of a bad node by a non-negative value as follows:
\begin{align}
0 \, > \, e_i \quad &\longrightarrow \quad \tilde{e}_i \, \geq \, 0 \,, \\
e_{i\pm1} \quad &\longrightarrow \quad e_{i\pm1}+\frac{e_i-\tilde{e}_i}{2} \,,
\end{align}
where
\begin{gather}
m \, \leq \, \tilde{e}_i \, \leq \, M \,, \\
m = e_i \mod 2 \,, \qquad M = \min(-e_i, F_i^{\text{eff}})
\end{gather}
with the effective flavor number $F_i^{\text{eff}} =N_{i-1}+N_{i+1}+F_i$.
If $-e_i \leq F_i^{\text{eff}}$, the range of the dual excess numbers is completely determined by the original excess number without referring to $F_i$, or equivalently $\sigma$; therefore, the local dualization and Higgsing commute. If $-e_i > F_i^{\text{eff}}$, on the other hand, the upper bound of the dual excess number depends on $F_i$. Nevertheless, we can pretend the upper bound is always given by $-e_i$, regardless of whether it is larger than $F_i^{\text{eff}}$ or not, and throw away the illegal dual frames violating
\begin{align}
\label{eq:positive rank condition}
\tilde{e}_i \quad \leq \quad F_i^{\text{eff}}
\end{align}
at the end. In principle, this condition must be checked at every gauge node after each dualization step, but as we will explain shortly, once this condition is satisfied for the final dual quiver, we are guaranteed that all the intermediate dual quivers satisfy it as well. Thus, it is enough to check this condition for the final quiver. 
The condition to be checked is nothing but the positivity of the rank given in \eqref{rankcond}, which, as we have seen before, must be satisfied by every $T_{\rho}^{\sigma}[SU(N)]$ theory.

Putting the above observations together we conclude that the operations of performing a local dualization and Higgsing commute, as long as the condition \eqref{eq:positive rank condition} is met: the final quiver does not depend on the order in which we perform them. We can therefore simplify the analysis by restricting to the case $\sigma=[1^N]$, in which the quiver just has $N$ flavors on the left charged under the first gauge group. The complete set of dual frames of the quiver we are interested in is obtained by determining the dual frames of the auxiliary quiver with $\sigma=[1^N]$ first and then performing the Higgsing on each dualization frame. It might be the case that some of the dual frames of the auxiliary theory with $\sigma=[1^N]$ we have introduced before correspond to $\rho$ partitions which do not satisfy \eqref{rankcond} for the $\sigma$ partition we are actually interested in. In this case we should simply drop these dual frames and restrict to those which do satisfy the constraint. The others are possible dual frames of the auxiliary theory but do not actually arise in the Higgsed theory. In summary, one should always enforce \eqref{rankcond} when working with the auxiliary theory and this is the only way the set of possible dual frames depends on $\sigma$. 

As we have mentioned earlier, a priori one should check after each local dualization operation that \eqref{rankcond} is satisfied, or equivalently that the rank of all the gauge nodes is above the bound imposed by \eqref{rankcond} (which is fixed once the partition $\sigma$ is known). However, since upon dualizing a node its rank always decreases, it is not possible that a quiver which does not satisfy the constraint \eqref{rankcond} becomes compatible with it after a local dualization move. It then follows that all final dual frames which satisfy \eqref{rankcond} can only arise from a sequence of local dualizations all compatible with \eqref{eq:positive rank condition}. This guarantees, as desired, that we only need to enforce  \eqref{rankcond} on the final set of dual frames of the auxiliary theory with $\sigma=[1^N]$.

One advantage of working with $\sigma=[1^N]$ is that \eqref{rank} simplifies to 
\be\label{rank2} N'_i=\sum_{j>i}\rho_j\ee 
and therefore the ranks of the various gauge groups now depend only on $\rho$. In short, this is telling us that the information encoded in $\sigma$ is inessential for the problem at hand and the full set of dual frames is obtained by specifying which $\rho$ partition correspond to a duality frame. As a result, we can immediately derive the following properties: 
\begin{itemize}
\item The $i$-th gauge group in the quiver is not good (bad or ugly) if and only if $\rho_i<\rho_{i+1}$, meaning that the partition is not ordered. 
\item If all the elements of $\rho$ are non negative, then $N_i\geq N_{i+1}$ $\forall i$ and therefore the rank of the gauge groups in the quiver form a decreasing sequence.
\item There are bad nodes with $N_f<N_c$, for which there is no dual frame without delta, only if some of the elements of the partition $\rho$ are negative. Later on we will see that also the converse is true: if an element of $\rho$ is negative then, with a sequence of local dualizations, we can get to a quiver having a node with $N_f<N_c$. 
\end{itemize}

We will now describe the maximal and minimal $\rho$ partitions we can get for a given quiver of the same form as in Figure \ref{newquiver}. 

\subsection{Local dualizations and moves on the $\rho$ partition} 

As we have seen, the quiver can be bad only if the initial partition $\rho$ is not ordered. In this situation we can identify the various dual frames by performing local dualizations at the bad nodes in the quiver, those such that the total number of flavors $N_f$ is less than twice the rank of the gauge group $N_c$. The maximal choice (largest rank for the dualized node) corresponds to replacing $N_c$ with $\lfloor N_f/2\rfloor$ while the minimal choice corresponds to replacing $N_c$ with $N_f-N_c$ when $N_f>N_c$ and with $0$ otherwise. We will now see that all possible local dualization moves can be neatly described in terms of simple manipulations of the partition $\rho$. 

As we have explained, the $i$-th gauge node is bad or ugly only if $\rho_i<\rho_{i+1}$. We can notice that switching the two elements $\rho_i$ and $\rho_{i+1}$ (i.e.~reordering them) corresponds to the minimal dualization choice $N_c\rightarrow N_f-N_c$ at the $i$-th node, without changing the rest of the quiver. This just follows from \eqref{rank2}. Similarly, the other dual frames can be obtained by considering a generalized switching move in which we first reorder the two elements and then we subtract a positive number $a$ from $\rho_{i+1}$ and add it to $\rho_i$, while preserving the inequality $\rho_{i+1}-a\geq \rho_i+a$. The maximum value we allow for $a$ is therefore 
$$a_{max}=\left\lfloor\frac{\rho_{i+1}-\rho_i}{2}\right\rfloor\;,$$
and corresponds to the maximal dual frame $N_c\rightarrow\lfloor N_f/2\rfloor$.
The value $a=0$ corresponds to the switching we have discussed before and all other values for $a$ correspond to other intermediate dual frames. 

The rule described above needs to be used with care in presence of bad nodes with $N_f<N_c$. In this case the minimal frame does not involve replacing $N_c$ with $N_f-N_c$, which is negative, but corresponds instead to setting to zero the rank of the gauge group. This can be implemented with a generalized switching move with $a>0$ and all we need to do is determine the correct value of $a$. This can be done as follows. The situation we are interested in arises when, after the switching, the new quiver includes a node of negative rank. This happens at the $i$-th gauge node when $\rho_i$ is negative and \be\label{cond22}-\rho_i>\sum_{j>i+1}\rho_j\;.\ee 
While in the initial quiver the rank of the $i$-th gauge group is $N_i=\rho_{i+1}+\sum_{j>i+1}\rho_j$ and is positive by assumption, after the switching the rank becomes negative if \eqref{cond22} holds. We can easily check that this happens only if the number of flavors (equal to $N_{i-1}+N_{i+1}$) is smaller than $N_i$. This immediately follows from \eqref{cond22}
\be N_{i-1}+N_{i+1}=\rho_i+\rho_{i+1}+2\sum_{j>i+1}\rho_j<\rho_{i+1}+\sum_{j>i+1}\rho_j=N_i\;.\ee
In this case the minimal dual frame is obtained by performing a generalized switching move with $a=-\rho_{i}-\sum_{j>i+1}\rho_j$. We should therefore always check after any generalized switching move that \eqref{cond22} is not satisfied. If it is, it means that the move we have just done is not physically sensible and should therefore be discarded. Our claim is that local dualizations can be entirely described in terms of manipulations of the corresponding $\rho$ partition and that with a sequence of generalized switching moves we can obtain all possible dual frames for our quiver. We are now going to determine the maximal and minimal dual frames for a generic bad quiver. 

\subsection{The maximal frame} \label{maxpart}

In this subsection we explain how to find the maximal dual frame for a given quiver. By maximal we mean that for every $i$ the rank of the $i$-th gauge node has the largest possible value. We can start by simply observing that, since all dual frames correspond to good theories, their associated $\rho$ partitions are ordered with non negative elements only. This condition can be rephrased in a simple geometrical way as we will now explain. 

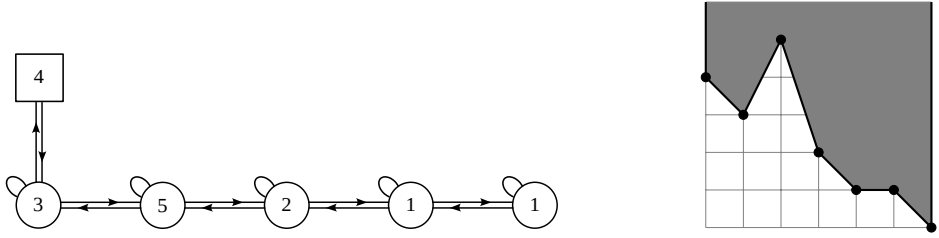
\begin{figure}[!ht]
\centering
\begin{subfigure}{.6\textwidth}
  \centering
  \includegraphics[width=.8\textwidth]{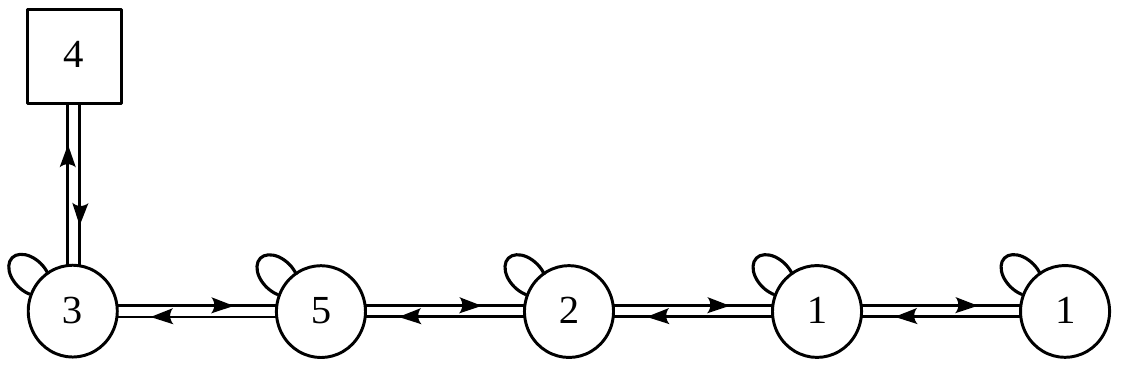}
\end{subfigure}
\begin{subfigure}{.3\textwidth}
  \centering
  \begin{tikzpicture}[x=.5cm,y=.5cm]
	\draw[step=.5cm,gray,very thin] (0,0) grid (6,6);
	\draw[ligne,black, fill=gray] (0,6)--(0,4)--(1,3)--(2,5)--(3,2)--(4,1)--(5,1)--(6,0)--(6,6);
	\node[bd] at (0,4) {};
	\node[bd] at (1,3) {};
	\node[bd] at (2,5) {};
	\node[bd] at (3,2) {};
	\node[bd] at (4,1) {};
	\node[bd] at (5,1) {};
	\node[bd] at (6,0) {};
\end{tikzpicture}
\end{subfigure}
\caption{On the left we have the linear quiver associated to the partition $\rho=[1,-2,3,1,0,1]$ of 4. On the right we have the corresponding diagram on the plane. The polygon is shaded in gray.}
\label{poly}
\end{figure}

Every quiver, good or bad, can be associated with a sequence of points on the plane with coordinates $(i,N_i)$. The point associated to the $N$ flavors on the left has coordinates $(0,N)$ and we also include the point $(k+1,0)$ if the quiver has length $k$. An example is provided in Figure \ref{poly}. 

\noindent For each such sequence of points we can consider the polygon shaded in gray in Figure \ref{poly}. It is easy to see that the quiver is good if and only if the corresponding polygon is convex. Given any initial quiver, not necessarily good, all its dual frames have an associated polygon with the following properties: 
\begin{enumerate}
\item The polygon is convex and all its edges have integral slope because each dual frame is good;  
\item It contains the polygon associated with the initial quiver because the rank never increases along a sequence of local dualizations.
\end{enumerate} 

 In order to describe the maximal dual frame, it turns out we need to consider the minimal polygon satisfying the two properties above. The $\rho$ partition associated with the minimal convex polygon, which we denote $\tilde\rho$, can be obtained as follows: 
\be\label{tilde1}\tilde\rho_{k+1}=\left\lfloor \text{Min}_{1\leq j\leq k+1}\left(\frac{N_{k+1-j}}{j}\right)\right\rfloor\;,\ee
and the other $k$ elements of the sequence are defined recursively by 
\be\label{tilde2}\tilde\rho_{\ell}=\left\lfloor \text{Min}_{1\leq j\leq \ell}\left(\frac{N_{\ell-j}-\sum_{i>\ell}\tilde\rho_i}{j}\right)\right\rfloor\;.\ee 
It can be shown that the partition $\tilde\rho$ as defined above is ordered since $\tilde\rho_{i}\geq\tilde\rho_{i+1}$ $\forall i$ and all the elements $\tilde\rho_i$ are non-negative, therefore the corresponding polygon satisfies property 1 above. Besides being ordered, the partition $\tilde{\rho}$ also satisfies the relation
$$\sum_{j\geq i}\tilde\rho_j\leq \sum_{j\geq i}\rho_j \quad \forall i\;,$$ 
where $\rho$ denotes the partition associated with the initial quiver. It follows that the polygon corresponding to $\tilde\rho$ satisfies also property 2 and is therefore associated with a possible dual frame. Moreover, due to minimality, every other convex polygon satisfying them contains the polygon associated with $\tilde{\rho}$. The proof that the quiver associated with the partition $\tilde\rho$ defined by \eqref{tilde1} and \eqref{tilde2} satisfies the two properties above and the associated polygon is minimal is presented in Appendix \ref{partitionsapp}.

\begin{figure}[!ht]
	\centering
	\includegraphics[width=.3\textwidth]{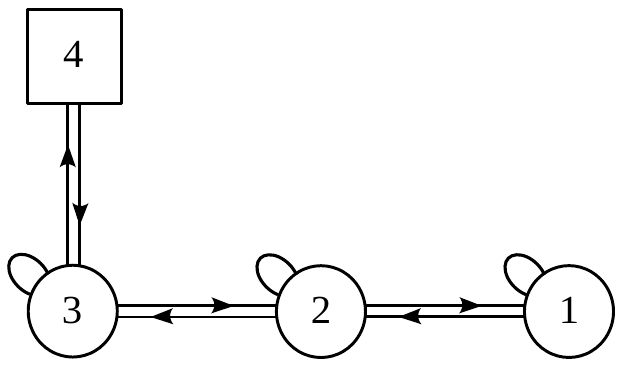}
	\caption{The quiver corresponding to the partition $\tilde{\rho}=[1,1,1,1,0,0]$. }
	\label{fig:quiver_3}
\end{figure}

Before proceeding with the discussion, let us determine the partition $\tilde\rho$ for the example in Figure \ref{poly} to illustrate how the procedure works. We have from \eqref{tilde1} and \eqref{tilde2}: 
$$\begin{array}{l} 
\tilde{\rho}_6=\left\lfloor \text{Min}\left( 1,\frac{1}{2},\frac{2}{3},\frac{5}{4},\frac{3}{5},\frac{4}{6}\right)\right\rfloor=0\;,\\
\tilde{\rho}_5=\left\lfloor \text{Min}\left( 1,\frac{2}{2},\frac{5}{3},\frac{3}{4},\frac{4}{5}\right)\right\rfloor=0\;,\\ 
\tilde{\rho}_4=\left\lfloor \text{Min}\left(\frac{2}{1},\frac{5}{2},\frac{3}{3},\frac{4}{4}\right)\right\rfloor=1\;,\\ 
\tilde{\rho}_3=\left\lfloor \text{Min}\left(\frac{5-1}{1},\frac{3-1}{2},\frac{4-1}{3}\right)\right\rfloor=1\;,\\
\tilde{\rho}_2=\left\lfloor \text{Min}\left(\frac{3-2}{1},\frac{4-2}{2}\right)\right\rfloor=1\;,\\
\tilde{\rho}_1=\left\lfloor \text{Min}\left(\frac{4-3}{1}\right)\right\rfloor=1\;,\\
\end{array}$$
and therefore $\tilde{\rho}=[1,1,1,1,0,0]$, corresponding to the quiver in Figure \ref{fig:quiver_3}. In other words, the maximal frame in this example is the $T[SU(4)]$ theory.
\\

Our claim is that the maximal dual frame can be obtained by performing a sequence of local dualizations in which we replace $N_c$ with $\lfloor N_f/2\rfloor$ for every bad node and the resulting polygon is the minimal one we have just described. We will now prove our claim using minimality of the polygon associated with $\tilde{\rho}$ (the proof of minimality is given in Appendix \ref{partitionsapp}). Let us consider a bad node of the initial quiver, whose corresponding point in the plane has coordinates $(i,N_i)$. After the dualization we replace $N_i$ with $\lfloor(N_{i-1}+N_{i+1})/2\rfloor$. Since both $(i-1,N_{i-1})$ and $(i+1,N_{i+1})$ are inside the minimal polygon, the whole segment connecting them will be inside the polygon as well by convexity. In particular the point $(i,(N_{i-1}+N_{i+1})/2)$ will be inside the polygon and since all its edges have integral slope we conclude that the point $(i,\lfloor(N_{i-1}+N_{i+1})/2\rfloor)$ as well is inside the polygon. As a result, we find that after a dualization move the polygon associated with the quiver is still inside the minimal polygon and by iterating the argument, we conclude this remains true also after an arbitrary sequence of dualizations. Since after finitely many dualizations the quiver becomes good and therefore the corresponding polygon is convex, by minimality it will necessarily coincide with the minimal polygon. We therefore conclude that the dual frame we get by iterating the dualization move $N_c\rightarrow \lfloor N_f/2\rfloor$ is described by the partition $\tilde{\rho}$ and this will be the maximal dual frame. 

Now we are in the position to prove a result we will need later on, namely that for every linear quiver without any flavors (or equivalently $N=0$ in Figure \ref{newquiver}) the dual frames obtained via a sequence of local dualizations are all trivial and correspond to a collection of (twisted) hypermultiplets.\footnote{This is consistent with our discussion in the previous section, where we considered the gluing of two $\mathcal{S}$-wall theories. In the deep IR, using the emergent symmetries, the result can be also expressed as a collection of free hypers.} This follows from the fact that the partition $\tilde{\rho}$ as defined in \eqref{tilde1} and \eqref{tilde2}, which describes the maximal frame, coincides with the trivial partition and consequently the gauge group has rank zero. Notice that this does not imply that the partition function is simply a product of delta functions times the hypermultiplet contributions. There might be several such contributions with a different number of deltas and we must sum over them, resultinging in a far from trivial expression for the partition function. The claim is that in each such term the interacting part is trivial.

\subsection{The minimal frame}\label{sixfour}

Let us now discuss the minimal dual frame. Since as we have seen the maximal dual frame can be obtained by replacing at each dualization step $N_c$ with $\lfloor N_f/2\rfloor$, which is the largest possible rank for the dualized node, one may guess that the minimal dual frame is instead obtained by replacing every time $N_c$ with $N_f-N_c$, which is the smallest possible rank. This is actually correct provided that the initial quiver is such that we never come across bad nodes with $N_f<N_c$ along the dualization sequence. Whenever such gauge nodes arise, we cannot replace $N_c$ with $N_f-N_c$, as we have already explained, and therefore the minimal dual frame cannot be obtained in this way. We would now like to understand for which quivers this issue arises and how to find the minimal dual frame in general. 

As we have discussed in Section \ref{sigmapar}, we can have a bad node with $N_f < N_c$ in the initial quiver only if the corresponding partition $\rho$ contains negative elements. Let us now look at this matter more in detail. Having a gauge node (say the $i$-th) with $N_f<N_c$ means that $N_{i-1}+N_{i+1}<N_{i}$ and this in terms of the elements of $\rho$ translates into the inequality 
\be\label{rhocond} \sum_{j\geq i}\rho_j+ \sum_{j\geq i+2}\rho_j<\sum_{j\geq i+1}\rho_j\;\rightarrow \rho_i+\sum_{j\geq i+2}\rho_j<0\;.\ee 
If \eqref{rhocond} is satisfied, we easily see that after switching $\rho_i$ and $\rho_{i+1}$ the rank of the $i$-th node becomes negative, which is precisely the expected issue. We therefore see that the initial quiver will have a gauge node with  $N_f<N_c$ only if \eqref{rhocond} is satisfied for some $i$ and is therefore not enough that one of the elements of $\rho$ is negative. On the other hand, if one of the elements of $\rho$ is negative, it is always possible with a sequence of dualizations to get to a quiver such that the corresponding partition satisfies \eqref{rhocond} and in this theory we will have a gauge node with $N_f<N_c$. This can be easily seen as follows: Take a negative element $\rho_i$ and apply a sequence of switching moves on $\rho$, bringing the element $\rho_i$ towards the end of the partition. This corresponds to dualizing the $i$-th node of the quiver first, then the gauge node $i+1$ and so on. After a finite number of steps we necessarily end up with a $\rho$ satisfying \eqref{rhocond}: This inequality is indeed obviously satisfied if we bring e.g. $\rho_i$ to the second last position of the partition. 

Once we get to a partition satisfying  \eqref{rhocond}, we can dualize the node with $N_f<N_c$  by setting its rank to zero. As a result of this operation the quiver will break into two decoupled quivers, one of which has no flavors (here we are exploiting the fact that the quiver has the form of Figure \ref{newquiver}) and therefore, thanks to the observation of the previous section that the interacting part of all its dual frames is trivial, we can simply drop it since it affects the dual frame just by adding a collection of free (twisted) hypermultiplets. The resulting quiver can be described by a new partition in which we fuse $\rho_i$ and all subsequent elements into a single element, without changing anything else. We call this operation on the partition collapse. 

The above discussion suggests a method to identify the minimal dual frame. The rule to find the corresponding partition works as follows: 
\begin{enumerate}
\item We first pick the last negative element of $\rho$, such that all subsequent elements in the partition are non negative. We focus on this subpartition and reorder it, putting the smallest element as last. We can simply ignore all the vanishing elements at the end of the resulting partition if any.
\item We then perform a sequence of local dualizations, bringing the last negative element towards the end of the partition until we satisfy the inequality  \eqref{rhocond} and collapse the partition. At this stage the negative element is gone. 
\item We repeat the above steps until we get to a ordered partition with positive elements only.
\end{enumerate}
Notice that if all elements of $\rho$ are positive, then only the first step is nontrivial and is equivalent to reordering the elements of the partition. This corresponds to a sequence of duality moves $N_c\rightarrow N_f-N_c$, in agreement with the expectation that we need to reduce the rank of the gauge groups as much as possible at each step.  

\begin{figure}[!ht]
	\centering
	\includegraphics[width=.25\textwidth]{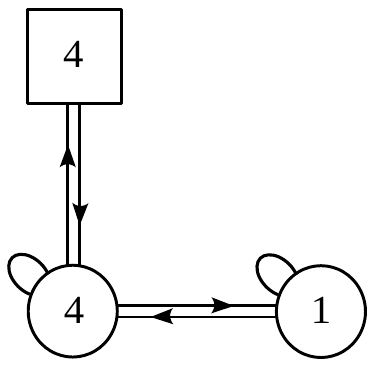}
	\hspace{4cm}
	\includegraphics[width=.25\textwidth]{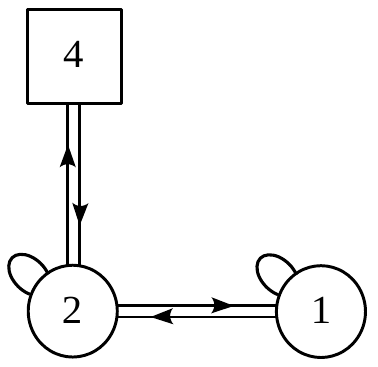}
	\caption{The bad quiver corresponding to the partition $\rho=[0,3,1]$ on the left and the associated good maximal frame with partition $\rho=[2,1,1]$ on the right.}
	\label{ex1}
\end{figure}

Before discussing the proof that the above procedure does indeed provide the minimal frame, let us discuss the example with $\rho=[0,3,1]$ shown on the left of Figure \ref{ex1}. In this case the minimal dual frame according to the above rule is described by the partition $\rho=[3,1]$ and therefore corresponds to a $U(1)$ theory with 4 flavors. The maximal frame is instead described by the partition $\rho=[2,1,1]$ and corresponds to the theory on the right of Figure \ref{ex1}. One can easily check that these are the only two possible dual frames.

\begin{figure}[!ht]
	\centering
	\includegraphics[width=.25\textwidth]{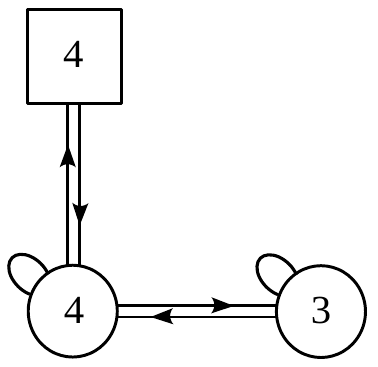}
	\caption{The quiver corresponding to the partition $\rho=[0,1,3]$.}
	\label{fig:quiver_6}
\end{figure}

We can now notice that the partition $\rho=[2,2]$, despite being smaller than $[3,1]$ and bigger than $[2,1,1]$, does not arise. We therefore find that, contrary to the case of the SQCD, not all partitions between the maximal and minimal appear as actual dual frames because we cannot reach them with a sequence of local dualizations. We would like to stress that this fact depends on the ordering of the elements of the initial partition. Indeed, if we consider instead from the theory in Figure \ref{fig:quiver_6} which just involves inverting the last two elements of $\rho$ in Figure \ref{ex1}, the maximal and minimal dual frames do not change but this time we do find a dual frame associated with the partition $[2,2]$. The corresponding sequence of generalized switching moves (or local dualizations) is 
$$[0,1,3]\rightarrow [0,2,2]\rightarrow [2,0,2]\rightarrow [2,2,0]=[2,2]\;.$$ 
In conclusion, all possible dual frames are labelled by partitions in between those associated with the minimal and maximal dual frames but not all such partitions actually correspond to a physical frame appearing in the partition function of the bad theory. 

Let us now come to the proof of minimality of the dual frame we have constructed above. If we start from the quiver associated with the partition $\rho$ and apply our rule, we obtain a new ordered partition we denote $\mathcal{I}(\rho)$. We want to prove that $\mathcal{I}(\rho)$ is larger than (or equal to) the partition associated with any of the other dual frames.  The derivation we provide is based on two technical lemmas. We will first illustrate the argument assuming their validity to explain the general idea and then give the proofs of the lemmas in Appendix \ref{partitionsapp}. We start by observing that the rule we have given to derive the minimal dual frame does not change good quivers (since all steps are trivial). In particular, if we apply it to any of the dual frames of our theory we simply recover the partition describing it. On the other hand, if we perform a generalized switching move (local dualization) on the partition $\rho$ obtaining $\rho’$ and apply our rule to it, we find another partition $\mathcal{I}(\rho’)$ which in general differs from $\mathcal{I}(\rho)$. If we can prove that $\mathcal{I}(\rho)\geq\mathcal{I}(\rho')$ for every duality move we obtain the desired result, since every dual frame arises as the outcome of a finite sequence of local dualizations and our rule, as we have noticed above, does not affect the corresponding partitions since in every dual frame the theory is good. Said differently, $\mathcal{I}(\rho')=\rho'$ if $\rho'$ is the partition associated with a dual frame.

We have reduced the problem to proving that $\mathcal{I}(\rho)\geq\mathcal{I}(\rho')$, with $\mathcal{I}(\rho')$ the partition obtained by applying the rule after a local dualization. This can be proven by induction on the length of the quiver. We first prove this is true for quivers of length one (SQCD), or equivalently partitions of length two, and then show that assuming the property holds for partitions of length $k-1$ implies it holds for partitions of length $k$. 

For quivers of length one the associated partition is $\rho=(N_f-N_c, N_c)$. If $N_f-N_c>N_c$, then the theory is good and we automatically have $\rho=\rho'=\mathcal{I}(\rho)=\mathcal{I}(\rho')$. If instead the theory is bad and $N_f-N_c>0$ then by applying our rule we find $\mathcal{I}(\rho)=(N_c, N_f-N_c)$, whereas any local dualization will lead to $\rho’=\mathcal{I}(\rho')=(N_c-a, N_f-N_c+a)$ with $a\geq0$ and we immediately conclude $\mathcal{I}(\rho)\geq\mathcal{I}(\rho')$. If instead $N_f<N_c$, by using our rule we simply collapse the partition and get $\mathcal{I}(\rho)=(N_f)$, which is larger than any other partition of $N_f$. We therefore conclude that $\mathcal{I}(\rho)\geq\mathcal{I}(\rho')$ in all cases. 

Let us now come to the inductive step, which is the most intricate. By induction we know that for every partition of length $k-1$ we have $\mathcal{I}(\rho)\geq\mathcal{I}(\rho')$ and we want to show this implies the same inequality for every partition of length $k$. Let us denote the initial partition as 
\be\label{initial} \rho=(\rho_1,\rho_2,\dots, \rho_k)\;.\ee 
If we now perform a local dualization at a gauge node which is not the first, the element $\rho_1$ is not affected and the new partition is 
\be \rho'=(\rho_1,\rho'_2,\dots, \rho'_k)\;.\ee 
As a preliminary step, let us apply our rule to the subpartitions of $\rho$ and $\rho'$ obtained by removing the first element $\rho_1$ and denote the resulting ordered partitions as $\mathcal{I}(\hat{\rho})$ and $\mathcal{I}(\hat{\rho}')$ respectively. By induction we know that $\mathcal{I}(\hat{\rho})\geq \mathcal{I}(\hat{\rho}')$. The next step is to apply our rule to 
$$\tilde{\rho}=(\rho_1,\mathcal{I}(\hat{\rho}))\;,$$ obtaining a partition of length $k$ $\mathcal{I}(\tilde{\rho})$ and to 
$$\tilde{\rho}'=(\rho_1,\mathcal{I}(\hat{\rho}'))\;,$$ obtaining another partition $\mathcal{I}(\tilde{\rho}')$. From the inequality $\mathcal{I}(\hat{\rho})\geq \mathcal{I}(\hat{\rho}')$ it is possible to show that $\mathcal{I}(\tilde{\rho})\geq\mathcal{I}(\tilde{\rho}')$. This will be proven in {\bf Lemma 1}. We can then conclude by observing that $\mathcal{I}(\tilde{\rho})=\mathcal{I}(\rho)$ and $\mathcal{I}(\tilde{\rho}')=\mathcal{I}(\rho')$ due to the way our rule is defined.

All is left to do is to analyze the case in which we dualize the first node of the quiver (which is possible only if $\rho_1<\rho_2$). In this case the partition $\rho'$ is of the form 
$$\rho'=(\rho_2-a,\rho_1+a,\rho_3,\dots, \rho_k)\; ,$$ 
with $a\geq0$. By applying our rule to $\rho$ we get a partition which is larger than that we find by applying it to $\rho'$. This is the content of {\bf Lemma 2} and can be proven directly without using the inductive step. Assuming the validity of the two lemmas, we have showed that our rule does provide the minimal dual frame. The proofs of the two lemmas are provided in Appendix \ref{partitionsapp}. 

\section{From unitary to special unitary gauge nodes}\label{sec:SU}

In this section we explain how the knowledge of a 3d mirror pair between quivers with unitary gauge groups and the associated mapping of the global symmetries that is provided by the dualization algorithm allow us to automatically find the mirror of a theory where some of the gauge nodes are replaced with special unitary groups (see also \cite{Dey:2020hfe} for related discussions). The same problem was addressed in \cite{Bourget:2021jwo} from the perspective of the brane webs.

The key idea is that we can turn a unitary gauge node into a special unitary one by gauging its $U(1)$ topological symmetry in an $\mathcal{N}=4$ preserving way.\footnote{Here we will be concerned with 3d $\mathcal{N}=4$ theories, but this applies also to less supersymmetric or non-supersymmetric theories; e.g.~3d $\mathcal N=2$ Seiberg-like dualities for a special unitary gauge group can be obtained from the unitary one using this method \cite{Park:2013wta}.} By knowing how this symmetry is mapped across the mirror duality we can then easily study this deformation in the dual. The topological symmetry is typically mapped to a flavor symmetry, so that the mirror dual that we get after the gauging is again Lagrangian and is obtained by suitably introducing some $U(1)$ gauge node. The precise mapping of which flavor symmetry the topological symmetry that we want to gauge maps to can be easily understood from the dualization algorithm perspective.

We will first explain this procedure for some good 3d $\mathcal{N}=4$ theories  and then discuss some examples of bad SQCD's. As usual we will show this procedure at the level of the $S^3_b$ partition function.

\subsection{Good SQCD}

Let us start considering the case of the good unitary SQCD, that is the $U(N_c)$ gauge theory with $N_f\geq 2N_c$ flavors. By gauging the only $U(1)$ topological symmetry we get the $SU(N_c)$ SQCD with $N_f\geq 2N_c$ flavors. Since we are gauging a topological symmetry, in order to do that in an $\mathcal{N}=4$ preserving way we should introduce a twisted vector multiplet rather than an ordinary vector multiplet. The difference between the two is that their representations under the $SU(2)_H$ and the $SU(2)_C$ factors of the $\mathcal{N}=4$ $SO(4)_R\cong SU(2)_H\times SU(2)_C$ R-symmetry are swapped. In terms of the $\mathcal{N}=2$ subalgebra, this means that the $\mathcal{N}=2$ adjoint chiral inside the twisted vector multiplet, which in our case is just a singlet since we are gauging a $U(1)$ symmetry, compared to an ordinary vector multiplet has the same canonical R-charge of 1 but opposite axial charge.\footnote{We remind that the canonical R-symmetry is defined as $U(1)_R=U(1)_H+U(1)_C$, where $U(1)_H$ and $U(1)_C$ are the Cartans of $SU(2)_H$ and $SU(2)_C$ respectively. This is not the parametrization that we use in our expressions for the $S^3_b$ partition function, where instead to recover the canonical R-symmetry we should replace $m_A=i\tfrac{Q}{4}$.} This in particular indicates that there is a new quadratic superpotential term between the added $U(1)$ adjoint chiral and the trace of the original $U(N_c)$ adjoint chiral
\be
\gd\mathcal{W}=\Phi_{U(1)}\,\mathrm{Tr}\,\Phi_{U(N_c)}\,.
\ee
Integrating them out we are left with a traceless adjoint chiral, as expected since now the gauge group is $SU(N_c)$.

At the level of the $S^3_b$ partition function, the contribution of the adjoint chiral for a twisted vector is obtained from the one for an ordinary vector that we have encouneterd so far by replacing $m_A\to i\tfrac{Q}{2}-m_A$. Hence, from the partition function of the $U(N_c)$ SQCD\footnote{Observe that we added the contact term $\mathrm{e}^{i\pi (Y_2-Y_1) \sum_{a=1}^{N_f}X_a}$. When considering the unitary theory, this can always be trivialized by setting $\sum_{a=1}^{N_f}X_a=0$ since the diagonal $U(1)$ part of the flavor symmetry can be re-absorbed with a gauge rotation, but it is instead crucial when gauging to go to the special unitary theory. The specific contact term we added is the one that appears in the expression for the partition function of the bad SQCD in \eqref{eq:3d_bad_SQCD}.}
\begingroup
\allowdisplaybreaks
\begin{align}
&\mathrm{e}^{i\pi (Y_2-Y_1) \sum_{a=1}^{N_f}X_a}\mathcal{Z}^{3d}_{\text{SQCD}_U(N_c,N_f)}\left(\vec{X};Y_1-Y_2\right)=\nn\\
&=\mathrm{e}^{i\pi (Y_2-Y_1) \sum_{a=1}^{N_f}X_a}\int\udl{\vec{Z}_{N_c}}\Gd_{N_c}^{3d}(\vec{Z};m_A)\mathrm{e}^{2\pi i(Y_1-Y_2)\sum_{j}Z_j}\prod_{j=1}^{N_c}\prod_{a=1}^{N_f}\sbfunc{i\frac{Q}{2}\pm(Z_j-X_a)-m_A}\,,
\end{align}
\endgroup
we can get that of the $SU(N_c)$ SQCD by integrating over $Y_2-Y_1$ and adding the singlet chiral
\begingroup
\allowdisplaybreaks
\begin{align}\label{eq:gauge3dSQCD}
&\sbfunc{i\frac{Q}{2}-2m_A}\int\udl{(Y_2-Y_1)}\mathrm{e}^{2\pi i (Y_2-Y_1) \left(\frac12 \sum_{a=1}^{N_f}X_a+B\right)}\mathcal{Z}^{3d}_{\text{SQCD}_U(N_c,N_f)}(\vec{X};Y_1-Y_2)=\nn\\
&=\sbfunc{i\frac{Q}{2}-2m_A}\int\udl{\vec{Z}_{N_c}}\Gd_{N_c}^{3d}(\vec{Z};m_A)\prod_{j=1}^{N_c}\prod_{a=1}^{N_f}\sbfunc{i\frac{Q}{2}\pm(Z_j-X_a)-m_A}\nn\\
&\qquad\qquad\qquad\qquad\quad\times\int\udl{(Y_2-Y_1)}\mathrm{e}^{2\pi i(Y_2-Y_1)\left(\frac{1}{2}\sum_{a=1}^{N_f} X_a+B-\sum_{j}Z_j\right)}\nn\\
&=\sbfunc{i\frac{Q}{2}-2m_A}\int\udl{\vec{Z}_{N_c}}\gd\left(\sum_{j}Z_j-B-\frac{1}{2}\sum_{a=1}^{N_f}X_a\right)\Gd_{N_c}^{3d}(\vec{Z};m_A)\nn\\
&\qquad\qquad\qquad\qquad\quad\times\prod_{j=1}^{N_c}\prod_{a=1}^{N_f}\sbfunc{i\frac{Q}{2}\pm(Z_j-X_a)-m_A}\nn\\
&=\int\udl{\vec{Z}_{N_c}}\gd\bigg(\sum_{j}Z_j\bigg)\widetilde{\Gd}_{N_c}^{3d}(\vec{Z};m_A)\prod_{j=1}^{N_c}\prod_{a=1}^{N_f}\sbfunc{i\frac{Q}{2}\pm\left(Z_j-X_a+B+\frac{1}{2}\sum_{a=1}^{N_f}X_a\right)-m_A}\nn\\
&=\mathcal{Z}^{3d}_{\text{SQCD}_{SU}(N_c,N_f)}(\vec{X};B)\,,
\end{align}
\endgroup
where in the last step we performed the change of variables $Z_j\to Z_j+B+\tfrac{1}{2}\sum_{a=1}^{N_f}X_a$ and we defined the $SU(N_c)$ vector multiplet contribution
\be
\widetilde{\Gd}_{N_c}^{3d}(\vec{Z};m_A)=\frac{\sbfunc{-i\frac{Q}{2}+2m_A}^{N-1}\prod_{j<l}^N s_b\left( -i\frac{Q}{2}+2m_A \pm (Z_j-Z_l) \right)}{\prod_{j<l}^N s_b\left( i\frac{Q}{2} \pm (Z_j-Z_l) \right)}
\ee
by first using the relation $\sbfunc{x}\sbfunc{-x}=1$ to simplify the contribution of the massive fields.
Notice that when integrating over $Y_2-Y_1$ we had the possibility to introduce the parameter $B$ which, together with the $U(1)$ diagonal flavor symmetry parameter $\tfrac{1}{2}\sum_{a=1}^{N_f}X_a$, appears as the real mass for the $U(1)$ baryonic symmetry in the final partition function of the $SU(N_c)$ SQCD. One can gauge this baryonic symmetry to go back to the $U(N_c)$ SQCD.

\begin{figure}[t]
	\includegraphics[width=\textwidth]{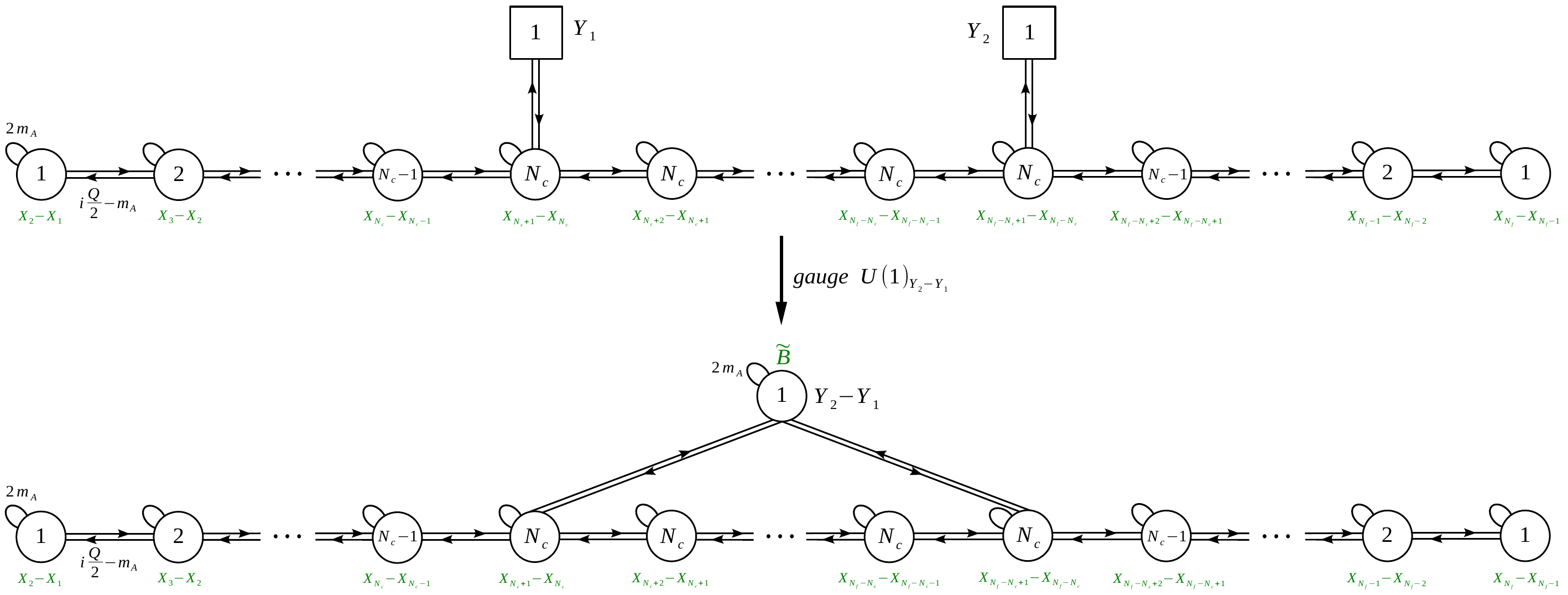}
	\caption{On top the mirror dual of the $U(N_c)$ SQCD with $N_f\geq 2N_c$ flavors. On the bottom the mirror dual of the $SU(N_c)$ SQCD with $N_f\geq 2N_c$ flavors obtained by gauging $U(1)_{Y_2-Y_1}$. We wrote the charges for one representative adjoint chiral and one bifundamental.}
		 \label{fig:3dmirrSQCD}
\end{figure}

Let us see what is the effect of the gauging in the mirror dual, which is depicted on the top of Figure \ref{fig:3dmirrSQCD}. The detailed mapping of the global symmetries has been worked out in \cite{Comi:2022aqo} and is reviewed in Appendix \ref{3dgoodres}. The symmetry that we want to gauge corresponds to the combination $Y_2-Y_1$, and the parameters $Y_1$, $Y_2$ only appear in the contribution to the partition function of the two flavors attached respectively to the leftmost and the rightmost $U(N_c)$ gauge nodes, which is
\begin{equation}
\prod_{j=1}^{N_c}\sbfunc{\pm(Z^{(N_c)}_j-Y_1)+m_A}\sbfunc{\pm(Z^{(N_f-N_c)}_j-Y_2)+m_A}\,,
\end{equation}
where the label $k=1,\cdots,N_f-1$ on the gauge parameters $Z^{(k)}_j$ increases going from the left to the right of the quiver. It is useful to perform the redefinition
\begin{equation}
\tilde{Y}_1=Y_2-Y_1\,,\,\,\,\, \tilde{Y}_2=Y_1+Y_2\quad\Longleftrightarrow\quad Y_1=\frac{\tilde{Y}_2-\tilde{Y}_1}{2}\,,\,\,\,\, Y_2=\frac{\tilde{Y}_1+\tilde{Y}_2}{2}\,,
\end{equation}
so to get the contribution
\begin{equation}
\prod_{j=1}^{N_c}\sbfunc{\pm\left(Z^{(N_c)}_j-\frac{\tilde{Y}_2-\tilde{Y}_1}{2}\right)+m_A}\sbfunc{\pm\left(Z^{(N_f-N_c)}_j-\frac{\tilde{Y}_1+\tilde{Y}_2}{2}\right)+m_A}\,.
\end{equation}
Notice that the parameter $\tilde{Y}_2$ is redundant since it can be removed by performing an overall change of variables $Z^{(k)}_j\to Z^{(k)}_j+\frac{\tilde{Y}_2}{2}$. The gauging that we performed in \eqref{eq:gauge3dSQCD} consists then in integrating over $\tilde{Y}_1$ with the phase $\mathrm{e}^{2\pi i B\tilde{Y}_1}$ and the twisted vector multiplet, which results in the quiver depicted on the bottom of Figure \ref{fig:3dmirrSQCD}. Notice that the parameter $B$ associated with the baryonic symmetry in the $SU(N_c)$ SQCD appears as the parameter for the topological symmetry of the new $U(1)$ gauge node, but shifted by some of the $X_a$ due to the contact terms in \eqref{eq:3d good mirror} (we call the resulting parameter $\tilde{B}$ in the figure). Moreover, adding a twisted vector multiplet as opposed to an ordinary vector multiplet when gauging a flavor symmetry in the mirror dual is the correct thing to do since we recall that under mirror symmetry the $SU(2)_H$ and the $SU(2)_C$ factors in the $\mathcal{N}=4$ $SO(4)_R\cong SU(2)_H\times SU(2)_C$ R-symmetry are swapped.

\subsection{A good linear quiver example}

Let us consider now a more complicated example. We start from the mirror pair between linear quiver theories with unitary gauge nodes only that are depicted in Figure \ref{fig:3dmirrUcomplicated}, where we also specify the mapping between flavor masses and FI parameters that can be obtained from the dualization algorithm. From this, we want to derive the mirror pair of Figure \ref{fig:3dmirrSUcomplicated} where on one side of the duality two $U(2)$ and one $U(5)$ gauge nodes have been replaced by $SU(2)$ and $SU(5)$ respectively. We consider this mirror pair to compare with the same example studied in \cite{Bourget:2021jwo} using the brane webs, see eqs.~(4.19)-(4.23) there; here we rederive it using the field theory technique we explained above for the SQCD.

\afterpage{
\begin{landscape}
	\begin{figure}
	\includegraphics[scale=.31]{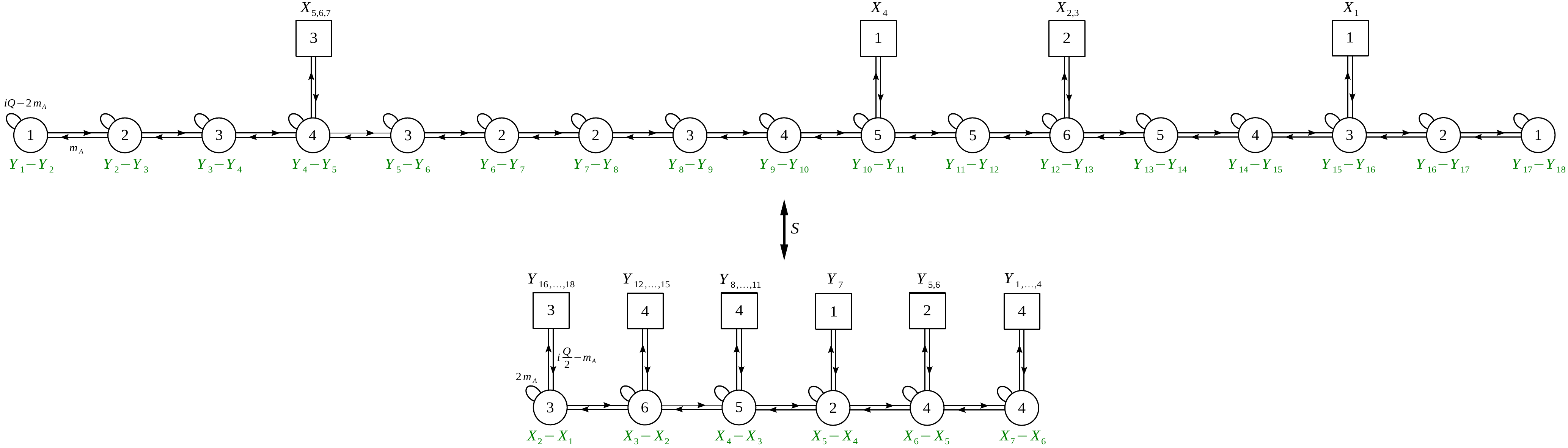}
	\caption{A 3d mirror duality between linear quiver theories with unitary nodes only, where we specify the mapping of symmetries obtained from the dualization algorithm.}
		 \label{fig:3dmirrUcomplicated}
\end{figure}
\begin{figure}
	\includegraphics[scale=.31]{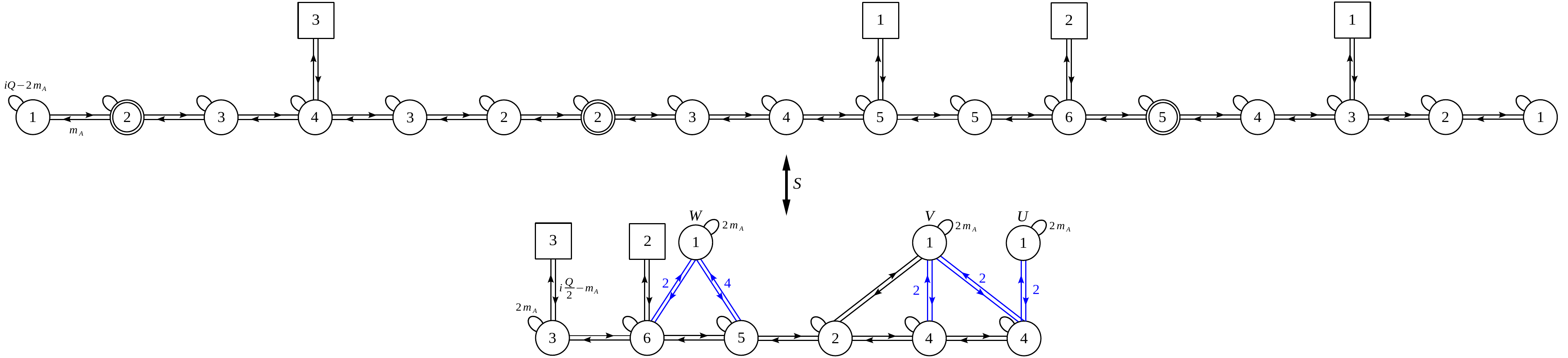}
	\caption{The 3d mirror duality between linear quiver theories with both unitary and special unitary nodes that can be obtained via gauging in the one of Figure \ref{fig:3dmirrUcomplicated}. The double circles denote special unitary gauge nodes. The numbers written close to the colored links denote the multiplicities of the corresponding chirals (which form hypers in pairs). In the bottom quiver we also label the $U(1)$ symmetries that were gauged to get this duality.}
		 \label{fig:3dmirrSUcomplicated}
\end{figure}
\end{landscape}
}

We can obtain the quiver on the top of Figure \ref{fig:3dmirrSUcomplicated} from the one on the top of Figure \ref{fig:3dmirrUcomplicated} by gauging three $U(1)$ symmetries that correspond to the following combinations of FI parameters:
\begin{equation}\label{eq:FIcomplicatedcase}
Y_2-Y_3\,,\quad Y_7-Y_8\,,\quad Y_{13}-Y_{14}\,,
\end{equation}
while leaving the others unchanged. This can be achieved by performing a redefinition of all the FI parameters
\begin{align}\label{eq:redefcomplcatedcase}
&Y_i=U+V+W+\tilde{Y}_i\,,\quad i=1,2\,,\nn\\
&Y_i=V+W+\tilde{Y}_i\,,\quad i=3,\cdots,7\,,\nn\\
&Y_i=W+\tilde{Y}_i\,,\quad i=8,\cdots,13\,,
\end{align}
where the $\tilde{Y}_i$ parameters are not all independent since we introduced the three extra parameters $U$, $V$, $W$. The gauging we are interested in corresponds in the $S^3_b$ partition function to integrating over $U$, $V$, $W$ since only the combination of FI's in \eqref{eq:FIcomplicatedcase} depend on them, while those of all the other gauge nodes do not. If we focus on the relevant part of the partition function we have\footnote{Here we ignore any contact term for simplicity. As we have seen in the SQCD case, these would just have the effect of shifting the parameters $B_i$ for the baryonic symmetries, so that we have some redefined parameters $\tilde{B}_i$.}
\begingroup\allowdisplaybreaks
\begin{align}
&\sbfunc{i\frac{Q}{2}-2m_A}^3\int\udl{U}\udl{V}\udl{W}\mathrm{e}^{-2\pi i(B_1U+B_2V+B_3W)}\mathrm{e}^{2\pi i(U+\tilde{Y}_2-\tilde{Y}_3)\sum_j Z^{(2)}_j}\nn\\
&\qquad\qquad\times\mathrm{e}^{2\pi i(V+\tilde{Y}_7-\tilde{Y}_8)\sum_j Z^{(7)}_j}\mathrm{e}^{2\pi i(W+\tilde{Y}_{13}-\tilde{Y}_{14})\sum_j Z^{(13)}_j}=\nn\\
&=\sbfunc{i\frac{Q}{2}-2m_A}^3\mathrm{e}^{2\pi i(\tilde{Y}_2-\tilde{Y}_3)\sum_j Z^{(2)}_j}\mathrm{e}^{2\pi i(\tilde{Y}_7-\tilde{Y}_8)\sum_j Z^{(7)}_j}\mathrm{e}^{2\pi i(\tilde{Y}_{13}-\tilde{Y}_{14})\sum_j Z^{(13)}_j}\nn\\
&\qquad\qquad\times\underbrace{\left(\int\udl{U}\mathrm{e}^{2\pi iU(\sum_j Z^{(2)}_j-B_1)}\right)}_{\gd(\sum_j Z^{(2)}_j-B_1)}\underbrace{\left(\int\udl{V}\mathrm{e}^{2\pi iV(\sum_j Z^{(7)}_j-B_2)}\right)}_{\gd(\sum_j Z^{(7)}_j-B_2)}\nn\\
&\qquad\qquad\times\underbrace{\left(\int\udl{W}\mathrm{e}^{2\pi iW(\sum_j Z^{(13)}_j-B_3)}\right)}_{\gd(\sum_j Z^{(13)}_j-B_3)}\nn\\
&=\sbfunc{i\frac{Q}{2}-2m_A}^3\mathrm{e}^{2\pi i(\tilde{Y}_2-\tilde{Y}_3)B_1}\mathrm{e}^{2\pi i(\tilde{Y}_7-\tilde{Y}_8)B_2}\mathrm{e}^{2\pi i(\tilde{Y}_{13}-\tilde{Y}_{14})B_3}\nn\\
&\qquad\qquad\times\gd\left(\sum_j Z^{(2)}_j-B_1\right)\gd\left(\sum_j Z^{(7)}_j-B_2\right)\gd\left(\sum_j Z^{(13)}_j-B_3\right)\,.
\end{align}
\endgroup
The three delta functions impose the conditions that turn the desired unitary nodes into special unitary, with the parameters $B_i$ playing the role of the three new baryonic masses. The three singlets correspond to the adjoint chirals inside the twisted vector multiplets associated with the gauged $U(1)$ topological symmetries and their role is again to give mass to the trace parts of the adjoint chirals for the three unitary nodes that we turned into special unitary.

Let us study now the effect of the gauging in the mirror dual depicted on the bottom of Figure \ref{fig:3dmirrUcomplicated}. From the redefinition \eqref{eq:redefcomplcatedcase} we can see for example that the $U(1)$ symmetry corresponding to the parameter $U$ we want to gauge acts with charge $1$ on two out of the four flavors attached to the rightmost $U(4)$ gauge node. The $U(1)$ symmetry associated with $V$ instead acts on the other two flavors of the rightmost $U(4)$ node, on both of the flavors attached to the $U(4)$ on its left and on the single flavor attached to the $U(2)$ node. Finally, the last $U(1)$ symmetry we gauge, which is associated with the parameter $W$, acts on all the four flavors attached to the $U(5)$ node and two out of the four flavors attached to the $U(6)$ node. Again for each of these three gaugings we add a twisted vector multiplet, as appropriate when gauging a flavor symmetry in the mirror dual. The result of the gauging can then be summarized with the quiver on the bottom of Figure \ref{fig:3dmirrSUcomplicated}. Again the parameters $B_i$ for the baryonic symmetry are mapped in the mirror dual to the topological symmetries of the new $U(1)$ nodes, up to possible shifts by $X_a$ due to contact terms. 

Any $U(N)$ node of good linear quivers can be turned into an $SU(N)$ node in this way, and the same gauging on the mirror dual side leads to the mirror dual of such $SU(N)$ quiver. In general, the freezing of the diagonal $U(1)$ allows a new baryonic symmetry, which appears as the topological symmetry of the new gauge node on the dual side. Furthermore, this manipulation is also applicable to a bad theory. The only difference is that the bad theory brings multiple mirror duals associated with some delta distributions of the FI parameters, some of which will be gauged to construct the $SU(N)$ nodes.

\subsection{Bad SQCD examples}

In this subsection we extend the procedure we just explained for gauging the topological symmetry to the case of  bad theories with unitary nodes  to obtain  bad theories with special unitary nodes. We will focus on the bad SQCD, but the same analysis can be equally applied to bad linear quiver theories.
The main difference compared to the previous section is in the effect of this gauging on the dual side.

As we have explained in this paper, for a bad theory we have both a mirror dual and an electric dual. In particular, the electric dual was defined starting from the mirror dual and taking  the
mirror dual of the interacting part of each frame given by a good theory. As we have anticipated in Section  \ref{sec:3}, in certain situations it might be useful to dualize also the free sector which is given by decoupled (twisted) hypers. We can do this by using the elementary mirror duality relating the SQED with one hyper to a free twisted hyper. At the level of partition function this corresponds to the following identity:
\begin{align}
\label{eq:SQED}
s_b\left(-\frac{iQ}{2}+2 m_A\right) \int \udl{Z} e^{2 \pi i \xi Z} \, s_b\left(\frac{iQ}{2}-m_A\pm Z\right) = s_b\left(\frac{iQ}{2}-\frac12 (iQ-2 m_A)\pm\xi\right).
\end{align}

The reason why this further dualization is important when studying special unitary quivers is that the twisted hypers are charged under the topological symmetry which we are gauging, so in principle they are no longer free.
However, even though this further dualization should be performed in each frame, this is actually trivial in the frames with a delta function. Indeed the presence of such a delta indicates, as explained in this paper and in \cite{Giacomelli:2023zkk}, that some monopole operator is taking a VEV that causes a spontaneous breaking of the topological symmetry. Hence, in these frames when we gauge the topological symmetry this gets Higgsed due to the VEV and in the IR the twisted hypers charged under it become free again.
In conclusion, this extra dualization of the free sector is important only for the case with no deltas and generic FI parameters.

Notice also that the effect of the gauging on the frames containing delta functions is just to implement the constraint they set.
The partition function of the special unitary SQCD  is then given by a sum of apparently good unitarity SQCDs with FI parameters tuned to the specific values set by the delta constraint. In particular, contrary to the case  of the bad unitary SQCD, the partition function will not be a distribution. Nevertheless because of the constraint on the FI parameters all the frames contributing to the partition function are still bad, indeed they contain monopole operators below the unitarity bound.

We will now discuss explicitly  two particular cases, the $SU(N_c)$ SQCD with $N_c=N_f=2$ and $N_c=6$, $N_f=7$.

\subsubsection{$N_c=N_f=2$}
\subsubsection*{Electric dual}
We begin by recalling the result for the  partition function of the electric dual of the 
 $U(N_c)$ SQCD with $N_c = N_f = 2$ from \cite{Giacomelli:2023zkk} which we review in \eqref{eq:3d_bad_SQCD}\footnote{Compared to \eqref{eq:3d_bad_SQCD} we actually used the constraint imposed by the delta to simplify some of the contributions, as explained in \cite{Giacomelli:2023zkk}.}
\begingroup
\allowdisplaybreaks
\begin{align}
	& 
	\mathcal{Z}^{3d}_{\text{SQCD}_U(2,2)}\left(\vec{X};Y_1-Y_2\right)
	\times
	e^{-\pi i (Y_1-Y_2) \sum_{a=1}^{2}X_a}
	= \nonumber\\
	& \quad = \,\,
	\Bigg[
	\delta\left(Y_1-Y_2 \right) \times
		e^{\pi i (iQ-2m_A)\sum_{a=1}^{2}X_a} 
	\nonumber\\
	& \qquad\qquad \times
	s_b\left(\frac{iQ}{2}-(iQ-2m_A)\right) 
	\mathcal{Z}^{3d}_{\text{SQCD}_U\left(1,2\right)}\left(\vec{X};Y_2-Y_1-(iQ-2m_A)\right)
	\Bigg]
	\nonumber\\[8pt]
	 & \qquad + 
	\Bigg[
	e^{-\pi i (Y_2-Y_1)\sum_{a=1}^{2}X_a} \times
	\prod_{j=1}^{2}s_b\left(\frac{iQ}{2}-\left(j-1\right)(iQ-2m_A)\pm (Y_1- Y_2)\right) 
	\Bigg]
	 \,.
	 \label{22}
\end{align}
\endgroup
Now we gauge the topological symmetry exactly in the same way as we did for the good SQCD. On the l.h.s.~the result is identical to the good case; we just get the $SU(2)$ SQCD with two flavor
\begingroup\allowdisplaybreaks
\begin{align}
	&\sbfunc{i\frac{Q}{2}-2m_A} \int \udl{(Y_2-Y_1)}\, e^{2\pi i B(Y_2-Y_1)}\mathcal{Z}^{3d}_{\text{SQCD}_U(2,2)}\left(\vec{X};Y_1-Y_2\right) e^{-\pi i (Y_1-Y_2) \sum_{a=1}^{2}X_a} = \nonumber\\[8pt]
	& =\sbfunc{i\frac{Q}{2}-2m_A}  
	\int \udl{\vec{Z}_{2}}\Delta_{2}^{(3d)}\left(\vec{Z};m_A \right) 
	\prod_{j=1}^{2}\prod_{a=1}^{2} s_b\left( i\frac{Q}{2}-m_A\pm(Z_j-X_a) \right)\nonumber\\
	& \quad\times \int \udl{(Y_2-Y_1)}\, 
	e^{2\pi i(Y_1-Y_2)\left(\sum_{j=1}^{2}Z_j-B -\frac{1}{2}\sum_{a=1}^{2}X_a \right)}
	\nonumber\\[8pt]
	& = 
	\int \udl{\vec{Z}_{2}}\,
	\delta\left( \sum_{j=1}^{2}Z_j-B -\frac{1}{2}\sum_{a=1}^{2}X_a \right)
	\widetilde{\Delta}_{2}^{(3d)}\left(\vec{Z};m_A \right) 
	\prod_{j=1}^{2}\prod_{a=1}^{2} s_b\left( i\frac{Q}{2}-m_A\pm(Z_j-X_a) \right)=
		\nonumber\\[8pt]
	& = \mathcal{Z}^{3d}_{\text{SQCD}_{SU(2,2)}}(\vec X,B)
	\,.
\end{align}
\endgroup
On the other hand, integrating over $Y_2-Y_1$ on the r.h.s.~of \eqref{22} gives the following contributions.
\begin{itemize}
\item Electric frame (A):
The first frame in \eqref{22} has an interacting part given by an SQED with two flavors and a delta function. The effect of the gauging is simply to implement the constraint set by the delta function
		\begingroup
	\allowdisplaybreaks
	\begin{align}
	& \Bigg[
	\int \udl{(Y_2-Y_1)}\,e^{2\pi i B(Y_2-Y_1)} \times
	\delta\left(Y_1-Y_2 \right) \nonumber\\
	& \qquad\quad\times
	e^{\pi i (iQ-2m_A)\sum_{a=1}^{2}X_a}  \times
	\mathcal{Z}^{3d}_{\text{SQCD}_U\left(1,2\right)}\left(\vec{X};Y_2-Y_1-(iQ-2m_A)\right) 
	\Bigg]
= \nonumber\\[10pt]
	& = 		e^{\pi i (iQ-2m_A)\sum_{a=1}^{2}X_a} \mathcal{Z}^{3d}_{\text{SQCD}_U\left(1,2\right)}\left(\vec{X};-(iQ-2m_A)\right) 
	\,.
	\end{align}
	\endgroup	
The electric frame (A) is then an  SQED with two flavors  with FI parameter frozen to the value $-(iQ-2m_A)$.

\item Electric frame (B): The second frame of theory \eqref{22} has only a free sector consisting of two twisted hypermultiplets.
As we have explained, to obtain a correct electric dual frame, we have to dualize each twisted hyper using \eqref{eq:SQED} as follows since  we will gauge the topological symmetry that they are charged under, which will make them interacting:
\begin{align}
&s_b\left(\frac{iQ}{2}-\frac12 (iQ-2m_A)\pm \left(Y_1- Y_2\pm\frac12(iQ-2m_A)\right)\right) \nn\\
&= s_b\left(-\frac{iQ}{2}+2 m_A\right) \int \udl{Z} e^{2 \pi i Z \left(Y_1- Y_2+\frac12(iQ-2m_A)\right)} \, s_b\left(\frac{iQ}{2}-m_A\pm Z\right) \nonumber \\
&\quad \times s_b\left(-\frac{iQ}{2}+2 m_A\right) \int \udl{W} e^{2 \pi i W \left(Y_1- Y_2-\frac12(iQ-2m_A)\right)} \, s_b\left(\frac{iQ}{2}-m_A\pm W\right)
\end{align}
where in the first line we have massaged the chirals in the second frame of \eqref{22}  such that they come in pairs correctly reconstructing the twisted hypermultiplets.
Now we proceed with gauging the topological symmetry with the usual prescription
\begin{align}
&\sbfunc{i\frac{Q}{2}-2m_A} \int \udl{(Y_2-Y_1)} \, e^{2\pi i B(Y_2-Y_1)} \times e^{-\pi i (Y_2-Y_1)\sum_{a=1}^{2}X_a} \nonumber \\
&s_b\left(-\frac{iQ}{2}+2 m_A\right) \int \udl{Z} e^{2 \pi i Z \left(Y_1- Y_2+\frac12(iQ-2m_A)\right)} \, s_b\left(\frac{iQ}{2}-m_A\pm Z\right) \nonumber \\
&\quad \times s_b\left(-\frac{iQ}{2}+2 m_A\right) \int \udl{W} e^{2 \pi i W \left(Y_1- Y_2-\frac12(iQ-2m_A)\right)} \, s_b\left(\frac{iQ}{2}-m_A\pm W\right)\,.
\end{align}
Performing the $Y_2-Y_1$ integration explicitly, we get
\begin{align}
\label{eq:SU22B2}
\sbfunc{-i\frac{Q}{2}+2m_A} \int \udl{Z} e^{2 \pi i Z (iQ-2m_A)} \, s_b\left(\frac{iQ}{2}-m_A\pm \left(Z\pm\left(\frac12 B-\frac14 \sum_{a = 1}^2 X_a\right)\right)\right)
\end{align}
where we have used that $\int \udl{(Y_2-Y_1)} e^{2 \pi i \xi (Y_2-Y_1)} = \delta(\xi)$ and shifted the integration variable $Z \rightarrow Z+\frac12 B-\frac14 \sum_{a = 1}^2 X_a$. Hence, the electric frame (B) is also given by the $\mathcal{N}=4$ SQED with two flavors, but compared to frame (A) it has  different frozen FI and mass parameters.
\end{itemize}

In both cases, we get an $\mathcal{N}=4$ SQED with two flavors, which is compatible with the findings of \cite{Ferlito:2016grh,Assel:2018exy,Bourget:2021jwo,Bourget:2023cgs}.
However, the mass parameter $B$ for the baryonic symmetry of the $SU(2)$ SQCD only appears in frame (B). In particular, from the partition function expression we see that the $U(1)$ baryonic symmetry of the $SU(2)$ SQCD is identified with the Cartan of the $SU(2)$ flavor symmetry of the SQED with two flavors of frame (B).
This indicates that this frame is capturing the baryonic branch of the $SU(2)$ theory.

\begin{figure}[t]
	\centering
	\includegraphics[width=0.55\textwidth]{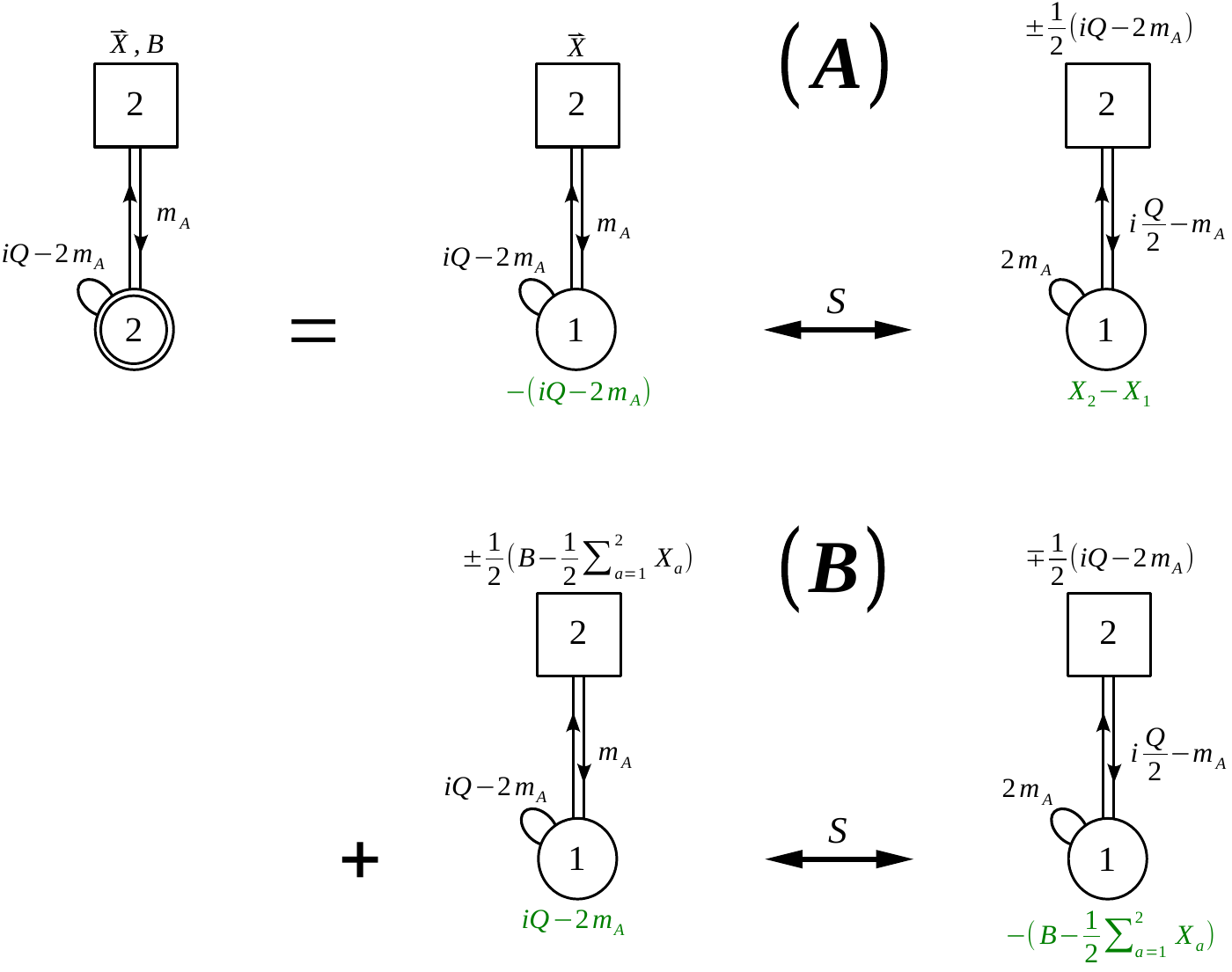}
	\caption{The electric and mirror dual frames of the $SU(2)$ theory with two flavors ($N_c = N_f = 2$). The electric frames have frozen FI parameters, whereas the mirror frames have frozen flavor mass parameters. Note that the baryonic mass parameter $B$ only appears in frame (B).}
	\label{fig:SU22}
\end{figure}

\subsubsection*{Mirror dual}

If we start instead from the mirror dual of the $U(2)$ SQCD  in  \eqref{22}  and perform the gauging, the situation is as follows.

\begin{itemize}
\item Mirror frame (A):
The first mirror frame is again  an SQED theory with two hypers (the mirror of the first frame in \eqref{22}) and comes with the same delta function. When we gauge the topological symmetry, which is now a flavor symmetry, we simply implement the delta constraint fixing the mass parameter.
 Both the electric frame (A) and the mirror frame (A) are depicted in the first line of Figure \ref{fig:SU22}.

\item Mirror frame (B):
In this case we directly gauge the topological symmetry in the second frame of \eqref{22} consisting of two twisted hypers
(without dualizing them first). We then obtain again an SQED theory with two hypers.
 Both the electric frame (B) and the mirror frame (B) are depicted in the second line of Figure \ref{fig:SU22}.
Notice that now the barionic  mass $B$ appears as an FI parameter in the mirror dual. In particular, the $U(1)$ baryonic symmetry is mapped to the $U(1)$ topological theory which is enhanced in the IR to $SU(2)$ in the SQED with two flavors.
\end{itemize}

As a nice consistency check we notice  that the mirror frames on the right column of Figure \ref{fig:SU22}, which we obtained by starting from the  mirror dual
of the $U(2)$ SQCD  in  \eqref{22}  and then gauging,  can also be obtained  by mirror dualizing directly the frames on the left column of Figure \ref{fig:SU22}.

\subsubsection{$N_c=6$, $N_f=7$}
\subsubsection*{Electric dual}
\begin{figure}[!ht]
	\centering
	\includegraphics[width=\textwidth]{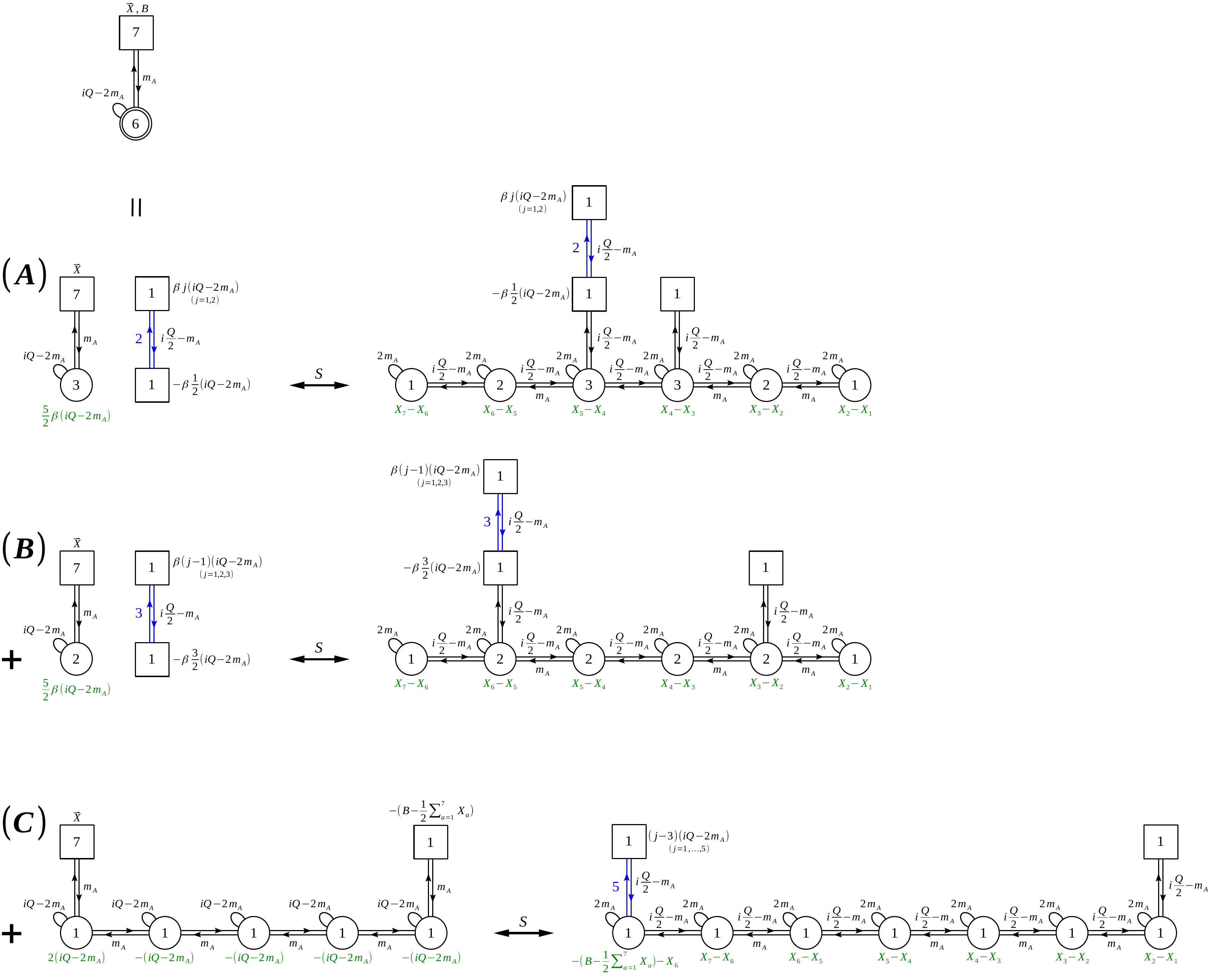}
	\caption{The electric and mirror dual frames of the $SU(6)$ theory with 7 hypers. }
	\label{fig:SU67}
\end{figure}

We begin by recalling the result for the   partition function of the electric dual of the 
 $U(N_c)$ SQCD with $N_6 = N_f = 7$ from \cite{Giacomelli:2023zkk}  which we review in \eqref{eq:3d_bad_SQCD}
\begingroup
\allowdisplaybreaks
\begin{align}
	&  
	\mathcal{Z}^{3d}_{\text{SQCD}_U(6,7)}\left(\vec{X};Y_1-Y_2\right) \times
	e^{-\pi i (Y_1-Y_2) \sum_{a=1}^{7}X_a}
	= \nonumber\\
	& \quad = 
	\sum_{\beta=\pm 1} \left[
	\delta\left(Y_1-\left(Y_2-\beta\frac{1}{2}(iQ-2m_A)\right) \right) \times
	e^{-\pi i (Y_2-Y_1+2 \beta(iQ-2m_A)) \sum_{a=1}^{7}X_a}
	\bigphantomspace\right.\nonumber\\
	& \qquad\qquad\times
	s_b\left(\frac{iQ}{2}-3(iQ-2m_A)\right)
	\prod_{j=1}^{2}s_b\left(\frac{iQ}{2}\pm j(iQ-2m_A)\right)
	\nonumber\\
	& \qquad\qquad \times 
	\mathcal{Z}^{3d}_{\text{SQCD}_U\left(3,7\right)}\left(\vec{X};Y_2-Y_1+2\beta(iQ-2m_A)\right)
	\left.\bigphantomspace\right]\nonumber\\[5pt]
	& \qquad
	+\sum_{\beta=\pm 1} \left[
	\delta\left(Y_1-\left(Y_2-\beta\frac{3}{2}(iQ-2m_A)\right) \right) \times
	e^{- \pi i (Y_2-Y_1+\beta(iQ-2m_A)) \sum_{a=1}^{7}X_a}
	\bigphantomspace\right.\nonumber\\
	& \qquad\qquad \times
	s_b\left(\frac{iQ}{2}-4(iQ-2m_A)\right)
	\prod_{j=1}^{3}s_b\left(\frac{iQ}{2}\pm j(iQ-2m_A)\right)
	\nonumber\\
	& \qquad\qquad \times
	\mathcal{Z}^{3d}_{\text{SQCD}_U\left(2,7\right)}\left(\vec{X};Y_2-Y_1+\beta(iQ-2m_A)\right)
	\left. \bigphantomspace \right]\nonumber\\[10pt]
	& \qquad + \left[
	e^{-\pi i (Y_2-Y_1)\sum_{a=1}^{7}X_a} \times
	\prod_{j=1}^{5}s_b\left(\frac{iQ}{2}-\left(j-\frac{5}{2}\right)(iQ-2m_A)\pm (Y_1- Y_2)\right)
	\bigphantomspace \right.\nonumber\\
	& \qquad\qquad \times
	\mathcal{Z}^{3d}_{\text{SQCD}_U\left(1,7\right)}\left(\vec{X};Y_2-Y_1\right)
	\left. \bigphantomspace \right]
	 \,.
	 \label{67}
\end{align}
\endgroup
Now we gauge the topological symmetry. Once again on the l.h.s.~we get the partition function of the special unitary SQCD
\begingroup\allowdisplaybreaks
\begin{align}
	& \sbfunc{i\frac{Q}{2}-2m_A} \int \udl{(Y_2-Y_1)}\, e^{2\pi i B(Y_2-Y_1)} \mathcal{Z}^{3d}_{\text{SQCD}_U(6,7)}\left(\vec{X};Y_1-Y_2\right) 
	e^{-\pi i (Y_1-Y_2) \sum_{a=1}^{7}X_a} = \nonumber\\[10pt]
	& = \sbfunc{i\frac{Q}{2}-2m_A} 
	\int \udl{\vec{Z}_{6}}\Delta_{6}^{(3d)}\left(\vec{Z};m_A \right) 
	\prod_{j=1}^{6}\prod_{l=1}^{7} s_b\left( i\frac{Q}{2}-m_A\pm(Z_j-X_l) \right)\nonumber\\
	& \quad\times \int \udl{(Y_2-Y_1)}\, 
	e^{2\pi i(Y_1-Y_2)\left(\sum_{j=1}^{6}Z_j-B -\frac{1}{2}\sum_{a=1}^{7}X_a \right)}
	\nonumber\\[8pt]
	& = 
	\int \udl{\vec{Z}_{6}}\,
	\delta\left( \sum_{j=1}^{6}Z_j-B -\frac{1}{2}\sum_{a=1}^{7}X_a \right)
	\widetilde{\Delta}_{6}^{(3d)}\left(\vec{Z};m_A \right) 
	\prod_{j=1}^{6}\prod_{a=1}^{7} s_b\left( i\frac{Q}{2}-m_A\pm(Z_j-X_a) \right) 
			\nonumber\\[8pt]
	& = \mathcal{Z}^{3d}_{\text{SQCD}_{SU(6,7)}}(\vec X,B)
	\,.
\end{align}
\endgroup
On the other hand, integrating over $Y_2-Y_1$ on the r.h.s.~gives the following contributions.
\begin{itemize}
	\item Electric frame (A):
	The first frame in \eqref{67} has an interacting part given by a $U(3)$ theory with 7 hypers, 2 free twisted hypers
	 and a delta function. The effect of the gauging is simply to implement the constraint set by the delta function
	\begingroup
	\allowdisplaybreaks
	\begin{align}
	\label{eq:SU67A}
	& \sum_{\beta=\pm 1} \left[
	\int \udl{(Y_2-Y_1)}\, e^{2\pi i B(Y_2-Y_1)}
	\delta\left(Y_1-\left(Y_2-\beta\frac{1}{2}(iQ-2m_A)\right) \right)
	\bigphantomspace\right.\nonumber\\
	& \qquad\qquad\times
	e^{-\pi i (Y_2-Y_1+2 \beta(iQ-2m_A)) \sum_{a=1}^{7}X_a} \nonumber\\
	& \qquad\qquad\times
	s_b\left(\frac{iQ}{2}-3(iQ-2m_A)\right)s_b\left(\frac{iQ}{2}+(iQ-2m_A)\right)
	s_b\left(\frac{iQ}{2}\pm 2(iQ-2m_A)\right) \nonumber \\
	& \qquad\qquad \times 
	\mathcal{Z}^{3d}_{\text{SQCD}_U\left(3,7\right)}\left(\vec{X};Y_2-Y_1+2\beta(iQ-2m_A)\right)
	\left.\bigphantomspace\right] = \nonumber\\
	& = \sum_{\beta=\pm 1} \left[
	e^{\pi i \beta(iQ-2m_A) B} \times
	e^{-\frac{5}{2}\pi i \beta(iQ-2m_A)\sum_{a=1}^{7}X_a} \bigphantomspace\right.\nonumber\\
	&\times
	\prod_{j = 1}^2 s_b\left(\frac{iQ}{2}-\frac12 (i Q-2 m_A)\pm\beta \frac{2 j+1}{2} (iQ-2 m_A)\right)
	\mathcal{Z}^{3d}_{\text{SQCD}_U\left(3,7\right)}\left(\vec{X}; \frac{5}{2}\beta(iQ-2m_A)\right)\,.
		\end{align}
	\endgroup
This result has the structure of an $\mathcal{N}=4$ SQCD with $U(3)$ gauge group and seven hypers, again with  frozen FI parameter, plus two twisted hypers.

	\item Electric frame (B): 
	The second frame in \eqref{67} has an interacting part given by a $U(2)$ theory with 7 hypers, 2 free twisted hypers
	 and a delta function. The effect of the gauging is simply to implement the constraint set by the delta function
\begingroup
	\allowdisplaybreaks
	\begin{align}
	& \sum_{\beta=\pm 1} \left[
	\int \udl{(Y_2-Y_1)}\, e^{2\pi i B(Y_2-Y_1)}
	\delta\left(Y_1-\left(Y_2-\beta\frac{3}{2}(iQ-2m_A)\right) \right)
	\bigphantomspace\right.\nonumber\\
	& \qquad\qquad\times
	e^{-\pi i (Y_2-Y_1+\beta(iQ-2m_A)) \sum_{a=1}^{7}X_a}
	\nonumber\\
	& \qquad\qquad \times
	s_b\left(\frac{iQ}{2}-4(iQ-2m_A)\right)s_b\left(\frac{iQ}{2}+(iQ-2m_A)\right)
	\prod_{j=2}^{3}s_b\left(\frac{iQ}{2}\pm j(iQ-2m_A)\right)	\nonumber\\
	& \qquad\qquad \times
	\mathcal{Z}^{3d}_{\text{SQCD}_U\left(2,7\right)}\left(\vec{X};Y_2-Y_1+\beta(iQ-2m_A)\right)
	\left. \bigphantomspace \right] = \nonumber\\
	& =
	\sum_{\beta=\pm 1} \left[
	e^{3\pi i \beta(iQ-2m_A) B} \times
	e^{-\frac{5}{2}\pi i \beta(iQ-2m_A)\sum_{a=1}^{7}X_a}
	\bigphantomspace\right.\nonumber\\
	& \qquad\qquad \times
	\prod_{j = 1}^3 s_b\left(\frac{iQ}{2}-\frac12 (i Q-2 m_A)\pm\left(\beta \frac{3}{2} (i Q-2 m_A)+\beta (j-1) (iQ-2 m_A)\right)\right)\nonumber\\
	& \qquad\qquad \times
	\mathcal{Z}^{3d}_{\text{SQCD}_U\left(2,7\right)}\left(\vec{X};\frac{5}{2}\beta(iQ-2m_A)\right)\,.
	\end{align}
	\endgroup	
This result has the structure of an $\mathcal{N}=4$ SQCD with $U(2)$ gauge group and seven flavors, again with a frozen FI parameter, plus three twisted hypers. 

	\item Electric frame (C):
		The third frame in \eqref{67} has an interacting part given by a $U(1)$ theory with 7 hypers, 5 free twisted hypers and no delta function. 
In this case, when gauging the topological symmetry the hypers become interacting and so, to have the correct electric dual, we should dualize each twisted hyper into an SQED using \eqref{eq:SQED} as follows:
\begin{align}
	&\prod_{j=1}^{5}s_b\left(\frac{iQ}{2}-\frac12 (iQ-2m_A)\pm (Y_1- Y_2-(j-3) (iQ-2m_A))\right)  \\
	&= s_b\left(-\frac{iQ}{2}+2 m_A\right)^5 \prod_{j=1}^{5} \int \udl{W_j} e^{2 \pi i W_j \left(Y_2- Y_1+(j-3) (iQ-2m_A)\right)} \, s_b\left(\frac{iQ}{2}-m_A\pm W_j\right) \,.\nonumber
\end{align}
Now we  gauge the topological symmetry with the usual prescription and obtain
	\begingroup
	\allowdisplaybreaks
	\begin{align}
	& \sbfunc{i\frac{Q}{2}-2m_A} \int \udl{(Y_2-Y_1)}\,
	e^{2\pi i B(Y_2-Y_1)} \times
	e^{-\pi i (Y_2-Y_1)\sum_{a=1}^{7}X_a} \nonumber\\
	& \qquad\times
	s_b\left(-\frac{iQ}{2}+2 m_A\right)^5 \prod_{j=1}^{5} \int \udl{W_j} e^{2 \pi i W_j \left(Y_2- Y_1+(j-3) (iQ-2m_A)\right)} \, s_b\left(\frac{iQ}{2}-m_A\pm W_j\right) \nonumber\\
	& \qquad\times
	\int \udl{Z}\Delta_{1}^{(3d)}\left(Z;m_A \right) \times 
	e^{2\pi i (Y_2-Y_1)Z} \times
	\prod_{a=1}^{7} s_b\left( \frac{iQ}{2}-m_A\pm(Z-X_a) \right) \,.
	\end{align}
	\endgroup
Then we perform the integration over $Y_2-Y_1$, getting $\delta(B-\frac12 \sum_{a = 1}^7 X_a+Z+\sum_{j = 1}^5 W_j)$, which becomes $\delta(W_5)$ with the following reparametrization:
\begingroup\allowdisplaybreaks
\begin{align}
Z &\rightarrow Z \,, \\
W_1 &\rightarrow W_1-Z \,, \\
W_j &\rightarrow W_j-W_{j-1} \,, \qquad\qquad j=2, \dots, 4 \,, \\
W_5 &\rightarrow W_5-W_4-B+\frac12 \sum_{a=1}^7 X_a \,.
\end{align}
\endgroup
Integrating over $W_5$, we obtain the electric frame (C):
\begingroup\allowdisplaybreaks
\begin{align}
	&e^{-4 \pi i \left(B-\frac12 \sum_{a=1}^7 X_a\right) (iQ-2m_A)} \nonumber \\
	&  \times \sbfunc{-i\frac{Q}{2}+2m_A} \int \udl{Z} 
	e^{4 \pi i Z (i Q-2 m_A)} \times
	\prod_{a=1}^{7} s_b\left( \frac{iQ}{2}-m_A\pm(Z-X_a) \right) \nonumber\\
	&  \times \sbfunc{-i\frac{Q}{2}+2m_A} \int \udl{W_1} e^{-2 \pi i W_1(iQ-2m_A)} \, s_b\left(\frac{iQ}{2}-m_A\pm (W_1-Z)\right) \nonumber \\
	&  \times \sbfunc{-i\frac{Q}{2}+2m_A}^3 \prod_{j=2}^{4} \int \udl{W_j} e^{-2 \pi i W_j (iQ-2m_A)} \, s_b\left(\frac{iQ}{2}-m_A\pm (W_j-W_{j-1})\right) \nonumber \\
	& \qquad\qquad\qquad\qquad\qquad\qquad \times s_b\left(\frac{iQ}{2}-m_A\pm \left(W_4+B-\frac12 \sum_{a=1}^7 X_a\right)\right)
\end{align}
\endgroup
whose corresponding quiver is shown in Figure \ref{fig:SU67}. Notice that in this linear quiver all the  FI parameters are frozen
but the baryonic symmetry acts non-trivially, so this frame describes the baryonic branch of the theory.
\end{itemize}

\subsubsection*{Mirror dual}

If we start instead from the mirror dual of the $U(6)$ SQCD with seven hypers in \eqref{67}  and perform the gauging the situation is as follows.

\begin{itemize}
\item Mirror frame (A): The first frame in the mirror of  \eqref{67} has an interacting part given by the mirror of the $U(3)$ theory with seven hypers, 2 free twisted hypers 	 and a delta function. The effect of the gauging is simply to implement the constraint set by the delta function which just fixes the flavor mass parameter.
 Both the electric frame (A) and the mirror frame (A) are depicted in the first line of Figure \ref{fig:SU67}.

\item Mirror frame (B): The second frame in the mirror of  \eqref{67} has an interacting part given by the mirror of the $U(2)$ theory with seven hypers, 3 free twisted hypers and a delta function. The effect of the gauging is simply to implement the constraint set by the delta function which just fixes the flavor mass parameter.
 Both the electric frame (B) and the mirror frame (B) are depicted in the first line of Figure \ref{fig:SU67}.

\item Mirror frame (C): The last frame in the mirror of \eqref{67}  is  the mirror of the $U(1)$ theory with seven hypers, which is a linear quiver with 6 $U(1)$ nodes, and 7 hypers  and  5 free twisted hypers (which now we  are not supposed to dualize).
Gauging the topological symmetry corresponds to gauging one the flavor nodes with the 5 hypers.
We obtain the last quiver in \ref{fig:SU67} where we see that the baryonic mass $B$ is now the FI of the new gauge node.
\end{itemize}

Again we notice  that the mirror frames on the right column of Figure  \ref{fig:SU67}, which we obtained by starting from the  mirror dual
of the $U(6)$ SQCD  in  \eqref{67}   and then gauging,  can also be obtained  by mirror dualizing directly the frames on the left column of Figure  \ref{fig:SU67}.

\acknowledgments
We would like to thank Riccardo Comi for discussions.
CH is supported by the National Natural Science Foundation of China under Grant No.~12247103.
FM is partially supported by STFC under Grant No.~ST/V507118 - Maths 2020 DTP. 
SP is partially supported by the MUR-PRIN grant No. 2022NY2MXY.
MS is partially supported by the ERC Consolidator Grant \#864828 “Algebraic Foundations of Supersymmetric Quantum Field Theory (SCFTAlg)” and by the Simons Collaboration for the Nonperturbative Bootstrap under grant \#494786 from the Simons Foundation.

\clearpage
\appendix
\section{Conventions for the 4d index and the 3d partition function}

\subsection{The 4d index}
We will here list our conventions for the 4d $\mathcal{N}=1$ supersymmetric index \cite{Romelsberger:2005eg,Kinney:2005ej,Dolan:2008qi}.

For a theory with gauge group $G$ and chiral  multiplets of R-charge $r$, in the representations $R_G$ and $R_F$ of the gauge and
flavor symmetry groups, the index is given by the following integral:
\begin{align}
	\CI_G(\vec{x}) = \oint \prod_{j=1}^{\mathrm{rk} G} \frac{dz_j}{|W_G| 2 \pi i z_j} 
	\frac{[(p;p)_\infty (q;q)_\infty]^{\mathrm{rk} G} }{ \prod_{\vec{\r} \in G} \Gamma(\vec{z}^{\, \vec{\r}} \, ) } 
	\prod_{\s_G \in R_G} \prod_{\s_F \in R_F} \Gamma \big( (pq)^{r/2} \vec{z}^{\, \vec{\s}_G} \vec{x}^{\, \vec{\s}_F} \big) \,,
\end{align}
where $\vec{\r}$ are the roots of the gauge group $G$, $\vec{\s}_G$ and $\vec{\s}_F$ are the weights of the representations $R_G$ and $R_F$, $|W_G|$ is the dimension of the Weyl group of $G$, and  $\vec{z}$ and  $\vec{x}$ are the gauge and flavor fugacities. 
It is convenient to define the integration measure for the $USp(2N)$ group 
\begin{equation}
	\udl{\vec{z}_{N}}=\frac{1}{2^N N!}\prod_{j=1}^N\frac{\udl{z_j}}{2\pi iz_j}\,,
\end{equation}
and the  combination
\begin{align}
\Gd_N(\vec{z};t) =\Gd_N(\vec{z}) A_N(\vec{z};t)\,, 
\end{align}
containing  the contributions from the vector multiplet $\Delta_N(\vec z)$ and the antisymmetric chiral multiplet $A_N(\vec{z};t)$
\begin{equation}
	\Gd_{N}(\vec z_{N})=\frac{\left[(p;p)_\infty (q;q)_\infty\right]^N}{\prod_{i=1}^N\Gpq{z_i^{\pm2}}\prod_{i<j}^N\Gpq{z_i^{\pm1} z_j^{\pm1}}}\,,
	\quad 
	A_N(\vec{z};t)=\Gpq{t}^N \prod_{i<j}^N\Gpq{t z_i^{\pm1} z_j^{\pm1}} \,.
\end{equation}

\subsection{The 3d partition function}
In this section we introduce the notation for the 3d $\mathcal{N}=2$ $S^3_b$ partition function \cite{Kapustin:2009kz,Jafferis:2010un,Hama:2011ea}.
For a theory with gauge group $G$ and chiral  multiplets of R-charge $r$, in the representations $R_G$ and $R_F$ of the gauge and
flavor symmetry groups, the $S^3_b$ partition function is given by the following integral:
\begin{align}
	\mathcal{Z} (Y,k,\vec{X}) = \frac{1}{|W_G|} \int \prod_{j=1}^{\mathrm{rk} G} d Z_j  & Z_{\text{cl}}(Y,k) 
	\frac{\prod_{\vec{\s}_G \in R_G} \prod_{\vec{\s}_F \in R_F} s_b \left( \frac{iQ}{2}(1-r) - \vec{\s}_G(\vec{Z}) - \vec{\s}_F(\vec{X}) \right)}{\prod_{\vec{\r} \in G} s_b \big( \frac{iQ}{2} - \vec{\r}(\vec{Z}) \big) }\,,
\end{align}
where $\vec{\r}$ are the roots of $G$, $\vec{\s}_G$ and $\vec{\s}_F$ are the weights of the representations $R_G$ and $R_F$, $|W_G|$ is the dimension of the Weyl group of the gauge group $G$, $\vec{Z}$ and $\vec{X}$ are parameters in the Cartan of the gauge and flavor groups and $Q=b+b^{-1}$ with $b$ the squashing parameter of $S^3_b$. 
The classical term
\begin{align}
	Z_{\text{cl}}(Y,k) = \exp \left[ 2\pi i Y \sum_{j=1}^{\mathrm{rk} G} Z_j + \pi i k \sum_{j=1}^{\mathrm{rk} G} Z_j^2 \right] 
\end{align}
contains the contribution of the FI parameter $Y$ and of the level $k$ Chern-Simons. It is convenient to define the integration measure for the $U(N)$ group 
\begin{equation}
	\udl{\vec{Z}_{N}}=\frac{1}{N!}\prod_{a=1}^{N}\udl{Z_a} \,,
\end{equation}
and the combination
\begin{align}
\Gd_N^{3d}(\vec{Z};m_A) =\Gd_N^{3d}(\vec{Z}\,) A_N^{3d}(\vec{Z};m_A)\,, 
\end{align}
containing the contributions from both the vector $\Gd_N^{3d}(\vec{Z}\,)$ and the adjoint chiral multiplet $A_N^{3d}(\vec{Z};m_A)$
\begin{equation}
	\Delta^{3d}_N\big(\vec Z\,\big)=\frac{1}{\prod_{a<b}^{N}s_b\left(i\frac{Q}{2}\pm(Z_a-Z_b)\right)} \,,
	\quad 
	A_N^{3d}(\vec{Z};m_A)=\prod_{a,b=1}^N\sbfunc{-i\frac{Q}{2}+2m_A+(Z_a-Z_b)}\,.
\end{equation}

\section{The SQCD }
\label{app:bad_SQCD}
\subsection{The 4d good SQCD}\label{The 4d mirror pair}
The 4d $USp(2N_c)$ SQCD with $N_f$ flavors, shown on top of Figure \ref{fig:good_SQCD_Sdualised_4d}, has the following index: 
\begin{align}
	& \mathcal{I}_{\text{SQCD}(N_c,N_f)}(\vec x;y_1,y_2;t;c) = \nonumber\\
	& \quad =
	\oint \udl{\vec{z}_{N_c}}\Delta_{N_c}\left(\vec{z};t\right)
	\prod_{j=1}^{N_c}\prod_{a=1}^{N_f}\Gamma_e\left( (pq)^{\frac{1}{2}}t^{-\frac{1}{2}} z_j^{\pm} x_a^{\pm} \right) \nonumber\\
	& \quad\qquad\times
	\prod_{j=1}^{N_c} \Gamma_e\left( c z_j^{\pm} y_1^{\pm} \right)
	\prod_{j=1}^{N_c} \Gamma_e\left( t^{\frac{N_f}{2}-N_c+1} c^{-1} z_j^{\pm} y_2^{\pm} \right) \nonumber\\
	& \quad\qquad\times
	\prod_{a=1}^{N_f} \Gamma_e\left((pq)^{\frac{1}{2}} t^{N_c-\frac{N_f}{2}-\frac{1}{2}}c x_a^{\pm}y_2^{\pm} \right)
	\prod_{j=1}^{N_c} \Gamma_e\left( pqt^{1-j}c^{-2} \right) \,.
\end{align}
If the SQCD is good (namely $N_f \geq 2N_c$), its mirror dual is unique and its index is the following:
\begingroup\allowdisplaybreaks
\begin{align}
	& \mathcal{I}_{\widehat{\text{SQCD}}(N_c,N_f\geq 2N_c)}(\vec x;y_1,y_2;t;c) = \nonumber\\
	& \quad =
	\oint 
	\prod_{k=1}^{N_c} \left( \udl{\vec{z}^{\,(k)}_{k}}\Delta_{k}\left(\vec{z}^{\,(k)};pqt^{-1}\right) \right) 
	\prod_{k=N_c+1}^{N_f-N_c-1} \left( \udl{\vec{z}^{\,(k)}_{N_c}}\Delta_{N_c}\left(\vec{z}^{\,(k)};pqt^{-1}\right) \right) 
	\nonumber\\
	& \quad\qquad\times
	\prod_{k=N_f-N_c}^{N_f-1} \left( \udl{\vec{z}^{\,(k)}_{N_f-k}}\Delta_{N_f-k}\left(\vec{z}^{\,(k)};pqt^{-1}\right) \right)
	\nonumber\\
	& \quad\qquad\times
	\prod_{k=1}^{N_c-1}\prod_{j=1}^{k}\prod_{l=1}^{k+1} \Gamma_e\left( t^{\frac{1}{2}} z_j^{(k)\,\pm}z_l^{(k+1)\,\pm} \right)
	\prod_{k=N_c}^{N_f-N_c-1}\prod_{j=1}^{N_c}\prod_{l=1}^{N_c} \Gamma_e\left( t^{\frac{1}{2}} z_j^{(k)\,\pm}z_l^{(k+1)\,\pm} \right)
	\nonumber\\
	& \quad\qquad\times
	\prod_{k=N_f-N_c}^{N_f-2}\prod_{j=1}^{N_f-k}\prod_{l=1}^{N_f-k-1} \Gamma_e\left( t^{\frac{1}{2}} z_j^{(k)\,\pm}z_l^{(k+1)\,\pm} \right)
	\nonumber\\
	& \quad\qquad\times
	\prod_{j=1}^{N_c}\Gamma_e\left( t^{\frac{1}{2}} z_j^{(N_c)\,\pm} y_1^{\pm} \right)\Gamma_e\left( t^{\frac{1}{2}} z_j^{(N_f-N_c)\,\pm} y_2^{\pm} \right)
	\nonumber\\
	& \quad\qquad\times
	\prod_{k=1}^{N_c}\prod_{j=1}^{k} \Gamma_e\left( (pq)^{\frac{1}{2}}t^{\frac{k-N_c}{2}}c^{-1} z_j^{(k)\,\pm}x_k^{\pm} \right)
	\prod_{k=N_c+1}^{N_f-N_c-1}\prod_{j=1}^{N_c} \Gamma_e\left( (pq)^{\frac{1}{2}}t^{\frac{k-N_c}{2}}c^{-1} z_j^{(k)\,\pm}x_k^{\pm} \right)
	\nonumber\\
	& \quad\qquad\times
	\prod_{k=1}^{N_f-N_c}\prod_{j=1}^{N_f-k} \Gamma_e\left( (pq)^{\frac{1}{2}}t^{\frac{k-N_c}{2}}c^{-1} z_j^{(k)\,\pm}x_k^{\pm} \right)
	\nonumber\\
	& \quad\qquad\times
	\prod_{k=1}^{N_c}\prod_{j=1}^{k} \Gamma_e\left( (pq)^{\frac{1}{2}}t^{\frac{N_c-k-2}{2}}c z_j^{(k)\,\pm}x_{k+1}^{\pm} \right)
	\prod_{k=N_c+1}^{N_f-N_c-1}\prod_{j=1}^{N_c} \Gamma_e\left( (pq)^{\frac{1}{2}}t^{\frac{N_c-k-2}{2}}c z_j^{(k)\,\pm}x_{k+1}^{\pm} \right)
	\nonumber\\
	& \quad\qquad\times
	\prod_{k=N_f-N_c}^{N_f-1}\prod_{j=1}^{N_f-k} \Gamma_e\left( (pq)^{\frac{1}{2}}t^{\frac{N_c-k-2}{2}}c z_j^{(k)\,\pm}x_{k+1}^{\pm} \right)
	\nonumber\\
	& \quad\qquad\times
	\prod_{k=1}^{N_c} \Gamma_e\left( (pq)^{\frac{1}{2}} t^{-\frac{1}{2}}c x_k^{\pm}y_1^{\pm} \right)
	\prod_{k=1}^{N_f-N_c} \Gamma_e\left( (pq)^{\frac{1}{2}} t^{N_c-\frac{N_f}{2}-\frac{1}{2}}c x_k^{\pm}y_2^{\pm} \right) 
	\nonumber\\
	& \quad\qquad\times
	\prod_{j=1}^{N_c}\Gamma_e\left( t^{N_f-N_c+2-j}c^{-2} \right) \,.
\end{align}
\endgroup
Therefore we have
\begin{equation}
	\mathcal{I}_{\text{SQCD}(N_c,N_f\geq 2N_c)}(\vec x;y_1,y_2;t;c) = \mathcal{I}_{\widehat{\text{SQCD}}(N_c,N_f\geq 2N_c)}(\vec x;y_1,y_2;t;c) \,.
	\label{eq:4d_SQCD_mirror_pair}
\end{equation}

\subsection{The 3d good SQCD}\label{3dgoodres}
The 3d $U(N_c)$ SQCD with $N_f$ flavors, shown on top of Figure \ref{fig:good_SQCD_Sdualised_3d}, has the following partition function: 
\begingroup\allowdisplaybreaks
\begin{align}
	\mathcal{Z}^{3d}_{\text{SQCD}_U(N_c,N_f\geq 2N_c)}\left(\vec{X};Y_1-Y_2\right) & = 
	\int \udl{\vec{Z}_{N_c}}\Delta_{N_c}^{(3d)}\left(\vec{Z};m_A \right) 
	\times e^{2\pi i (Y_1-Y_2) \sum_{j=1}^{N_c}Z_j} 
	\nonumber\\
	& \qquad\quad\times
	\prod_{j=1}^{N_c}\prod_{a=1}^{N_f} s_b\left( \frac{iQ}{2}-m_A\pm(Z_j-X_a) \right)
	\label{eq:def_Z3d_SQCD_electric}
\end{align}
\endgroup
If the SQCD is good (namely $N_f \geq 2N_c$), its mirror dual is unique and its partition function is the following:
\begingroup\allowdisplaybreaks
\begin{align}
	&\mathcal{Z}^{3d}_{\widehat{\text{SQCD}}_U(N_c,N_f\geq 2N_c)}\bigg(Y_1,Y_2;X_{k+1}-X_{k}\,(k=1,\dots,N_f-1)\bigg) =\nonumber\\
	& \quad = 
	\int 
	\prod_{k=1}^{N_c}\left(\udl{\vec{Z}_{k}^{(k)}}\Delta_k^{(3d)}\left(\vec{Z}^{(k)};i\frac{Q}{2}-m_A \right) \right)
	\prod_{k=N_c+1}^{N_f-N_c-1}\left(\udl{\vec{Z}_{N_c}^{(k)}}\Delta_{N_c}^{(3d)}\left(\vec{Z}^{(k)};i\frac{Q}{2}-m_A \right) \right)
	\nonumber\\
	& \qquad\times
	\prod_{k=N_f-N_c}^{N_f-1}\left(\udl{\vec{Z}_{N_f-k}^{(k)}}\Delta_{N_f-k}^{(3d)}\left(\vec{Z}^{(k)};i\frac{Q}{2}-m_A \right) \right)
	\prod_{k=1}^{N_f-1}e^{-2\pi i(X_k-X_{k+1})\sum_{j=1}^{N_k}Z_j^{(k)}}
	\nonumber\\
	& \qquad\quad\times
	\prod_{k=1}^{N_c-1}\prod_{j=1}^{k}\prod_{l=1}^{k+1} s_b\left( m_A \pm\left(Z_j^{(k)}-Z_l^{(k+1)}\right) \right)
	\nonumber\\
	& \qquad\quad\times
	\prod_{k=N_c}^{N_f-N_c-1}\prod_{j=1}^{N_c}\prod_{l=1}^{N_c} s_b\left( m_A \pm\left(Z_j^{(k)}-Z_l^{(k+1)}\right) \right)
	\nonumber\\
	& \quad\qquad\times
	\prod_{k=N_f-N_c}^{N_f-2}\prod_{j=1}^{N_f-k}\prod_{l=1}^{N_f-k-1} s_b\left( m_A \pm\left(Z_j^{(k)}-Z_l^{(k+1)}\right) \right)
	\nonumber\\
	& \quad\qquad\times
	\prod_{j=1}^{N_c}s_b\left( m_A \pm\left(Z_j^{(N_c)}- Y_1\right)\right)s_b\left( m_A \pm\left(Z_j^{(N_f-N_c)}- Y_2 \right)\right) \,,
\end{align}
\endgroup
where the rank of the $U(N_k)_{Z^{(k)}}$ gauge node is given by
\begin{align}
	N_k = 
	\begin{cases}
		k \qquad\qquad\,\,\, \text{for } k=1,\dots,N_c-1 \text{ (increasing ramp)}\\
		N_c \qquad\qquad\! \text{for } k=N_c,\dots,N_f-N_c \text{ (plateau)} \\
		N_f-k \qquad \text{for } k=N_f-N_c+1,\dots,N_f-1 \text{ (decreasing ramp)}
	\end{cases}
	\,.
\end{align}

Therefore we have
\begin{align}
	& \left(e^{2\pi i Y_2 \sum_{a=1}^{N_f}X_a}\right) 
	\mathcal{Z}^{3d}_{\text{SQCD}_U(N_c,N_f\geq 2N_c)}\left(\vec{X};Y_1-Y_2\right) 
	= \nonumber\\
	& \qquad =
	\left(
	e^{2\pi i Y_1 \sum_{a=1}^{N_c} X_a} \times
	e^{2\pi i Y_2 \sum_{a=1}^{N_f-N_c} X_a}
	\right)
	\mathcal{Z}^{3d}_{\widehat{\text{SQCD}}_U(N_c,N_f\geq 2N_c)}\left(Y_1,Y_2;X_{k+1}-X_{k}\right) \,,
\end{align}
which is the 3d reduction of \eqref{eq:4d_SQCD_mirror_pair}.
This identity is represented in Figure \ref{fig:good_SQCD_Sdualised_3d}.

Moreover, we find it convenient to shift all of the magnetic gauge variables by $Y_1$, so that also on this mirror side the integral depends on the combination $Y_2-Y_1$. We get
\begin{align}
\label{eq:3d good mirror}
	& \left(e^{\pi i (Y_2 -Y_1)\sum_{a=1}^{N_f}X_a}\right) 
	\mathcal{Z}^{3d}_{\text{SQCD}_U(N_c,N_f\geq 2N_c)}\left(\vec{X};Y_1-Y_2\right) 
	= \nonumber\\
	& \qquad =
		e^{2\pi i (Y_2-Y_1) \left(\sum_{a=1}^{N_f-N_c} X_a-\frac{1}{2}\sum_{a=1}^{N_f}X_a\right)}
	\mathcal{Z}^{3d}_{\widehat{\text{SQCD}}_U(N_c,N_f\geq 2N_c)}\left(0,Y_2-Y_1;X_{k+1}-X_{k}\right) \,.
\end{align}

\subsection{The 4d bad SQCD}
The 4d  $USp(2N_c)$ SQCD with $N_f < 2N_c-1$ corresponding the bad case (or with $N_f=2N_c-1$  corresponding to ugly case) has, as its electric dual, a sum of frames, as represented in Figure \ref{fig:bad_SQCD_Sdualised_4d}. The corresponding index identity is the following \cite{Giacomelli:2023zkk}:
\begingroup\allowdisplaybreaks
\begin{align}\label{eq:4d_bad_SQCD}
    & 
    \mathcal{I}_{\text{SQCD}\left(N_c,\,N_c \leq N_f \leq 2N_c-1\right)}(\vec x;y_1,y_2;t;c) = 
    \nonumber\\[10pt]
    & = \quad 
    \sum_{n=0}^{M+\epsilon} 
    \sum_{\alpha=\pm 1}
    \sum_{\substack{\beta = 1 \text{ if } n-\epsilon = 0, \\ \beta = \pm1 \text{ otherwise}}}
    \left\{
    \tilde\delta\left(y_1,\left(y_2 t^{-(n-\epsilon)\beta}\right)^{\alpha}\right)
    \frac{\prod_{j=1}^{M+\epsilon-n}\Gamma_e\left(t^{-j}\right)}{\prod_{j=0}^{M+\epsilon-n} \Gamma_e\left(t^{\, j-2M-1}\right)} 
    \right. \nonumber\\
    & \qquad\qquad\qquad\quad \times 
    \prod_{j=0}^{M-\epsilon+n}\Gamma_e\left(c^{-2} t^{\, j-2M}\right)
    \prod_{j=1}^{M-\epsilon+n+1}\Gamma_e\left(t^{\, j-M-1} w_1^\pm w_2^\pm\right)
    \nonumber\\
    & \qquad\qquad\qquad\quad \times \left.\left.
    \mathcal{I}_{\text{SQCD}\left(\frac{N_f}{2}+\epsilon-n,\,N_f\right)}\left(\vec x;w_2,w_1;t;t^{\frac{M-\epsilon+n+1}{2}}c\right) \right|_{w_{1,2}=y_{1,2} t^{\mp \frac{1}{2}\left(M+\epsilon-n+1\right)\beta}} 
    \right\} \nonumber\\[10pt]
    & \qquad\quad + \left\{
    \prod_{j=0}^{2M+1}\Gamma_e\left(c^{-2} t^{\, j-2M}\right) 
    \prod_{j=1}^{2M+2}\Gamma_e\left(t^{\, j-M-1} y_1^\pm y_2^\pm\right) 
    \phantom{\frac{\frac{\frac{1}{1}}{1}}{\frac{\frac{1}{1}}{1}}}
    \right. \nonumber\\
    & \qquad\qquad\qquad \left.
    \phantom{\frac{\frac{\frac{1}{1}}{1}}{\frac{\frac{1}{1}}{1}}}
    \times
    \mathcal{I}_{\text{SQCD}\left(N_f-N_c,\,N_f\right)}\left(\vec x;y_2,y_1;t;t^{\frac{2N_c-N_f}{2}}c\right) \right\}\,,
    \end{align}
\endgroup
where we defined $M=N_c-\frac{N_f}{2}-1$ and 
\be
\epsilon=\begin{cases}0&N_f\text{ even}\,,\\-\frac{1}{2}&N_f\text{ odd}\,.\end{cases}
\ee
We also defined 
\begin{align}
	\tilde{\delta}\left(x,y\right) = \frac{2 \pi i x}{(p;p)_\infty (q;q)_\infty}\delta\left(x-y\right)\,.
	\label{eq:def_delta_tilde}
\end{align}
Notice that each frame but the last one is associated with a delta function enforcing a constraint on the $y_i$ fugacities.
Notice also that for $N_f = 2 N_c-1$, i.e.~the ugly case, only the last term with no delta constraint contributes.

On the other hand for $N_f<N_c$, i.e.~the evil SQCD case, we have
\begingroup\allowdisplaybreaks
\begin{align}\label{eq:4d_evil_SQCD}
    & 
    \mathcal{I}_{\text{SQCD}\left(N_c,\,N_f < N_c\right)}(\vec x;y_1,y_2;t;c) = 
    \nonumber\\[10pt]
    & = \quad 
    \sum_{n=0}^{\frac{N_f}{2}+\epsilon} 
    \sum_{\alpha=\pm 1}
    \sum_{\substack{\beta = 1 \text{ if } n-\epsilon = 0, \\ \beta = \pm1 \text{ otherwise}}}
    \left\{
    \tilde\delta\left(y_1,\left(y_2 t^{-\left(n-\epsilon\right)\beta}\right)^{\alpha}\right)
    \frac{\prod_{j=1}^{M+\epsilon-n}\Gamma_e\left(t^{-j}\right)}{\prod_{j=0}^{M+\epsilon-n} \Gamma_e\left(t^{j-2M-1}\right)} 
    \right. \nonumber
    \\
    & \qquad\qquad\qquad \times 
    \prod_{j=0}^{M-\epsilon+n}\Gamma_e\left(c^{-2} t^{j-2M}\right)
    \prod_{j=1}^{M-\epsilon+n+1}\Gamma_e\left(t^{j-M-1} w_1^\pm w_2^\pm\right)
    \nonumber\\
    & \qquad\qquad\qquad \times \left.\left.
    \mathcal{I}_{\text{SQCD}\left(\frac{N_f}{2}+\epsilon-n,\,N_f\right)}\left(\vec x;w_2,w_1;t;t^{\frac{M-\epsilon+n+1}{2}}c\right) 
    \right|_{w_{1,2}=y_{1,2} t^{\mp \frac{1}{2}\left(M+\epsilon-n+1
    \right)\beta}} 
    \right\}\,.
\end{align}
\endgroup
In this case each frame comes with a delta function and there is no frame with generic fugacities.

\subsection{The 3d bad SQCD}
The 3d  $U(N_c)$ SQCD with $N_f < 2N_c-1$ corresponding the bad case (or with $N_f=2N_c-1$  corresponding to ugly case) has, as its electric dual, a sum of frames, as represented in Figure \ref{fig:bad_SQCD_Sdualised_3d}, whose partition function reads \cite{Giacomelli:2023zkk}
\begingroup\allowdisplaybreaks
\begin{align}\label{eq:3d_bad_SQCD}
	& 
	\mathcal{Z}^{3d}_{\text{SQCD}_U(N_c,N_c \leq N_f \leq 2N_c-1)}\left(\vec{X};Y_1-Y_2\right) \times
	e^{\pi i (Y_2-Y_1) \sum_{a=1}^{N_f}X_a}
	= \nonumber\\[10pt]
	& \quad = 
	\sum_{n=0}^{M+\epsilon} 
    \sum_{\substack{\beta = 1 \text{ if } n-\epsilon = 0, \\ \beta = \pm1 \text{ otherwise}}}
	\left[
	\delta\Big(Y_1-\left(Y_2-\beta(n-\epsilon)(iQ-2m_A)\right) \Big) 
	\times 
	\bigphantomspace\right.
	\nonumber \\
	&\qquad \quad \times
	\frac{\prod_{j=1}^{M+\epsilon-n}s_b\left(\frac{iQ}{2}+j(iQ-2m_A)\right)}{\prod_{j=0}^{M+\epsilon-n} s_b\left(\frac{iQ}{2}-(j-(2N_c-N_f-1))(iQ-2m_A)\right)} 
	\times 
	e^{\pi i (W_1-W_2)\sum_{a=1}^{N_f}X_a}
	\nonumber\\ 
	&\qquad \quad \times
	\prod_{j=1}^{M-\epsilon+n+1}s_b\left(\frac{iQ}{2}-\left(j-N_c+\frac{N_f}{2}\right)(iQ-2m_A)\pm (W_1- W_2)\right) 
	\nonumber\\
	&\left.\qquad \quad \times
	\mathcal{Z}^{3d}_{\text{SQCD}_U\left(\frac{N_f}{2}+\epsilon-n,N_f\right)}\left(\vec{X};W_2-W_1\right)\Bigg|_{W_{1,2}=Y_{1,2}\mp\frac{1}{2}\beta\left(M+\epsilon-n+1\right)(iQ-2m_A)}
	 \right] \nonumber\\[10pt]
	 &\qquad+
	e^{\pi i (Y_1-Y_2)\sum_{a=1}^{N_f}X_a} 
	 \times
	\prod_{j=1}^{2 N_c-N_f}s_b\left(\frac{iQ}{2}-\left(j-N_c+\frac{N_f}{2}\right)(iQ-2m_A)\pm (Y_1- Y_2)\right) 
	\nonumber\\
	&\left.\qquad \quad \times
	\mathcal{Z}^{3d}_{\text{SQCD}_U\left(N_f-N_c,N_f\right)}\left(\vec{X};Y_2-Y_1\right)
	 \right] \,.
\end{align}
\endgroup

On the other hand for $N_f<N_c$, i.e.~the evil SQCD case, we have
\begingroup\allowdisplaybreaks
\begin{align}\label{eq:3d_evil_SQCD}
	& 
	\mathcal{Z}^{3d}_{\text{SQCD}_U(N_c,N_f < N_c)}\left(\vec{X};Y_1-Y_2\right) \times
	e^{\pi i (Y_2-Y_1) \sum_{a=1}^{N_f}X_a}
	= \nonumber\\[10pt]
	& \quad = 
	\sum_{n=0}^{\frac{N_f}{2}+\epsilon} 
	\sum_{\substack{\beta = 1 \text{ if } n-\epsilon = 0, \\ \beta = \pm1 \text{ otherwise}}}
	\left[
	\delta\Big(Y_1-\left(Y_2-\beta(n-\epsilon)(iQ-2m_A)\right) \Big) 
	\times
	\bigphantomspace\right. \nonumber \\
	&\qquad \quad \times
	\frac{\prod_{j=1}^{M+\epsilon-n}s_b\left(\frac{iQ}{2}+j(iQ-2m_A)\right)}{\prod_{j=0}^{M+\epsilon-n} s_b\left(\frac{iQ}{2}-(j-(2N_c-N_f-1))(iQ-2m_A)\right)} 
	\times
	e^{\pi i (W_1-W_2)\sum_{a=1}^{N_f}X_a}
	\nonumber\\ 
	&\qquad \quad \times
	\prod_{j=1}^{M-\epsilon+n+1}s_b\left(\frac{iQ}{2}-\left(j-N_c+\frac{N_f}{2}\right)(iQ-2m_A)\pm (W_1- W_2)\right) 
	\nonumber\\
	&\left.\qquad \quad \times
	\mathcal{Z}^{3d}_{\text{SQCD}_U\left(\frac{N_f}{2}+\epsilon-n,N_f\right)}\left(\vec{X};W_2-W_1\right)\Bigg|_{W_{1,2}=Y_{1,2}\mp\frac{1}{2}\beta\left(M+\epsilon-n+1\right)(iQ-2m_A)}
	 \right] .
\end{align} 
\endgroup
In this case each frame comes with a delta function and there is no frame with generic fugacities.

\afterpage{
\begin{landscape}
\begin{figure}[!ht]
\centering
    \includegraphics[scale=.4,center]{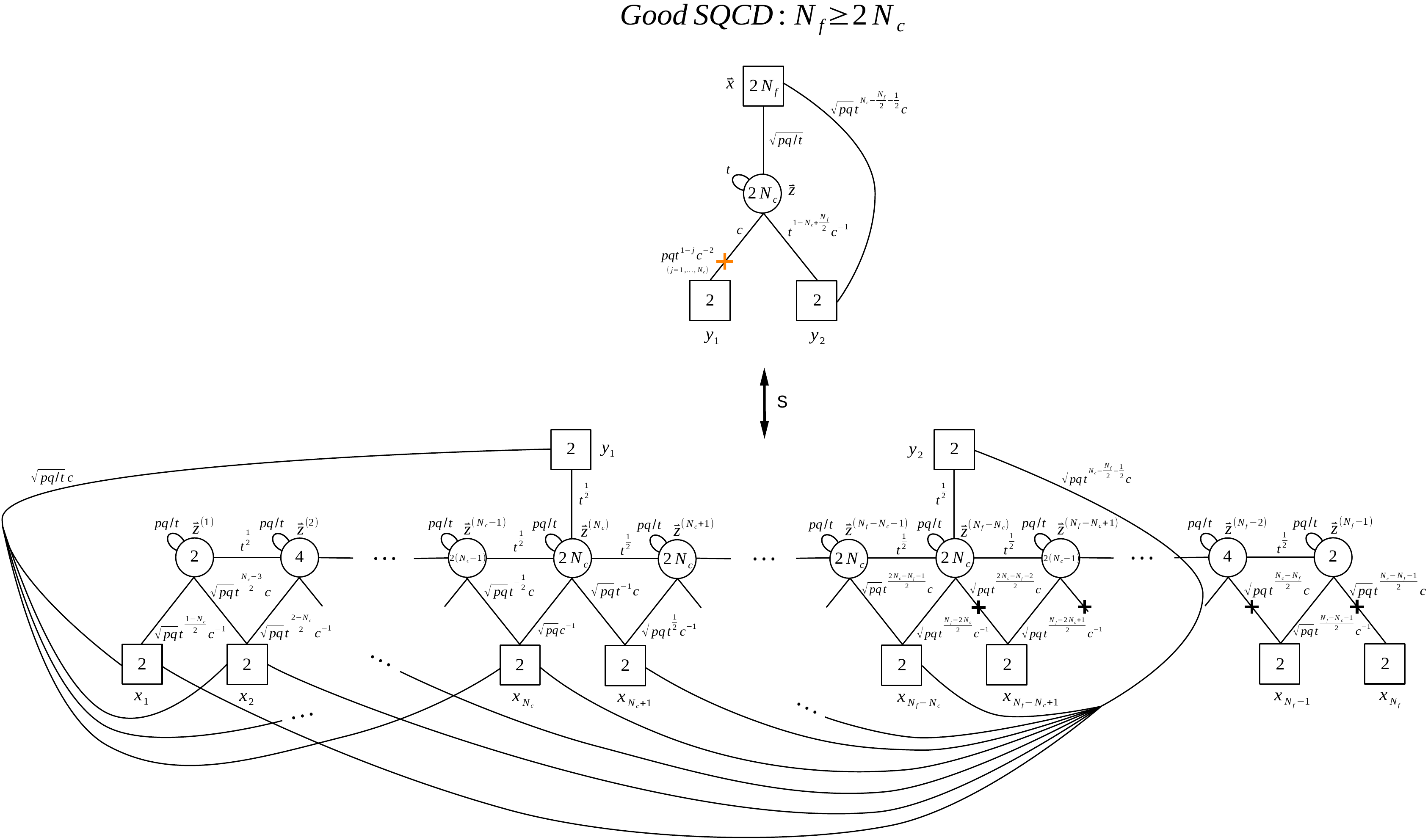}
    \caption{The good 4d SQCD and its  mirror.}
    \label{fig:good_SQCD_Sdualised_4d}
\end{figure}
\end{landscape}
}

\afterpage{
\begin{landscape}
\begin{figure}[!ht]
\centering
    \includegraphics[scale=.45,center]{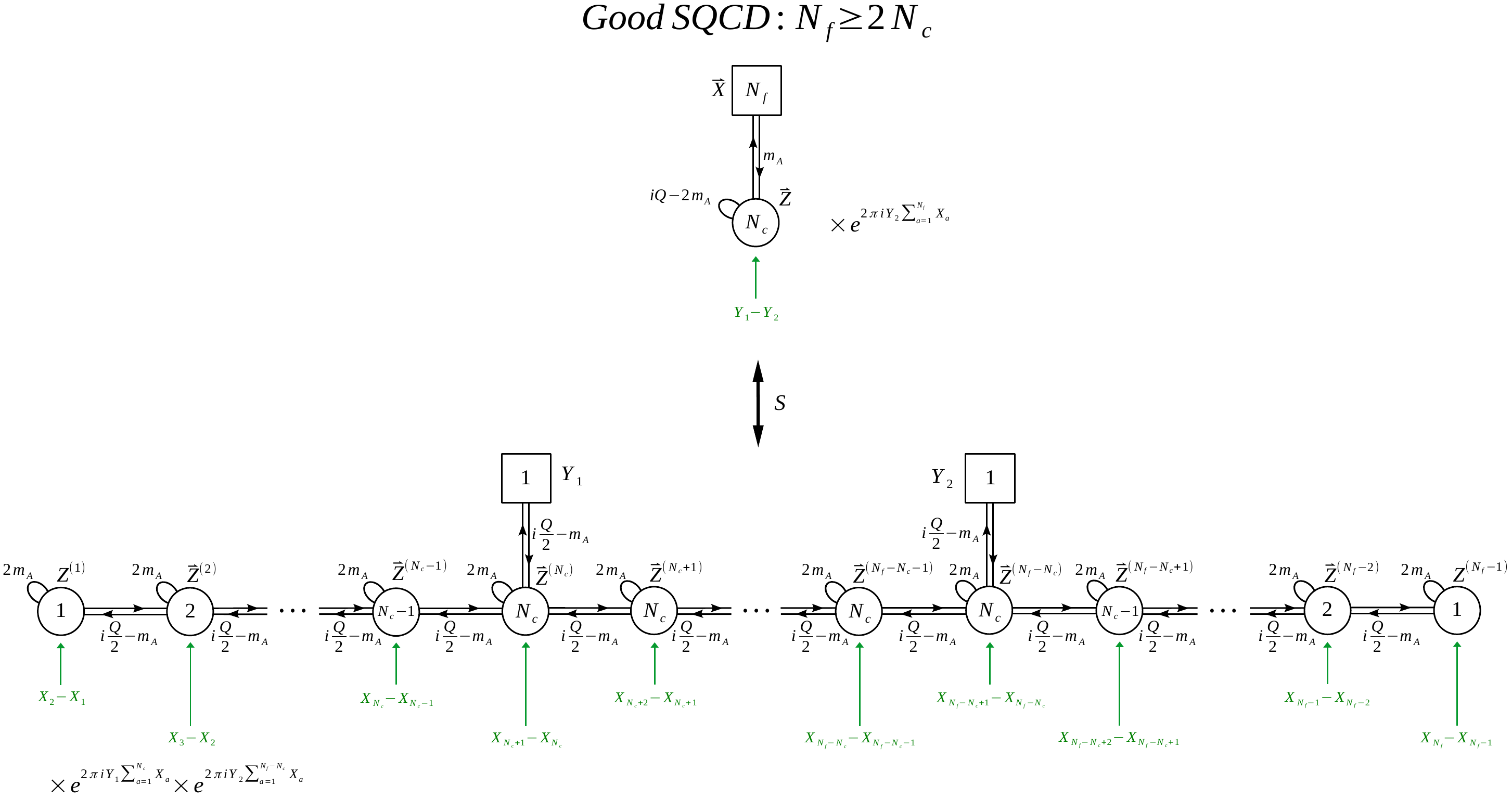}
    \caption{The good 3d SQCD and its  mirror. FI parameters are denoted in green (with an arrow that shows which gauge node they are referred to). Mixed background CS couplings are also indicated.}
    \label{fig:good_SQCD_Sdualised_3d}
\end{figure}
\end{landscape}
}

\afterpage{
\begin{landscape}
\begin{figure}[!ht]
\centering
    \includegraphics[scale=.41,center]{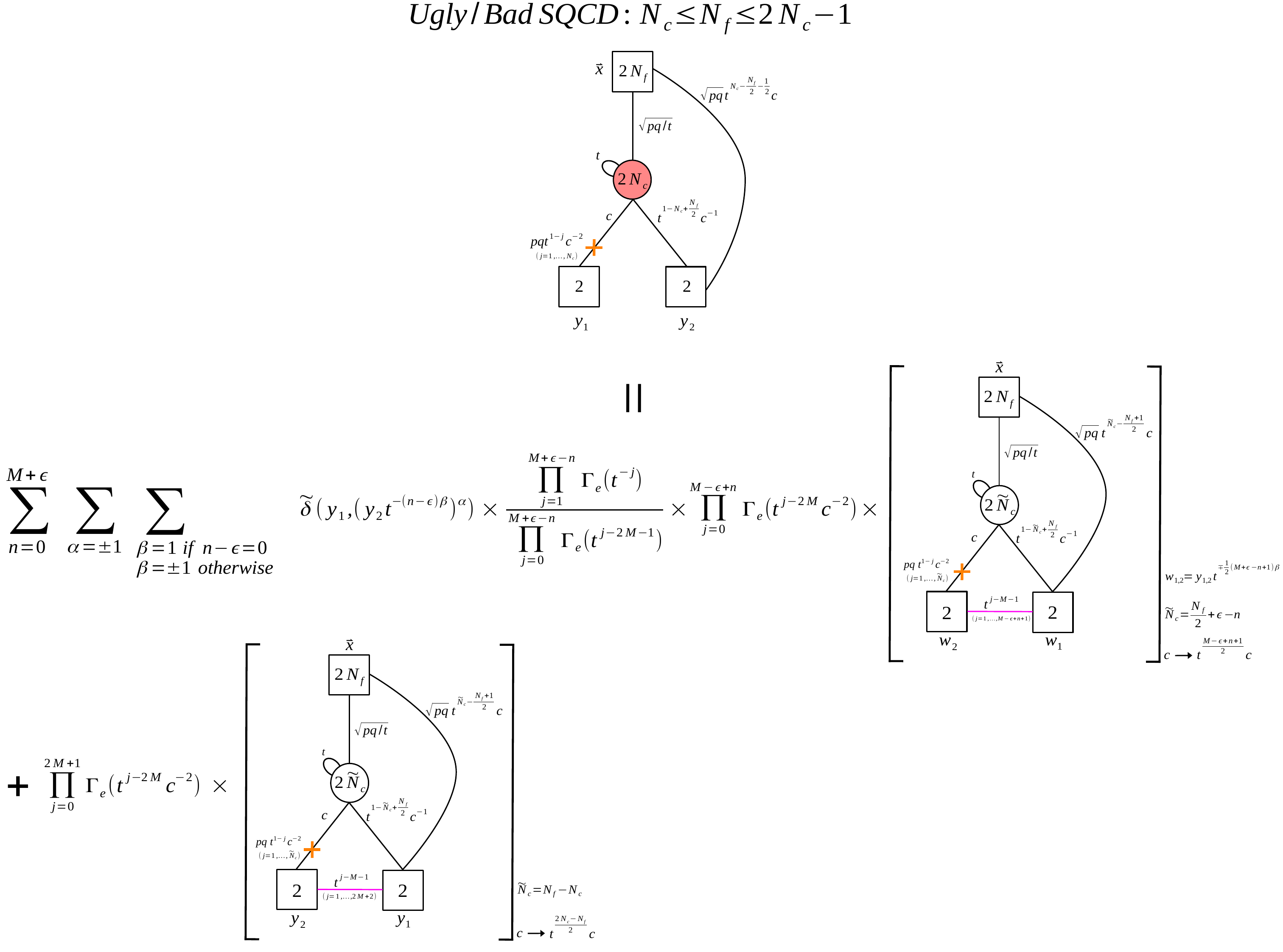}
    \caption{The 4d ugly/bad SQCD  and its mirror dual given by  a sum of frames.}
    \label{fig:bad_SQCD_Sdualised_4d}
\end{figure}
\end{landscape}
}

\afterpage{
\begin{landscape}
\begin{figure}[!ht]
\centering
    \includegraphics[scale=.41,center]{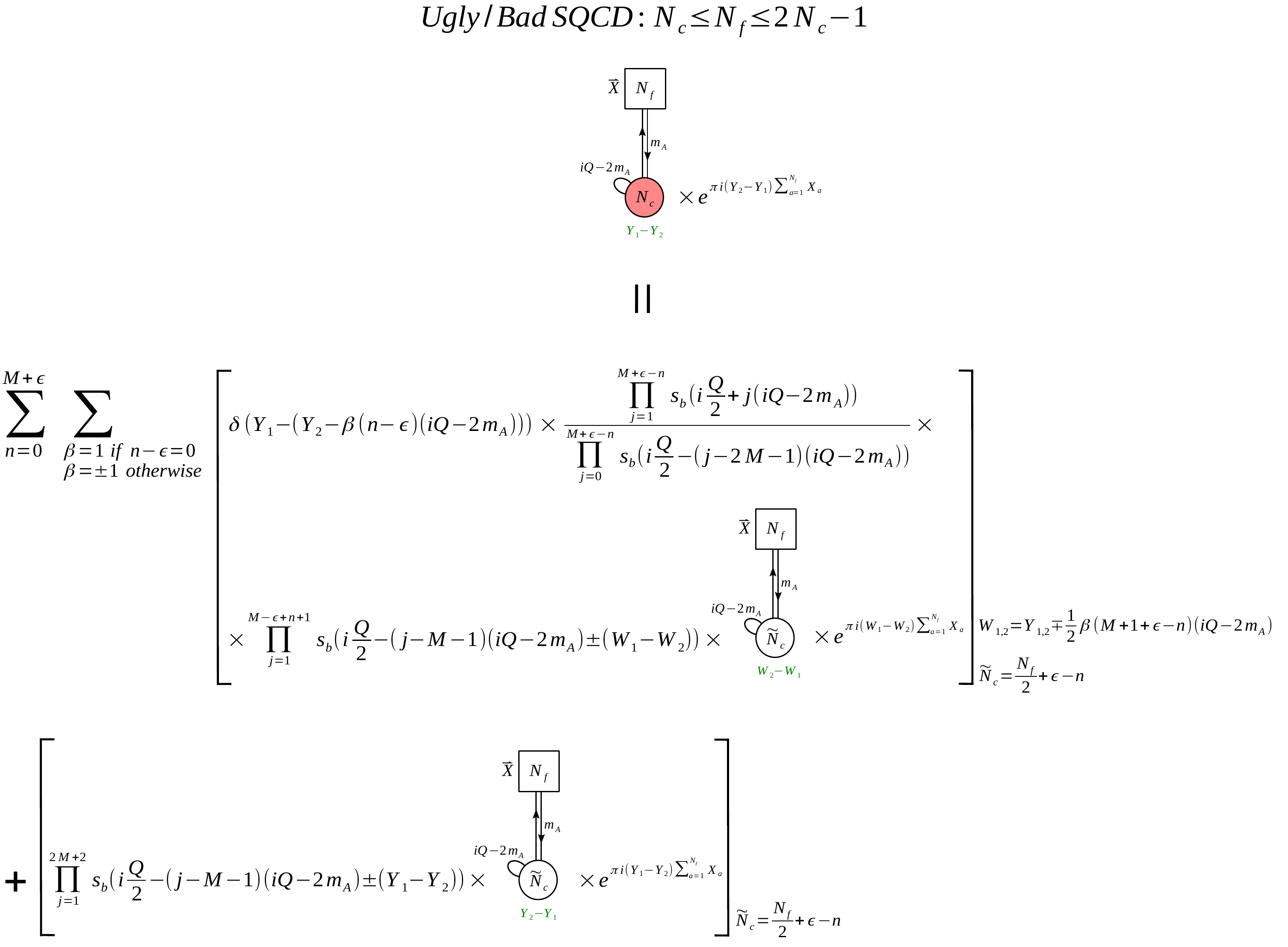}
    \caption{The 3d ugly/bad SQCD and  its mirror dual given by  a sum of frames.}
    \label{fig:bad_SQCD_Sdualised_3d}
\end{figure}
\end{landscape}
}

\clearpage
\section{\texorpdfstring{$E_{\rho}^{\sigma}[USp(2N)]$ and $T_{\rho}^{\sigma}[SU(N)]$ theories}{}}
\label{app:Erhosigma_Trhosigma}
The description of the $E_{\rho}^{\sigma}[USp(2N)]$ and the $T_{\rho}^{\sigma}[SU(N)]$ theories requires some preliminary definitions.

Given an integer $N$, we define the two partitions
\begin{align}
	\rho & = \left[ \rho_1,\rho_2,\dots,\rho_L \right] = \left[ N^{l_N},(N-1)^{l_{N-1}},\dots,1^{l_1} \right] \,,\\
	\sigma & = \left[ \sigma_1,\sigma_2,\dots,\sigma_L \right] = \left[ N^{k_N},(N-1)^{k_{N-1}},\dots,1^{k_1} \right] \,,
\end{align}
with $\sigma^T<\rho$ and where some of the $l_n, k_m$ integers can be zero and must satisfy the conditions
\begin{gather}
	N = \sum_{n=1}^{N} n \cdot l_n = \sum_{m=1}^{N} m \cdot k_m \,,\label{eq:N}\\
	L = \sum_{n=1}^{N} l_n \,,\qquad 
	K = \sum_{m=1}^{N} k_m \,.\label{eq:LM}
\end{gather}
The gauge ranks $N_i$ and the flavor ranks $M_i$ are given by
\begin{align}
	N_{L-n} & = \sum_{j=n+1}^{L} \rho_j - \sum_{m=n+1}^{N} (m-n)k_m \,,\\
	M_{L-m} & = k_m \,.
\end{align}

The above definitions apply to the case in which the linear quivers are good theories. However, by abuse of notation, in the main text we still referred with $E_{\rho}^{\sigma}[USp(2N)]$ and $T_{\rho}^{\sigma}[SU(N)]$ to the bad 4d and 3d linear quivers, respectively. In such a case, $\rho$ is not a partition, but just a sequence of integers summing up to $N$, which is not necessarily ordered nor of positive integers
\begin{equation}
\rho = \left[ \rho_1,\rho_2,\dots,\rho_L \right]\,,\qquad \sum_{i=1}^L\rho_i=N\,.
\end{equation}
All of the above formulas still apply to the bad case, except for the first equalities in \eqref{eq:N} and \eqref{eq:LM}. Instead, in this case $L$ is just the number of entries in the sequence $\rho$.

\subsection{The 4d $E_{\rho}^{\sigma}[USp(2N)]$ family}\label{ersdef}

\begin{figure}[!ht]
\centering
    \includegraphics[width=\textwidth]{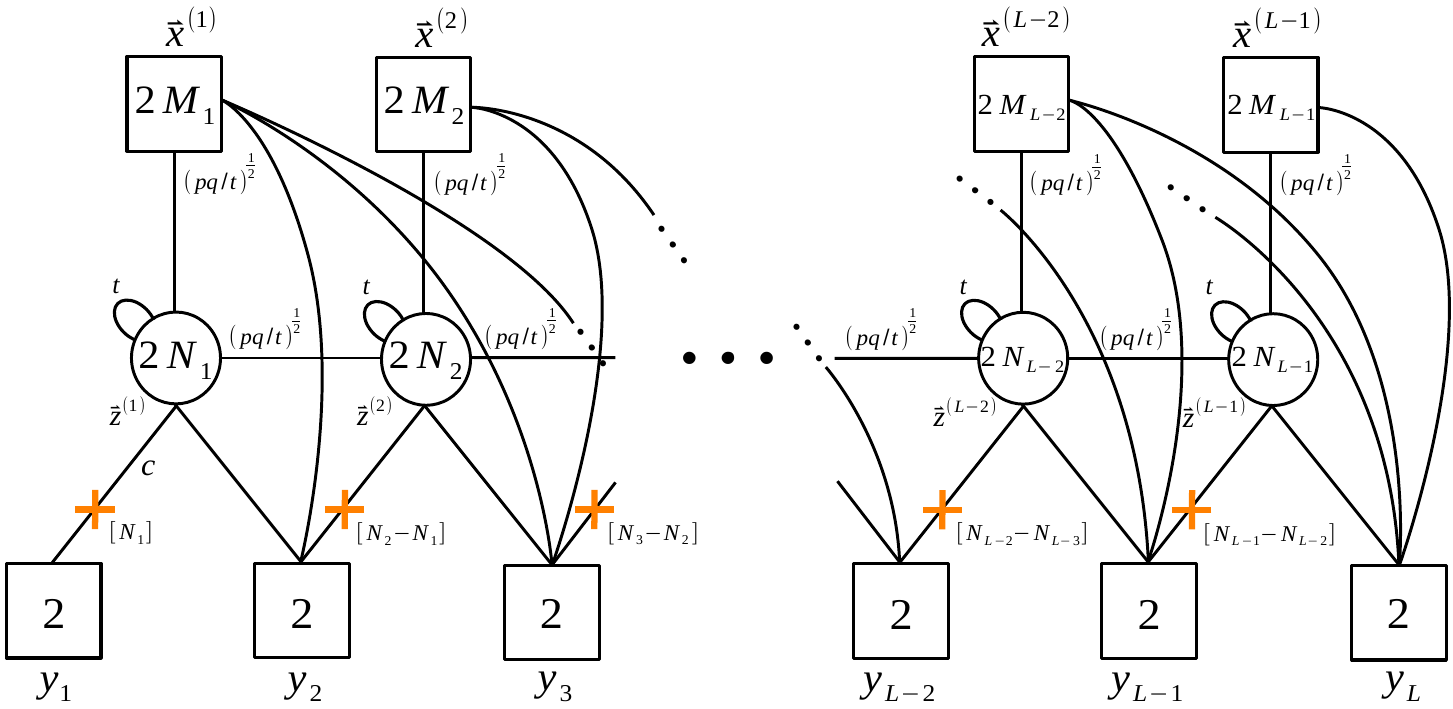}
    \caption{The theory $E_{\rho}^{\sigma}[USp(2N)]$. The crosses represent the singlets flipping the chiral on which they are placed (possibly dressed with some power of the antisymmetric). Their number is written in square brackets: if $N_j-N_{j-1}\leq 0$, the $SU(2)_{y_j}\times USp(2N_j)$ chiral has no flipping field.}
    \label{fig:Generic_Erhosigma_LtoR}
\end{figure}

The index corresponding to the generic $E_{\rho}^{\sigma}[USp(2N)]$ theory in Figure \ref{fig:Generic_Erhosigma_LtoR} is
\begingroup\allowdisplaybreaks
\begin{align}
	& \mathcal{I}_{E_{\rho}^{\sigma}[USp(2N)]}(\vec{x};\vec{y};t;c) = \nonumber\\
	& \quad = \oint 
	\prod_{j=1}^{L-1} \udl{\vec{z}^{\,(j)}_{N_j}}\Delta_{N_j}\left(\vec{z}^{\,(j)};t\right)
	\prod_{j=1}^{L-2}\prod_{a=1}^{N_j}\prod_{b=1}^{N_{j+1}} \Gamma_e\left( (pq)^{\frac{1}{2}}t^{-\frac{1}{2}} z_a^{(j)\pm}z_b^{(j+1)\pm} \right) \nonumber\\
	& \qquad\quad\times
	\prod_{j=1}^{L-1}\prod_{a=1}^{N_j}\prod_{b=1}^{M_j} \Gamma_e\left( (pq)^{\frac{1}{2}}t^{-\frac{1}{2}} z_a^{(j)\pm}x_b^{(j)\pm} \right) \nonumber\\
	& \qquad\quad\times
	\prod_{j=1}^{L-1}\prod_{a=1}^{N_j} \Gamma_e\left( \xi[\vec{y}_j,\vec{z}^{\,(j)}] z_a^{(j)\pm}y_j^{\pm} \right)
	\prod_{j=1}^{L-1}\prod_{a=1}^{N_j} \Gamma_e\left( \xi[\vec{y}_{j+1},\vec{z}^{\,(j)}] z_a^{(j)\pm}y_{j+1}^{\pm} \right)\nonumber\\
	& \qquad\quad\times
	\prod_{j=1}^{L-1}\prod_{k=j+1}^{L}\prod_{a=1}^{M_j} \Gamma_e\left( \xi[\vec{y}_{k},\vec{x}^{\,(j)}] y_k^{\pm} x_a^{(j)\pm} \right)
	\prod_{j=1}^{L-1}\prod_{k=1}^{N_j-N_{j-1}} \Gamma_e\left( pq \, \xi[\vec{y}_{j},\vec{z}^{\,(j)}]^{-1} t^{1-k} \right) \,,
	\label{eq:Index_Generic_Erhosigma_LtoR}
\end{align}
\endgroup
where $\xi[\vec{u},\vec{v}]$ is the (uniquely determined) charge of the field transforming in the bifundamental representation of the groups parametrized by the Cartans $\vec{u}$ and $\vec{v}$. Indeed in the saw-like structure formed by the $SU(2)_{y_i}\times USp(2N_j)_{z_j}$ chirals, the only fixed charge is that of the first one (the one transforming under $SU(2)_{y_1}\times USp(2N_1)_{z_1}$), and its value is $c$. The other charges in the saw are uniquely fixed by the NSVZ condition at each gauge node and by the requirement that the fields forming a triangle form a superpotential term. Therefore they cannot be written in a generic way, and we employ the notation $\xi[\vec{u},\vec{v}]$ just explained.
The last term in \eqref{eq:Index_Generic_Erhosigma_LtoR} encodes the flipping fields (represented in Figure \ref{fig:Generic_Erhosigma_LtoR} as crosses): they flip the $SU(2)_{y_{j}}\times USp(2N_j)$ chirals dressed with $k=0,\dots,N_j-N_{j-1}$ powers of the antisymmetric. If $N_j-N_{j-1}\leq 0$ for a certain $j$, that chiral has no flipping field attached.
The superpotential of $E_{\rho}^{\sigma}[USp(2N)]$ theories is discussed in Section \ref{sec:QFT_Ingredients}.

\subsection{The 3d $T_{\rho}^{\sigma}[SU(N)]$ family}\label{trsdef}

\begin{figure}[!ht]
\centering
    \includegraphics[width=.8\textwidth]{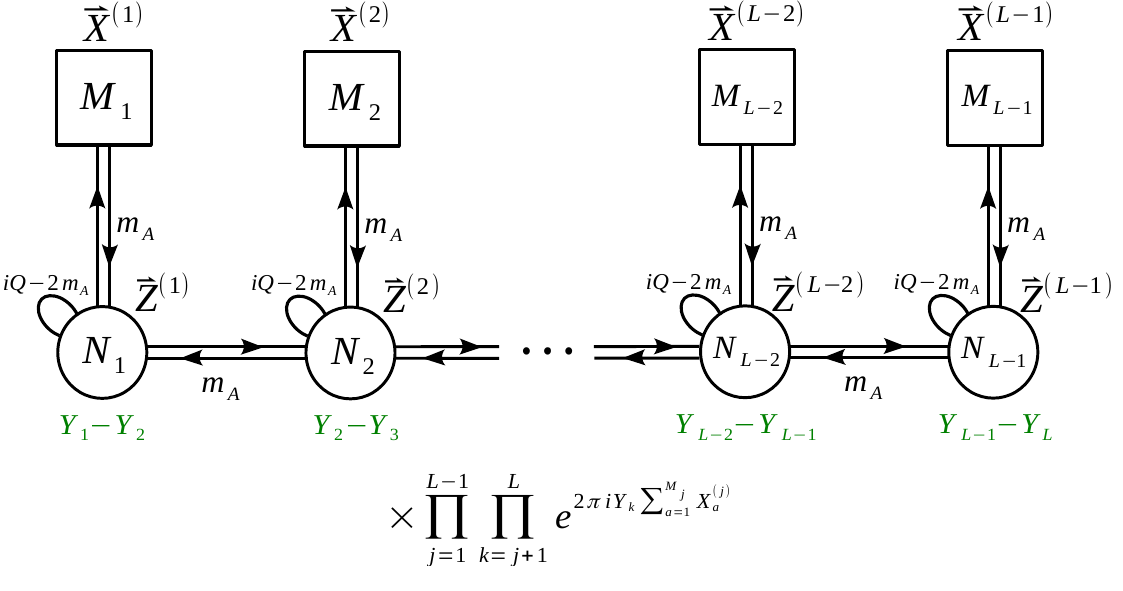}
    \caption{The theory $T_{\rho}^{\sigma}[SU(N)]$.}
    \label{fig:Generic_Trhosigma}
\end{figure}

The partition function corresponding to the generic $T_{\rho}^{\sigma}[SU(N)]$ theory in Figure \ref{fig:Generic_Trhosigma} is
\begingroup\allowdisplaybreaks
\begin{align}
	& \mathcal{Z}^{3d}_{T_{\rho}^{\sigma}[SU(N)]}\left(\vec{X};\vec{Y};m_A\right) = \nonumber\\
	& \qquad = 
	\int \prod_{j=1}^{L-1} \left(\udl{\vec{Z}_{N_j}^{(j)}}\Delta_{N_j}^{(3d)}\left(\vec{Z}_{N_j}^{(j)};m_A \right)\right)
	\prod_{j=1}^{L-1} e^{2 \pi i (Y_j-Y_{j+1})\sum_{a=1}^{N_j}Z_a^{(j)}}
	\prod_{j=1}^{L-1}\prod_{k=j+1}^{L}e^{2\pi i Y_k \sum_{a=1}^{M_j}X_a^{(j)}}
	\nonumber\\
	& \qquad\quad \times
	\prod_{j=1}^{L-2}\prod_{a=1}^{N_j}\prod_{b=1}^{N_{j+1}} s_b\left( i\frac{Q}{2}-m_A\pm\left(Z_a^{(j)}-Z_b^{(j+1)}\right) \right)
	\nonumber\\
	& \qquad\quad \times
	\prod_{j=1}^{L-1}\prod_{a=1}^{N_j}\prod_{b=1}^{M_{j}} s_b\left( i\frac{Q}{2}-m_A\pm\left(Z_a^{(j)}-X_b^{(j)}\right) \right) \,.
\end{align}
\endgroup

\section{Derivation of the $(1,0)$-$(1,0)$ swap move}
\label{app:5branes_swap_proofs}
In this appendix we will derive the $(1,0)$-$(1,0)$ blocks swap moves in 4d (namely \eqref{eq:B10B10_swap_4d} and \eqref{eq:D5_swap_4d}). Their 3d counterpart can be demonstrated analogously, or can just be obtained by taking the 3d limit of the 4d identities.

\begin{figure}[!ht]
\centering
    \includegraphics[width=\textwidth]{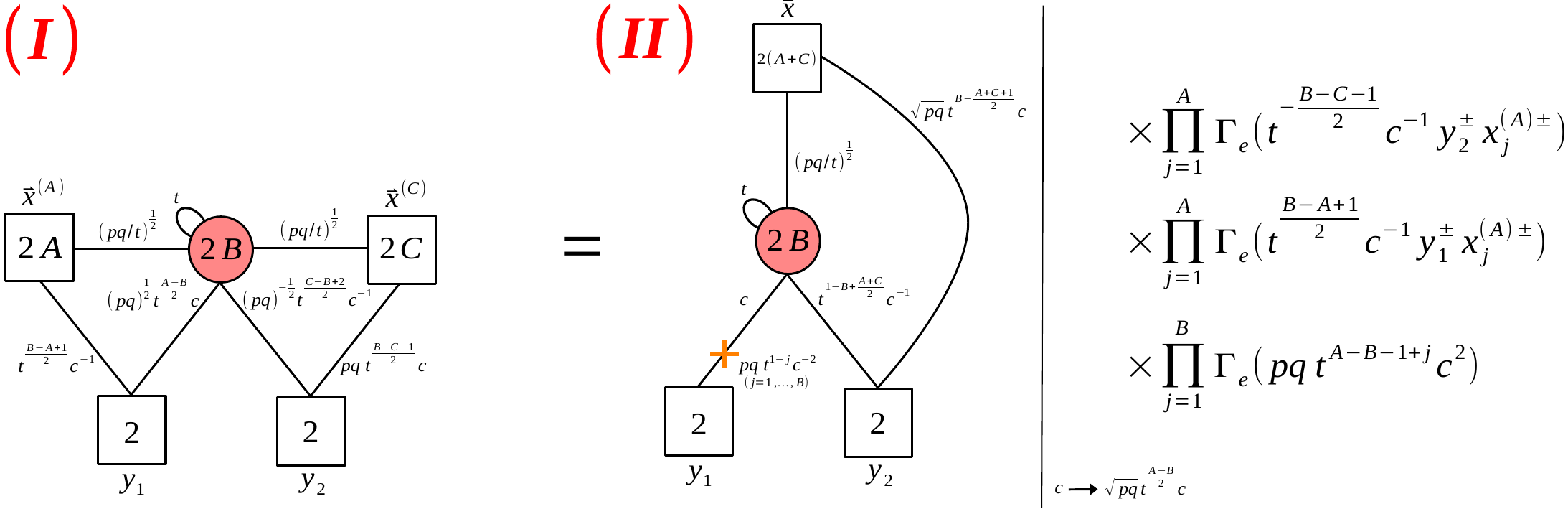}
    \caption{We express the two joint $(1,0)$-blocks as a bad SQCD with extra singlets.}
    \label{fig:I_to_III}
\end{figure}

Let us start considering the 4d theory (I) at the top of Figure \ref{fig:I_to_III}, built as a concatenation of two $(1,0)$ blocks. The ranks of the groups are taken such that they satisfy $A \ge B \ge C$ and $A+C<2B$. The proof for the other ranges discussed in Section \ref{sec:B01-swap}, namely $C \ge B \ge A$ and $B\ge A,\,B\ge C$, is analogous.

We can merge the $USp(2A)_{x^{(A)}}$ and the $USp(2C)_{x^{(C)}}$ flavor symmetries into a larger $USp(2(A+C))_x$ flavor symmetry,
add and remove singlets (which we write aside) to recognise the  $USp(2B)$ SQCD with $A+C$ flavors with the standard choice of singlets in (II).
It is also conveint to shift the $c$ fugacities so we can apply  formula \eqref{eq:4d_bad_SQCD} to get 
the electric dual of the SQCD as sum of frames.  Notice that we are allowed to apply the bad SQCD result only because we assumed $A+C<2B$ from the very beginning. 

\begin{figure}[!ht]
\centering
    \includegraphics[width=\textwidth,center]{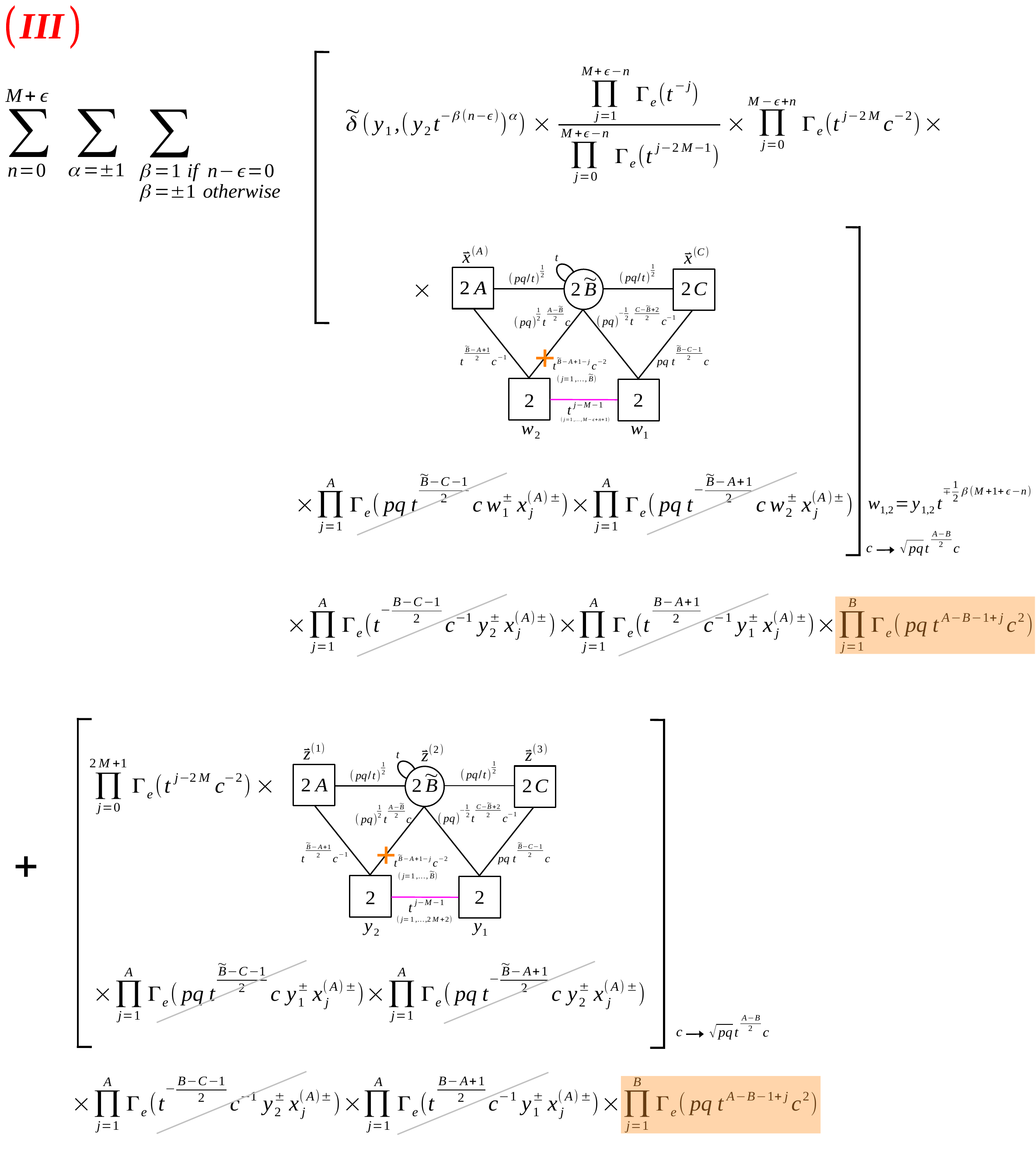}
    \caption{After taking the electric dual of the bad SCQD we  replace each good SQCD frame with two joint $(1,0)$-blocks.
    Keeping into account the delta constraint we can simplify the contribution of the massive singlets.
    The highlighted singlets (the same contribution in each frame) need to  be flipped and moved to (I)  to obtain the identity we wanted to prove.}
    \label{fig:V}
\end{figure}

Finally we can rewrite each SQCD of fame as two joint $(1,0)$ blocks by splitting the $USp(2(A+C))_x$ group into the original $USp(2A)_{x^{(A)}}$ and $USp(2C)_{x^{(C)}}$ flavor symmetries, multiplying and dividing by the missing $USp(2A)_{x^{(A)}} \times USp(2)_{w_2}$ and $USp(2C)_{x^{(C)}} \times USp(2)_{w_1}$ singlets needed for this operation (or the $USp(2A)_{x^{(A)}} \times USp(2)_{y_2}$ and $USp(2C)_{x^{(C)}} \times USp(2)_{y_1}$ singlets for the frame outside the summation).
 This process, together with the implementation of the $c$ redefinitions (explicitly written in the picture), leads to sum of quivers  (III) in Figure \ref{fig:V}
corresponding  to the  $(1,0)$-$(1,0)$block swap \eqref{eq:B10B10_swap_4d} we wanted to demonstrate. 

\clearpage
\section{Properties of maximal and minimal frames}\label{partitionsapp} 

\subsection{Properties of the maximal frame} 

The partition associated with the maximal frame is defined by the equations \eqref{tilde1} and \eqref{tilde2}:
\be\label{max1}\tilde\rho_{k+1}=\left\lfloor \text{Min}_{1\leq j\leq k+1}\left(\frac{N_{k+1-j}}{j}\right)\right\rfloor\;,\ee
and
\be\label{max2}\tilde\rho_{\ell}=\left\lfloor \text{Min}_{1\leq j\leq \ell}\left(\frac{N_{\ell-j}-\sum_{i>\ell}\tilde\rho_i}{j}\right)\right\rfloor\;.\ee 
Let us first prove this is a partition, namely that all the $\tilde\rho_i$ are positive and they are ordered $\tilde\rho_i\geq\tilde\rho_{i+1}$. The ordering can be proven as follows. From the definition \eqref{max2} we know there is at least a value of $j$ such that 
\be\label{xxx} \tilde\rho_i+1>\frac{N_{i-j}-\sum_{\ell>i}\tilde\rho_\ell}{j}\;.\ee 
For later convenience let us introduce the quantity $A_i=\sum_{\ell>i+1}\tilde\rho_\ell$. The equation above can therefore be written as 
$$ \tilde\rho_i+1>\frac{N_{i-j}-\tilde\rho_{i+1}-A_i}{j}.$$ 
We also know that 
\be\label{yyy} \tilde\rho_{i+1}\leq \frac{N_{i+1-m-1}-A_i}{m+1}=\frac{N_{i-m}-A_i}{m+1},\ee 
which holds for arbitrary $0 \leq m \leq i$, and in particular for $m=j$. 
Plugging this relation with $m=j$ inside \eqref{xxx} we find 
$$\tilde\rho_i+1>\frac{N_{i-j}-A_i-(N_{i-j}-A_i)/(j+1)}{j}=(N_{i-j}-A_i)\frac{j+1-1}{j(j+1)}=\frac{N_{i-j}-A_i}{j+1}\;.$$ 
We can now notice that the r.h.s.~of this expression and of \eqref{yyy} are identical and therefore by combining them we find $\tilde\rho_i+1>\tilde\rho_{i+1}$ and since all $\tilde\rho_i$'s are integer by definition we conclude that \be\label{zzz}\tilde\rho_i\geq\tilde\rho_{i+1}\;,\ee 
as desired. The positivity of the integers $\tilde{\rho}_i$ now follows immediately by noticing that \eqref{max1} clearly implies that $\tilde\rho_{k+1}$ is nonnegative. The inequality \eqref{zzz} then tells us the same is true for all other elements $\tilde\rho_i$ with $i\leq k$. 

What we have just shown is that \eqref{tilde1} and \eqref{tilde2} define a good unitary quiver with $k$ gauge groups of rank $\tilde{N}_i=\sum_{j>i}\tilde\rho_j$. By setting $m=0$ in \eqref{yyy} we find $\tilde\rho_{i+1}\leq N_i-A_i$ and therefore 
\be\label{www} \tilde{N}_i=\sum_{j>i}\tilde\rho_j=\tilde\rho_{i+1}+A_i\leq N_i\;,\ee 
meaning that the rank of each gauge group is smaller that that of the original quiver we started from. Now all we need to do to fully justify the argument of Section \ref{maxpart} is to prove maximality of the quiver defined by the partition $\tilde\rho$. This means showing that for every ordered partition $\Pi$ of $N$ with $k+1$ elements $\Pi_i$ (defining a good unitary quiver with $k$ gauge groups of rank $\hat{N}_i=\sum_{j>i}\Pi_j$), if the condition $\sum_{j>i}\Pi_j\leq \sum_{j>i}\tilde\rho_j$ is not satisfied $\forall i$, then it is not possible that $\hat{N}_i\leq N_i$ for every $i$ and therefore the associated polygon cannot contain that of the original quiver. Since our local duality moves can only decrease the rank of the gauge groups, such a quiver cannot correspond to any of the dual frames we get via a sequence of local dualizations. This result then implies that  \eqref{tilde1} and \eqref{tilde2} define the maximal dual frame. We will prove this result by induction. 

We first argue that $\Pi_{k+1}$ cannot be greater than $\tilde\rho_{k+1}$. If this was the case, that is if we assume $\Pi_{k+1}\geq\tilde\rho_{k+1}+1$, then we would conclude that \be\label{ineq1}\hat{N}_{k+1-j}\geq j\Pi_{k+1}\geq j\tilde\rho_{k+1}+j\;\; \forall j>0\;,\ee 
where we have used  the definition $\hat{N}_i=\sum_{j>i}\Pi_j$ and the fact that the partition $\Pi$ is ordered and therefore the element $\Pi_{k+1}$ is the smallest. The contradiction now comes from \eqref{max1}, which implies that there is at least a value of $j$ such that 
$$\tilde\rho_{k+1}+1>\frac{N_{k+1-j}}{j}\Rightarrow j\tilde\rho_{k+1}+j>N_{k+1-j}\;.$$ 
Combining this with \eqref{ineq1} we conclude that there is at least a $j$ such that $\hat{N}_{k+1-j}>N_{k+1-j}$ and therefore the quiver cannot correspond to a dual frame of the original one. Said differently, for every partition associated with one of the dual frames the $(k+1)$-th element cannot be larger than $\tilde\rho_{k+1}$. 

For the inductive step we have to show that the conditions $\sum_{j>i+1}\Pi_j\leq \sum_{j>i+1}\tilde\rho_j$ and $\hat{N}_\ell\leq N_{\ell}$ $\forall \ell$ imply that also  $\sum_{j>i}\Pi_j\leq \sum_{j>i}\tilde\rho_j$ must hold. As a preliminary remark, we observe that for every $j>0$ 
\be\label{ineq2}\hat{N}_{i+1-j}=\sum_{\ell>i+1-j}\Pi_{\ell}\geq \sum_{\ell>i+1}\Pi_{\ell}+j\Pi_{i+1}=\sum_{\ell>i}\Pi_{\ell}+(j-1)\Pi_{i+1}\;,\ee 
where we have used the fact that the partition $\Pi$ is ordered.
The argument now proceeds as before: we assume by contradiction that the inequality we want to derive does not hold. Since we assume $\sum_{j>i}\Pi_j > \sum_{j>i}\tilde\rho_j$ while $\sum_{j>i+1}\Pi_j\leq \sum_{j>i+1}\tilde\rho_j$ by induction, we conclude that $\Pi_{i+1}\geq\tilde\rho_{i+1}+1$. From \eqref{ineq2} we then find 
\begin{align}
\hat{N}_{i+1-j}\geq\sum_{\ell>i}\Pi_{\ell}+(j-1)\Pi_{i+1}>\sum_{\ell>i}\tilde\rho_{\ell}+(j-1)\Pi_{i+1}\geq \sum_{\ell>i}\tilde\rho_{\ell}+(j-1)\tilde\rho_{i+1}+j-1\;,
\end{align}
where we exploit the assumption that the inequality we want to derive is not satisfied. The above expression can be rewritten as 
\be\label{ineq3} \hat{N}_{i+1-j}\geq \sum_{\ell>i}\tilde\rho_{\ell}+(j-1)\tilde\rho_{i+1}+j= \sum_{\ell>i+1}\tilde\rho_{\ell}+j\tilde\rho_{i+1}+j\;\; \forall j>0\;.\ee 
We now conclude by observing that the definition \eqref{max2} implies that there is at least a $j>0$ such that the r.h.s.~of \eqref{ineq3} is strictly larger than $N_{i+1-j}$ and therefore the partition $\Pi$ we are considering does not correspond to a dual frame of the original quiver. This completes our proof.

\subsection{Proof of the lemmas for the minimal frame} 

Let us now prove the lemmas needed in deriving the extremality of the minimal dual frame.

\paragraph{Proof of Lemma 1.} 

The statement is that $\mathcal{I}(\rho)$, the partition we get by applying our rule to $[\rho_1,\mu]$, is larger than (or equal to) $\mathcal{I}(\rho')$, namely the partition we find by applying our rule to $[\rho_1,\mu']$, if  $\mu\geq \mu'$. Notice that in Section \ref{sixfour} the partitions $\mu$ and $\mu'$ are denoted $\mathcal{I}(\hat{\rho})$ and $\mathcal{I}(\hat{\rho}')$ respectively and are themselves obtained by applying our rule to the sequences of integers $\hat{\rho}$ and $\hat{\rho}'$. On the other hand, the inequality we are after holds true for arbitrary partitions $\mu$ and $\mu'$ such that  $\mu\geq \mu'$. In order to prove this, let us consider separately the cases $\rho_1<0$ and $\rho_1\geq0$. 

{\bf \boldmath For $\rho_1<0$} all we have to do is to collapse the partitions $\mu$ and $\mu'$. It is convenient to introduce the two sequences 
\be a_n=\sum_{i>n}\mu_i;\quad b_n=\sum_{i>n}\mu'_i;\quad 0\leq n<k-1\;.\ee 
Since  $\mu\geq \mu'$, we have $a_n\leq b_n$ $\forall n$ and therefore  $\rho_1+a_n\leq \rho_1+b_n$. If we denote as $n'$ the last value of $n$ such that $\rho_1+a_n>0$, the effect of the collapse is to leave the first $n'$ elements from $\mu_1$ to $\mu_{n'}$ of the partition unchanged and to replace the element $n'+1$ with $\rho_1+a_{n'}$. This will be the last element of the resulting partition. Due to the inequality $\rho_1+a_n\leq \rho_1+b_n$, we conclude that also the collapse of $\mu'$ will leave the first $n'$ elements unchanged and therefore the inequality we are looking for follows directly from  $\mu\geq \mu'$. 

{\bf \boldmath For $\rho_1\geq0$} we have instead to reorder the two partitions, which is equivalent to putting $\rho_1$ in the right position along the partitions. If $\rho_1$ ends up at position $i$ for the partition $\mu$, we have $\mu_{i-1}>\rho_1\geq\mu_i$. We call the resulting partition $\mathcal{I}(\rho)$. Analogously, for the second partition $\rho_1$ ends up in position $j$ if we have $\mu'_{j-1}>\rho_1\geq\mu'_j$ and we call the resulting partition $\mathcal{I}(\rho')$. Notice that $i$ and $j$ are not necessarily equal. Similarly to the previous case, define 
 \be a_n=\sum_{m<n}I(\rho)_m;\quad b_n=\sum_{m<n}I(\rho')_m;\quad 2\leq n\leq k\;.\ee 
If $i=j$ then we have $a_n\geq b_n$ due to $\mu\geq \mu'$ and therefore $\mathcal{I}(\rho)\geq \mathcal{I}(\rho')$ as desired. If instead $i<j$ we still have $a_n\geq b_n$ for $n\leq i$ and $n > j$ directly from  $\mu\geq \mu'$, whereas in the range $i< n<\leq j$ the inequality $a_n\geq b_n$ follows from $$a_n=\sum_{m<n-1}\mu_m+\rho_1\geq \sum_{m<n}\mu_m\geq \sum_{m<n}\mu'_m=b_n.$$ 
In the last case $j<i$ we again find $a_n\geq b_n$ for $n\leq j$ and $n> i$ as a result of the inductive step, whereas in the intermediate range $j<n\leq i$ the inequality follows from $$a_n=\sum_{m<n-1}\mu_m+\mu_{n-1} > \sum_{m<n-1}\mu_m+\rho_1\geq \sum_{m<n-1}\mu'_m+\rho_1=b_n\;.$$ 
This concludes the proof of Lemma 1. 

\paragraph{Proof of Lemma 2.} 

The statement is that if we apply our rule to the partition $\rho$ we get a partition $\mathcal{I}(\rho)$ which is larger than (or equal to) the $\mathcal{I}(\rho')$ we get by applying it to 
\be \rho'=[\rho_2-a,\rho_1+a,\rho_3,\dots, \rho_k]\;,\ee 
where $\rho_1<\rho_2$ and $a$ is a non negative integer such that $\rho_1+a\leq\rho_2-a$. As before, we can first apply our rule to the subpartition obtained by deleting the first two elements of $\rho$ and $\rho'$ and then to the entire partitions without affecting the final result. We can therefore assume without loss of generality that the elements $\rho_3,\dots, \rho_k$ are ordered and all positive. We divide the discussion into cases according to the sign of $\rho_{1,2}$. 

{\bf \boldmath If both $\rho_1$ and $\rho_2$ are negative} then also $\rho_1+a$ and $\rho_2-a$ will be negative and therefore the effect of our rule will be in both cases to collapse twice the tail of the partition $[\rho_3,\dots, \rho_k]$. We can now notice that this double collapse leads to the same answer we would get by starting from the partition $[\rho_1+\rho_2,\rho_3,\dots, \rho_k]$, where we have ``fused'' the first two elements together. Since after this fusion operation $\rho$ and $\rho'$ become identical, the outcome is the same and therefore we conclude $\mathcal{I}(\rho)=\mathcal{I}(\rho')$ in this case. 

{\bf \boldmath If both $\rho_1$ and $\rho_2$ are non negative} then also $\rho_1+a$ and $\rho_2-a$ will be non negative and the effect of applying our rule will simply be to put these elements in the correct position along the ordered partitions. More precisely, we find\footnote{Notice that $\rho_2$ can be at the beginning of the partition ($I=3$) and/or $\rho_1$ can be at the end of the partition ($J=k+1$). Our argument applies in this case as well. The same remark applies to the partition $\mathcal{I}(\rho')$.}
\be \mathcal{I}(\rho)=[\rho_3,\dots, \rho_{I-1},\rho_2,\rho_I,\dots,\rho_{J-1},\rho_1,\rho_J,\dots]\;,\ee 
\be \mathcal{I}(\rho')=[\rho_3,\dots, \rho_{L-1},\rho_2-a,\rho_L,\dots,\rho_{M-1},\rho_1+a,\rho_M,\dots]\;.\ee 
Indeed we have the inequalities $I\leq L\leq M\leq J$. It is convenient in this case to introduce the sequence of integers $$\rho''=\mathcal{I}(\rho)-\mathcal{I}(\rho')=[\mathcal{I}(\rho)_i-\mathcal{I}(\rho')_i]\,.$$ By inspection the first $I-1$ elements are zero, those in the range $(I,L-1)$ are positive, those in the range $(L,M-1)$ are zero again, those in the range $(M,J-1)$ are negative and the rest are zero. We now introduce the sequence of partial sums of $\rho''$ 
$$c_n=\sum_{m\leq n}\rho''_m.$$ 
Since both $\mathcal{I}(\rho)$ and $\mathcal{I}(\rho')$ are partitions of $N$, we necessarily have $c_k=0$. Due to the structure of the sequence $\rho''$, we find that the sequence $c_n$ is initially zero, then it starts increasing until it reaches the value $a$ (if $a > 0$), it is constant for a while and then decreases back to zero. Overall, we have $c_n\geq0$ $\forall n$. Since $c_n$ is also equal to the sum of the first $n$ elements of $\mathcal{I}(\rho)$ minus the sum of the first $n$ elements of $\mathcal{I}(\rho')$, the positivity of $c_n$ directly implies $\mathcal{I}(\rho)\geq\mathcal{I}(\rho')$ as desired.

{\bf \boldmath If $\rho_1<0$ and $\rho_2\geq 0$} the outcome of our rule depends on the value of $\rho_1$, $\rho_2$ and $a$. We will therefore discuss all possible cases separately. Let us examine first of all the effect on the partition $\rho$. We first put the element $\rho_2$ in the correct position, say it ends up in position $I$ (namely the position between $\rho_{I-1}$ and $\rho_I$). This implies that $\rho_i$ with $i<I$ is larger than $\rho_2$ (with the exception of $\rho_1$ of course). After this we collapse the partition using $\rho_1$. There are two cases to be considered: either the element $\rho_2$ disappears in the collapse (it is included in the last element of the partition $\mathcal{I}(\rho)$) or it does not. 

Let us consider the first case. In this situation the partition $\mathcal{I}(\rho)$ will have the form\footnote{$I(\rho)$ can also be just $[\hat \rho]$ with $\hat \rho = \sum_{n \geq 0} \rho_n=N$, in which case, $I(\rho) \geq I(\rho')$ trivially.}
$$\mathcal{I}(\rho)=[\rho_3,\dots, \rho_J,\hat{\rho}]\;;\quad \hat{\rho}=\rho_1+\rho_2+\sum_{n>J}\rho_n\;,$$ 
where necessarily $J<I$. In the partition $\rho'$ we have the elements $\rho_1+a$ and $\rho_2-a$, which are smaller than or equal to $\rho_2$. Regardless of their sign, the resulting partition $\mathcal{I}(\rho')$ will be of the form 
$$\mathcal{I}(\rho')=[\rho_3,\dots, \rho_{J-1},\dots]\;.$$ 
If there is a collapse, it cannot involve $\rho_{J-1}$ due to the inequality  $\hat{\rho}\geq0$ and the fact that $\rho_J>\rho_2$, implying that $\rho_1+a+\rho_J+\sum_{n>J}\rho_n>0$. If both $\rho_1+a$ and $\rho_2-a$ are non negative (and therefore no collapse is involved), we know they will end up in the partition in position $I$, which is larger than $J$, or after it since they are smaller than (or equal to) $\rho_2$. Overall, we conclude that the first $J-3$ elements of $\mathcal{I}(\rho)$ and $\mathcal{I}(\rho')$ are equal and $\mathcal{I}(\rho)$ has length $J-1$. If $\mathcal{I}(\rho')$ also includes the element $\rho_J$, then we immediately reach the desired inequality $\mathcal{I}(\rho)\geq\mathcal{I}(\rho')$. Otherwise it means that in $\mathcal{I}(\rho')$ the element $\rho_J$ is removed by the collapse and both $\rho_2-a$ and $\rho_1+a+\rho_J+\sum_{n>J}\rho_n$ are strictly smaller than $\rho_J$. These observations lead again to  $\mathcal{I}(\rho)\geq\mathcal{I}(\rho')$.

Let us now move to the discussion of the second option. In this case $\mathcal{I}(\rho)$ will have the form 
$$\mathcal{I}(\rho)=[\rho_3,\dots, \rho_{I-1},\rho_2, \rho_I,\dots,\rho_J, \tilde{\rho}]\;;\quad \tilde{\rho}=\rho_1+\sum_{n>J}\rho_n\;.$$ 
We have indeed the inequality $\rho_{I-1}>\rho_2\geq\rho_I$ and also $\rho_{J+1}>\tilde{\rho}$ due to our rule. 
If both $\rho_1+a$ and $\rho_2-a$ are negative the effect of applying the rule on $\rho'$ will be to collapse the partition twice, which is equivalent (as we have explained) to collapsing the partition once with $\rho_1+\rho_2$. Since $\tilde \rho \geq 0$ and $\rho_2\geq\rho_{J+1}$, we have $\rho_1+\rho_2+\sum_{n>J+1}\rho_n\geq 0$ and this implies $\mathcal{I}(\rho')$ will be of the form 
$$\mathcal{I}(\rho')=[\rho_3,\dots, \rho_{J+1},\dots]\;.$$ 
This is a consequence of the inequality $\rho_1+\rho_2+\sum_{n>J+1}\rho_n\geq 0$ which tells us the collapse cannot involve $\rho_{J+1}$. If we now consider the difference 
$$\mathcal{I}(\rho)-\mathcal{I}(\rho')=[0,\dots, 0, \rho_2-\rho_I, \rho_I-\rho_{I+1},\dots, \rho_J-\rho_{J+1}, \dots]\;,$$ 
we see that all the $J-1$ elements displayed are non negative and since $\mathcal{I}(\rho)$ has exactly length $J$ we conclude that $\mathcal{I}(\rho)\geq\mathcal{I}(\rho')$. 

\noindent When instead both $\rho_1+a$ and $\rho_2-a$ are non negative, then $\mathcal{I}(\rho')$ has the form 
$$\mathcal{I}(\rho')=[\rho_3,\dots, \rho_{I-1}, \dots, \rho_{\ell}, \rho_2-a, \dots, \rho_K, \rho_1+a, \dots]\;.$$ 
it is now convenient to consider again 
$$a_n=\sum_{m\leq n}\mathcal{I}(\rho)_m\;;\quad b_n=\sum_{m\leq n}\mathcal{I}(\rho')_m\;.$$ 
The difference $a_n-b_n$ can be checked to be non negative for $n\leq \ell-2$, is equal to $a$ for $n=\ell-1$ and also for all $n$ up to $\text{Min}\{J-1,K-1\}$. If $K\geq J$ this directly implies the desired result, otherwise for $K\leq n\leq J-1$ we find $a_n-b_n=\rho_{n+1}-\rho_1>0$. Since in any case $a_n-b_n\geq 0$ for every $n<J$, we conclude again that $\mathcal{I}(\rho)\geq\mathcal{I}(\rho')$.

\noindent Finally, in the case in which $\rho_2-a\geq 0$ and $\rho_1+a<0$ we have that $\mathcal{I}(\rho')$ is produced by a collapse induced by $\rho_1+a$ and then we put $\rho_2-a$ in the correct position. All the elements from $\rho_3$ to $\rho_J$ will necessarily be present in $\mathcal{I}(\rho')$ since $\rho_1+a+\sum_{n>J}\rho_n=\tilde{\rho}+a$ is non negative and therefore $\rho_J$ cannot be involved in the collapse. If $\rho_2-a$ is larger than $\rho_J$ the partition $\mathcal{I}(\rho')$ will be of the form 
$$\mathcal{I}(\rho')=[\rho_3,\dots, \rho_2-a, \dots, \rho_J, \dots]\;.$$ 
If instead $\rho_2-a<\rho_J$ the element following $\rho_J$ in $\mathcal{I}(\rho')$ can be either $\rho_2-a$ or $\rho_{J+1}$ or $\tilde{\rho}+a$ and as a result $\mathcal{I}(\rho')$ will be respectively of the form 
\begin{itemize} 
\item  $\mathcal{I}(\rho')=[\rho_3,\dots, \rho_J, \rho_2-a, \dots]\;,$
\item  $\mathcal{I}(\rho')=[\rho_3,\dots, \rho_J, \rho_{J+1}, \dots]\;,$
\item  $\mathcal{I}(\rho')=[\rho_3,\dots, \rho_J, \tilde{\rho}+a, \rho_2-a]\;.$
\end{itemize}
In each of these four possibilities, by considering the difference $\mathcal{I}(\rho)-\mathcal{I}(\rho')$, we easily find again that $\mathcal{I}(\rho)\geq\mathcal{I}(\rho')$.  This concludes our argument. 

\clearpage
\bibliographystyle{JHEP}
\bibliography{ref}

\end{document}